\documentclass[letterpaper, 10pt]{book}

\usepackage[left=4.7cm,right=4.7cm,top=5.04cm, bottom=5.04cm]{geometry}

\usepackage[tbtags,intlimits]{amsmath}
\usepackage{graphicx}
\usepackage{appendix}
\usepackage[rflt]{floatflt}

\usepackage[square,comma,sort&compress]{natbib}
\usepackage{txfonts}

\usepackage[Ale]{fncychap}

\usepackage{fancyhdr}

\fancypagestyle{plain}{
                        \fancyfoot{}
                        \fancyhead{}
                        \renewcommand{\headrulewidth}{0pt}
                        \renewcommand{\footrulewidth}{0pt}
                      }

\newcommand{\startTechnicalDetails}{
        \vspace{.1cm}
        \begin{center}
        \Large{*\hspace{.5cm} *\hspace{.5cm} *}
        \end{center}
        \begin{footnotesize}}
\newcommand{\stopTechnicalDetails}{
        \end{footnotesize}
        \begin{center}
        \Large{*\hspace{.5cm} *\hspace{.5cm} *}
        \end{center}
        \vspace{.1cm}}

\newcommand{\rmd}[2]{\ifnum #1 = 1
                       \text{d} #2 \
                     \else \text{d}^#1 #2 \
                     \fi}

\newcommand{\threeVec}[1]{{\textbf #1}}

\newcommand{\MQQ}{\mathcal{O} \left( M^2 / Q^2 \right) }

\newcommand{\phaseFactor}[1]{ \frac{\rmd{3}{ \threeVec{#1} }}
{(2\pi)^3 2 E_{\threeVec{#1}}}}

\newcommand{\myBox}[1]{\vspace{.2cm}
		       \\*
                       \vspace{.2cm}
                       \fbox{
                             \parbox{.96\columnwidth}{
                             #1
                             \vspace{-.3cm}}
                            }
                      }

\newcommand{\eqnIndent}{\hspace{.5cm}}

\def\slashi#1{\rlap{\sl/}#1}
\def\slashii#1{\setbox0=\hbox{$#1$}             % set a box for #1
   \dimen0=\wd0                                 % and get its size
   \setbox1=\hbox{\sl/} \dimen1=\wd1            % get size of /
   \ifdim\dimen0>\dimen1                        % #1 is bigger
      \rlap{\hbox to \dimen0{\hfil\sl/\hfil}}   % so center / in box
      #1                                        % and print #1
   \else                                        % / is bigger
      \rlap{\hbox to \dimen1{\hfil$#1$\hfil}}   % so center #1
      \hbox{\sl/}                               % and print /
   \fi}                                         %
\def\slashiii#1{\setbox0=\hbox{$#1$}#1\hskip-\wd0\hbox to\wd0{\hss\sl/\/\hss}}
\def\slashiv#1{#1\llap{\sl/}}
\DeclareMathOperator{\tr}{Tr}

\title{Single spin asymmetries and gauge invariance\\ in hard scattering
processes}

\author{Fetze Pijlman}

\begin{document}

%\maketitle

%\thispagestyle{empty}
%\phantom{boktor}
%{}

%\pagebreak

\thispagestyle{empty}

\begin{center}
        {VRIJE UNIVERSITEIT}
        \\
        \vspace{2cm} {\bf
        {\LARGE Single spin asymmetries and gauge invariance\\ 
                in hard scattering processes\par}}
        \vspace{2cm}
        {ACADEMISCH PROEFSCHRIFT}
        \\
        \vspace*{1cm}
        ter verkrijging van de graad Doctor aan\\
        de Vrije Universiteit Amsterdam,\\
        op gezag van de rector magnificus\\
        prof.dr.\ T.~Sminia,\\
        in het openbaar te verdedigen \\
	ten overstaan van de promotiecommissie \\
        van de faculteit der Exacte Wetenschappen\\
        op donderdag 12~januari~2006 om 13.45 uur\\
        in de aula van de universiteit, \\
        De Boelelaan 1105 %}
        \vspace*{2.5cm}

        door \\
        \vspace{1cm}

        {\large \bf Fetze Pijlman} \\
        \vspace{1cm}
        geboren te Leiden
\end{center}

\newpage

\noindent
\thispagestyle{empty}
promotor: prof.dr. P.J.G. Mulders

\pagebreak

\indent

\thispagestyle{empty}

\noindent
This thesis is partly based on the following publications:
\\
%\begin{itemize}
\\
%\item
B.~L.~G.~Bakker, M.~van Iersel, and F.~Pijlman,\newline
\emph{Comparison of relativistic bound-state calculations in front-form and
      instant-form dynamics,}
Few Body Syst.\  {\bf 33}, 27 (2003).
\newline
\\
%\item
D.~Boer, P.~J.~Mulders, and F.~Pijlman,\newline
\emph{Universality of T-odd effects in single spin and azimuthal asymmetries,}\newline
Nucl.\ Phys.\ B {\bf 667}, 201 (2003).
\newline
\\
%\item
A.~Bacchetta, P.~J.~Mulders, and F.~Pijlman,\newline
\emph{New observables in longitudinal single-spin asymmetries in semi-inclusive
      DIS,}\newline
Phys.\ Lett.\ B {\bf 595}, 309 (2004).
\newline
\\
%\item
C.~J.~Bomhof, P.~J.~Mulders, and F.~Pijlman,\newline
\emph{Gauge link structure in quark-quark correlators in hard processes,}\newline
Phys.\ Lett.\ B {\bf 596}, 277 (2004).
\newline
\\
%\item
F.~Pijlman,\newline
\emph{Color gauge invariance in hard processes,}\newline
Few Body Syst.\  {\bf 36}, 209 (2005).
\newline
\\
%\item
F.~Pijlman,\newline
\emph{Factorization and universality in azimuthal asymmetries,}\newline
Proceedings of the 16th international spin physics symposium 2004, hep-ph/0411307.
\newline
\\
%\item
A.~Bacchetta, C.~J.~Bomhof, P.~J.~Mulders, and F.~Pijlman,\newline
\emph{Single spin asymmetries in hadron-hadron collisions,}\newline
Phys.\ Rev.\ D {\bf 72}, 034030 (2005).
%\end{itemize}

\indent

\newpage

\thispagestyle{empty}

\phantom{boktor}

\vspace{\stretch{1}}
Printed by Universal Press - Veenendaal, The Netherlands.

\tableofcontents

\chapter{Introduction\label{Chapter1}}

%An overview is given of the research field to which the work reported on
%in this thesis belongs. 
%Together with an enumeration of a few successes in elementary particle
%physics, the theory of quantum chromodynamics 
%will be introduced, and 
%the phenomena of single spin asymmetries will be discussed.
%The last section contains an outline of the
%thesis.

\section{Particle physics, it is amazing!}

Progress in the understanding of elementary particles is amazing. 
For more than a century the smallest building
blocks of nature have been studied,
and discoveries are still being made today. While studying
the ingredients of nature, fundamental and inspiring theories have been
developed making this field of increasing interest.

At the beginning of the last century, Einstein studied the concept
of time in order to explain the discrepancy between the theories of Newton
and Maxwell. This led to the publication of his theory
of special relativity in
1905~\cite{Einstein:1905ve}. Combining this theory and the quantum
theory, Dirac predicted the
existence of the antiparticle of both the electron and
proton~\cite{Dirac:1931kp}.
The antiparticle of the electron, the positron, was discovered in 1933 by
Anderson~\cite{Anderson:1933mb} and the antiproton was found
in \nolinebreak 1955 \nolinebreak by \nolinebreak \mbox{Chamberlain}
\nolinebreak et \nolinebreak al.\cite{Chamberlain:1955ns}

Around the 1930's,
explanations were sought for $\beta$-decay, which is one particular
form of nucleus-disintegration.
Experimental studies of this phenomenon seemed to show that the
energy before and after the decay were not the same:
some energy was missing. To circumvent the potential violation of
energy conservation (Newton's law),
Pauli suggested between 1930 and 1933 at several
conferences
a new kind of particle\footnote{Pauli publicized this new particle at
several conferences among which the Solvay Congresses in 1930 and 1933.},
which would be produced during
radioactive decay without
notice. This new particle, called neutrino by Fermi, was observed
in 1956 by Reines and Cowan~\cite{Reines:1953pu}.

Around 1960 particle accelerators discovered new kinds of hadronic matter.
In order to classify the observed hadrons
Gell-Mann in Ref.~\cite{Gell-Mann:1964nj}
and Zweig in Ref.~\cite{Zweig:1964jf}
introduced, independently,
a substructure with three types of quarks.
Since then several other quark-types have been discovered and
just a decade ago the last
quark with a mass of almost 200 times the proton
mass, the top quark,
was discovered at Fermilab~\cite{Abe:1995hr,Abachi:1995iq}.
Since this quark
and its mass were already predicted on the basis
of data taken by the large electron-positron collider at CERN,
this was once again a stunning success for particle physics.

In the last century particles have been found which were predicted by theory
and theories have been developed on the basis of
experimental observations. It is expected that during this
century some of the predicted particles, such as
the ones responsible for spontaneous symmetry breaking (the Higgs-sector),
will be observed. The interplay
between experiment and theory in this field is a guaranteed
success to explore
what nature will offer us next.

\section{QCD and single spin asymmetries}

Quantum chromodynamics (QCD) describes the interactions between quarks and
gluons and is constructed from powerful theories and concepts.
The main ingredients
are: the theory of relativity, the quantum theory, and the concept of
gauge invariance.
The first gauge theory
was developed more than a century ago. Around 1865 Maxwell wrote down
his equations describing the interactions between electromagnetic
fields and matter. The equations are a set of differential equations which
also raised some questions, one of which was that the potentials obeying the
equations are not unique. This
point became clarified towards the end of the nineteenth century; it was
considered as a mathematical symmetry which was apparently
left in the equations.
This mathematical symmetry allowed for a set of
transformations of the potentials which would not
affect physical observables. Nowadays this symmetry is named
\emph{gauge invariance} and the potentials are often called the
gauge fields.

In the beginning of the
twentieth century gauge invariance was considered more seriously.
While incorporating the symmetry in quantum mechanics, Fock discovered that,
besides the gauge fields, the
wave function of the electron
should transform as well to maintain consistency with
the theory of relativity. In order to preserve invariance of observables under
gauge transformations, the wave function of the electron must obtain
a space-time-dependent phase.
However, the question
remained whether the gauge fields were to be considered as fundamental fields
or whether they just alleviated complex calculations. For a review on
the historical roots of gauge invariance the reader is referred to Jackson,
Okun~\cite{Jackson:2001ia}.

\begin{figure}
\begin{center}
\includegraphics[width=6cm]{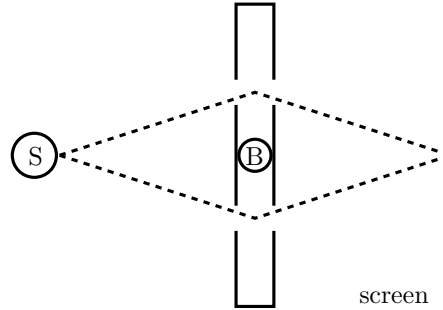}
\end{center}
\caption{The schematic setup of an Aharonov-Bohm like experiment.
S represents a source of electrons producing an interference pattern
on the screen owing to the two slits.
The interference pattern shifts in a particular
direction if the solenoid in B, pointing out
of the plane, is given a current.
If the screen is far away from the two slits then
the shift is proportional to
$\oint \rmd{1}{\threeVec{x}}\cdot \threeVec{A}(\threeVec{x})$.
The path of the integral is the closed path formed by the two dashed lines
and $\threeVec{A}(\threeVec{x})$ is the potential field. 
One can call this shift of interference pattern 
an asymmetry because the direction of the shift 
depends on the direction of the magnetic field.
\label{intro1}}
\end{figure}

In the second half of the twentieth century the question
on whether potentials are more fundamental than electric and magnetic fields
was finally addressed.
Aharonov and Bohm apparently\footnote{It seems that Ehrenberg and Siday
already pointed out that enclosed magnetic fluxes could
cause phase shifts. Their
work~\cite{Ehrenberg} has been cited in the subsequent paper of Aharonov
and Bohm~\cite{Aharonov2}.} rediscovered that an electron can obtain a
phase shift from its interaction with the potential even if it
only travels in
regions in which there is no electric or magnetic field~\cite{Aharonov:1959fk}.
The experiment carried
out by Chambers showed that instead of the electric and magnetic
fields, the non-uniquely defined potentials should be
considered as the fundamental fields in quantum electrodynamics~\cite{Chambers}.
A schematic setup of the experiment is given in Fig.~\ref{intro1}.

As compared to electrons and photons, the situation of quarks and gluons
is much more involved. In contrast to electrons and photons,
free quarks and gluons have for instance never
been observed. They only seem to exist in
a hadron (confinement)
which indicates a strong interaction.
On the other hand,
%when hadrons collide at high energies,
perturbation theory turns out to provide a satisfactory description
for collisions involving hadrons at high energies,
showing that
the interaction at high energies must be weak.
This particular scale dependence of the interaction strength
confronted physicists with a big challenge.

\begin{figure}
\begin{center}
\includegraphics[width=5cm]{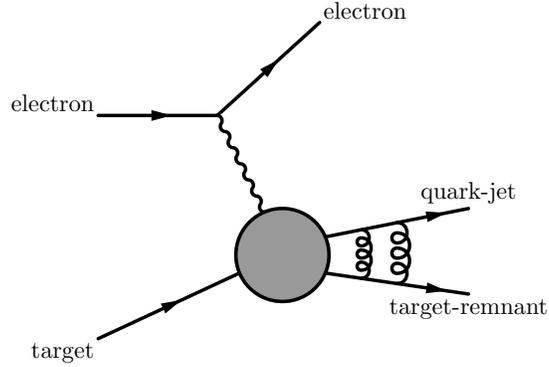}
\end{center}
\caption{An illustration of the interactions between the quark-jet
and the target-remnant (Sivers effect).
These interactions, which are on the amplitude level
and lead to phases as we will see in chapter~\ref{chapter3},
give rise to interference contributions
in the cross section and could produce
single spin asymmetries
in the idealized jet-production in semi-inclusive deep-inelastic
scattering.\label{intro2a}}
\end{figure}
\begin{figure}
\begin{center}
\includegraphics[width=8cm]{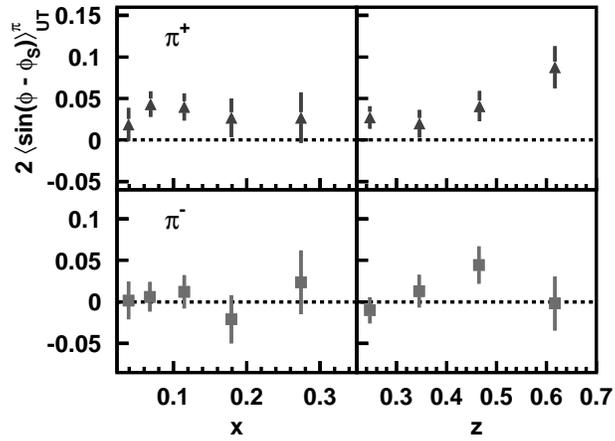}
\end{center}
\caption{A nonzero transverse 
single target-spin asymmetry for $\pi^+$ and $\pi^-$
in electron-proton scattering
measured by the HERMES collaboration.
As will be explained in chapter~\ref{chapter3}, the asymmetry is
in the scaling limit
proportional to the phase picked up by the outgoing
quark and is also called the Sivers effect. The plot
was taken from Ref.~\cite{Airapetian:2004tw}.
\label{intro2b}}
\end{figure}

%In contrast to the common belief t
The solution
came from a quantum field theory.
In quantum field theories infinities often appear.
In the 1940's Dyson,
Feynman, Schwinger, and Tomonaga showed that in quantum electrodynamics such
infinities can be handled by renormalizing the observables.
In contrast to the general expectation,
\mbox{'t Hooft} and Veltman showed in 1972
that this was also possible for non-Abelian
gauge theories~\cite{'tHooft:1972fi,'tHooft:1972ue}. Using the machinery
of 't Hooft and Veltman, Gross, Politzer, and Wilczek derived
in Ref.~\cite{Gross:1973id,Politzer:1973fx} \nolinebreak that
a scale-dependent interaction strength
appears when quarks and gluons are characterized
with a color.
The constructed non-Abelian gauge theory,
called quantum chromodynamics, has a vanishing interaction strength at
large momentum transfers - as those occurring
in high energy collision experiments -
 which is called asymptotic freedom.

%In the end it turned out that
%by introducing a color for the quarks and gluons one could obtain an asymptotic
%free theory; quantum chromodynamics was born.
%Quantum chromodynamics can be regarded as a non-Abelian version of Maxwell's
%theory in which relativistic and quantum effects are incorporated.

Not being able to apply perturbation theory, most of the low energy regime of
quantum chromodynamics is at present not calculable from first principles.
Bjorken and
Feynman introduced the idea of absorbing the nonperturbative part in
parton probability functions
which could be measured in several experiments.
These functions describe how quarks are distributed in hadrons
(distribution functions) or how they ``decay'' into a hadron and accompanying
jet (fragmentation functions).
The functions introduced by Feynman depend only on
the longitudinal momentum fraction because at
high energies the
transverse momenta of quarks in hadrons
can often be neglected. This somewhat ad hoc procedure
of absorbing the nonperturbative part in functions,
called the parton model, can be translated into more rigorous QCD and
is very successful in describing various kinds of data.

One observation, studied in this thesis, cannot be explained
by the parton model, namely the observation of
single spin asymmetries.
In single spin asymmetries, one of the participating particles in a scattering
process carries or acquires a certain
polarization. If the scattering cross section
depends on the direction of this polarization, one has a
single spin asymmetry.
Large single spin asymmetries in inelastic collisions
were discovered in hyperon-production in
hadron-hadron scattering at Fermilab~\cite{Bunce:1976yb}. Since then,
single spin asymmetries have been observed in various processes.

Several explanations for single spin asymmetries at large scales
were developed over the last two decades. One of the most important
ideas, proposed by
Sivers in Ref.~\cite{Sivers:1989cc,Sivers:1990fh},
was to allow for a nontrivial
correlation between the transverse momentum of the quark and its
polarization. After incorporating the transverse momenta of quarks
in an extended version of the parton model,
it is at present understood that there are
two sources for single spin asymmetries. The first is the presence of 
interactions within a fragmentation process
(see Collins~\cite{Collins:1993kk}). The other source,
%One of them,
appearing
in the idealized single jet-production in semi-inclusive deep-inelastic
scattering, see Fig.~\ref{intro2a},
turns out to be
the phase which the struck quark picks up due to its interaction with the
target-remnant (also exists for fragmentation). 
This particular phase
shows up as a gauge link, which is a mathematical operator,
in the definition of transverse momentum dependent distribution functions.
Since 
these functions are defined in terms of 
nonlocal operators inside matrix elements,
the presence of this gauge link is also needed for invariance under local
gauge transformations.
Note that the obtained
phase of the outgoing quark has similarities with the phases of
the electrons in the Aharonov-Bohm experiment.
Having the same origin of the effect for both asymmetries is surprising.
The first nonzero experimental data which directly
measures this phase was obtained by the HERMES collaboration in 2004
and is
given in Fig.~\ref{intro2b}.

\newpage

\section{Outline of the thesis}

The appearance and treatment of
phases in several
hard scattering processes will be studied in this thesis. 
In 2002 this particular topic became popular after
Brodsky, Hwang, and Schmidt showed that such phases, leading to single
spin asymmetries, could be generated within a
model calculation~\cite{Brodsky:2002cx}. The obtained phases were attributed
by
Collins~\cite{Collins:2002kn}, Belitsky, Ji, and Yuan~\cite{Belitsky:2002sm}
to the presence of a fully closed gauge link in the definition of
parton distribution functions. In this thesis these ideas are
implemented in a diagrammatic expansion which is a field theoretical
description of hard scattering processes. The effect of the gauge link is
studied
in several hard processes like hadron-hadron and lepton-hadron
scattering. Although only QCD is studied in this thesis,
gauge links also appear in other gauge theories like
quantum electrodynamics. It is therefore to be expected that gauge links
could provide a description of single spin asymmetries in pure electromagnetic
scattering as well.

%\sloppy
For a full appreciation of the contents of this thesis 
familiarity with particle
physics and quantum field theory is needed. Some excellent textbooks or reviews
have been
written by Anselmino, Efremov, Leader~\cite{Anselmino:1994gn},
Barone, Ratcliffe~\cite{Barone}, Halzen, Martin~\cite{Halzen:1984mc},
Leader~\cite{leader}, Peskin, Schroeder~\cite{Peskin:1995ev},
Ryder~\cite{Ryder:1985wq}, Weinberg~\cite{Weinberg:1995mt,Weinberg:1996kr}.

%\fussy
Chapter~\ref{chapter2} will begin with an introduction of
the kinematics of some processes in which the hard
scale is set by an electromagnetic interaction.
A discussion of the diagrammatic approach
will be given together with the definitions of distribution
and fragmentation functions. This chapter contains some new results from
Ref.~\cite{Boer:2003cm,Bacchetta:2004zf}.

In chapter~\ref{chapter3}, the diagrammatic approach will be applied to
obtain cross sections of
some electromagnetic processes assuming factorization. 
The gauge link inside the definition of
parton distribution and fragmentation functions
will be derived,
showing the consistency of the applied approach
at leading order in $\alpha_S$ (tree-level).
The presence of the gauge link will lead to
the interesting prediction of Collins~\cite{Collins:2002kn}
that T-odd distribution functions in the Drell-Yan process
appear with a different sign
compared to semi-inclusive deep-inelastic scattering.
This chapter is based on Ref.~\cite{Boer:2003cm,Bacchetta:2004zf}.

Chapter~\ref{chapter4} will begin by considering gauge links
in more complicated processes (and beyond tree-level). Besides the gauge links
which are found in the electromagnetic processes new gauge links
will be encountered which is quite a surprise.
We will also see that the appearance of these
new structures is an essential ingredient in the discussion
of factorization.
A set of tools will be
developed which allows for a quick determination of the gauge link for
arbitrary scattering processes. This chapter
contains unpublished material of which some results were given in
Ref.~\cite{Bomhof:2004aw, Pijlman:2004wb, Pijlman:2004mr}.

In chapter~\ref{chapter5} the physical
effect of the new gauge links will be
illustrated in almost back-to-back hadron-production in
hadron-hadron scattering. A new observable is constructed
which is directly sensitive to the intrinsic
transverse momenta of partons. In the same way as T-odd distribution
functions change sign in the electromagnetic processes, the T-odd
distribution functions receive a gauge link dependent factor
in the studied asymmetries of
hadron-hadron scattering.
This chapter is based on Ref.~\cite{Bacchetta:2005rm}.

%\section*{content}
%
%\begin{itemize}
%
%\item What is the field of subatomic physics? atoms consist of
%nuclei and electrons, proton consist of quarks  ...etc .. Aim is to
%understand the the smallest building blocks, although that will certainly
%not be enough to understand macroscopic physics. Since the building blocks are
%universal they appear everywhere, beginning over universe, on earth and in
%stars, etc. To understand those processes better we need to understand the
%building blocks.
%
%\item Elementary particle physics has always given surprises and noble prices%
%
%\item Subatomic physics is much different from the environment around us. There
%are two sources for this. 1 quantum mechanics and relativity. Without any doubt
%the two major discoveries in physics of the last century. The combination of
%the two is formulated by quantum field theory and will be the theory used in
%this thesis. The two mechanisms give rise to effects which do not exist in
%our world and make physics interesting. Theory is far from being perfect.
%Many things are ill defined and makes life hard.

%\item Give list of successes in particle physics.
%(number of colours of quarks)

%\item Treat Aharonov Bohm effect and say that we discovered a similar effect
%      (second section)

%\item Treat an experimental SIDIS and make connection with Sivers and Aharonov
%Bohm effect. Give matrix element of Sivers function.

%\item Outline of thesis.

%\end{itemize}

\chapter{%Scattering theory and the parametrizations
High energy scattering and quark-quark correlators
\label{chapter2}}

\vspace{-.4cm}

The formalism will be introduced
which was
initiated by Collins, Ralston, and Soper in
Ref.~\cite{Soper:1976jc,Collins:1977iv,Soper:1979fq,Ralston:1979ys}.
The formalism
carefully considers the role of intrinsic
transverse momenta of partons in hard scattering processes.
Although the first part of the chapter
is already present in the literature (and partly based on Ref.~\cite{pietsNotes}),
it remains
worthwhile to look at some
parts in more detail to
elaborate upon the approximations
and the philosophy behind certain
approaches.
Since this part is meant as an introduction for the
non-experts, the reader can skip those sections
which are familiar to him or her.

Starting from section~\ref{sectionDistr},
the second part
contains new elements
which have been developed over the last few years.
These new
elements, which are one of the highlights in QCD-phe\-no\-me\-no\-lo\-gy,
originate from the
presence of Wilson lines or gauge links
in parton distribution and fragmentation
functions. These functions will be defined and they turn out to
provide valuable information of partons inside hadrons and parton decay into
hadrons.
As we will see in the following chapters, the presence of these Wilson lines
in the definitions of these functions
lead to interesting predictions. A summary
will be given at the end of this chapter.

\newpage

\section{Kinematics of electromagnetic scattering processes}

Physicists use often the
Lorentz-invariance of the theory to choose
the most convenient frame
for their purposes.
This has produced several frame definitions and frame-dependent
interpretations.
In principle the definitions and results can be compared by making
the appropriate coordinate
transformations
but in practice this has often led to confusion.
In this section several scattering processes will be introduced and
their kinematics will be set up such that theoretical
predictions and experimental results can be compared frame-independently.

%In order to clarify this situation one should express
%all results in easy-to-compare
%frame-independent observables which is possible owing to the
%Lorentz-invariance of the theory .

As was advocated at the Transversity workshop
in Trento 2004 (see also Ref.~\cite{Bacchetta:2004jz}),
one can, in order to clarify this
situation, express all results in easy-to-compare
frame-independent observables which is possible owing to the
Lorentz-invariance of the theory.
For example,
the variables in the invariant
cross section are usually the momentum and spin vectors
which have specific transformation properties.
A much better choice would be to express the
cross section in terms of the possible invariants. The invariants are
frame-independent and can therefore be directly calculated in any frame!

A drawback of this approach is that equations become rather lengthy and that
we are not used to think in a frame-independent
manner. To aid our
intuition and to support the frames which are already in use,
a Cartesian basis
will be employed which will serve as an interface. Such bases were
already introduced before, see for instance Lam, Tung~\cite{Lam:1978pu}
and
Meng, Olness, Soper~\cite{Meng:1991da}.
This will result in short
expressions while maintaining manifest frame-independence.

In this thesis the metric tensor of
Bjorken and Drell~\cite{Bjorken} will be employed, reading
\begin{align}
g^{00} &= - g^{11} = - g^{22} = -g^{33} = 1,&  g^{ij} &= 0\ \ \text{for }
i \neq j,
\end{align}
and the antisymmetric tensor is normalized such
that $\epsilon^{0123} = 1$.
The Einstein summation convention will also be used,
meaning that if a certain
index appears twice in a product it is automatically summed over all its
values unless stated otherwise.
Furthermore, natural units with $\hbar=c=1$ will be used.

\subsection{Semi-inclusive deep-inelastic lepton-hadron scattering}

\begin{figure}
\begin{tabular}{cp{.5cm}c}
\begin{minipage}[b][4cm][c]{4cm}
\includegraphics[width=4cm]{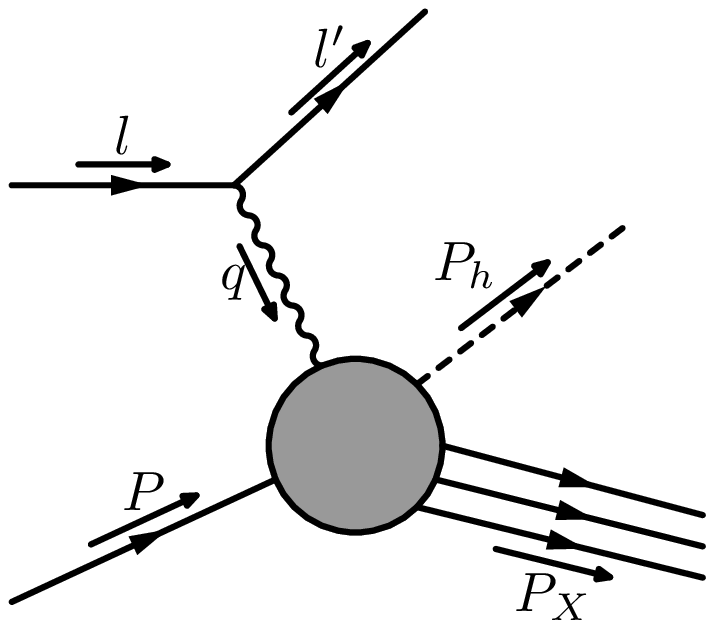}
\end{minipage}
&      &
\begin{minipage}[b][4cm][t]{6cm}
\includegraphics[width=6cm]{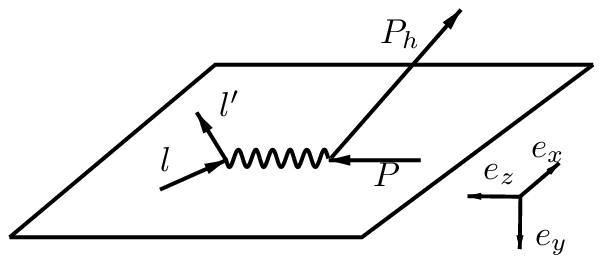}
\end{minipage}
\\
(a)&  & (b)
\end{tabular}
\caption{Semi-inclusive DIS. Figure~(a) illustrates
the process in leading order in $\alpha_\text{e.m.}$. Figure~(b)
presents the definition of the Cartesian
basis, except for the vectors $e_y$ and $P_h$ all vectors lie in the
plane shown.\label{figTheory1}}
\end{figure}

In the deep-inelastic scattering (DIS) process, a lepton
with momentum $l$ and mass $m_e$,
strikes with a large momentum difference ($l\cdot P \gg M^2$)
a hadron (sometimes called target), with momentum $P$ and mass $M$.
The interaction, mediated through the exchange of a highly virtual
photon with momentum $q$ (and $-M^2 \gg q^2$),
causes the hadron to break up into all kinds of
particles being most often other hadrons. The measurement is called:
\emph{inclusive} if only the scattered electron is measured,
\emph{semi-inclusive} if an
additional particle (or more) with momentum $P_h$ and
mass $M_h$
is measured, and \emph{exclusive} if all (but one) final-state particles
are detected. The semi-inclusive process is illustrated in
Fig.~\ref{figTheory1}a.

Traditionally the following
Lorentz-invariants have been introduced to characterize
experimental events
\myBox{
\vspace{-.3cm}
\begin{align}
Q^2 &\equiv - q^2,& W^2 &\equiv (P+q)^2,&
s &\equiv (P+l)^2,& x_B& \equiv \frac{Q^2}{2P\cdot q},
\nonumber
\\
y &\equiv \frac{P\cdot q}{P\cdot l}, &
z_h &\equiv \frac{-2 P_h \cdot q}{Q^2},&
z &\equiv \frac{P\cdot P_h}{P\cdot q}, & &
\end{align}
\vspace{-.6cm}
\begin{flushright}
\emph{several Lorentz-invariants in semi-inclusive DIS}
\end{flushright}
}
where $z = z_h (1+\mathcal{O}(M^2/Q^2))$ and $y = Q^2 (1+\mathcal{O}(M^2/Q^2))/(x_B s)$.

The Cartesian basis
is defined through an orthogonal
set of basis vectors $e_i$
(see Fig.~\ref{figTheory1}b).
The space-like vector $e_z$ is chosen such that it is
pointing in the opposite direction
of $q$. The time-like
vector $e_t$
is constructed
from $P$ subtracting its projection along $q$,
and
the transverse directions are fixed by choosing
$e_x$ along the components of the sum of the lepton momenta
which are
perpendicular to $e_z$ and $e_t$. The definition of $e_y$
follows from demanding a right-handed coordinate system. This Cartesian
basis is by construction
frame independent
and is mathematically defined as
\myBox{
\vspace{-.3cm}
\begin{align}
e_z^\mu &\equiv \frac{-q^\mu}{Q},&
e_t^\mu &\equiv \frac{q^\mu + 2 x_B P^\mu}{Q\sqrt{1 + \tfrac{4 x_B^2 M^2}{Q^2}}},
&&
\nonumber\\
g_{\perp}^{\mu\nu} &\equiv g^{\mu\nu} - e_t^\mu e_t^\nu + e_z^\mu e_z^\nu,&
A_{\perp}^\mu &\equiv g_{\perp}^{\mu\nu} A_\nu\ (\text{for any }A), &
\epsilon_{\perp}^{\rho\nu} &\equiv
 \epsilon^{\sigma\mu\rho\nu} {e_z}_\sigma {e_t}_\mu,
\nonumber\\
e_x^\nu &\equiv \frac{ l_\perp^\nu + {l'_\perp}^\nu}
{\left| \threeVec{l}_\perp + \threeVec{l}'_\perp \right|},&
e_y^\rho &\equiv \epsilon_\perp^{\rho\nu} {e_x}_\nu,&
-1 &= \epsilon_{\mu\nu\rho\sigma} e_t^\mu e_x^\nu e_y^\rho e_z^\sigma.
\label{hiro13}
\end{align}
\vspace{-.8cm}
\begin{flushright}
\emph{Cartesian basis for semi-inclusive DIS}
\end{flushright}
}
Note that the antisymmetric tensor,
$\epsilon_\perp^{\rho\nu}$, has been defined.

\sloppy
In a Cartesian basis a general vector can be easily decomposed
into a linear combination
of the basis vectors having frame-independent coefficients, for example
\begin{equation}
P_h = (P_h \cdot e_t)\ e_t - (P_h \cdot e_x)\ e_x - (P_h \cdot e_y)\ e_y
- (P_h \cdot e_z)\ e_z.
\end{equation}
By using such decompositions \emph{head-on}
cross sections\footnote{Cross sections are generally
not invariant under
Lorentz transformations in contrast to head-on
cross sections (see for instance Ref.~\cite{Peskin:1995ev}).
In the latter the initial particles are aligned.}
can be written in terms of invariants only. Those cross sections
can be indicated with the following notation:
$\sigma ( \text{Inv:}\ p_1, p_2, \ldots )$ $=
\sigma(\text{all possible invariants of }p_1,p_2,\ldots)$.

\begin{figure}
\begin{center}
\begin{tabular}{cp{.5cm}c}
\includegraphics[width=5cm]{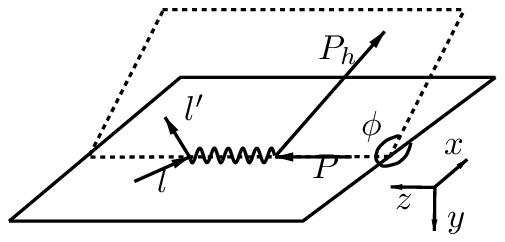} & &
\includegraphics[width=5cm]{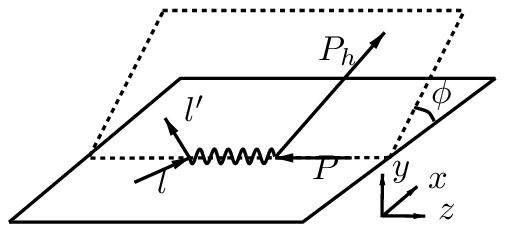} \\
(a) & & (b)
\end{tabular}
\end{center}
\caption{Two examples of fixed frames. Only the
momentum $P_h$ has a nonzero $y$-component and lies in the
dashed planes.
\label{expFrames}}
\end{figure}

\fussy
The advantage of this approach becomes clear by the following example.
Suppose we are interested in azimuthal asymmetries in semi-inclusive DIS
 and we would like to predict
or measure the quantity
\begin{equation}
A = \int \rmd{2}{{\threeVec{P}_h}_{\perp}}
\ {\threeVec{P}_h}_{\perp} \cdot \threeVec{e}_y\
\frac{ \rmd{2}{ \sigma ( \text{Inv:}\ l,l',P,S,P_h ) } }
     { \rmd{2}{{\threeVec{P}_h}_\perp} }.
\end{equation}
If our frame would be defined such that our proton is moving along the $+z$-axis
and our $+x$-axis is proportional to $l^\perp + {l'}^\perp$
(see Fig.~\ref{expFrames}a), then $A$ would read
in that frame
\begin{align}
\text{frame a: }A &= \int \rmd{2}{{\threeVec{P}_h}_{\perp}}\
| {\threeVec{P}_h}_{\perp} |  \sin{\phi_{{\threeVec{P}_h}_{\perp}}}\
\frac{ \rmd{2}{ \sigma ( \text{Inv:} l,l',P,S,P_h ) } }
     { \rmd{2}{{\threeVec{P}_h}_\perp} },\\
\intertext{
while if our $z$-axis would lie in the opposite direction
and keeping the same $x$-axis (see Fig.~\ref{expFrames}b),
$A$ would read}
\text{frame b: }A &= -\int \rmd{2}{{\threeVec{P}_h}_{\perp}}\
| {\threeVec{P}_h}_{\perp} |  \sin{\phi_{{\threeVec{P}_h}_{\perp}}}\
\frac{ \rmd{2}{ \sigma ( \text{Inv:} l,l',P,S,P_h ) } }
     { \rmd{2}{{\threeVec{P}_h}_\perp} }.
\end{align}
Although the appearance of the expressions differs by a sign
(due to a different definition
of the azimuthal angle $\phi_{\threeVec{P}_{h\perp}}$),
the quantity $A$ is the same in the two different
coordinate systems.

\subsection{The Drell-Yan process}

\begin{figure}
\begin{tabular}{cp{.5cm}c}
\includegraphics[width=4cm]{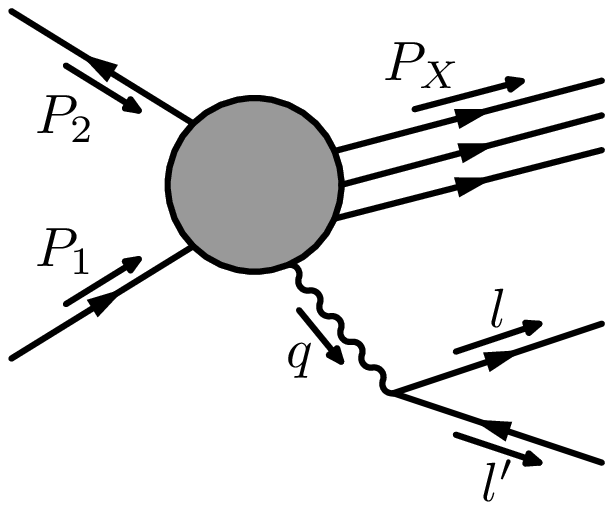}& &
\includegraphics[width=6cm]{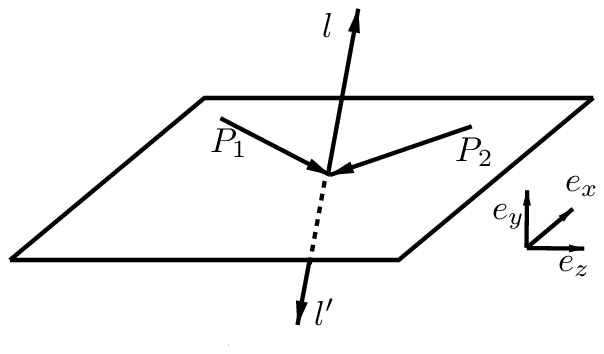}\\
(a) & &(b)
\end{tabular}
\caption{The Drell-Yan process.
Figure~(a) illustrates the process in leading order in $\alpha_\text{e.m.}$.
Figure~(b) presents the Cartesian basis in which
only $l$, $l'$, and $e_y$ are out
of the drawn plane.\label{drellYan}}
\end{figure}

In the Drell-Yan process two hadrons with momentum $P_1$ and $P_2$ collide
and produce a virtual photon with a large invariant momentum squared
$q^2 \gg M^2$.
This virtual photon decays into an antilepton and lepton
which are measured in the final state. The process
has been illustrated in Fig.~\ref{drellYan}a.

Characterizing invariants for this process are
\myBox{
\vspace{-.3cm}
\begin{align}
Q^2 &\equiv q^2, & s &\equiv (P_1+P_2)^2,&
%&&
%\nonumber\\
x_1& \equiv \frac{Q^2}{2P_1\cdot q},& x_2& \equiv \frac{Q^2}{2P_2\cdot q},&
y & \equiv \frac{l \cdot P_1}{q\cdot P_1},
\end{align}
\vspace{-.6cm}
\begin{flushright}
\emph{several Lorentz-invariants in Drell-Yan}
\end{flushright}
}
and
the Cartesian basis is chosen to be (see also Fig.~\ref{drellYan}b)
\myBox{
\vspace{-.3cm}
\begin{align}
e_t^\mu &\equiv \frac{q^\mu}{Q},&
e_z^\mu &\equiv \frac{ \sqrt{\tfrac{x_1}{x_2}} P_1 -
                       \sqrt{\tfrac{x_2}{x_1}} P_2}
                     { \sqrt{ s\! -\! M_1^2 \left(1\! -\! \tfrac{x_1}{x_2} \right)
                              \!  -\! M_2^2 \left(1\! -\! \tfrac{x_2}{x_1} \right)}},
\\
g_{\perp}^{\mu\nu} &\equiv g^{\mu\nu} - e_t^\mu e_t^\nu + e_z^\mu e_z^\nu, &
A_{\perp}^\mu &\equiv g_{\perp}^{\mu\nu} A_\nu\ (\text{for any }A),
\qquad
\epsilon_{\perp}^{\rho\nu} \equiv
 \epsilon^{\sigma\mu\rho\nu} {e_z}_\sigma {e_t}_\mu,
\nonumber\\
e_x^\mu &\equiv \frac{-\left( P_1 + P_2 \right)_{\perp}^\mu }
                                {\sqrt{-\left( P_1 + P_2 \right)_{\perp}^2}},
&
e_y^\rho & \equiv \epsilon_{\perp}^{\rho\nu} {e_x}_\nu,
\qquad \qquad \qquad \ \ \,
-1 = \epsilon_{\mu\nu\rho\sigma}e_t^\mu e_x^\nu e_y^\rho e_z^\sigma
\nonumber .
\end{align}
\vspace{-1cm}
\begin{flushright}
\emph{Cartesian basis for Drell-Yan}
\end{flushright}
}
Since $q$ is now time-like, $e_t$ is chosen along $q$. The
vector $e_z$ is chosen
perpendicular to $q$ and such that $P_1$ has a large
positive component. The sum of the incoming hadron momenta perpendicular
to $e_t$ and $e_z$ sets the direction of $e_x$, and $e_y$ follows from $e_x$.

\subsection{Semi-inclusive electron-positron annihilation}

\begin{figure}
\begin{tabular}{cp{.5cm}c}
\includegraphics[width=4cm]{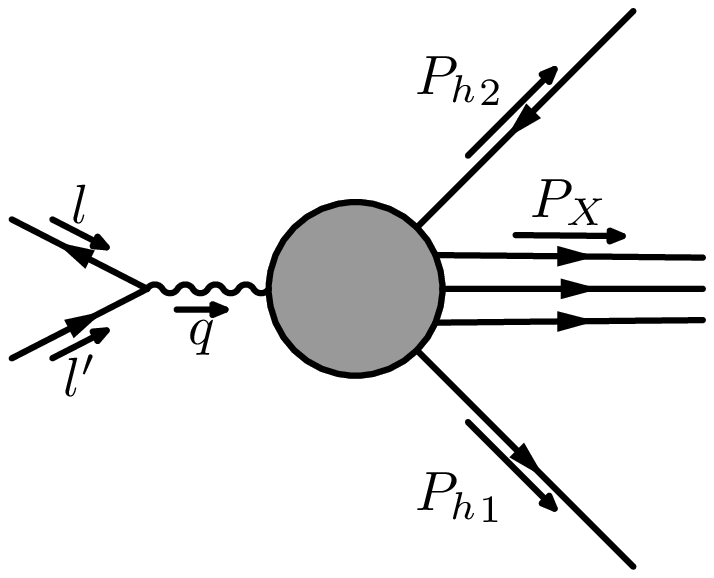}& &
\includegraphics[width=6cm]{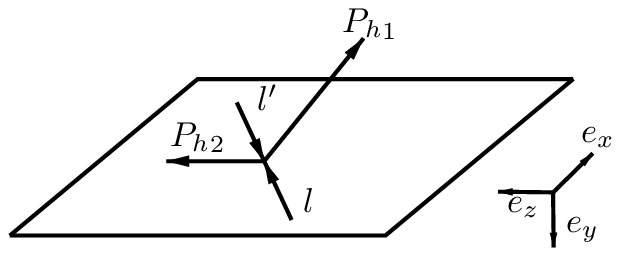}\\
(a) & &(b)
\end{tabular}
\caption{The electron-positron annihilation process.
Figure~(a) illustrates the process in leading order in $\alpha_\text{e.m.}$.
Figure~(b) presents
the Cartesian basis in which only $P_{h_1}$ and $e_y$ are out of the plane.
\label{figePeM}}
\end{figure}

In electron-positron annihilation an electron and positron collide
with a large momentum difference, producing at
leading order in $\alpha_S$ two
jets (see Fig.~\ref{figePeM}a). We will assume here
that in both jets one hadron is detected and that $Q^2>M^2$.

The variables which characterize this process are (see Fig.~\ref{figePeM}b)
\myBox{
\begin{align}
Q^2 &\equiv q^2,& s &\equiv (l_1+l_2)^2,&
z_{h1} &\equiv \frac{2 P_{h1} \cdot q}{Q^2},&
z_{h2}&\equiv \frac{2 P_{h2} \cdot q}{Q^2}.
\end{align}
\vspace{-.5cm}
\begin{flushright}
\emph{several Lorentz-invariants in electron-positron annihilation}
\end{flushright}
}
The Cartesian basis is defined through
\myBox{
\begin{align}
e_t^\mu & \equiv \frac{q^\mu}{Q},&
e_z^\mu &\equiv
        \frac{ \frac{2}{z_{h2}} P_{h2}^\mu - q^\mu}
             {Q\sqrt{1-\frac{4 M_{h2}^2}{z_{h2}^2 Q^2}}},&
& \nonumber\\
g_{\perp}^{\mu\nu} &\equiv g^{\mu\nu} - e_t^\mu e_t^\nu + e_z^\mu e_z^\nu,&
A_{\perp}^\mu &\equiv g_{\perp}^{\mu\nu} A_\nu\ (\text{for any }A),&
\epsilon_{\perp}^{\rho\nu} &\equiv
 \epsilon^{\sigma\mu\rho\nu} {e_z}_\sigma {e_t}_\mu,
\nonumber\\
e_x^\mu &\equiv \frac{l_{\perp}^\mu}{\sqrt{- {l_{\perp}}^2}},&
e_y^\rho &\equiv \epsilon_{\perp}^{\rho\nu} {e_x}_\nu, &
-1 &= \epsilon_{\mu\nu\rho\sigma}e_t^\mu e_x^\nu e_y^\rho e_z^\sigma .
\end{align}
\vspace{-.5cm}
\begin{flushright}
\emph{Cartesian basis for electron-positron annihilation}
\end{flushright}
}

\section{Cross sections\label{theoryMaster}}

In this section the cross section formula for semi-inclusive DIS
will be derived and
results for Drell-Yan and electron-positron annihilation will be stated.
First, some conventions will be given.

The helicity of a parton with momentum $p$ and spin $s$ is defined here to be
\begin{equation}
\lambda \equiv
\frac{\threeVec{s}\cdot \threeVec{p}}{|\threeVec{s}\cdot \threeVec{p}|}.
\end{equation}
Dirac spinors and particle states are normalized such that
\begin{align}
\bar{u}(k,\lambda)\ u(k,\lambda') &= 2 m\ \delta_{\lambda\lambda'},
\\
\langle \threeVec{P}, \lambda\ | \ \threeVec{P}', \lambda' \rangle &=
2 E_\threeVec{P}\ (2\pi)^3 \ \delta^3(\threeVec{P}' - \threeVec{P})\
\delta_{\lambda\lambda'}.
\end{align}

The standard cross section
for semi-inclusive DIS is
(see for example Ref.~\cite{Peskin:1995ev})
\begin{multline}
\mathrm{d} \sigma = \frac{1}{F}
\phaseFactor{P_h}\ \phaseFactor{l'} \\
\times
\sum_X \int \phaseFactor{P_X}
| \mathcal{M} |^2 \ (2\pi)^4 \delta^4 \left( l+P- P_X - P_h - l'
\right).\label{cross}
\end{multline}
As we can see from this equation, the cross section is built up
out of: several phase-space factors, a sum over all
possible final states, an invariant amplitude, a delta-function which expresses
momentum conservation, and a flux factor $F$ which is given by
\begin{equation}
F \equiv 4\ E_{\threeVec{l}} E_\threeVec{P}\ |v_l - v_P|,
\end{equation}
where $v_l$ and $v_P$ are the velocities. The phase-space factors together with
the delta-function are Lorentz invariant. Since the invariant
amplitude is also Lorentz invariant the transformation properties of the
cross section are set by the flux. In this thesis only
\emph{head-on} collisions will be considered,
meaning that the motion of the initial particles is
aligned. In that case the flux takes the form
\begin{equation}
F = 2s \left(1+ \MQQ \right).
\end{equation}

For nucleons, consisting of strongly interacting quarks and gluons,
the interaction between
the electrons and the hadrons is
at lowest order in $\alpha\equiv e^2/(4\pi)$
mediated through the exchange of a virtual
photon.
Owing to the presence of charged quarks,
the virtual photon feels the electromagnetic
current between the incoming and outgoing hadrons.
The process is illustrated in Fig.~\ref{currentSidis} and leads for
the invariant amplitude to
\begin{equation}
i \mathcal{M} = (-ie)\ \overline{u}(l',\lambda') \gamma_\rho
u(l,\lambda)\
\frac{-i}{q^2}\ \
%\times
{}_\text{out}\langle P_h, P_X |
(-ie) J^\rho (0) | P,S \rangle_\text{in,c} + \mathcal{O}(e^3),
\label{amplitudeSIDIS}
\end{equation}

\begin{floatingfigure}{3.3cm}
\begin{center}
\includegraphics[width=3.3cm]{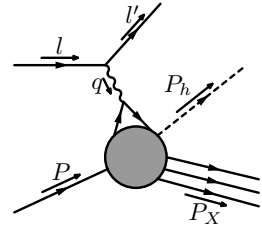}
\end{center}
\caption{The virtual photon coupling to the quark current in semi-inclusive
DIS.\label{currentSidis}\vspace{-1cm}}
\end{floatingfigure}

\noindent
where the subscript c indicates that only connected matrix elements should
be considered.
The blob in Fig.~\ref{currentSidis}
expresses that all kinds of interactions are
present.
Mathematically this means that
the currents in Eq.~\ref{amplitudeSIDIS} are in the Heisenberg
picture. A derivation of this equation is often omitted in textbooks but gives
considerable insight into the approximations made. Therefore, a
schematic derivation for the interested reader is provided in appendix
\ref{eersteAfleiding}.

\indent

The square
of the amplitude, needed for the cross section, can be written
as a contraction between the leptonic tensor and a hadronic part
in leading order of $\alpha$, giving (neglecting lepton masses)
\begin{align}
\left| \mathcal{M} \right|^2 &= \frac{e^4}{Q^4}\ L^{(lH)}_{\mu\nu}
H_{(lH)}^{\mu\nu}\left( 1 + \mathcal{O}(e^2) \right),\\
L^{(lH)}_{\mu\nu} &= \delta_{\lambda \lambda'} \left(
2 l_\mu l'_\nu {+} 2 l_\nu l'_\mu {-} Q^2 g_{\mu\nu} 
{+} 2 i \lambda\ \epsilon_{\mu\nu\rho\sigma} q^\rho l^\sigma \right),\\
H_{(lH)}^{\mu\nu} &=
{}_\text{in}\langle P,S | J^\mu (0) | P_X; P_h,S_h \rangle_\text{out,c}
\nonumber\\
& \eqnIndent \eqnIndent \eqnIndent \eqnIndent\times
{}_\text{out}\langle P_X; P_h,S_h | J^\nu (0) | P,S \rangle_\text{in,c},
\end{align}
where the $\lambda$ is the helicity of the  incoming electron and
$\lambda'$ is the helicity of the outgoing electron.
Defining now the hadronic tensor $W$ to be\newline
\begin{equation}
2M\ W_{(lH)}^{\mu\nu}
 = \frac{1}{(2\pi)^4} \sum_X \int \phaseFactor{P_X}
(2\pi)^4 \delta^4(l' + P_h + P_X - P - l)\ H_{(lH)}^{\mu\nu},
\end{equation}
enables us to write the cross section as
\myBox{
\begin{equation}
E_{\threeVec{P}_h} E_{ \threeVec{l}'}\
\frac{  \rmd{6}{\sigma} }
{\rmd{3}{\threeVec{l}'} \rmd{3}{\threeVec{P}_h}} =
\frac{M}{s}\ \frac{ \alpha^2 }{Q^4}\ L^{(lH)}_{\mu\nu}\ W_{(lH)}^{\mu\nu}
\left(1+ \mathcal{O}(\alpha)\right).
\label{crossSIDIS}
\end{equation}
\begin{flushright}
\emph{cross section for semi-inclusive DIS}
\end{flushright}
}
The interesting information on the distribution and fragmentation of quarks
is captured in the hadronic tensor $W^{\mu\nu}_{(lH)}$.

The inclusive cross section can be obtained by summing over all
observed final-state hadrons and
integrating over their phase space. This leads for the hadronic tensor to
\begin{multline}
2M\ W_{(DIS)}^{\mu\nu}
 = \frac{1}{2\pi} \sum_X \int \phaseFactor{P_X}
(2\pi)^4 \delta^4(l' + P_X - P - l)\\
\times {}_\text{in}\langle P,S| J^\mu(0) | P_X \rangle_\text{out,c}\
{}_\text{out}\langle P_X | J^\nu (0) | P,S \rangle_\text{in,c},
\label{W_dis}
\end{multline}
and gives for the cross section
\myBox{
\begin{equation}
E_{ \threeVec{l}'}\
\frac{   \rmd{3}{\sigma} }
{\rmd{3}{\threeVec{l}'} } =
\frac{2M}{s} \frac{ \alpha^2 }{Q^4} L^{(lH)}_{\mu\nu} W_{(DIS)}^{\mu\nu}
\left(1+\mathcal{O}(\alpha) \right).
\end{equation}
\begin{flushright}
\emph{cross section for inclusive DIS}
\end{flushright}
}

The cross sections for Drell-Yan and electron-positron annihilation can be
derived similarly.
One obtains for Drell-Yan in terms of
\begin{align}
L^{(DY)}_{\mu\nu} &= \delta_{\lambda \lambda'} \left(
2 l_\mu l'_\nu + 2 l_\nu l'_\mu - Q^2 g_{\mu\nu} + 2 i \lambda \epsilon_{\mu\nu
\rho\sigma} q^\rho l^\sigma \right),\\
H^{\mu\nu}_{(DY)} &=
{}_\text{in}\langle P_A,S_A; P_B,S_B | J^\mu (0) | P_X \rangle_\text{out,c}\
{}_\text{out}\langle
P_X | J^\nu (0) | P_A,S_A; P_B,S_B \rangle_{\text{in,c}}, \\
W_{(DY)}^{\mu\nu}
 & = \frac{1}{(2\pi)^4} \sum_X \int \phaseFactor{P_X}
(2\pi)^4 \delta^4(P_1 + P_2 - l-l'-P_X)
H^{\mu\nu}_{(DY)},
\end{align}
the following cross section (a factor $2$ was included for summing
over lepton polarizations)
\myBox{
\begin{equation}
E_\threeVec{l} E_{\threeVec{l}'}\
\frac{  \rmd{6} \sigma }
     { \rmd{3}{\threeVec{l}} \rmd{3}{\threeVec{l}'} }
=
\frac{ \alpha^2 }{s Q^4} L^{(DY)}_{\mu\nu} W^{\mu\nu}_{(DY)}
\left(1+ \mathcal{O}(\alpha) \right).
\end{equation}
\begin{flushright}
\emph{cross section for Drell-Yan}
\end{flushright}
}
For electron-positron annihilation
the cross section for producing two almost back-to-back hadrons
reads (a factor $\frac{1}{2}$ was included for averaging
over lepton polarizations)
\myBox{
\begin{equation}
E_{\threeVec{P}_{h1}} E_{\threeVec{P}_{h2}}\
\frac{  \rmd{6}{\sigma}}
     { \rmd{3}{\threeVec{P}_{h1}} \rmd{3}{\threeVec{P}_{h2}}}
     = \frac{\alpha^2}{4 Q^6} L^{(e^+e^-)}_{\mu\nu} W_{(e^+e^-)}^{\mu\nu}
     \left(1+ \mathcal{O}(\alpha) \right),
\end{equation}
\begin{flushright}
\emph{cross section for electron-positron annihilation}
\end{flushright}
}
where (with $| \Omega \rangle $ representing the physical vacuum)
\begin{align}
L^{(e^+e^-)}_{\mu\nu} &= \delta_{\lambda\lambda'}                     \
\left(
2 l_\mu l'_\nu + 2 l_\nu l'_\mu - Q^2 g_{\mu\nu}
+ 2 i \lambda\ \epsilon_{\mu\nu\rho\sigma} l^\rho {l'}^\sigma
\right),\\
H_{(e^+e^-)}^{\mu\nu} &=
\langle \Omega | J^\mu (0) | P_X;P_{h1},S_{h1}; P_{h2},S_{h2}
\rangle_\text{out,c}
\nonumber
\\
& \phantom{=} \eqnIndent \eqnIndent \eqnIndent \eqnIndent
\eqnIndent \eqnIndent \eqnIndent \eqnIndent\times
{}_\text{out}\langle P_X;P_{h1},S_{h1}; P_{h2},S_{h2} | J^\nu (0) | \Omega \rangle_\text{c},
\\
W_{(e^+e^-)}^{\mu\nu} &= \frac{1}{(2\pi)^4} \int \phaseFactor{P_X}\ (2\pi)^4
\delta^4 (q-P_X-P_{h1}-P_{h2})\                                         \
H_{(e^+e^-)}^{\mu\nu}.
\end{align}

\section{Operator product expansion}

There are two methods to gain more information from the hadronic tensor. In 1968
the
first method was proposed by Wilson
in Ref.~\cite{wilson}
 and is called the \emph{operator product
expansion}. The second method, \emph{the diagrammatic expansion},
was proposed by Politzer in Ref.~\cite{Politzer:1980me} in 1980
and will be introduced in the next section.

The operator product expansion is useful for inclusive measurements.
As an illustration let us consider the inclusive DIS process.
Having no hadrons observed in
the final state, the sum over all final QCD-states is complete. Together with
the fact that the proton is a stable particle
one can rewrite the hadronic
tensor in Eq.~\ref{W_dis}
into\footnote{A necessary condition to have a stable particle is that
the sum over energies of its possible decay products
is larger than the energy of the
considered particle. Together with the fact that the zeroth momentum
component of the virtual
photon in DIS is positive, it can be shown that
the hadronic tensor vanishes if $q$ is replaced by $-q$. This enables
one to obtain the commutator.}
\begin{equation}
2 M W^{\mu\nu}_{(DIS)} = \frac{1}{2\pi}
\int \rmd{4}{x}  e^{iqx}
\langle P,S | \left[ J^\mu (x), J^\nu (0) \right]
 | P,S \rangle_\text{c} .
\end{equation}
According to the Einstein causality principle the commutator of two
physical operators should vanish for space-like separations. In our case
this means that only the area $x^2 > 0$ gives a contribution. Under the
assumption
that the hadronic tensor is well behaving for $x^2 > 0$ one can show that
the main contribution comes from
$x^2 \approx 0$ in the Bjorken limit (fixed $x_B$ and $Q\rightarrow \infty$),
implying light-cone dominance.

The idea of Wilson,
which was later proven in perturbation theory
in 1970 by Zimmerman\footnote{For reference, please consider chapter 20 of
Ref.~\cite{Weinberg:1996kr}.},
is that for small
separations one can make a Taylor expansion for operators. This expansion
is called the operator product expansion and reads
\begin{equation}
\mathcal{O}_A (x) \mathcal{O}_B (0) \underset{x \approx 0}{\approx}
\sum_n C_{AB}^n \mathcal{O}_n (0) .
\end{equation}
This above relation
also holds for commutators and by using dispersion relations the 
short distance expansion
can be applied for inclusive DIS.

In Drell-Yan
%the situation
%is similar. T
the sum over final QCD-states is complete which enables one to
obtain a product of current operators. One can
gain insight in the various structure functions in which the cross
section can be decomposed but since these structure functions
will
depend on the two hadrons they
are inconvenient for the study of the structure
of a single nucleon. In addition, the process is not light-cone dominated
which complicates the application of the operator product expansion.

Another situation is encountered in electron-positron annihilation.
When summing
over all final states the commutator can be obtained, but
if we are interested in how quarks decay into hadrons we
would not be able to sum over such a complete set. The use of the operator
product expansion is therefore limited here as well.

Summarizing, the operator product expansion is a useful approach
that finds applications
in inclusive DIS and
electron-positron annihilation.
At the same time,
it is also limited to those processes. Parton distribution functions
which are measured
in DIS cannot be compared to more complex or less inclusive 
processes and quark decay cannot be studied within this approach.
We will
proceed by applying an extended form of Feynman's parton
model. In the original parton
 model it is assumed that the underlying process is a partonic
scattering
process multiplied by distribution and fragmentation functions. These functions
describe the probability of finding on shell constituents in a hadron or
how a quark decays into a particular hadron.
In the next section the diagrammatic approach as
an extension of this model will be discussed.

%By assuming this model it is conjectured that
%one can go beyond the formal operator product
%expansion by applying a diagrammatic expansion for some kinematical
%regimes. This approach will be used
%in the rest of this thesis and will be introduced in the next section.

\section{The diagrammatic expansion and the parton model\label{diagram}}

\subsubsection{Background of the diagrammatic expansion}

The operator product expansion is of limited use for the
study of the nucleon's structure.
In the case of Drell-Yan we were not able to study the
structure of a single nucleon although one could imagine that the chance of
producing a virtual photon should just be proportional to the chances of finding
a quark in the nucleon and an antiquark in the other nucleon. This idea
of expressing the cross section in terms of probability functions which
are then convoluted with some parton scattering cross section was suggested
by Feynman and is nowadays called the \emph{parton model}.

The parton model had already lots of successes. It gave for instance
an intuitive explanation for the
approximate Bjorken scaling which was observed at SLAC.
In an
QCD-improved version of the parton model one could even predict the
scaling violation with a set of equations
called \emph{evolution equations}
(for example see Altarelli, Parisi~\cite{Altarelli:1977zs}).
Another
success of the parton model is the observation of jets. A jet is a
set of particles of which their momentum differences can be characterized
with a hadronic size. By
assuming that these jets are produced by partons which ``decay'' into
these jets one is able to predict the number of jets appearing in scattering
processes. However, the appearance of
jets also creates a problem with color. Since partons carry color charges,
they should somehow loose this color when decaying into a set of colorless
hadrons. It appears that this issue
does not influence the scattering cross sections
at large momentum transfers.

\begin{floatingfigure}{3.3cm}
\begin{center}
\includegraphics[width=3.3cm]{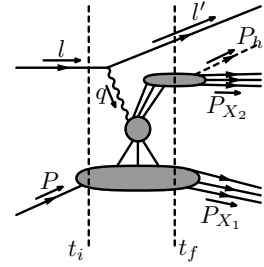}
\end{center}
\caption{The diagrammatic approach illustrated for two
jet-production and an observed hadron in semi-inclusive DIS.\vspace{.3cm}
\label{diagramSIDIS}}
\end{floatingfigure}

The success of the parton model relies on the asymptotic freedom property of
QCD~\cite{Gross:1973id,Politzer:1973fx}. This property
allows one to apply perturbation theory for elementary particle
scattering in strong interaction physics in the presence of
large scales.
Strictly speaking, it remains, however, 
to be proven whether one can apply perturbation
theory in hadronic scattering processes as well.
The present idea is that suitable hadronic scattering processes
can be described in terms of short-distance physics,
the hard scattering part, and the long-distance nonperturbative
physics which is captured in probability and decay
functions. Since the latter are nonperturbative
in nature,
the approach should at least be self-consistent to all orders in perturbation
theory. This description in separated terms is called \emph{factorization}.

The diagrammatic approach is an extended form of Feynman's parton model. 
Originating from field theory, the
approach includes the possibility that several
parton-fields from a hadron can participate in a scattering process
(involving multi-parton correlators),
whereas the parton model only considers the possibility of hitting 
a single parton.
The approach agrees with the operator product expansion
when applicable.
In 1980 it
was suggested by Politzer in Ref.~\cite{Politzer:1980me} in order to
describe the subleading orders in $M/Q$, involving ``higher twist'' operators
in matrix elements (to be discussed in chapter~\ref{chapter3}).
Subsequently, it was applied by Ellis, Furmanski,
and Petronzio in Ref.~\cite{Ellis:1982wd,Ellis:1982cd}.
Although similar assumptions as in the parton model
are made, the starting point is
more general because it allows for more possible interactions.
As we will see later, some of these interactions will provide an explanation for
single spin asymmetries 
(see the work of Qiu and Sterman~\cite{Qiu:1990xx,Qiu:1990xy,Qiu:1998ia}).
The diagrammatic approach was further developed and used in several 
applications, some of which to be discussed later in this thesis.
%In Ref.~\cite{Ellis:1982wd,Ellis:1982cd} and
%also in the works of Qiu and Sterman~\cite{Qiu:1990xx,Qiu:1990xy,Qiu:1998ia}
%the small transverse momenta in the hard
%part are usually expanded with respect to the big momenta. In the processes
%presented here such an expansion can be postponed.

\sloppy
The assumption in the diagrammatic approach is that interactions
between the incoming hadrons and outgoing jets can be described in perturbation
theory
and hence can be diagrammatically expanded
with in the hard part a sufficiently small coupling constant.
With respect to asymptotic freedom this requires the
incoming hadrons and outgoing jets to be well separated
in momentum space. We will therefore
impose
that the products of external momenta are large 
($P_i \cdot P_j \gg M^2$ for $i \neq j$) and assume that  
interactions between outgoing
jets can be neglected. Non-perturbative physics
inside the jets and hadrons is maintained.
Together with the
assumption of adiabatically switching on and off the interactions, the
applied assumptions are sufficient to describe general QCD-scattering processes.

\fussy
As an example, in the case of semi-inclusive DIS
a hadron is detected
which is well separated from the incoming nucleon in momentum
space. Hence, there must have
been a partonic scattering. The virtual photon has struck a quark which after
interacting with the photon decays in a separate jet
including the observed hadron.
This process is
illustrated in Fig.~\ref{diagramSIDIS}. There is also
the chance that more jets are being produced which may not be observed.
However,
that possibility is expected to be subleading in $\alpha_s$.
In section \ref{theoryMaster} an expression for the invariant amplitude
was obtained in
Eq.~\ref{amplitudeSIDIS}. Making the assumption of
adiabatically switching on and off the
interactions more explicit (see also Fig.~\ref{diagramSIDIS}),
this result can be rewritten into (see also appendix~\ref{eersteAfleiding})
\begin{equation}
\begin{split}
&
\langle e',P_X,P_h | i T | P, e \rangle
\\
&
\
= \bar{u}(k_e',\lambda') (-ie) \gamma_\mu u(k_e,\lambda)\ \frac{-i}{q^2}
\\
&
\ \ \phantom{=}
\times
\lim_{\substack{t_i \rightarrow -\infty \\ t_f \rightarrow \infty}}
\int\!\! \rmd{4}{z}\! e^{-iqz}
{}_\text{out}\langle P_h,P_X | U(t_0,t_f)
\biggl[ U(t_f,t_i) (-ie) J_I^\mu (z) \biggr]
 U(t_i,t_0) | P,S \rangle_\text{in,c},
\end{split}
\end{equation}
where the subscript $I$ \nolinebreak \mbox{denotes}
\nolinebreak the \nolinebreak \mbox{interaction} \nolinebreak \mbox{picture}
 \nolinebreak and \nolinebreak $t_0$ \nolinebreak \mbox{defines}
the \nolinebreak \mbox{quantization}  plane.

%\pagebreak

%\phantom{boktor}

\begin{floatingfigure}{3.3cm}
\begin{center}
\includegraphics[width=3.3cm]{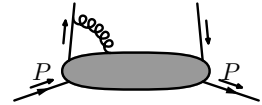}
\end{center}
%\vspace{-1cm}
\caption{An interactions
%between the elementary scattering
%cross section
%and an incoming jet
which can be absorbed in the jet definition
\label{reducible}}
\end{floatingfigure}

%\vspace{-.6cm}
\sloppy
The assumption of the diagrammatic expansion can now be used to expand
the bracketed term in the above equation.
In general a complete expansion will connect an
arbitrary number of lines to the several jets. Such a complete expansion,
however, is
not necessary. The interactions between lines which are connected to one jet
as illustrated in Fig.~\ref{reducible}
can be absorbed in the matrix elements. Therefore, only those
parts will be expanded which cannot be absorbed in one of the
participating jets. In general this leads to
matrix elements in which the interaction picture fields become
Heisenberg fields.

\subsubsection{Applying the diagrammatic expansion}

\begin{figure}
\begin{center}
\vspace{.3cm}
\begin{tabular}{cp{.5cm}cp{.cm}c}
\includegraphics[width=3cm]{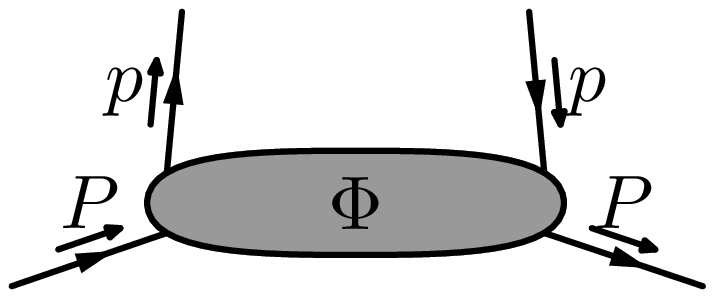}& &
\includegraphics[width=3cm]{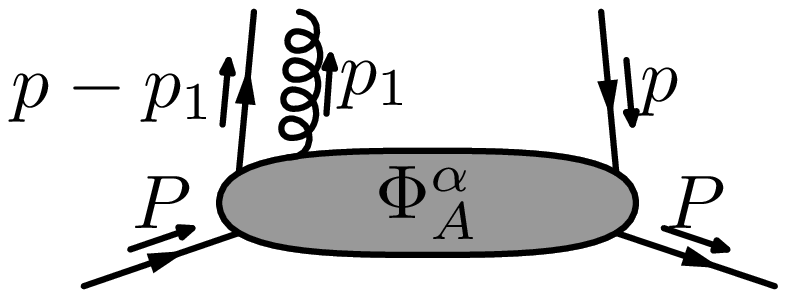}& &
\includegraphics[width=3cm]{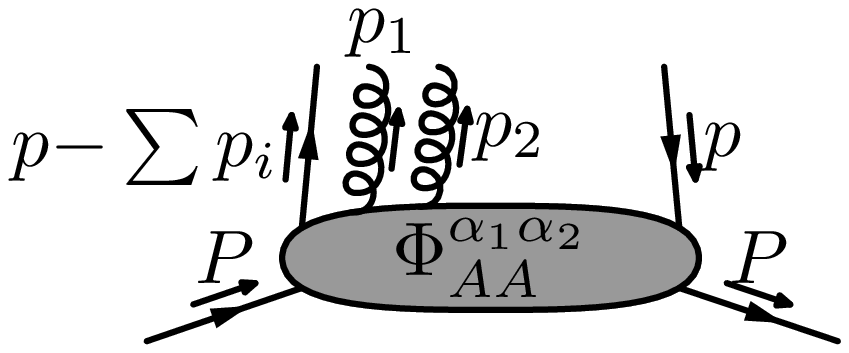}
\end{tabular}
\end{center}
\caption{Some of the distribution correlators containing two quark-fields
\label{figDistr}}
\end{figure}

\begin{figure}
\begin{center}
\vspace{.3cm}
\begin{tabular}{cp{.4cm}cp{.4cm}c}
\includegraphics[width=3cm]{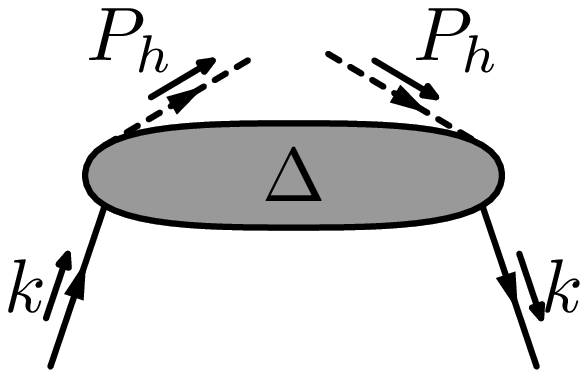}& &
\includegraphics[width=3cm]{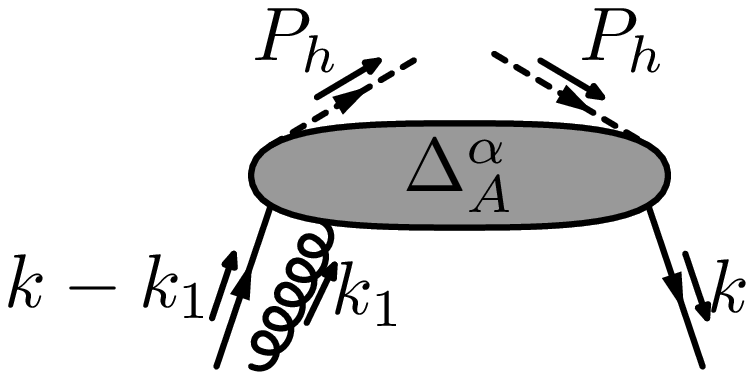}& &
\includegraphics[width=3cm]{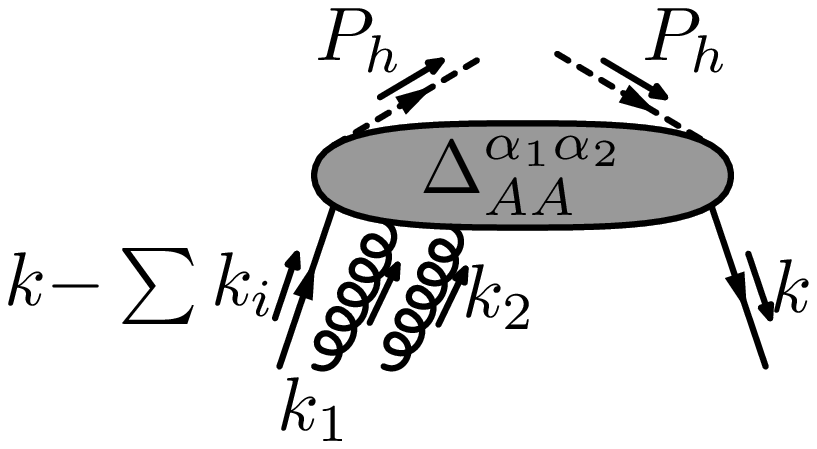}
\end{tabular}
\end{center}
\caption{Some of the fragmentation correlators containing two quark-fields
\label{figFrag2}}
\end{figure}

\fussy
Using the diagrammatic approach the cross section for a general scattering process
can be calculated in a number of steps.  
An outline of its derivation for two examples 
is given in appendix~\ref{theoryAppendix2}.
\begin{enumerate}
\item Write down all squared
Feynman diagrams with an arbitrary amount of external
parton-lines
and connect them in all possible ways to the external jets and particles.
Each external parton-line carries
an independent momentum variable (for example $p_i$ or $k_i$).
Any interaction
which can be absorbed in one of the participating jets should not be included.
\item Replace the external spinors or polarization vectors of step
1 by
an appropriate correlator as will be defined below.
For instance, $u\bar{u} \rightarrow \Phi$, $u \epsilon^\mu \bar{u}
\rightarrow \Phi_A^\alpha$, etc.
\item Integrate over all parton momenta and
impose total
momentum conservation by adding $(2\pi)^4 \delta^4(\text{incoming }-
\text{ outgoing parton momenta}$).
\item If there is any QED part in the diagram, calculate that part with
ordinary Feynman rules.
\item Divide by the flux factor and multiply by the phase-space factors of the
produced particles,
$\rmd{3}{\threeVec{k}_i} / ( (2\pi)^3 2 E_{\threeVec{k}_i} )$.
\end{enumerate}
Some of the correlators which appear in cross sections
are (see also Fig.~\ref{figDistr} and Fig.~\ref{figFrag2})\nopagebreak
\myBox{
\begin{align}
\Phi_{ij,ab}(p) &{=} \int \! \frac{ \rmd{4}{\xi} }{(2\pi)^4}\ e^{ip\xi}
\langle P,S | \bar{\psi}_{j,b}(0) \
\psi_{i,a} (\xi) | P,S \rangle_\text{c},
\label{defPhi}\\
{\Phi_{A_l}^\alpha}_{ij,ab}(p,p_1) &{=}
 \int\! \frac{\rmd{4}{\xi} \rmd{4}{\eta}}{(2\pi)^8}\
e^{ip\xi} e^{ip_1 (\eta-\xi)}
\langle P,S | \bar{\psi}_{j,b}(0)\  A_l^\alpha(\eta)\
\psi_{i,a} (\xi) | P,S \rangle_\text{c},
\label{defPhiA}\\
&\nonumber
\\
{\Phi_{A_{l_1} \dots A_{l_n}}^{\alpha_1 \ldots \alpha_n}}_{ij,ab}&(p,p_1,\ldots,p_n)
\nonumber\\
& =
\int
\frac{\rmd{4}{\xi} \rmd{4}{\eta_1}\ldots \rmd{4}{\eta_n}}{(2\pi)^{4(n+1})}
\ e^{ip\xi} e^{ip_1 (\eta_1-\xi)} \ldots e^{ip_n(\eta_n-\xi)}
\nonumber\\
&\eqnIndent \eqnIndent\phantom{=} \times
\langle P,S | \bar{\psi}_{j,b}(0)\  A_{l_1}^{\alpha_1}(\eta_1)
\ldots A_{l_n}^{\alpha_n}(\eta_n)\
\psi_{i,a} (\xi) | P,S \rangle_\text{c},
\end{align}
\begin{flushright}
\emph{some parton distribution correlators containing two quark-fields
\label{figFrag}}
\end{flushright}
}
\myBox{
\begin{align}
\Delta_{ij,ab}(k) &{=}
\sum_{\substack{X}}\! \int\! \phaseFactor{P_X}\!
\int\! \frac{ \rmd{4}{\xi} }{(2\pi)^4} e^{ik\xi}
\langle \Omega | \psi_{i,a} (\xi) | P_h,S_h;P_X \rangle_{\text{out,c}}
\nonumber\\
&\eqnIndent \phantom{=}\times {}_{\text{out}}
\langle P_h,S_h;P_X | \bar{\psi}_{j,b}(0) | \Omega \rangle_\text{c},
\label{defDelta}
\\
{\Delta_{A_l}^\alpha}_{ij,ab}(k,k_1)
&{=}
\sum_{\substack{X}}\! \int\! \phaseFactor{P_X}\!
\int\! \frac{ \rmd{4}{\xi} \rmd{4}{\eta} }{(2\pi)^8}
e^{ik\xi} e^{-i k_1 \eta}
\langle \Omega | \psi_{i,a} (\xi) | P_h,S_h;P_X \rangle_{\text{out,c}}
\nonumber\\
&\eqnIndent \phantom{=}\times
{}_{\text{out}}
\langle P_h,S_h;P_X | \bar{\psi}_{j,b}(0) A_l^\alpha(\eta)| \Omega \rangle_\text{c},
\label{defDeltaA}\\
& \nonumber\\
{\Delta_{A_{l_1} \ldots A_{l_n}}^{\alpha_1 \ldots \alpha_n}}_{ij,ab}&(k,k_1,\ldots,k_n)
\nonumber\\
&{=}
\sum_{\substack{X}}\! \int\! \phaseFactor{P_X}\!
\int\! \frac{ \rmd{4}{\xi}\! \rmd{4}{\eta_1}\! \ldots\! \rmd{4}{\eta_n} }
{(2\pi)^{4(n+1)}} e^{ik\xi}e^{-i k_1 \eta_1}\! \ldots\! e^{-i k_n \eta_n}
\nonumber\\
&\eqnIndent \phantom{=} \times
\langle \Omega | \psi_{i,a} (\xi) | P_h,S_h;P_X \rangle_{\text{out,c}}
\nonumber\\
&\eqnIndent \phantom{=} \times
{}_{\text{out}}\langle P_h,S_h;P_X |
\bar{\psi}_{j,b}(0)
A_{l_1}^{\alpha_1}(\eta_1) \ldots A_{l_n}^{\alpha_n}(\eta_n)|
\Omega \rangle_\text{c}.
\end{align}
\begin{flushright}
\emph{some parton fragmentation correlators containing two quark-fields}
\end{flushright}
}
All fields, so $\psi$ and $A$,
carry a color index ($a,b \in 1,2,3,\ l_i \in 1,\ldots, 8$)
over which it is summed
in the cross section (no averaging for initial states),
and the indices $\{i,j\}$ denote Dirac indices. 
%Since only the
%connected parts of the matrix elements are considered, there is no
%time-ordering present
Time-ordering is not present because the matrix elements are connected
and the correlators will be integrated over the small parton's momentum 
component, putting the fields on the light-front
(see Jaffe~\cite{Jaffe:1983hp,Jaffe:1985je},
Diehl, Gousset~\cite{Diehl:1998sm}).


\begin{figure}
\begin{center}
\begin{tabular}{cp{.5cm}c}
\includegraphics[width=5cm]{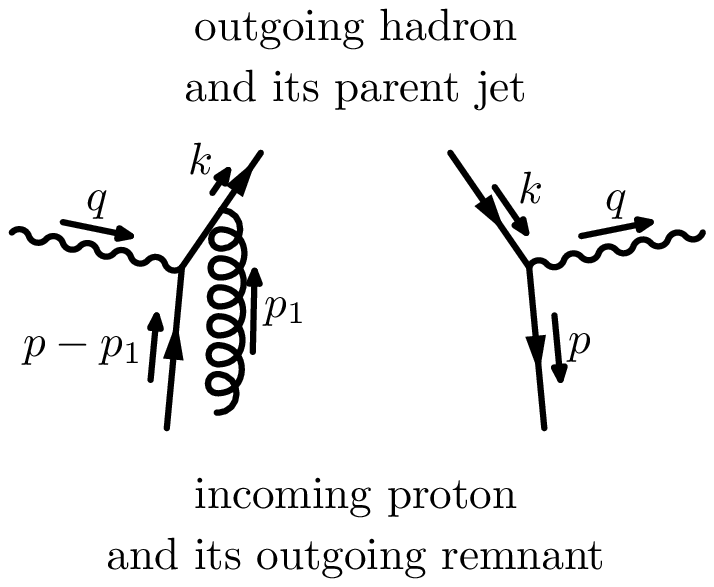}& &
\includegraphics[width=5cm]{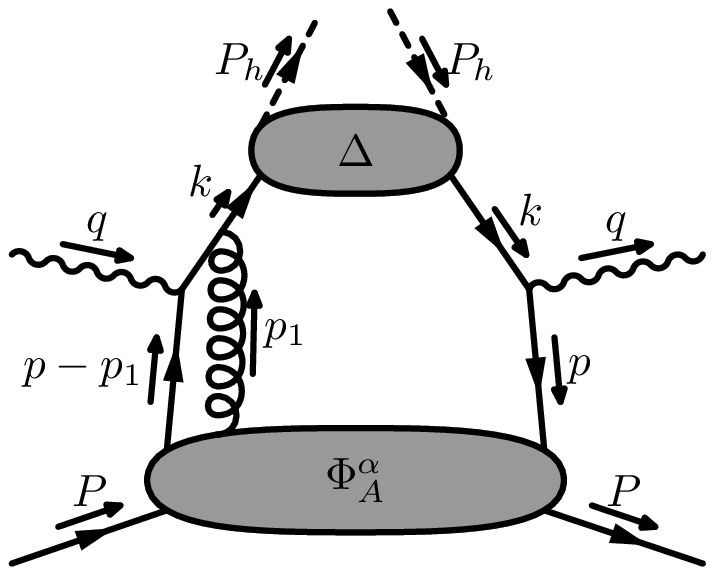}  \\
&&\\
(a) & & (b)
\end{tabular}
\end{center}
\caption{Application of the diagrammatic approach in semi-inclusive
DIS. Figure~(a) represents
the two jets and the interactions
between them, and in figure~(b) the correlators describing the jets are included.
\label{examplediagram}}
\end{figure}

To illustrate these rules we return to our example in which
there are only two jets produced (see the amplitude diagram in
Fig.~\ref{diagramSIDIS}).
In step 1 we have to
write down for the cross section
\emph{all} squared Feynman diagrams which contribute to this process.
Let us consider one of the contributing diagrams as given
in
Fig.~\ref{examplediagram}a.
Expressing the colors very explicitly by
giving the spinors and polarization vector a
color charge (superscript) and a color index (subscript), one finds for
the diagram
%which is mathematically expressed as
%(the spinors and polarization vector have been given an explicit
%color charge (superscript) and color index (subscript),
(note also that $\bar{u}^{q_1}_a \sim \delta_{q_1a}$)
\begin{multline}
\sum_{\begin{subarray}{c} q_1,q_2,\\q_3,q_4,g_1 \end{subarray}}\!\!\!\!
\bar{u}^{(q_1)}_a(k) i \gamma_\alpha t^l_{ab} i
\frac{\slashi{k} - \slashiv{p}_1 + m}{(k{-}p_1)^2 {-} m^2 {+} i \epsilon}
\gamma^\nu u^{(q_2)}_b(p-p_1) g \epsilon_l^{(g_1)\alpha}(p_1) \times
\bigg[
\bar{u}^{(q_3)}_c(k) \gamma^\mu \delta_{cd} u^{(q_4)}_d(p) \bigg]^\dagger
\\
{=}
\tr^{D,C}\! \bigg(
\big[ u_c(k) \bar{u}_a(k) \big]
i \gamma_\alpha t^l_{ab} i
\frac{\slashi{k} - \slashiv{p}_1 + m}{(k{-}p_1)^2 {-} m^2 {+} i \epsilon}
\gamma^\nu
\big[ u_b(p{-}p_1)g\epsilon_l^\alpha(p_1) \bar{u}_c(p) \big]
\gamma^\mu
\bigg),
\end{multline}
where $\tr^{D,C}$ stands for a trace in color and Dirac space, and
$u_c(k) \equiv \sum_q u^{(q)}_c$ and similarly for $\epsilon_l$.
According to step 2 we replace the bracketed terms by $\Delta$ and
$\Phi_{A_l}^\alpha$ and in step 3 we integrate
over $p$, $k$, and $p_1$, and multiply
this by $(2\pi)^4 \delta(p+q-k)$. In step 4 we multiply the result
of step 3 with the
leptonic tensor and the photon propagator. Applying step 5 this diagram
contributes to the cross section as
\begin{equation}
\begin{split}
\rmd{1}{\sigma} &{=} \frac{1}{2s} \phaseFactor{P_h}\phaseFactor{l'}
L^{(lH)}_{\mu\nu} \frac{e^4}{Q^4}
\int \rmd{4}{p} \rmd{4}{k} \rmd{4}{p_1} (2\pi)^4 \delta^4(p+q-k)\\
&\phantom{=} \times
\tr^{D,C}\left(
%\frac{1}{3}\tr^C
\left[ \Delta(k)  \right]
% \tr^C \left(
i \gamma_\alpha t_l i
\frac{\slashi{k} - \slashiv{p}_1 + m}{(k{-}p_1)^2 {-} m^2 {+} i \epsilon}
\gamma^\nu
\big[ \Phi_{A_l}^\alpha(p,p_1) \big]
\gamma^\mu \right)
%\right)
%\\
%&\phantom{=}
+ \ldots,
\end{split}
\label{theory43}
\end{equation}
%where it was used that $\Delta_{ab}(k) = \delta_{ab} \tr^C \Delta(k)/3$ and
where the dots denote contributions from other diagrams.\nopagebreak

The previous example illustrates that the diagrammatic approach
enables one to express any cross section in a set of correlators.
The result is in general an \emph{infinite sum} of
partonic scattering diagrams connected to \emph{all kinds} of correlators.
That result is fairly exact; it relies on the possibility
of applying perturbation theory between the external jets and the possibility
of defining the jets.
However, the cross section is not yet a product of probability
functions as it is in the parton
model. In order to obtain this product
similar assumptions as
in the parton model will be made here.

These assumptions are that the parton-lines connecting the soft blobs
are approximately
on their mass-shell and collinear
with their \emph{parent hadron}, the hadron
which is connected to the blob of the parton.
This assumption is less strict than
the assumptions of the
successful parton model in which partons were treated as in essence
free, collinear, and
on the mass-shell. Since the assumptions made here are similar to those made
in the
parton model, we will refer to them as the parton model assumptions.

To exploit these assumptions a set of light-like vectors
will be chosen such that
$P\sim Q\ n_+ + \mathcal{O}(M^2/Q)\ n_-$
and $P_h \sim Q\ n_- + \mathcal{O}(M^2/Q)\ n_+$ and
\begin{equation}
n_- \cdot n_+ = 1,\qquad \text{and }\quad \bar{n}_- \sim n_+,
\end{equation}
where the bar on a vector denotes reversal
of the spatial vector components.
This introduction of vectors, called a
\emph{Sudakov-decomposition}, has the advantage that it seems that our
target-hadron is now moving very fast in a particular direction without actually
having
boosted the target. The frame in which the proton is moving very fast is also
called the \emph{infinite momentum frame} and if its momentum is opposite to the
momentum of the photon one also refers to it as the
\emph{Breit frame}.
Using the parton model assumptions for the integration over $p\cdot n_-$ and
$k\cdot n_+$
and applying the relation $\Delta_{ab}(k) = \delta_{ab} \tr^C \Delta(k)/3$,
one obtains for Eq.~\ref{theory43} ($p^\pm \equiv p\cdot n_\mp$,
etc.)
\begin{equation}
\begin{split}
\rmd{1}{\sigma} &{=} \frac{1}{2s} \phaseFactor{P_h}\phaseFactor{l'}
L^{(lH)}_{\mu\nu} \frac{e^4}{Q^4}
\int \rmd{2}{p_T} \rmd{2}{k_T}
\rmd{2}{{p_1}_T} \rmd{1}{p_1^+} (2\pi)^4 \delta^2(p_T+q_T-k_T)
\\
& \eqnIndent \phantom{=}\times
\tr^D \Bigg[ \frac{1}{3}
\tr^C \Bigg(
\int \rmd{1}{k^+} \Delta(k) \Bigg) \tr^C \Bigg(
i \gamma_\alpha t_l i
\frac{\slashi{k} - \slashiv{p}_1 + m}{(k-p_1)^2 - m^2 + i \epsilon}
\Big|_{\substack{p_1^- = 0\\ k^+ = 0}}
\gamma^\nu
\\
& \eqnIndent \phantom{=}\times
\int \rmd{1}{p^-} \rmd{1}{p_1^-} \Phi_{A_l}^\alpha(p,p_1) \big)
\gamma^\nu \Bigg)
\Bigg] \Bigg|_{\substack{
p^+ = -q^+\\ k^- = q^-}} \left(1+
\MQQ \right) + \ldots,
\end{split}
\raisetag{26pt}
\end{equation}
where the subscript $T$ denotes transverse components with respect to
$n_-$ and $n_+$.

The applied Sudakov-decomposition can be used for general
scattering processes as long as the scalar product of the observed momenta
is large. For each observed hadron one can introduce a light-like vector
along which the hadron is moving. Since the parton-lines are approximately
collinear and on shell, one of the components of the parton momenta,
$p\cdot n$, must appear to be very small. In general one should be able
to neglect these components
in the hard scattering part such that one can integrate the considered
correlator over this variable.
In the next section we will see that correlators which are integrated
over the small momentum components are probabilities in leading
order in $M/Q$.

\section[Quark distribution functions for spin-$\frac{1}{2}$ hadrons]
{Quark distribution functions for spin-$\frac{\mathbf{1}}{\mathbf{2}}$
hadrons\label{sectionDistr}}

The various functions for spin-$\frac{1}{2}$ hadrons will be
introduced and their relevance will be pointed out. For spin-$1$ targets the
reader is referred to Bacchetta, Mulders~\cite{Bacchetta:2000jk}.
To define the parton distributions a set of light-like vectors is constructed
such that
\myBox{
\vspace{-.2cm}
\begin{align}
1 &= n_- \cdot n_+,&  n_- &\sim \bar{n}_+,
\nonumber\\
P &= \frac{M^2}{2P^+} n_- + P^+ n_+,&
\epsilon_{T}^{\mu\nu} &\equiv \epsilon^{\rho\sigma\mu\nu} {n_+}_\rho {n_-}_\sigma,
\nonumber\\
g_{T}^{\mu\nu} &= g^{\mu\nu} - n_+^\mu n_-^\nu - n_+^\nu n_-^\mu,&
A_{T}^\mu &= g_{T}^{\mu\nu} A_\nu,\ \text{for any }A.
\label{frameDistr}
\end{align}
\begin{flushright}
\emph{the basis in which parton distribution functions are defined}
\end{flushright}
}
For any vector $A$ we also define $A^{\pm} \equiv A \cdot  n_\mp$,
which means 
that $P^+$ is defined to be a Lorentz-invariant.
To describe the spin of the hadron one usually introduces
\begin{align}
S &= -S_L \frac{M}{2P^+} n_- + S_L \frac{P^+}{M} n_+ + S_T,&  \text{with }&
S_L^2 + \threeVec{S}_T^2 = 1,
\end{align}
which satisfies the necessary constraints: $P\cdot S = 0$, $S^0 = 0$
if $\threeVec{P}=0$, and $S^2 = -1$.

In the next subsections we will parametrize an expansion in $M/P^+$ of
quark-quark correlators, but $M/P^+$ does not have to be small.
However, in order to make use of the truncated expansion,
calculations in the next chapters will be performed such that
$P^+ \gg M$. Since the light-like vectors $n_-$ and $n_+$ are defined
up to a rescaling ($n_+ {\rightarrow} \alpha n_+$,
$n_- {\rightarrow} \alpha^{-1} n_-$),
this does not put any constraint on $P$ or the frame.
In fact, for a target at
rest one has for instance $n_+ = (M / 2P^+)\ (1,0,0,1)$ and
$n_- = (P^+/M)\ (1,0,0,-1)$ where $P^+ \gg M$ can still be chosen.

As discussed in the previous section,
one encounters in the
diagrammatic expansion
an infinite set of
correlators which can all be integrated over the small momentum components.
As we will see in the next chapter, this infinite set can be rewritten
into a single new correlator containing the gauge link (to be defined below).
In the discussed electromagnetic processes
basically two kinds of correlators appear in the final result.

\begin{figure}
\begin{center}
\begin{tabular}{cp{.5cm}c}
\includegraphics[width=5cm]{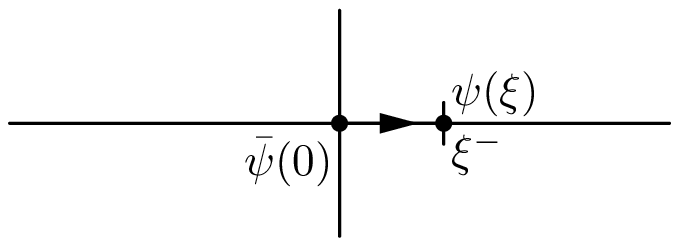} &  &
\includegraphics[width=5cm]{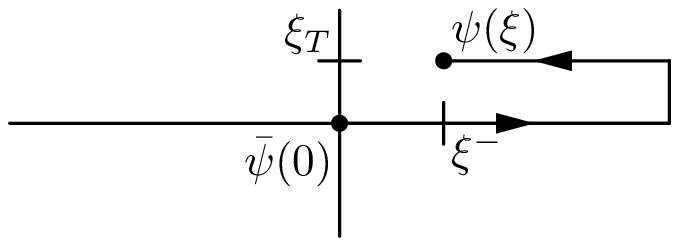} \\
a: $\mathcal{L}^{0_T\!,\ \xi^+} (0^-,\xi^-) $ &   &
b: $\mathcal{L}^{[+]} (0,\xi^-) $
\\
& \\
\includegraphics[width=5cm]{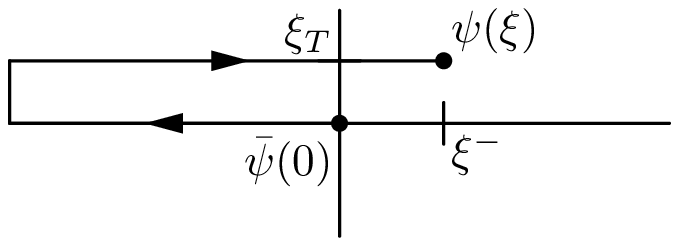} &       &
\includegraphics[width=5cm]{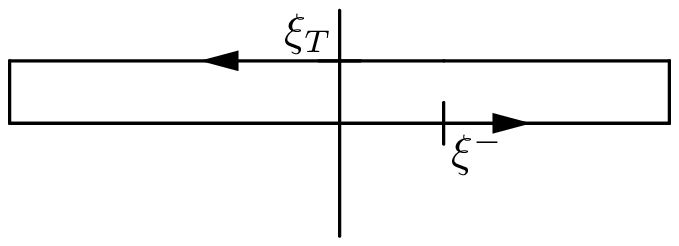}
\\
c: $\mathcal{L}^{[-]} (0,\xi^-) $ &      &
d: $\mathcal{L}^{[ \square ]} (0,\xi^-) $
\end{tabular}
\end{center}
\caption{The paths of the various gauge links which connect
the two quark-fields in the correlator.
\label{plaatjeLinks} }
\end{figure}

The first
kind appears in cross sections which are not sensitive to the transverse
momenta of the constituents and is the
so-called \emph{integrated} correlator. Including
a \emph{Wilson line operator} $\mathcal{L}$, it reads
\begin{equation}
\Phi_{ij}(x,P,S) = \int \frac{\rmd{1}{\xi^-}}{2\pi} e^{ixP^+\xi^-}
\langle P,S | \bar{\psi}_j (0) \mathcal{L}^{0_T\!,\ \xi^+}(0^-,\xi^-) \psi_i(\xi)
|P,S \rangle_\text{c} \big|_{\substack{\xi^+ = 0\\ \xi_T = 0}},
\label{integratedCorrelator}
\end{equation}
where over the color indices was summed and
where $x$ is the longitudinal momentum fraction of the quark with respect
to its parent hadron, $x \equiv p^+ / P^+$.
The Wilson line operator, or also called \emph{gauge link},
is a $3\times 3$ color-matrix-operator and makes the bilocal
operator $\bar{\psi}_{j,b}(0)\ \psi_{i,a}(\xi)$ invariant under color gauge transformations. A Wilson
line along a path $\Xi^\mu(\lambda)$ with $\Xi^\mu(0) = a^\mu$ and
$\Xi^\mu(1) = b^\mu$
is defined as
\begin{multline}
\mathcal{L}(a,b) \equiv 1 - ig \int_0^1 \rmd{1}{\lambda} \frac{\rmd{1}{\Xi^\mu}}
{\rmd{1}{\lambda}} A_\mu (\Xi(\lambda))
\\+
(-ig)^2 \int_0^1 \rmd{1}{\lambda_1} \frac{\rmd{1}{\Xi^\mu}}
{\rmd{1}{\lambda_1}} A_\mu (\Xi(\lambda_1)) \int_{\lambda_1}^1 \rmd{1}{\lambda_2}
\frac{\rmd{1}{\Xi^\mu}}
{\rmd{1}{\lambda_2}} A_\mu (\Xi(\lambda_2)) + \ldots,
\end{multline}
where $A^\mu = A^\mu_l t^l$.
In Eq.~\ref{integratedCorrelator}
an abbreviation was introduced for links along straight paths.
In the abbreviation it is indicated
which variables are constant
along the path ($0_T$ and $\xi^+$) and which coordinates are running
(the minus components).
The path for this case is illustrated in Fig.~\ref{plaatjeLinks}a.
The integrated correlator will be parametrized in the next
subsection.

The second kind of correlator is encountered in cross sections which are
sensitive to the transverse momenta of the quarks.
Calling it the \emph{unintegrated correlator},
it also contains a Wilson
line operator and reads
\begin{equation}
\Phi_{ij}^{[\pm]}(x,p_T,P,S) =
\int \frac{ \rmd{1}{\xi^-} \rmd{2}{\xi_T} }{(2\pi)^3}
e^{ip\xi}
\langle P,S | \bar{\psi}_j (0)\ \mathcal{L}^{[\pm]} (0,\xi^-)\
\psi_i (\xi) | P,S \rangle_\text{c} \bigg|_{\substack{\xi^+ = 0\\p^+ = xP^+}},
\end{equation}
where
\begin{equation}
\mathcal{L}^{[\pm]} (0,\xi^-)
\equiv
\mathcal{L}^{0_T\!,\ \xi^+}(0^-,\pm \infty^-)
\mathcal{L}^{\pm \infty^-\!,\ \xi^+} (0_T,\xi_T)
\mathcal{L}^{\xi_T\!,\ \xi^+}(\pm \infty^-,\xi^-),
\end{equation}
and similarly for $\{ \xi^+,\xi^-,\infty^- \} \leftrightarrow
\{ \xi^-,\xi^+,\ \infty^+ \}$.
In these correlators the link runs via a nontrivial path
like the ones drawn in Fig.~\ref{plaatjeLinks}b and
Fig.~\ref{plaatjeLinks}c.
After an integration over $p_T$ the unintegrated
correlator reduces to the integrated correlator.
The nontrivial paths in the unintegrated correlator are
a source for interesting phenomena like single spin asymmetries. These
correlators will be parametrized in the second subsection.

As a last remark, in chapter~\ref{chapter4} and \ref{chapter5} we will also
encounter a gauge link of which its path is closed, see 
Fig.~\ref{plaatjeLinks}d. This ``closed''
gauge link or Wilson loop is defined as
\begin{equation}
\mathcal{L}^{[ \Box ]}(0,\xi^\pm) \equiv  \left[ \mathcal{L}^{[+]}(0,\xi^\pm)\right]
                                        \left[
                                         \mathcal{L}^{[-]}(0,\xi^\pm)
                                        \right]^\dagger  .
\end{equation}

\subsection{Integrated distribution functions}

The specific form of the integrated distribution correlator
%, as defined in
(Eq.~\ref{integratedCorrelator})
%\begin{equation}
%\Phi_{ij}(x,P,S) = \int \frac{\rmd{1}{\xi^-}}{2\pi} e^{ixP^+\xi^-}
%\langle P,S | \bar{\psi}_j (0)\ \mathcal{L}^{0_T\!,\ \xi^+}(0,\xi^-)\ \psi_i(\xi)
%|P,S \rangle_\text{c} \Big|_{\begin{subarray}{l} \xi^+ = 0\\ \xi_T = 0
%\end{subarray}},
%\end{equation}
has the following analytical properties
\begin{alignat}{2}
{\Phi}^\dagger (x,P,S) &=
\gamma^0\ \Phi (x,P,S)\ \gamma^0
& \qquad &\text{(hermiticity)},\\
\Phi (x,P,S) &=
\gamma^0\ \Phi (x,\bar{P},-\bar{S})\ \gamma^0
&  &\text{(parity)},\\
{\Phi}^* (x,P,S) &=
i \gamma_5 C\ \Phi (x,\bar{P},\bar{S})\ i \gamma_5 C
&\qquad & \text{(time-reversal)},
\end{alignat}
where $C$ denotes charge conjugation and $\gamma_5 \equiv i \gamma^0 \gamma^1
\gamma^2 \gamma^3$.
Note that $P\rightarrow \bar{P}$ implies that $n_- \leftrightarrow n_+$.
Using these constraints the correlator has been
decomposed on a basis of Dirac
structures~\cite{Ralston:1979ys,Jaffe:1991ra,Mulders:1995dh}.
In the notation
of Ref.~\cite{Mulders:1995dh}
this gives\footnote{T-odd integrated distribution functions are
in this approach zero and
have therefore been discarded.}
(for conventions on names see for instance Ref.~\cite{Boer:1998im})
\myBox{
\begin{align}
\Phi(x,P,S)\ =\
                &
                 \frac{1}{2} \biggl( f_1\ \slashii{n}_+ +
                 S_L\ g_1\ \gamma_5 \slashii{n}_+ +
                 h_1\ \gamma_5 \slashi{S}_T \slashii{n}_+ \biggr)
                 &\qquad & (\text{twist 2}) \nonumber\\
                 &
                 + \frac{M}{2P^+} \biggl(
                 e + g_T\ \gamma_5 \slashi{S}_T +
                 S_L\ h_L\ \gamma_5 \frac{[\slashii{n}_+, \slashii{n}_-]}{2} \biggr)
                 & & (\text{twist 3})
                 \nonumber\\
                 &+ \text{higher twist}. &&
                 %\\
                 %&
                 %\frac{M}{2P^+} \biggl(
                 %-\lambda\ e_L\ i \gamma_5 - f_T\ \epsilon_{T}^{\rho\sigma}
                 %\gamma_\rho {S_T}_\sigma +i\ h  \frac{[\slashii{n}_+, \slashi{n_-}]}{2}
                 %\biggr)
                 %& & (\text{twist 3}),
\label{intDistr}
\end{align}
\begin{flushright}
\emph{parametrization of the integrated correlator}
\end{flushright}
}
where all functions depend on $x$
%. The functions also depend 
and on a renormalization
scale. The scale dependence can be calculated by applying evolution
equations (for example see Ref.~\cite{Brock:1993sz,thesisAlex}).
The above expansion in $M/P^+$ is not to all orders. Calculations of
cross sections should therefore be constructed such that
$M/P^+$ is small in order to employ the above parametrization.

In the parametrization we have indicated the twist of the distribution
functions. The twist of a distribution function defines at which order
in $M/P^+$
the function appears in the parametrization. This definition
of ``operational twist'' for nonlocal operators
was introduced
by Jaffe in Ref.~\cite{Jaffe:1996zw}. The standard
definition of twist, which counts the dimensions of local operators,
agrees with this definition for local operators appearing in the Taylor
expansion of the nonlocal matrix element.

As will be shown in the next subsection,
the leading twist functions allow for a probability
interpretation.
In order to obtain a parton interpretation for the functions,
the Heisenberg field operators, which are present in the correlators,
can be expanded in
creation and annihilation operators
at particular point in time.
Therefore,
the functions which are integrated over $x$ and are local in time
allow for a parton
interpretation. Moreover,
in those matrix elements
the gauge link has vanished, simplifying the interpretation.
The
function $f_1$ describes the chance of hitting an unpolarized
quark (red, green, or blue)
in an unpolarized nucleon, the
function $g_1$ describes longitudinally
polarized quarks - or actually chirally left or right handed quarks -
in a longitudinally polarized nucleon, and $h_1$ or
\emph{transversity} describes it for transverse polarizations. This last
function is at present the only leading twist integrated function which has
not yet been measured.
Recently, it has been suggested in
Ref.~\cite{Bakker:2004ib}
by Bakker, Leader, and Trueman that this function
appears together with orbital angular momentum of the partons in a simple
sum-rule. This would provide for the first time
in semi-inclusive DIS access to the orbital angular momentum of partons
inside a nucleon, making the transversity function of increasing interest.
For more information on the interpretation of these
functions the reader is referred to Ref.~\cite{Barone,Bacchetta:2002xd}.

\subsection{Transverse momentum dependent distribution
            functions\label{subsecUnintDistr}}

\subsubsection{Gauge Invariant correlators and T-odd behavior}

In electromagnetic processes one encounters at first sight
two kinds of unintegrated correlators in the final result of the diagrammatic
expansion.
They are defined
as 
%(with an $\mathcal{L}^{[+]}$ or $\mathcal{L}^{[-]}$)
\begin{equation}
\Phi_{ij}^{[\pm]}(x,p_T,P,S) =
\int \frac{ \rmd{1}{\xi^-} \rmd{2}{\xi_T} }{(2\pi)^3}
e^{ip\xi}
\langle P,S | \bar{\psi}_j (0)\ \mathcal{L}^{[\pm]} (0,\xi^-)\
\psi_i (\xi) | P,S \rangle_\text{c} \bigg|_{\substack{\xi^+ = 0\\p^+=xP^+}},
\end{equation}
%where $\mathcal{L}^{[\pm]} (0,\xi)$ is a \emph{Wilson line operator},
%making the
%structure invariant under color gauge transformations. This Wilson line
%operator, also called \emph{gauge link}, is defined as
%\begin{align}
%\mathcal{L}^{[\pm]} (0,\xi) &=\mathcal{L}^{0_T}(0^-,\pm \infty^-)
%                              \mathcal{L}^{\pm \infty^-} (0_T,\xi_T)
%                              \mathcal{L}^{\xi_T}(\infty^-,\xi^-),\\
%\intertext{where}
%\mathcal{L}^{a}(b,c) &= \mathcal{P} \exp{-ig \int_b^c \rmd{1} \eta \cdot
%                         A(\text{at}\ a, \text{running}\ \eta)} \nonumber\\
%&\equiv
%1-ig \int_b^c \rmd{1} \eta \cdot
%                         A(\text{at}\ a, \text{running}\ \eta) + \nonumber\\
%&(-ig)^3 \int_b^c \rmd{1} \eta \cdot
%                         A(\text{at}\ a, \text{running}\ \eta)
%                         \int_\eta^c \rmd{1}{\eta'}\cdot
%                         A(\text{at}\ a, \text{running}\ \eta')
%                         + \mathcal{O}(g^3).
%\end{align}
where the two different paths of the gauge links $\mathcal{L}^{[\pm]}$ are
indicated in Fig.~\ref{plaatjeLinks}b,c.
In general also other paths will appear in link operators,
but those will be discussed in chapter \ref{chapter4} and \ref{chapter5}.
The analytical structure of the correlator has the following properties
\begin{alignat}{2}
{\Phi^{[\pm]}}^\dagger (x,p_T,P,S) &=
\gamma^0\ \Phi^{[\pm]} (x,p_T,P,S)\ \gamma^0
& \quad &\text{(hermiticity)},\\
\Phi^{[\pm]} (x,p_T,P,S) &=
\gamma^0\ \Phi^{[\pm]} (x,-p_T,\bar{P},-\bar{S})\ \gamma^0
&  &\text{(parity)},\\
{\Phi^{[\pm]}}^* (x,p_T,P,S) &=
i \gamma_5 C\ \Phi^{[\mp]} (x,-p_T,\bar{P},\bar{S})\ i \gamma_5 C
&\qquad & \text{(time-reversal)}.
\end{alignat}
The last equation shows that the time-reversal operation relates
 the two different paths of the gauge links. This enables us to
decompose $\Phi$ in two classes, one which
is the average and is called T-even,
and one which is the difference and is called
T-odd (T-odd does not mean breaking of time-reversal in QCD, the name is
similar to P-odd)
\begin{align}
\Phi^{[\text{T-even}]} (x,p_T,P,S)&= \frac{1}{2} \left( \Phi^{[+]}(x,p_T,P,S) + \Phi^{[-]}(x,p_T,P,S) \right),
\label{theoryTeven}
\\
\Phi^{[\text{T-odd}]}(x,p_T,P,S) &= \frac{1}{2} \left( \Phi^{[+]}(x,p_T,P,S) - \Phi^{[-]}(x,p_T,P,S) \right).
\label{theoryTodd}
\end{align}

\sloppy
Although the phenomenology of
T-odd distribution functions was studied, see for
instance Anselmino, D'Alesio, Boglione,
Murgia~\cite{Anselmino:1994tv,Anselmino:1998yz,Anselmino:2002pd} and
Boer, Mulders~\cite{Boer:1998nt}, \mbox{T-odd}
distribution functions were not really
believed to
exist as separate distributions
(for example see Collins in Ref.~\cite{Collins:1993kk}). After Brodsky, Hwang,
and Schmidt
showed in Ref.~\cite{Brodsky:2002cx}
that unsuppressed T-odd effects could be generated in a particular model,
the existence of T-odd distribution functions was taken as
a serious possibility (see also Collins~\cite{Collins:2002kn}
and Belitsky, Ji, Yuan~\cite{Belitsky:2002sm}).
At present the data of HERMES as shown in chapter~\ref{Chapter1},
Fig.~\ref{intro2b}, indicate that
T-odd functions might really exist.

\fussy
Another mechanism for T-odd functions has also been suggested
by  Anselmino, Barone, Drago, and Murgia in Ref.~\cite{Anselmino:2002yx}
based on nonstandard time-reversal. This mechanism is not taken into account
here but could be included at a later stage. If this mechanism is realized in
nature it would lead to
universality problems for the distribution functions similar
to those already appearing for fragmentation functions. The latter problem
will be discussed in the next section.

The T-even and T-odd parts of $\Phi^{[\pm]}$ have identical
parity and hermiticity properties and they can be parametrized
in a set of functions.
Before going to the explicit parametrizations it is interesting to note that
a distribution correlator can now be given by a T-even and a sign
dependent T-odd part
\begin{equation}
\Phi^{[\pm]}(x,p_T,P,S) = \Phi^{[\text{T-even}]}(x,p_T,P,S)
\pm \Phi^{[\text{T-odd}]}(x,p_T,P,S).
\end{equation}
This means that T-odd distribution functions enter with a
sign~\cite{Collins:2002kn,Brodsky:2002rv} depending
on the path of the
gauge link. In the next chapters we will see
that this path is set by the process
or subprocess.

\subsubsection{The introduction of the distribution functions}

The correlator can be parametrized as
follows~\cite{Levelt:1993ac,Tangerman:1994eh,Mulders:1995dh,Boer:1997nt,
Bacchetta:2004zf,Goeke:2005hb}
\myBox{
\small
\vspace{-.2cm}
\begin{align}
\Phi^{[\text{T-even}]}&(x,p_T,P,S) &&
\nonumber\\
=
                   & \frac{1}{2} \left(
                   f_1\ \slashii{n}_+
                     + \left( S_L\ g_{1L} -\tfrac{p_T\cdot S_T}{M}\
                            g_{1T}  \right)
                       \gamma_5 \slashii{n}_+ \right)
                     &\ & (\text{twist 2})
                     \nonumber\\
                   &+ \frac{1}{2}
                   \left( h_{1T}\ \gamma_5 \slashi{S_T} \slashii{n}_+
                     + \left( S_L\ h_{1L}^\perp -\tfrac{p_T\cdot S_T}{M}\
                              h_{1T}^\perp  \right)
                       \frac{\gamma_5 \slashiv{p}_T \slashii{n}_+}{M} \right)
                     && (\text{twist 2})
                     \nonumber\\
                   &+ \left( \frac{M}{2P^+} \right) \left(
                     e + f^\perp\ \frac{\slashiv{p}_T}{M} \right)
                     && (\text{twist 3, unpolarized})
                     \nonumber\\
                   &+ \left( \frac{M}{2P^+} \right)
                   \left(
                    \left( S_L\ g^\perp_{L}\ -\tfrac{p_T\cdot S_T}{M}\
                            g^\perp_{T} \right)
                            \frac{\gamma_5 \slashiv{p}_T}{M}
                            + g_T'\ \gamma_5 \slashi{S}_T
                    \right)
                     && (\text{twist 3, polarized})
                     \nonumber\\
                   &+ \left( \frac{M}{2P^+} \right)
                     \left(
                     h_T^\perp\ \frac{ \gamma_5 [ \slashi{S}_T, \slashiv{p}_T ]}
                     {2M}
                     \right)
                     && ( \text{twist 3, polarized})
                     \nonumber\\
                   &+ \left( \frac{M}{2P^+} \right)
                      \left(
                      \left( S_L\ h_{L}\ -\tfrac{p_T\cdot S_T}{M}\
                            h_{T}  \right)
                     \frac{\gamma_5 [ \slashii{n}_+, \slashii{n}_- ]}{2}
                     \right)
                     && ( \text{twist 3, polarized})
                     \nonumber\\
                   &+\text{higher twist},
                   &&
\nonumber\\
\Phi^{[\text{T-odd}]}&(x,p_T,P,S) &&
\nonumber\\
=
                   & \frac{1}{2} \left(
                   f_{1T}^\perp\
                     \frac{ \epsilon_{T}^{\mu\nu} {S_T}_\mu {p_T}_\nu
                     \slashii{n}_+ }{M} +
                     h_1^\perp\ \frac{i \slashiv{p}_T \slashii{n}_+ }{M} \right)
                     && (\text{twist 2})
                     \nonumber\\
                   &+ \left( \frac{M}{2P^+} \right)
                     \left( h\ \frac{i [ \slashii{n}_+, \slashii{n}_- ]}{2}
                     + g^\perp\ \frac{\epsilon_{T}^{\mu\nu} {p_{T}}_\mu
                       \gamma_\nu \gamma_5}{M} \right) \qquad
                     && (\text{twist 3, unpolarized})
                     \nonumber\\
                   & +\left( \frac{M}{2P^+} \right)
                   \left(
                     \frac{\epsilon_T^{\mu\nu} {S_T}_\mu {p_T}_\nu}{M} e_T^\perp
                     -f_T\ \epsilon_{T}^{\mu\nu} \gamma_\mu {S_T}_\nu
                     \right)
                     && (\text{twist 3, polarized})
                     \nonumber\\
                   & -\left( \frac{M}{2P^+} \right) \left( \big(
                     S_L\ f_L^\perp - \tfrac{p_T\cdot S_T}{M}\ f_T^\perp
                     \big) \frac{ \epsilon_{T}^{\mu\nu}
                     \gamma_\mu {p_T}_\nu}{M} \right)
                     &\qquad& (\text{twist 3, polarized})
                     \nonumber\\
                   &-\left( \frac{M}{2P^+} \right)
                   \left(
                     \left( S_L\ e_{L}\ -\tfrac{p_T\cdot S_T}{M}\
                            e_{T}  \right) i \gamma_5 \right)
                     && (\text{twist 3, polarized})
                     \nonumber\\
                   &+\text{higher twist}.
                   &&
\label{unintDistr}
\end{align}
\vspace{-1cm}
\begin{flushright}
\emph{parametrization of the distribution quark-quark correlator}
\end{flushright}
}
All functions have the arguments $x$ and $p_T^2$
and also depend on a renormalization scale. In contrast to the integrated
distribution functions the scale dependence is not known for transverse
momentum dependent functions (see for instance Henneman~\cite{thesisAlex}).

In the T-odd correlator the new functions $g^\perp,\ e_T^\perp,$
and $f_T^\perp$ are included.
The function $g^\perp$ (as defined in Ref.~\cite{Bacchetta:2004zf})
and the existence of the others were discovered in
Ref.~\cite{Bacchetta:2004zf}.
Subsequently, a complete parametrization was given by
Goeke, Metz, and Schlegel in Ref.~\cite{Goeke:2005hb}.
The fact that $\int \rmd{2}{p_T}\ \Phi^{[\text{T-odd}]}(x,p_T) = 0$ leads to
constraints for the T-odd functions $h$, $f_T$, $e_L$, and $f_T^\perp$
(see for example Ref.~\cite{Boer:2003cm,Goeke:2005hb}).

The first transverse moment of the correlator $\Phi$ and 
some function $f_i$ is defined as
\begin{align}
\Phi_\partial^\alpha (x,P,S) 
&
\equiv \int \rmd{2}{p_T} p_T^\alpha\ \Phi(x,p_T,P,S)\label{hiro3}.
\\
f_i^{(1)}(x)
&\equiv \int \rmd{2}{p_T} \frac{\threeVec{p}_T^2}{2M^2}\ f_i (x,p_T^2).
\end{align}

The introduced functions describe how the quarks are distributed in the
nucleon. The leading twist functions (twist 2) are again probability
functions
and therefore contain valuable information. For instance, the
functions
$f_1(x,p_T^2)$ and $g_{1L}(x,p_T^2)$ are generalizations of $f_1(x)$  and $g_{1}(x)$.
For more information on the interpretation of \mbox{T-even}
functions the reader is referred to Ref.~\cite{Boer:1998im, Boer:2003xz}.

For the interested reader the proof for the probability
interpretation of $f_1$ is given here
(see for instance Ref.~\cite{Jaffe:1985je,Bacchetta:1999kz}, and
Ref.~\cite{Brodsky:2002ue} for related work)
\begin{equation}
\begin{split}
f_1 &= \frac{1}{2} \tr \slashii{n}_- \Phi^{[ \text{T-even}] }\\
    &=\!\!\! \sum_X \!\!\! \int \!\!\!
       \phaseFactor{P_X}\!\! \int\!\!\! \frac{\rmd{2}{\xi_T}\! \rmd{1}{\xi^-}}
       {(2\pi)^3} e^{ip\xi}\!
       \langle P,S| \!
       \left(\tfrac{\gamma^- \gamma^+}{2} \psi (0) \right)^\dagger \!\!
       \mathcal{L}^{0_T\!,\ \xi^+}\!(0^-,\infty^-)\!
       \mathcal{L}^{\infty^-,\ \xi^+}\!(0_T,\infty_T)
       | P_X \rangle_\text{c}
    \\
    & \eqnIndent \times \frac{1}{\sqrt{2}}
       \langle P_X |
       \mathcal{L}^{\infty^-,\ \xi^+}(\infty_T,\xi_T)
       \mathcal{L}^{\xi_T\!,\ \xi^+} (\infty^-,
       \xi^-) \tfrac{\gamma^- \gamma^+}{2} \psi(\xi) | P,S \rangle_\text{c}
       \Big|_{\xi^+ = 0}
    \\
    &= \frac{1}{\sqrt{2}} \sum_X  \int \phaseFactor{P_X}\
     \delta(p^+ {+} P_X^+ {-} P^+)\ \delta^2
       (p_T {-} {P_X}_T) \\
    &   \eqnIndent \times \left| \langle P,S |
        \left(\tfrac{\gamma^- \gamma^+}{2} \psi (0) \right)^\dagger
       \mathcal{L}^{0_T\!,\ \xi^+}(0^-,\infty^-)
       \mathcal{L}^{\infty^-,\ \xi^+}(0_T,\infty_T)
       | P_X \rangle_\text{c}  \right|^2 \Big|_{\xi^+ = 0} > 0.
\end{split}
\raisetag{18pt}
\end{equation}
For the other leading twist functions (T-even and T-odd)
the proof is analogous. The fact
that the leading twist functions are probabilities leads to several
positivity bounds, see for instance
Soffer~\cite{Soffer:1994ww}, and Bacchetta et al.~\cite{Bacchetta:1999kz}.

The leading twist T-odd functions
(such as the \emph{Sivers function} $f_{1T}^\perp$)
do not have a parton interpretation in terms of quarks
although they are probabilities as well. To obtain an interpretation we
consider the transverse moment of a T-odd correlator (see also Eq.~\ref{hiro3})
and
rewrite the quark transverse momentum as a derivative acting on the
gauge links alone
\begin{multline}
\Phi_\partial^{[\text{T-odd}]\alpha}
(x,P,S)
 =\! \frac{1}{2}\!\! \int\!\! \rmd{2}{p_T}\!\!\!
\int\! \frac{ \rmd{1}{\xi^-} \rmd{2}{\xi_T}}{(2\pi)^3} e^{ip\xi}
\langle P,S | \bar{\psi}_j (0)
\\
\times
 \left( i \partial_{\xi_T}^\alpha\!
\left[ \mathcal{L}^{[+]}(0,\xi^-) {-} \mathcal{L}^{[-]}(0,\xi^-) \right] \right)
\psi_i (\xi) | P,S \rangle_\text{c} \Big|_{\begin{subarray}{l}
\xi^+=0\\p^+=xP^+ \end{subarray}}\!\!.
\label{calina22}
\end{multline}
Using identities of Ref.~\cite{Boer:2003cm} (the second identity can be
proven by using that
$G^{+\alpha} \sim [iD^+,iD^\alpha]$,
$i\partial_\xi^+ \mathcal{L}^{\xi^+,\xi_T}(\eta^-,\xi^-) =
\mathcal{L}^{\xi^+,\xi_T}(\eta^-,\xi^-) iD^+(\xi)$,
and shifting
$iD^+$ to the side)
\myBox{
\vspace{-.3cm}
\begin{align}
&
i \partial_{\xi_T}^\alpha \mathcal{L}^{\pm \infty^-\!,\ a^+} (0_T,\xi_T)
\nonumber\\
&
\eqnIndent
= \mathcal{L}^{\pm \infty^-\!,\ a^+} (0_T,\xi_T) i D_T^\alpha(\pm \infty,a^+,\xi_T),
\nonumber \\
&
\nonumber\\
&
i D_T^\alpha(\zeta^-,a^+,\xi_T)
\mathcal{L}^{\xi_T\!,\ a^+} (\zeta^-,\xi^-)
\nonumber
\\
&
\eqnIndent
=
\mathcal{L}^{\xi_T\!,\ a^+}(\zeta^-,\xi^-)
i D_T^\alpha(\xi^-,a^+,\xi_T)
\nonumber
\\
&
\eqnIndent \eqnIndent
-g\! \int_{\zeta^-}^{\xi^-}\! \rmd{1}{\eta^-}
\mathcal{L}^{\xi_T\!,\ a^+} (\zeta^-, \eta^-) G^{+\alpha}_T(\eta^-,a^+,\xi_T)
\mathcal{L}^{\xi_T\!,\ a^+} (\eta^-,\xi^-),
\label{theoryIdentity}
\end{align}
\vspace{-1cm}
\begin{flushright}
\emph{identities concerning gauge links}
\end{flushright}
}
where $a^+$ is some constant and
$iD_T^\alpha(\xi^-,a^+,\xi_T) \equiv i \partial_{\xi_T}^\alpha +
g A_T^\alpha(\xi^-,a^+,\xi_T)$,
the derivative on the difference of the links can be written as
\begin{multline}
\Phi_\partial^{[\text{T-odd}]\alpha}
(x,P,S) =
\frac{g}{2}\!
\int\! \frac{ \rmd{1}{\xi^-}\! \rmd{1}{\eta^-}\! }{2\pi} e^{ixP^+\xi^-}
\langle P,S | \bar{\psi}_j (0)
\\
\times
\mathcal{L}^{0_T\!,\ \xi^+}(0^-,\eta^-) G^{+\alpha}_T(\eta)
\mathcal{L}^{0_T\!,\ \xi^+}(\eta^-,\xi^-)
\psi_i (\xi) | P,S \rangle_\text{c}
\big|_{\substack{\eta^+ = \xi^+ = 0\\ \eta_T=\xi_T=0}}.
\label{theoryGluonicPole}
\end{multline}
This matrix element is also called a \emph{gluonic pole matrix element}. In
1991 such a matrix element was suggested by Qiu and Sterman in
Ref.~\cite{Qiu:1991pp,Qiu:1991wg} as an explanation
for single spin asymmetries in photon-production in
hadron-hadron scattering and it has been subsequently studied
in several other
articles~\cite{Hammon:1996pw,Boer:1997bw,Qiu:1998ia,Kanazawa:2000cx,Boer:2001tx,Burkardt:2004ur}.
We see here that the same matrix element appears in the
two separately suggested mechanisms for \mbox{T-odd}
effects~\cite{Boer:2003cm}
(the soft gluon effects of Qiu and Sterman and the gauge link).

%Using the identities in Eq.~\ref{theoryIdentity} one can also write
%Eq.~\ref{theoryGluonicPole}
%as
%\begin{multline}
%\Phi_\partial^{[\text{T-odd}]\alpha}
%(x,P,S)
%= \frac{1}{2} \int \rmd{2}{p_T}
%\int \frac{ \rmd{1}{\xi^-} \rmd{2}{\xi_T}}{(2\pi)^3} e^{ip\xi}
%\langle P,S | \bar{\psi}_j (0)
%\\
%\times
%\left( i \partial_{\xi_T}^\alpha \mathcal{L}^{[ \Box ]}(0,\xi^-)\right)
%\mathcal{L}^{\xi^+\!,\ \xi_T}(0^-,\xi^-)
%\psi_i (\xi) | P,S \rangle_\text{c} \Big|_{\substack{\xi^+=0\\p^+=xP^+}},
%\end{multline}
%where the Wilson loop, $\mathcal{L}^{[ \Box ]}(0,\xi^-)$, is a closed Wilson
%line of which its path is given in Fig.~\ref{plaatjeLinks}d.
%Its definition is
%\begin{equation}
%\mathcal{L}^{[ \Box ]}(0,\xi^\pm) \equiv  \left[ \mathcal{L}^{[+]}(0,\xi^\pm)\right]
%                                        \left[
%                                         \mathcal{L}^{[-]}(0,\xi^\pm)
%                                        \right]^\dagger  .
%\end{equation}

At first sight it seems strange that the nontrivial path is the origin of T-odd
functions. One might think that such functions should
vanish,
since in the light-cone gauge, $A^+ = 0$, the T-odd distribution
functions become proportional to transverse gauge links at infinity
(see Eq.~\ref{calina22} and Fig.~\ref{plaatjeLinks}c, d) and
shouldn't those matrix elements vanish? Since we are at present not
able to calculate those matrix elements in QCD we do not really
know the answer,
but we do know that nonzero $A$-fields can give effects
in areas where the physical fields,
like
$G^{\alpha\beta}$,
are required to vanish. An example of such an effect is
the Aharanov-Bohm experiment which was discussed in
chapter~\ref{Chapter1}.

\subsubsection{The Lorentz invariance relations and
$\boldsymbol{g}^{\pmb{\perp}}$}

Before the paper of Brodsky, Hwang, and
Schmidt~\cite{Brodsky:2002cx} appeared,
physical effects from the gauge link were assumed to be absent
which led to several interesting observations. Not
only did T-odd effects disappear, but relations between the
various functions in the correlator were also obtained.
By arguing that the correlator $\Phi(x,p_T,P,S)$
could be written in terms
of only fermion fields, the starting point then was
another object $\Phi(p,P,S)$ which is defined by
\begin{equation}
\Phi_{ij}(p,P,S) = \int \frac{\rmd{4}{\xi}}{(2\pi)^4}\ e^{ip\xi}\
\langle P,S |\ \bar{\psi}_j (0)\ \psi_i(\xi)\ | P,S \rangle_\text{c}.
\end{equation}
This quantity $\Phi(p,P,S)$ is the fully unintegrated correlator,
from which
$\Phi(x,p_T,P,S)$ (without gauge link) is obtained via
\begin{equation}
\Phi(x,p_T,P,S) = \int \rmd{1}{p^-} \Phi(p,P,S).
\end{equation}
Without the gauge link the parametrization of the
object $\Phi(p,P,S)$ contains less functions
than $\Phi(x,p_T,P,S)$. This led to the so-called
\emph{Lorentz-invariance
relations}\footnote{Based on Lorentz
invariance, relations
were also derived by Bukhvostov, Kuraev, and Lipatov
in Ref.~\cite{Bukhvostov:1983,Bukhvostov:1983te,Bukhvostov:1984as}.
It is at present unclear why their relations are different from the relations
given here.}
(see Boer, Jakob, Henneman, Mulders, Tangerman~\cite{Jakob:1997wg,Mulders:1995dh,Boer:1997nt,Henneman:2001ev})
\begin{align}
g_T(x) &= g_1(x) + \frac{\rmd{1}{}}{\rmd{1}{x}} g_{1T}^{\perp(1)} (x),&
g_L^\perp (x) &= - \frac{\rmd{1}{}}{\rmd{1}{x}} g_T^{\perp(1)}(x),
\quad h_T(x) = - \frac{\rmd{1}{}}{\rmd{1}{x}} h_{1T}^{\perp(1)}(x),
\nonumber\\
h_L(x) &= h_1(x) - \frac{\rmd{1}{}}{\rmd{1}{x}} h_{1L}^{\perp(1)}(x),&
h_{1L}^\perp(p^+,\threeVec{p}_{T}^2) &= h_T (p^+,\threeVec{p}_{T}^2)
- h_T^\perp (p^+, \threeVec{p}_{T}^2).
\label{theory45}
\end{align}
Similar relations for T-odd distribution functions were obtained as well.

Taking the gauge link into account affects these relations. Since the
gauge link runs in the $n_-$-direction
via infinity (see Fig.~\ref{plaatjeLinks}), it is not clear
how to construct an $n_-$-independent $\Phi(p,P,S)$ which after
an integration over $p^-$ leads to $\Phi(x,p_T,P,S)$
containing the gauge link.
In 2003, this was made explicit by
Goeke, Metz, Pobylitsa, and Polyakov
in Ref.~\cite{Goeke:2003az}.
Taking the $n_-$-dependence into account, they showed that the
former
proof of the Lorentz-invariance relations failed.

It seems to be impossible to maintain the Lorentz-invariance relations
for T-odd functions since the involved matrix elements
are intrinsically nonlocal
(see for instance Eq.~\ref{theoryGluonicPole}).
This is in contrast to
the first transverse moment of a T-even function which is local in the
light-cone gauge. This could imply that the Lorentz-invariance relations
might still
hold for the T-even functions.

The $n_-$-dependence of the gauge link
implies not only the need to revisit the Lorentz-invariance
relations~\cite{Goeke:2003az}, but also leads to new functions in the
parametrizations as discovered in Ref.~\cite{Bacchetta:2004zf}
and confirmed in Ref.~\cite{Goeke:2005hb}.
Since the origin of $g^\perp$ is connected to the non-validity of the
Lorentz-invariance relations, a measurement of
$g^\perp$ or checking the Lorentz-invariance relations
(Eq.~\ref{theory45}) would be interesting.
In the next chapter it will be pointed out how $g^\perp$ can be accessed.

\section[Quark fragmentation functions into
spin-$\frac{1}{2}$ hadrons]
{Quark fragmentation functions into spin-$\frac{\mathbf{1}}{\mathbf{2}}$ hadrons
\label{sectFragm}}

\vspace{-.5cm}

The introduction of the fragmentation functions proceeds analogously to the
introduction of the distribution functions.
A set of light-like vectors is introduced such that
\myBox{
\vspace{-.4cm}
\begin{align}
1 &= n_- \cdot n_+,&  n_- &\sim \bar{n}_+,   \nonumber\\
P_h &= P_h^-\ n_- + \frac{M_h^2}{2P_h^-}\ n_+,&
\epsilon_{T}^{\mu\nu} &\equiv \epsilon^{\rho\sigma\mu\nu} {n_+}_\rho
{n_-}_\sigma,
\nonumber\\
g_{T}^{\mu\nu} &= g^{\mu\nu} - n_+^\mu n_-^\nu - n_+^\nu n_-^\mu,&
A_{T}^\mu &= g_{T}^{\mu\nu} A_\nu,\ \text{for any }A.
\label{frameFragm}
\end{align}
\vspace{-.8cm}
\begin{flushright}
\emph{the basis in which parton fragmentation functions are defined}
\end{flushright}
}
Also here we define for every vector $A$ the Lorentz-invariants
$A^\pm \equiv A \cdot n_\mp$.
The spin of the observed hadron is decomposed as
\begin{equation}
S_h = S_{hL} \frac{P_h^-}{M_h}\ n_- - S_{hL} \frac{M_h}{2P_h^-}\ n_+ + S_{hT}
\qquad \text{with } S_{hL}^2 + \threeVec{S}_{hT}^2 = 1.
\end{equation}

Similarly as for distribution functions, $M_h/P_h^-$ does not
have to be small. However, calculations of cross sections
will be constructed in such a way that $P_h^- \gg M_h$.

In general one encounters integrated and unintegrated
fragmentation functions. The integrated fragmentation functions
will be discussed in the next subsection.
For the unintegrated functions we will see that
the two different link structures (as for the distribution
functions) produce problems with universality~\cite{Boer:2003cm}.
In the second subsection,
these functions will be parametrized.

\subsection{Integrated fragmentation functions}

The integrated correlator which appears in cross sections is
expressed as\footnote{In order to write the gauge link in one of the matrix 
elements the light-cone gauge was chosen in an intermediate step.}
\begin{multline}
\Delta_{ij}(z^{-1},P_h,S_h) \equiv
\frac{1}{3} \sum_X \int \phaseFactor{P_X}
\int \frac{ \rmd{1}{\eta^+} }{2\pi} e^{iz^{-1}P_h^-\eta^+}
{}_\text{out}\langle P_h,S_h;P_X | \bar{\psi}_j(0)
| \Omega \rangle_\text{c}
\\
\times
\langle \Omega | \mathcal{L}^{0_T\!,\ \eta^-}(0^+,\eta^+)\psi_i (\eta) | P_h,S_h;P_X
\rangle_\text{out,c} \big|_{\substack{\eta^- = 0\\ \eta_T = 0}},
\end{multline}
where a factor $1/3$ is introduced in the definition
to average over the initial quark's color,
and where the color indices are contracted.
Its analytical structure satisfies the following constraints
\begin{alignat}{2}
{\Delta}^\dagger (z^{-1},P_h,S_h) &=
\gamma^0\ \Delta (z^{-1},P_h,S_h)\ \gamma^0
& \quad &\text{(hermiticity)},\\
\Delta (z^{-1},P_h,S_h) &=
\gamma^0\ \Delta (z^{-1},\bar{P}_h,-\bar{S}_h)\ \gamma^0
&\qquad  &\text{(parity)}.
\end{alignat}
For fragmentation functions there is no constraint from time-reversal. In the
case of distribution functions we have that $\mathcal{T} |P \rangle_\text{in}
= |P \rangle_\text{out} =
| P \rangle_{\text{in}}$, but for fragmentation functions there are several
particles in the out-state which might interact with each other. Therefore,
such an identity does not hold and integrated T-odd
fragmentation functions appear (in contrast to the integrated distribution
functions). This mechanism to generate T-odd functions
was introduced by Collins in Ref.~\cite{Collins:1993kk}
and was shown to exist in model calculations by
Bacchetta, Kundu, Metz, and Mulders in
Ref.~\cite{Bacchetta:2001di,Bacchetta:2002tk}.

We continue by defining T-odd and T-even for fragmentation and write
\begin{align}
\Delta(z^{-1},P_h,S_h) &\equiv \Delta^{[\text{T-even}]}(z^{-1},P_h,S_h)
+ \Delta^{[\text{T-odd}]}(z^{-1},P_h,S_h),\\
\intertext{where $\Delta^{[\text{T-even}]}(z^{-1},P_h,S_h)$ and 
$\Delta^{[\text{T-odd}]}(z^{-1},P_h,S_h)$ obey}
{\Delta^{[\text{T-even}]}}^*(z^{-1},P_h,S_h) & = 
\phantom{(-)}(i \gamma_5 C)\ \Delta^{[\text{T-even}]}(z^{-1},P_h,S_h)\ (i \gamma_5 C),\\
{\Delta^{[\text{T-odd}]}}^*(z^{-1},P_h,S_h) & = 
{(-)}(i \gamma_5 C)\ \Delta^{[\text{T-odd}]}(z^{-1},P_h,S_h)\ (i \gamma_5 C).
\end{align}
This gives the following
parametrization of fragmentation into a spin-$\frac{1}{2}$ hadron\nopagebreak
\myBox{
\small
\vspace{-.2cm}
\begin{align}
\Delta(z^{-1},P_h,S_h) = \
                &
                \frac{1}{z} \biggl(
                D_1\ \slashii{n}_- - S_{hL}\ G_1\ \slashii{n}_- \gamma_5
                + H_1\ {\slashi{S}_h}_T \slashii{n}_- \gamma_5 \biggr)
                & \quad & (\text{twist 2, T-even})
                \nonumber\\
                &+ \frac{M_h}{z P_h^-} \biggl(
                -G_T\ {\slashi{S}_{h}}_T \gamma_5 + S_{hL}\ H_L\
                \frac{ [ \slashii{n}_-, \slashii{n}_+]}{2} \gamma_5
                + E \bigg)
                && (\text{twist3, T-even})
                \nonumber\\
                &+\frac{M_h}{z P_h^-} \biggl(
                D_T\ \epsilon_{T}^{\rho\sigma}
                \gamma_\rho {{S_h}_T}_\sigma -
                S_{hL}\ E_L\ i \gamma_5 + i\ H\ \frac{[\slashii{n}_-, \slashii{n}_+]}{2}\biggr)
                && (\text{twist3, T-odd})
                \nonumber\\
                &+ \text{higher twist}. &&
\label{intFrag}
\end{align}
\vspace{-1cm}
\begin{flushright}
\emph{parametrization of the integrated fragmentation correlator}
\end{flushright}
}
where all functions depend on $z$ and a renormalization scale.

The functions $D_1$, $G_1$, and $H_1$ have similar interpretations as the analogous
distribution functions. The function $D_1$ describes for instance
how an unpolarized quark
(being either red, green, or blue) decays into a hadron plus jet (over final
state colors is summed).
T-odd effects appear at subleading twist for fragmentation into
spin-$\frac{1}{2}$ hadrons,
in contrast to fragmentation into spin-$1$ hadrons where T-odd effects
already appear at leading twist
(see for example Bacchetta, Mulders~\cite{Bacchetta:2001rb}).

\subsection{Transverse momentum dependent fragmentation functions\label{hiro2}}

\subsubsection{Gauge invariant correlators and T-odd behavior}

Similarly as for the unintegrated distribution correlators one encounters
fragmentation correlators with two different links, defined through
%(for economics of notation we have written the transverse gauge link
%in one of the matrix elements)
\begin{equation}
\begin{split}
&
\Delta_{ij}^{[\pm]}(z^{-1},k_T,P_h,S_h)
\\
& \ \equiv
\frac{1}{3} \sum_X \int \phaseFactor{P_X}
\int \frac{\!\rmd{1}{\eta^+}\! \rmd{2}{\eta_T}\! }{(2\pi)^3} e^{ik\eta}
\\
& \eqnIndent
\times
{}_\text{out}\langle P_h,S_h;P_X | \bar{\psi}_j(0)
\mathcal{L}^{0_T\!,\ \eta^-}(0,\pm \infty^+)
\mathcal{L}^{\pm \infty^+,\ \eta^-}(0_T,\infty_T)
| \Omega \rangle_\text{c}
\\
& \eqnIndent
\times
\langle \Omega |
\mathcal{L}^{\pm\infty^+\!,\ \eta^-}(\infty_T,\eta_T)
\mathcal{L}^{0_T\!,\ \eta^-}(\pm \infty^+,\eta^+)
 \psi_i (\eta) | P_h,S_h;P_X
\rangle_\text{out,c}\Big|_{\substack{\eta^- = 0\\k^-=z^{-1}P_h^-}}.
\end{split}
\end{equation}
When studying the analytical structure of this correlator one finds
\begin{alignat}{2}
{\Delta^{[\pm]}}^\dagger (z^{-1},k_T,P_h,S_h) &=
\gamma^0\ \Delta^{[\pm]} (z^{-1},k_T,P_h,S_h)\ \gamma^0
& \quad &\text{(hermiticity)},\\
\Delta^{[\pm]} (z^{-1},k_T,P_h,S_h) &=
\gamma^0\ \Delta^{[\pm]} (z^{-1},-k_T,\bar{P}_h,-\bar{S}_h)\ \gamma^0
&\qquad  &\text{(parity)}.
\end{alignat}
As for the integrated correlators,
the time-reversal operation
does not lead to additional constraints.
This means that the functions appearing in the two different fragmentation
correlators, $\Delta^{[\pm]}$, cannot be related. That holds for the T-odd
functions as
well as for the T-even functions. Since
the functions become universal after an integration over
$k_T$, there could be a universality relation for the unintegrated
functions but at present such a QCD-relation
is unknown. This forms a problem for the by Collins~\cite{Collins:1993kk}
suggested method of accessing transversity via the
\emph{Collins function} $H_1^\perp$
(\emph{Collins effect}).

The problem with universality comes from
the interplay of two effects,
the final-state interactions (in and out-states) and
the gauge link. If one of the two mechanisms would be suppressed
the situation is simplified. For instance,
if final-state interactions in the out-states are
suppressed, then T-odd functions will enter with a sign depending
on the gauge link or process similar to the situation for the distribution
functions.
On the other hand, if gauge links do not influence the
expectation value of matrix elements,
then fragmentation functions will be the same in all
processes (no gauge link means no process-dependence from that source).
The latter scenario has been observed by Metz in a model
calculation~\cite{Metz:2002iz} and has subsequently been advocated
by Metz and Collins in Ref.~\cite{Collins:2004nx}. In the
next chapter the discussion on universality will be continued.

\subsubsection{The parametrizations}

Transverse moments of fragmentation functions are often encountered. For
a function $D_i (z,z^2 k_T^2)$, it is defined as
\vspace{-.2cm}
\begin{equation}
D_i^{(1)} (z) \equiv z^2 \int \rmd{2}{k_T}
\frac{\threeVec{k}_T^2}{2M_h^2}\ D_i(z,z^2k_T^2).
\end{equation}

The correlator is decomposed as (see also
Ref.~\cite{Levelt:1994np,Boer:1997mf,Bacchetta:2004zf,Goeke:2005hb})
\myBox{
\small
\vspace{-.2cm}
\begin{align}
\Delta^{[\pm]}(z^{-1},\hspace{.06cm}&\hspace{-.06cm}k_T,P_h,S_h)= 
\Delta^{[\pm,\text{T-even}]}(z^{-1},k_T,P_h,S_h) + 
\Delta^{[\pm,\text{T-odd}]}(z^{-1},k_T,P_h,S_h) 
\nonumber\\
\Delta^{[\pm,\text{T-even}]}&(z^{-1},k_T,P_h,S_h) \nonumber\\
&= \begin{alignedat}[t]{2}
                   & z \left( D_1^{[\pm]}\ \slashii{n}_-
                     + \left( S_{hL}\ G_{1L}^{[\pm]}\ -
            \tfrac{k_T\cdot {S_h}_T}{M_h}\  G_{1T}^{[\pm]} \right)
                       \gamma_5 \slashii{n}_- \right)
                     &\ & (\text{twist 2})\\
                   &+ z \left(
                   H_{1T}^{[\pm]}\ \gamma_5 \slashi{{S_h}_T} \slashii{n}_-
                     {+} \left( S_{hL}\ H_{1L}^{\perp [\pm]} {-}
                         \tfrac{k_T\cdot {S_h}_T}{M_h}\ H_{1T}^{\perp [\pm]} \right)
                       \frac{\gamma_5 \slashi{k}_T \slashii{n}_-}{M_h} \right)
                     && (\text{twist 2})\\
                   &+ \left( \frac{zM_h}{P_h^-} \right) \left(
                     E^{[\pm]} + D^{\perp [\pm]}\ \frac{\slashi{k}_T}{M_h} \right)
                     && (\text{twist 3, unpolarized})\\
                   &+ \left( \frac{zM_h}{P_h^-} \right) \left(
                    \left( S_{hL} G_{L}^{\perp [\pm]} {-}
                        \tfrac{k_T\cdot {S_h}_T}{M_h}\ G_{T}^{\perp [\pm]}  \right)
                            \frac{\gamma_5 \slashi{k}_T}{M_h}
                         {+}   G_T^{'[\pm]}\ \gamma_5 {\slashi{S}_h}_T
                            \right)
                     && (\text{twist 3, polarized})\\
                   &+ \left( \frac{zM_h}{P_h^+} \right)
                   \left(
                     H_T^{\perp [\pm]}\ \frac{ \gamma_5 [ {\slashi{S}_h}_T, \slashi{k}_T ]}
                     {2M_h}\right) && ( \text{twist 3, polarized})\\
                   &+ \left( \frac{zM_h}{P_h^-} \right)
                      \left( S_{hL}\ H_{L}^{[\pm]}\ -
                          \tfrac{k_T\cdot {S_h}_T}{M_h}\  H_{T}^{[\pm]} \right)
                     \frac{\gamma_5 [ \slashii{n}_-, \slashii{n}_+ ]}{2}
                     && ( \text{twist 3, polarized})\\
                   &+\text{higher twist},
                   &&
                   \end{alignedat}
                   \nonumber\\
\Delta^{[\pm,\text{T-odd}]}&(z^{-1},k_T,P_h,S_h) \nonumber\\
&= \begin{alignedat}[t]{2}
                   & z \left( D_{1T}^{\perp [\pm]}\
                     \frac{ \epsilon_{T}^{\mu\nu} {k_T}_\mu {S_{hT}}_\nu
                     \slashii{n}_- }{M_h} +
                     H_1^{\perp [\pm]}\ \frac{i \slashi{k}_T \slashii{n}_- }{M_h} \right)
                     && (\text{twist 2})\\
                   &+ \left( \frac{zM_h}{P_h^-} \right)
                     \left( H^{[\pm]}\ \frac{i [ \slashii{n}_-, \slashii{n}_+ ]}{2}
                     + G^{\perp [\pm]}\ \frac{\epsilon_{T}^{\mu\nu} {k_{T}}_\mu
                       \gamma_\nu \gamma_5}{M_h} \right)
                     &\qquad \quad \ \ & (\text{twist 3, unpolarized})\\
                   & +\left( \frac{zM_h}{P_h^-} \right) \left(
                     \frac{\epsilon_T^{\mu\nu} {k_T}_\mu {S_{hT}}_\nu }{M_h}
                     E_T^{\perp[\pm]} +
                     D_T^{[\pm]}\ \epsilon_{T}^{\mu\nu} \gamma_\mu {S_h}_\nu
                     \right)
                     && (\text{twist 3, polarized})\\
                   & +\left( \frac{zM_h}{P_h^-} \right) \left( \big(
                     S_{hL}\ D_L^{\perp [\pm]}
                     - \tfrac{k_T\cdot S_{hT}}{M_h}\ D_T^{\perp [\pm]} \big)
                      \frac{ \epsilon_{T}^{\mu\nu}
                     \gamma_\mu {k_T}_\nu}{M_h} \right)
                     && (\text{twist 3, polarized})\\
                   &-\left( \frac{zM_h}{P_h^-} \right)
                     \left( S_{hL}\ E_{L}^{[\pm]} -
                          \tfrac{k_T\cdot {S_h}_T}{M_h}\  E_{T}^{[\pm]} \right) i \gamma_5
                     && (\text{twist3, polarized})\\
                   & + \text{higher twist}. &&
                   \end{alignedat}
\raisetag{13pt}
\label{unintFrag}
\end{align}
\vspace{-.5cm}
\begin{flushright}
\emph{parametrization of the fragmentation quark-quark correlator}
\end{flushright}
}
where all functions depend on $z$ and $z^2 k_T^2$. Also here a renormalization
scale is involved.
In order to address the universality issue, 
it is important to measure T-even and T-odd fragmentation
functions in different processes.
For fragmentation functions a set of Lorentz invariance
relations has also been put forward. For the validity of these relations, the same
issues as discussed in the previous section play a role.

\newpage

\section{Summary and conclusions}

We introduced several Cartesian bases and
expressed head-on cross sections
in terms of Lorentz-invariants. It was shown how
frame-inde\-pen\-dent observables can be defined, simplifying
comparisons between theoretical predictions,
experimental observations, and
fixed frame definitions which are already
in use in the literature.

Two approaches were discussed
to access the parton distribution functions. The
first approach is the operator product expansion. Although this method has a firm
theoretical basis its applicability turns out to be limited. The other
approach, the diagrammatic expansion, is a field theoretical extension of the
parton model. Although slightly less rigorous than the operator product
expansion, it can be applied to most hadronic scattering processes
and agrees with the operator product expansion when applicable. The
diagrammatic approach will be used in the rest of the thesis.

It was indicated that an
infinite set of diagrams, appearing in the diagrammatic ex\-pan\-sion,
can be rewritten such that
transverse momentum
dependent correlators including a gauge link appear.
These correlators underlie the definition of parton distributions,
and provide important information on the partonic structure of hadrons.
The definition of these correlators contains
a bilocal operator and a gauge link
which ensures invariance under local color gauge transformations.
Since the path of the gauge link introduces a directional dependence,
new parton distribution
and fragmentation functions  were discovered at twist three
(for related work, see Goeke, Metz, Schlegel~\cite{Goeke:2005hb}).
A measurement of these new functions would contribute to the understanding of 
the theoretical description.

Using time-reversal, the functions can be divided in two classes,
called T-even and T-odd.
It was
shown that the first transverse moment of T-odd distribution functions
corresponds to the gluonic pole matrix element, which Qiu and Sterman suggested
to explain T-odd effects~\cite{Qiu:1991pp,Qiu:1991wg}.
As discovered by Collins~\cite{Collins:2002kn}, it
was also found that T-odd distribution
functions appear with opposite signs in the parametrization of correlators
which have a gauge link via plus or minus infinity.
In chapter~\ref{chapter4} we will see that
other link structures in correlators can appear, giving more complex factors.

For fragmentation functions
a universality problem was encountered due to the two possible mechanisms
to produce T-odd effects, the final-state interactions and the presence
of a gauge link.
This means that the value of unintegrated fragmentation
functions and their transverse moments
can be different for different experiments, forming a potential problem for
extracting transversity via the Collins effect. It is therefore important
to compare fragmentation functions which are measured in different
processes (for example their $z$-dependences).
If for instance one of the two effects is suppressed, a simple
sign relation (plus or minus) should appear
between functions which have gauge links via plus or minus infinity.
The discussion on the universality of fragmentation functions
will be continued in the next chapter.

% and
%this forms a potential problem  to
%measure the transversity function, $h_1$ by extracting in semi-inclusive DIS
%the product of $h_1$ and $H_1^\perp$ (Collins function) of which the
%latter can also be accessed separately in
%electron-positron annihilation. In the next chapter we
%will see that the gauge link in the fragmentation functions in semi-inclusive
%DIS and electron-positron annihilation are not the same, meaning that the
%Collins effect in electron-positron annihilation
%would be different from the Collins effect in semi-inclusive
%DIS. This particular suggestion to extract transversity might therefore be
%complicated,
%but the measurements will remain very interesting. One could for instance verify
%whether the $z$-dependence of the Collins function in semi-inclusive DIS
%compared to the Collins function in
%electron-positron annihilation has a similar behavior. If it would, then
%that would be a clear indication for universality up to a factor.
%A discussion on the universality of fragmentation functions
%will be given in the next chapter.

\newpage

\begin{subappendices}

\section{Outline of proof of Eq.~\ref{amplitudeSIDIS}\label{eersteAfleiding}}

For the interested reader a derivation of Eq.~\ref{amplitudeSIDIS}
will be presented here. The derivation is based on section 4.2 and 7.2
of Peskin and Schroeder~\cite{Peskin:1995ev}.

The invariant amplitude can be obtained via the S-matrix
\begin{align}
S &\equiv \textbf{1} + i T, \label{theory233}\\
{}_\text{free}\langle P_X, P_h, l' | iT | P,l \rangle_\text{free}
 &\equiv (2\pi)^4 \delta^4 \left(
l+P-l'-P_h-P_X \right) i \mathcal{M}, \label{iM}
\end{align}
and the \emph{LSZ reduction formula}
\begin{equation}
\begin{split}
&{}_\text{free}\langle P_X, P_h, l' | S | P,l \rangle_\text{free}
\\
&\times \frac{i Z_e}{l^2 - m_e^2 + i \epsilon}
\frac{i Z_P}{P^2 - M^2 + i \epsilon}
\frac{i Z_e}{l'^2 - m_e^2 + i \epsilon}
\frac{i Z_h}{P_h^2 - M_h^2 + i \epsilon}
\frac{i Z_X}{P_X^2 - M_X^2 + i \epsilon}
\\
&\eqnIndent \thicksim
\int \rmd{4}{x_1} \rmd{4}{x_2} \rmd{4}{y_1}\rmd{4}{y_2}\rmd{4}{y_3}
e^{i( l'y_1 + P_h y_2 + P_X y_3 - P x_1 - l x_2)}
\\
& \phantom{\eqnIndent \thicksim} \eqnIndent
\times
\langle \Omega | \mathcal{T}
\left[ \tfrac{\overline{u}(l',\lambda') \psi_e (y_1)}{2 m_e} \right]
\Phi_h (y_2) \Phi_X (y_3)
\left[ \tfrac{\bar{\psi}_P (x_1) U(P,S)}{2 M} \right]
\left[ \tfrac{\bar{\psi}_e (x_2) u(l,\lambda)}{2 m_e} \right] | \Omega \rangle,
\end{split}
\label{LSZ1}
\end{equation}
where the $Z_i$'s represent the field-strength renormalizations,
$| \Omega \rangle $ represents the physical vacuum,
$\mathcal{T}$ is the time-ordering operator, and $ \thicksim $ means that the
two expressions agree in the
vicinity of the poles. These poles are generated by
the unbounded interval of the integrals and correspond to the asymptotic
incoming and outgoing states.
The task is now to work out the right-hand-side and to identify these poles.

The time-ordered product can be calculated in the interaction picture. In the
interaction picture the Heisenberg fields are decomposed
in creation and annihilation operators
at some point in
time, $t_0$, which also defines the \emph{quantization plane}.
Evolving these operators in time by the Hamiltonian defines the 
\emph{interaction picture fields}. Since
the creation and annihilation operators can be interpreted as
 asymptotic states it
is a priory not clear whether this expansion is valid for quarks and gluons
in QCD. Since our final result in this section does not depend on the explicit 
expansion of these creation and annihilation operators, 
we will skip here this technical point. Assuming the vacuum
structure of QCD to be simple and indicating the fields in the interaction picture
with a superscript $I$, we obtain
\begin{equation}
\begin{split}
\text{Eq.~\eqref{LSZ1}} & \thicksim 
\lim_{\substack{T \rightarrow \infty(1-i \epsilon)}}
\int \rmd{4}{x_1} \rmd{4}{x_2} \rmd{4}{y_1}\rmd{4}{y_2}\rmd{4}{y_3}
e^{i( l'y_1 + P_h y_2 + P_X y_3 - P x_1 - l x_2)}
\\
&\eqnIndent \times \langle 0 | \mathcal{T} 
\left[ \tfrac{\overline{u}(l',\lambda') \psi^I_e (y_1)}{2 m_e} \right]
\Phi^I_h (y_2) \Phi^I_X (y_3) U(T,-T)
\left[ \tfrac{\bar{\psi}^I_P (x_1) U(P,S)}{2 M} \right]
\left[ \tfrac{\bar{\psi}^I_e (x_2) u(l,\lambda)}{2 m_e} \right] | 0 \rangle
\\
& \eqnIndent \times
\left[ \langle 0 | U(T,-T) | 0 \rangle \right]^{-1} ,
\end{split}
\raisetag{12pt}
\label{lsz2}
\end{equation}
where
\begin{align}
\Phi^I_i (x) &\equiv U(x^0,t_0)\ \Phi(x)\ U^\dagger(x^0,t_0),\\
U(t_b,t_a) &\equiv \mathcal{T} \exp \left[ -i \int_{t_a}^{t_b} \rmd{1}{t'} 
H_I (t') \right], \label{theory67}
\end{align}
 and where $H_I$ is the interacting part of the Hamiltonian in the interaction picture,
and $|0\rangle $ is the bare vacuum.
The interaction Hamiltonian contains the
electromagnetic and QCD interactions. The electromagnetic
interactions can be treated in perturbation
theory while the QCD interactions in general cannot.

To produce nonforward matrix elements there should be at least two
electromagnetic interactions,
once between the incoming and outgoing electron and once between the
incoming and outgoing hadrons, giving
\begin{equation}
\begin{split}
&
{}_\text{free}\langle P_X, P_h, l' | iT | P,l \rangle_\text{free}
\\
& \phantom{=}\times \frac{i Z_e}{l^2 - m_e^2 + i \epsilon}
\frac{i Z_P}{P^2 - M^2 + i \epsilon}
\frac{i Z_e}{l'^2 - m_e^2 + i \epsilon}
\frac{i Z_h}{P_h^2 - M_h^2 + i \epsilon}
\frac{i Z_X}{P_X^2 - M_X^2 + i \epsilon} 
\\
& \thicksim
\int \rmd{4}{x_1} \rmd{4}{x_2} \rmd{4}{y_1} \rmd{4}{y_2} \rmd{4}{y_3}
e^{i( l'y_1 + P_h y_2 + P_X y_3 - P x_1 - l x_2)}(-i)^2
\lim_{T\rightarrow \infty (1-i \epsilon)}
 \int \rmd{4}{z_1} \rmd{4}{z_2}
 \\
& \eqnIndent
\times 
\langle 0 | \mathcal{T} 
\left[ \tfrac{\overline{u}(l',\lambda') \psi^I_e (y_1)}{2 m_e} \right]
\Phi^I_h (y_2) \Phi^I_X (y_3) U(T,z_1) A^I_\rho (z_1) e J_{e,I}^\rho (z_1)
U(z_1,z_2) A^I_\sigma (z_2)
\\
& \eqnIndent
\times e J_{q,I}^\sigma (z_2) U(z_2,-T)
\left[ \tfrac{\bar{\psi}^I_P (x_1) U(P,S)}{2 M} \right]
\left[ \tfrac{\bar{\psi}^I_e (x_2) u(l,\lambda)}{2 m_e} \right] | 0 \rangle
\left[ \langle 0 | U(T,-T) | 0 \rangle  \right]^{-1},
\label{lsz3}
\end{split}
\raisetag{16pt}
\end{equation}
where $J_{e/q,I}^\rho (z) = Q_{e/q}\  \overline{\psi}_{e/q}^I (z) \gamma^\rho \psi_{e/q}^I (z)$. Using
Wick's theorem the lowest nontrivial order in $e$ 
can be worked out, giving explicitly
\begin{align}
&\text{Eq.~\eqref{lsz3}}
\nonumber\\
&{=}
\int \rmd{4}{x_1} \rmd{4}{x_2} \rmd{4}{y_1} \rmd{4}{y_2} \rmd{4}{y_3}
e^{i( l'y_1 + P_h y_2 + P_X y_3 - P x_1 - l x_2)} (-i)^2
\lim_{T \rightarrow \infty (1- i \epsilon)} \int \rmd{4}{z_1} \rmd{4}{z_2} 
\nonumber \nopagebreak\\
& \phantom{=} \times
\langle 0 | \mathcal{T}
\left[ \tfrac{\overline{u}(l',\lambda') \psi^I_e (y_1)}{2 m_e} \right]
e J_{e,I}^\rho (z_1)
\left[ \tfrac{\bar{\psi}^I_e (x_2) u(l,\lambda)}{2 m_e} \right]
 | 0 \rangle
\langle 0 | \mathcal{T} A^I_\rho (z_1) A^I_\sigma (z_2)
| 0 \rangle  
\nonumber \nopagebreak\\
&\phantom{=} \times
\langle 0 | \mathcal{T} 
\Phi^I_h (y_2) \Phi^I_X (y_3) U(T,z_1)
U(z_1,z_2) e J_{q,I}^\sigma (z_2) U(z_2,-T)
\left[ \tfrac{\bar{\psi}^I_P (x_1) U(P,S)}{2 M} \right]
 | 0 \rangle 
\nonumber \nopagebreak\\
& \phantom{=}
 \times \left[ \langle 0 | U(T,-T) | 0 \rangle \right]^{-1} + 
\mathcal{O}(e^3)
\nonumber \displaybreak[0]\\
& 
{=}  \frac{i}{l^2 - m_2^2 + i \epsilon}
\frac{i}{{l'}^2 - m_2^2 + i \epsilon}(-ie) 
\overline{u}(l',\lambda') \gamma_\rho
u(l,\lambda)
\int \rmd{4}{x_1} \rmd{4}{y_2}\rmd{4}{y_3}
e^{i( P_h y_2 + P_X y_3  - P x_1)}
\nonumber\\
& \phantom{\eqnIndent =} \times (-i)
\lim_{\substack{T\rightarrow \infty(1-i\epsilon)}} 
\int \rmd{4}{z_2} e^{-iq z_2} \frac{-i}{q^2}
\langle 0 | \mathcal{T} 
\Phi^I_h (y_2) \Phi^I_X (y_3) U(T,z_1)
U(z_1,z_2) e J_I^\sigma (z_2) 
\nonumber \nopagebreak\\
& \phantom{\eqnIndent =} \times  U(z_2,-T)
\left[ \tfrac{\bar{\psi}^I_P (x_1) U(P,S)}{2 M} \right]
 | 0 \rangle
\left[ \langle 0 | U(T,-T) | 0 \rangle \right]^{-1} +
\mathcal{O}(e^3).
\label{lsz4}
\end{align}
In order to obtain the singularities in the hadron energies it is assumed
 that the incoming
hadron only starts interacting after some point in time, $t_i$,
and  that
the outgoing hadrons are well-separated wave-packets
after some point in time called $t_f$. 
These assumptions, known as 
\emph{adiabatically switching on and off the
interactions}, restricts $z_2^0$ to be $t_i < z_2^0 < t_f$.
By doing so, the
poles of the 
incoming nucleon and outgoing hadrons can be identified 
after which the limits
$t_i \rightarrow -\infty$ and $t_f \rightarrow \infty$ can be taken. 
One obtains
\begin{align}
\text{Eq.~\eqref{lsz4}}&
=  \frac{i}{l^2 - m_2^2 + i \epsilon}
\frac{i}{{l'}^2 - m_2^2 + i \epsilon}(-ie) \overline{u}(l',\lambda') \gamma_\rho
u(l,\lambda) \frac{-i}{q^2}
\nonumber \\
&
\phantom{\thicksim} \times
\int \rmd{4}{x_1} \rmd{4}{y_2}\rmd{4}{y_3}
e^{i( P_h y_2 + P_X y_3  - P x_1)} (-i)
\lim_{\substack{t_i \rightarrow -\infty \\ t_f \rightarrow \infty}}
\lim_{T\rightarrow(\infty - i \epsilon)}
\int \rmd{4}{z_2} e^{-iq z_2} 
\nonumber\\
& \phantom{\thicksim} \times
\langle 0 | \mathcal{T} 
\Phi^I_h (y_2) \Phi^I_X (y_3) U(T,t_f) \bigg[
U(t_f,t_i) e J_I^\sigma (z_2) \bigg]   U(t_i,-T)
\left[ \tfrac{\bar{\psi}^I_P (x_1) U(P,S)}{2 M} \right]
 | 0 \rangle
\nonumber\\
& \phantom{\thicksim} \times
\left[ \langle 0 | U(T,-T) | 0 \rangle \right]^{-1} + 
\mathcal{O}(e^3)
\nonumber \displaybreak[0]\\
&=
\frac{i}{l^2 - m_2^2 + i \epsilon}
\frac{i}{{l'}^2 - m_2^2 + i \epsilon}
\frac{iZ_P}{P^2 - M^2 + i \epsilon}
\frac{iZ_X}{P_X^2 - M_X^2 + i \epsilon}
\frac{iZ_h}{P_h^2 - M_h^2 + i \epsilon}
\nonumber \\
&\phantom{\thicksim} \times (-ie)
\overline{u}(l',\lambda') \gamma_\rho
u(l,\lambda) (-i)
\lim_{\substack{t_i \rightarrow -\infty \\ t_f \rightarrow \infty}}
\int \rmd{4}{z_2} e^{-iq z_2} \frac{-i}{q^2}
\nonumber\\
& \phantom{\thicksim} \times
{}_\text{out}\langle P_h,P_X | \mathcal{T}
U(t_0,t_f) \bigg[
U(t_f,t_i) e J_I^\sigma (z_2) \bigg]   U(t_i,t_0) | P,S \rangle_\text{in,c}
+ \mathcal{O}(e^3).
\end{align}
Note that all the necessary poles
including the field-strength renormalizations
 at order $e^2$ (for nonforward matrix elements
$Z_e = 1$ at this order) have been produced. 
After shifting the current, $J^\sigma (z_2)$,
yielding the delta-function which expresses momentum conservation,
we can read off the invariant amplitude
$\mathcal{M}$ from Eq.~\ref{iM},
giving
\begin{equation}
\begin{split}
i \mathcal{M} &= (-ie)\overline{u}(l',\lambda') \gamma_\rho
u(l,\lambda) \\
& \phantom{=} \times
\lim_{\substack{t_i \rightarrow -\infty \\ t_f \rightarrow \infty}}
\frac{-i}{q^2}
{}_\text{out}\langle P_h, P_X | U(t_0,t_f) \bigg[ U(t_f,t_i)
(-ie)J_I^\rho (0) \bigg] U(t_i,t_0) | P,S \rangle_\text{in,c} + \mathcal{O}(e^3)
\\
&= (-ie)\ \overline{u}(l',\lambda') \gamma_\rho
u(l,\lambda)\
\frac{-i}{q^2}\
{}_\text{out}\langle P_h, P_X | 
(-ie)J^\rho (0) | P,S \rangle_\text{in,c} + \mathcal{O}(e^3).
\end{split}
\raisetag{18pt}
\end{equation}
Note that the current, $J^\rho$, is here in the Heisenberg picture.

\section{The diagrammatic expansion\label{theoryAppendix2}}

For the interested reader
a short outline of the derivation of two diagrams in the
diagrammatic approach will be given here. It
will be pointed out how the results can be generalized.

Following the same procedure as applied for semi-inclusive DIS
in the previous appendix,
 one can derive that the invariant amplitude for general
processes reads
\begin{equation}
\langle \text{free states }P | i T | \text{free states }Q \rangle
{=}
\lim_{\substack{
t_i \rightarrow -\infty \\ t_f \rightarrow \infty }}
{}_\text{out}\langle P | U(t_0,t_f)
\biggl[ U(t_f,t_i) \biggr]
 U(t_i,t_0) | Q \rangle_\text{in,c},
\label{hier1}
\end{equation}
where the $U$'s are defined in Eq.~\ref{theory67}.
When applying the
diagrammatic expansion one expands the bracketed term.

At tree-level the result is expressed in terms of fields inside matrix elements.
These fields can always be transported to the origin, yielding
momentum
conservation for the incoming and outgoing particles. The lowest order result
is the tree-level squared
Feynman diagram with spinors or polarization vectors
replaced by Heisenberg fields in matrix elements, multiplied by
a delta-function expressing momentum conservation.

\begin{figure}
\begin{center}
\begin{tabular}{cp{.5cm}c}
\includegraphics[width=5cm]{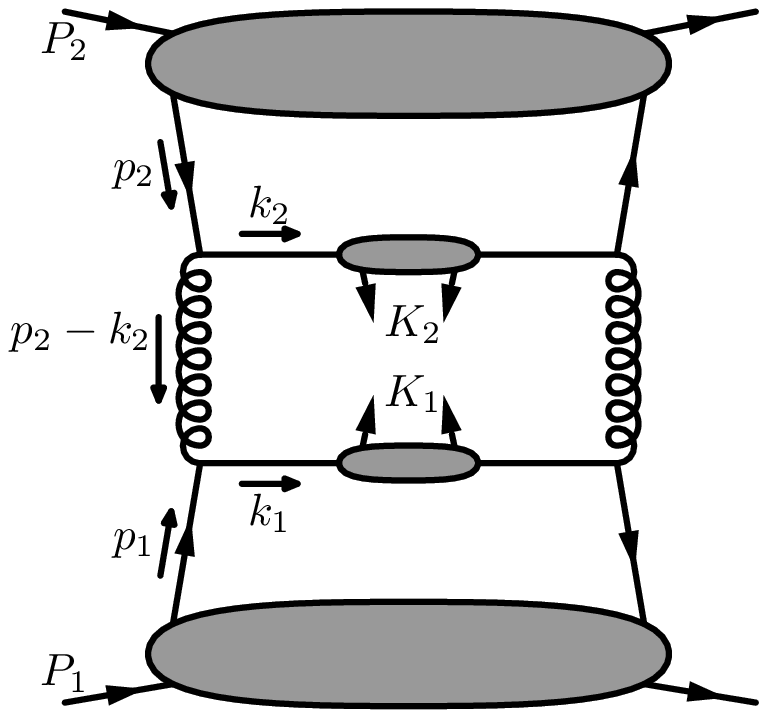}  & &
\includegraphics[width=5cm]{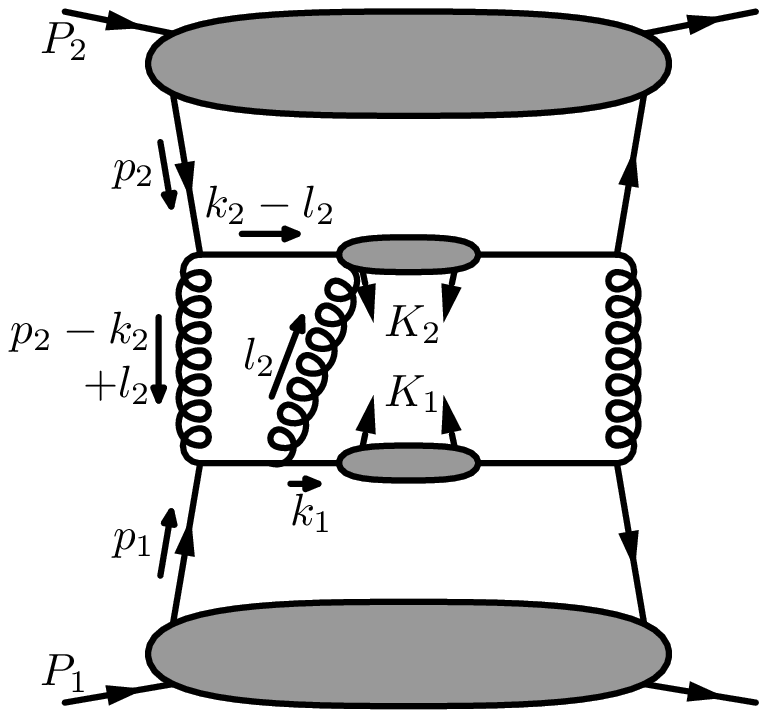}\\
(a) & &(b)
\end{tabular}
\end{center}
\caption{Two contributions which appear in the diagrammatic approach.
\label{theoryApp3}}
\end{figure}

For example, consider hadron-hadron scattering with momenta $P_1$ and
$P_2$ producing two hadrons
with momenta $K_1$ and $K_2$
 approximately
back-to-back in the azimuthal plane.
It is assumed that the outgoing hadrons are well separated from
the incoming hadrons. If this is not the case then fracture functions
are needed to describe
the combined process of distribution and fragmentation.
Although the incoming
and outgoing partons
can be quarks or gluons, we will restrict ourselves
to (anti)quarks as external partons, the gluons can be included later on.
In that case the
incoming quarks or antiquarks interact via a gluon giving
quarks and antiquarks in the final state. Expanding
the bracketed term of Eq.~\ref{hier1} to lowest order
this contribution is
\begin{align}
&\hspace{-.2cm}
\langle K_1, K_2, X | iT | P_1, P_2 \rangle \nonumber\\
&\hspace{-.2cm}
=\! (ig)^2 \!\!\!
\int\!\! \rmd{4}{x}\! \rmd{4}{y}
{}_\text{out}\langle K_1, K_2, X |
\mathcal{T} \bar{\psi} (x) \slashii{A}(x) \psi(x)
\bar{\psi} (y) \slashii{A}(y) \psi(y) | P_1, P_2 \rangle_\text{in,c}
{+} \mathcal{O}(g^3).
\label{hier2}
\end{align}
Overall momentum conservation can be simply obtained by making an overall
shift of $x$ and subsequently redefining $y$. One finds
\vspace{-.1cm}
\begin{multline}
\langle K_1, K_2, X | iT | P_1, P_2 \rangle
=
\int \rmd{4}{y}
{}_\text{out}\langle K_1, K_2, X |
\mathcal{T} \bar{\psi} (0) \slashii{A}(0) \psi(0)
\bar{\psi} (y) \slashii{A}(y) \psi(y) | P_1, P_2 \rangle_\text{in,c}
\\
\vspace{-.2cm}
\times
(2\pi)^4 \delta^4 \big(P_1 {+} P_2 {-} K_1 {-} K_2 {-} \sum_i P_{X_i} \big)
+ \mathcal{O}(g^3).
\label{hier3}
\end{multline}
Since no gluons appear in the initial or final state,
those fields need to be contracted with each other yielding the gluon
propagator
(Wick's theorem).
Each of the quark-fields can be connected to one of the incoming
or outgoing jets. All
possibilities should be included. Picking one of the contributions
as displayed in Fig.~\ref{theoryApp3}a
and working out the contractions gives
($k_1 \equiv P_{X_3}{+}K_1$, $k_2 \equiv P_{X_4}{+}K_2$, 
$p_i \equiv P_i {-} P_{x_i}$)
%\begin{equation}
%\begin{split}
%\text{Contribution}&\text{ from Fig.~\eqref{theoryApp3}a}
%\\
%&=
%\int \rmd{4}{x} \rmd{4}{y} \frac{\rmd{4}{l}}{(2\pi)^4} e^{-il(x-y)}
%\frac{-ig^2}{l^2 + i \epsilon} \\
%& \eqnIndent \times
%{}_\text{out}\langle K_1, P_{X3} | \bar{\psi}(x) | \Omega \rangle_\text{c}\
%(it_b \gamma^\mu) \
%{}_\text{out}\langle P_{X1} | \psi(x) | P_1 \rangle_\text{in,c} \\
%&\eqnIndent \times \
%{}_\text{out}\langle K_2, P_{X4} | \bar{\psi}(y) | \Omega \rangle_\text{c}\
%(it_b \gamma_\mu) \
%{}_\text{out}\langle P_{X2} | \psi(y) | P_2 \rangle_\text{in,c}.
%\end{split}
%\label{hier3}
%\end{equation}
%By shifting the matrix elements one obtains explicit momentum conservation
\begin{equation}
\begin{split}
\text{Eq.~\eqref{hier3}} =&
(2\pi)^4 \delta^4 \big(P_1 {+} P_2 {-} K_1 {-} K_2 {-} \sum_i P_{X_i} \big)
\frac{-ig^2}{(p_2-k_2)^2 + i \epsilon}
\\
& \eqnIndent \times
{}_\text{out}\langle K_1,P_{X3} | \bar{\psi}(0) | \Omega \rangle_\text{c}\
(it_b \gamma^\mu)  \
{}_\text{out}\langle P_{X1} | \psi(0) | P_1 \rangle_\text{in,c}
\\
& \eqnIndent \times
{}_\text{out}\langle K_2, P_{X4} | \bar{\psi}(0) | \Omega \rangle_\text{c}\
(it_b \gamma_\mu ) \
{}_\text{out}\langle P_{X2} | \psi(0) | P_2 \rangle_\text{in,c}.
\end{split}
\label{hier4}
\raisetag{12pt}
\end{equation}

If more interactions are included these steps can be performed
in the same manner. Each interaction in $U$
consists of fields which are fully
integrated over their independent coordinate variables. 
As an example we will consider one higher order
contribution which will yield Fig.~\ref{theoryApp3}b. The interaction 
can be expressed as
\begin{multline}
\langle K_1, K_2, X | iT | P_1, P_2 \rangle 
\\
= 
(ig)^3 \int \rmd{4}{z} \rmd{4}{y}
{}_\text{out}\langle K_1, K_2, X |
\mathcal{T} 
\bar{\psi} (z) \slashii{A}(z) \psi(z)\bar{\psi} (0)
\slashii{A}(0) \psi(0)
\bar{\psi} (y) \slashii{A}(y) \psi(y) | P_1, P_2 \rangle_\text{in,c}
\\
\times
(2\pi)^4 \delta^4(P_1 {+} P_2 {-} K_1 {-} K_2 {-} \sum P_{X_i})
 + \ldots,
\label{dinsdag11}
\end{multline}
where the dots denote contributions from other interactions.
We need to include all possible contractions but we will consider now one
of the possibilities. Contracting $\psi(z)$ with $\bar{\psi}(0)$ and
$A(0)$ with $A(y)$ and considering the remaining fields to be connected to the
jets in a certain way, we find
\begin{equation}
\begin{split}
&
\text{One contribution of Eq.~\eqref{dinsdag11}}
\\
&
=
(2\pi)^4 \delta^4(P_1 {+} P_2 {-} K_1 {-} K_2 {-} \sum P_{X_i})
g^3 \int \rmd{4}{z} \rmd{4}{y} \frac{\rmd{4}{l_1} \rmd{4}{l_2}}{(2\pi)^8}
e^{il_1y} e^{-il_2z} \frac{-i}{l_1^2}
\\
& 
\times
{}_\text{out}\langle K_1,P_{X_3}| \bar{\psi}(z) | \Omega \rangle_\text{c}
i \gamma_\alpha t_a \frac{i(\slashi{l}_2 + m)}{l_2^2 - m^2 + i \epsilon}
i \gamma^\mu t_b 
{}_\text{out}\langle P_{X_1} | \psi(0) | P_1 \rangle_\text{in,c}
\\
& 
\times
{}_\text{out}\langle K_2,P_{X_4}| \bar{\psi}(y) A^\alpha_a(z) | \Omega 
\rangle_\text{c} i \gamma_\mu t_b 
{}_\text{out}\langle P_{X_2} | \psi(0) | P_2 \rangle_\text{in,c}
\end{split}
\label{hier6}
\end{equation}
We would like the
fields which already appeared at ``tree-level''
to be at the origin. Shifting
those fields and redefining the other integration variables, one finds
\begin{equation}
\begin{split}
\text{Eq.~\eqref{hier6}} =&
(2\pi)^4 \delta^4 \big(P_1 {+} P_2 {-} K_1 {-} K_2 {-} \sum_i P_{X_i}\big)
g^3 \int \rmd{4}{z}  \frac{\rmd{4}{l_2}}{(2\pi)^4}
e^{-il_2z} \frac{1}{(l_2{+}p_2{-}k_2)^2}
\\
& 
\times
{}_\text{out}\langle K_1,P_{X_3}| \bar{\psi}(0) | \Omega \rangle_\text{c}
i \gamma_\alpha t_a 
\frac{i(\slashi{l}_2 + \slashi{k}_1 + m)}{(l_2{+}k_1)^2 {-} m^2 + i \epsilon}
i \gamma^\mu t_b 
{}_\text{out}\langle P_{X_1} | \psi(0) | P_1 \rangle_\text{in,c}
\\
& 
\times
{}_\text{out}\langle K_2,P_{X_4}| \bar{\psi}(0) A^\alpha_a(z) | \Omega \rangle_\text{c}
i \gamma_\mu t_b 
{}_\text{out}\langle P_{X_2} | \psi(0) | P_2 \rangle_\text{in,c}
\end{split}
\end{equation}
%\frac{-ig^2}{(P_2 - P_{X_2} - k_2)^2 + i \epsilon}\\
%&\times
%{}_\text{out}\langle K_1, P_{X3} | \bar{\psi}(0) | \Omega \rangle_\text{c}
%(it_b\gamma^\mu)
%{}_\text{out}\langle P_{X1} | \psi(0) | P_1 \rangle_\text{in,c}
%\\
%&\times
%\int \rmd{4}{z} \frac{\rmd{4}{l}}{(2\pi)^4}
%e^{-ilz}
%{}_\text{out}\langle K_2, P_{X4} | \bar{\psi}(0)A_a^\alpha(z) | \Omega
% \rangle_\text{c} (it_b\gamma_\mu)
%\\
%&\times
%\frac{i(\slashi{K}_1 + \slashi{P}_{X3} +\slashi{l}+m)}{(K_1 + P_{X3} + l)^2-m^2+ i \epsilon}
%ig(\gamma_\alpha t_a)
%{}_\text{out}\langle P_{X2} |
% \psi(0) | P_2 \rangle_\text{in,c}.
%\end{split}
%\end{equation}
The result is the original expression plus an additional interaction which
contains integrals over $z$ and $l_2$. We are actually deriving the Feyman
rules.
If we would have considered a virtual
correction there would have been an additional contraction and one would
remain with one integral over $l_2$.

In general one should have the following situation:
the tree-level result consists of momentum conservation and matrix elements
containing fields at the origin; if more external partons\footnote{
External partons are incoming or outgoing partons connected to correlators
or decaying as a jet.} are
present then the corresponding fields 
are integrated over their coordinates and momenta, and their
interactions are described by the Feynman rules.
Virtual corrections are similar
except that the integral over the coordinates can be carried out.

The cross section is obtained by removing the delta-function related
to momentum conservation, taking the square,
and multiplying with the removed delta-function
(see Eq.~\ref{cross}, \ref{theory233}, \ref{iM}).
Except for the phase-space
and flux factors this is the cross section.
In general the total momentum conservation can be rewritten by giving
each matrix element a number ($i$) and naming the sum of the outgoing momenta
(also $P_{Xi}$ here)
${K_i}_\text{out}$ and incoming momentum ${P_i}_\text{in}$.
The total momentum conservation can then be written as
\begin{multline}
(2\pi)^4 \delta^4 \left(\sum_{i=1} {K_i}_\text{out} - \sum_{j=1} {P_j}_\text{int}
\right)
= \prod_n \left[
\int \rmd{4}{t_n} \delta^4(t_n \pm {P_n}_\text{in} \mp {K_n}_\text{out})
\right]\\
\times (2\pi)^4 \delta^4(\text{external parton momenta}),
\end{multline}
where $t_i$ are the momenta of the external partons and where
the last delta-function expresses the momentum conservation of the
external partons.
The first $n$ delta-functions are used
to shift the fields of the matrix elements by introducing another
integral over the coordinates
\begin{equation}
\delta^4(t_n \pm {P_n}_\text{in} \mp {K_n}_\text{out})
= \int \frac{\rmd{4}{\xi_n}}{(2\pi)^4}
e^{i \left( t_n \pm {P_n}_\text{in} \mp {K_n}_\text{out} \right) \xi_n }.
\end{equation}
After an integration over the unobserved momenta one
obtains the correlators as introduced in section \ref{diagram}.

\newpage

\thispagestyle{empty}

\end{subappendices}

%\section*{contents}
%
%\begin{itemize}
%
%\item Warn the reader for boring stuff and send him to the next chapter if
%he does not like it.
%
%
%\item kinematics
%
%\item introduce leptonic and hadronic tensor. Show which asymmetries
%are possible and what angular dependences that gives.
%
%\item (LSZ reduction formulae) Operator product expansion and its limits
%
%\item diagrammatic approach derived for more complex processes,
%its approximations (factorization but not too much), LSZ reduction formulae,
%this should end with a list of rules for drawing diagrams (including
%blobs with more then two quark leggs)
%
%Give an explicit tree-level example of SIDIS in terms of PHI and Delta
%
%
%\item Introduce parton distributions and fragmentation functions and provide
%as much information as possible. Just mention the gauge link but not derive it.
%
%Lorentz-invariance relations and new functions which violate them.
%
%\end{itemize}

\chapter{Electromagnetic scattering processes at leading order in
$\boldsymbol{\alpha}_{\boldsymbol{S}}$\label{chapter3}}

\markboth{\thechapter\ Electromagnetic scattering processes at leading order in
$\alpha_S$}{}

\vspace{-.5cm}

The diagrammatic approach, introduced
in the previous chapter, will be employed
to describe at leading order in $\alpha_S$ cross sections
in which the hard scale is set by an electromagnetic interaction.
In the discussion on the
meaning of $\alpha_S$ it will be
argued that even at leading order there is an infinite amount of interactions
that should be considered.
Including next-to-leading order corrections in inverse powers of the hard
scale,
these leading-order-$\alpha_S$-interactions
will be evaluated in order to obtain the cross sections for semi-inclusive DIS,
electron-positron annihilation, and Drell-Yan
in terms of distribution and
fragmentation correlators. These correlators contain gauge links of which
their paths are different for different processes.
Using the parametrizations of these
correlators some explicit asymmetries for semi-inclusive DIS will be given.

In the second part of this chapter, the universality of
fragmentation functions
will be discussed and
the link structures appearing in deeply-virtual Compton
scattering will be studied. The calculation of the latter illustrates
the wide-ranging applicability of the diagrammatic
approach. A summary is provided at the end of this chapter.

\newpage

\section[Leading order in $\alpha_S$]
{Leading order in $\boldsymbol{\alpha}_{\boldsymbol{S}}$}

In the previous chapter the diagrammatic
expansion was developed to handle scattering
processes. In order to apply this expansion,
the various jets need to be
well separated. This can be further
translated into the demand to have large
momentum transfers, implying that the inner products of momenta connected to
different jets are large.
Besides an expansion in $\alpha_s$, this
presence of large energy-scales allows for a possible
expansion of physical observables in the inverse powers of that energy-scale
($M/Q$). Both expansions will be made
throughout the chapter.

\begin{figure}
\begin{center}
\begin{tabular}{cp{.001cm}cp{.001cm}c}
\includegraphics[width=3.3cm]{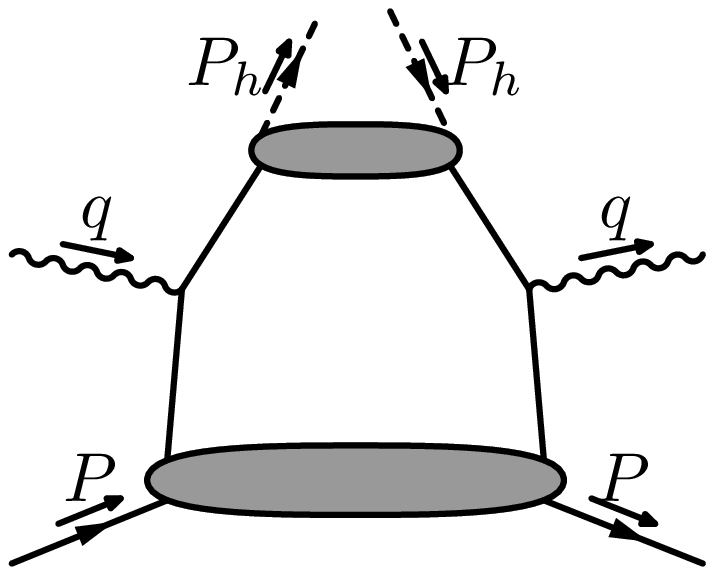} &&
\includegraphics[width=3.3cm]{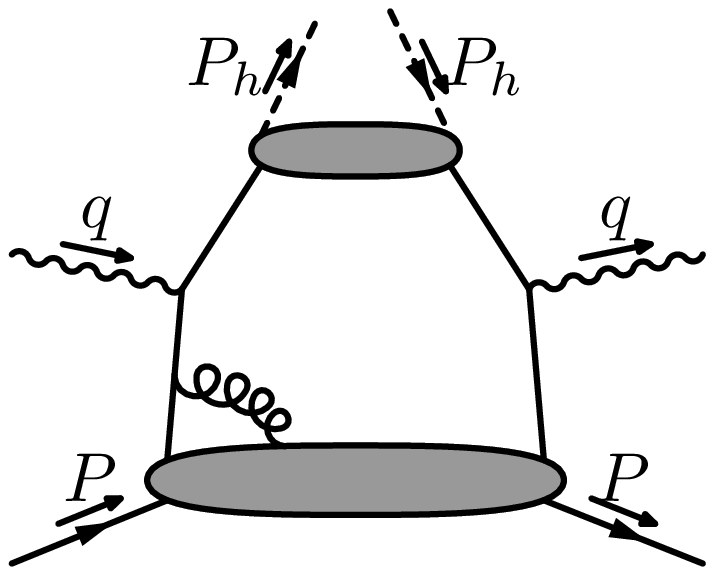} &&
\includegraphics[width=3.3cm]{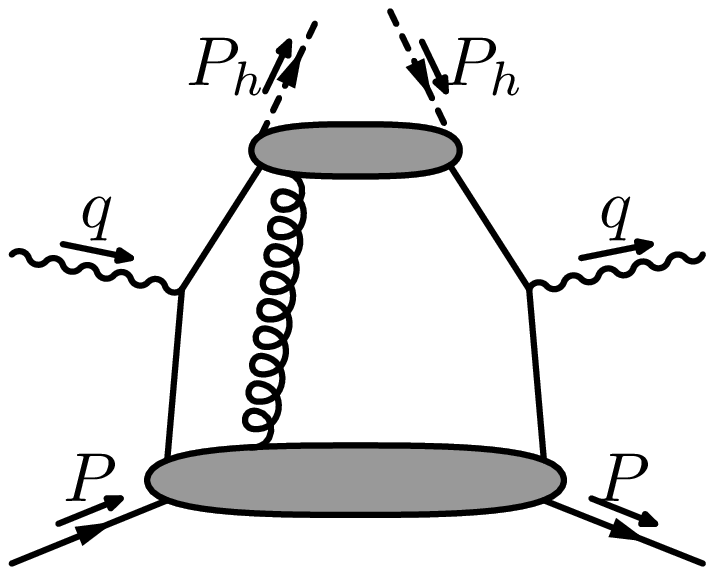}\\
(a) && (b) && (c)\\
\includegraphics[width=3.3cm]{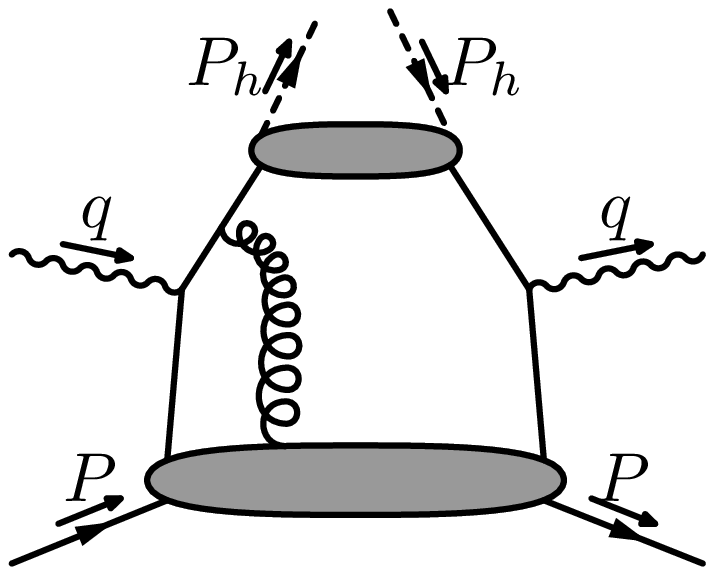} &&
\includegraphics[width=3.3cm]{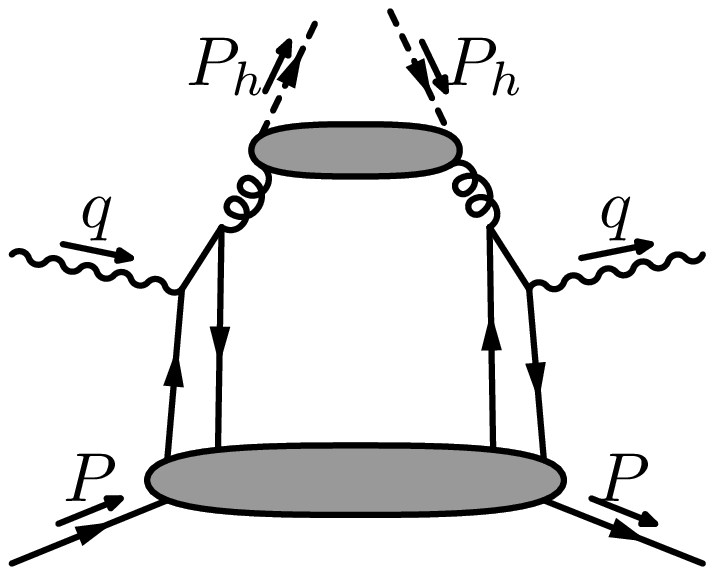} &&
\includegraphics[width=3.3cm]{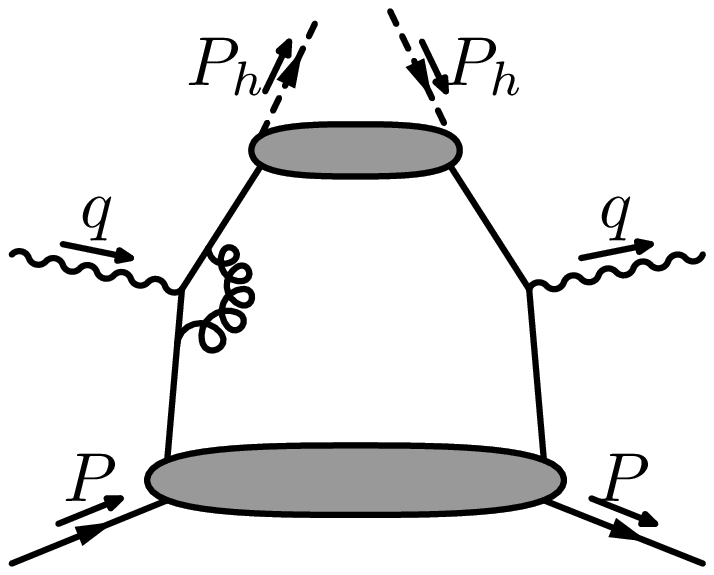}\\
(d) && (e) && (f)\\
\includegraphics[width=3.3cm]{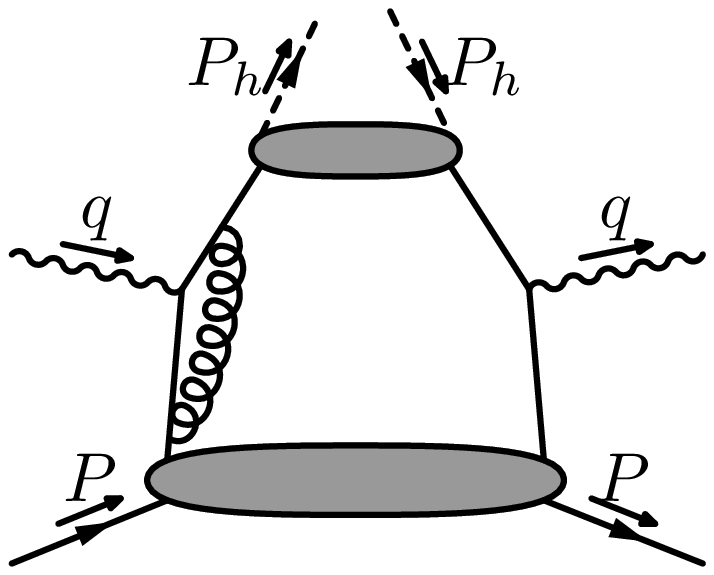} &&
\includegraphics[width=3.3cm]{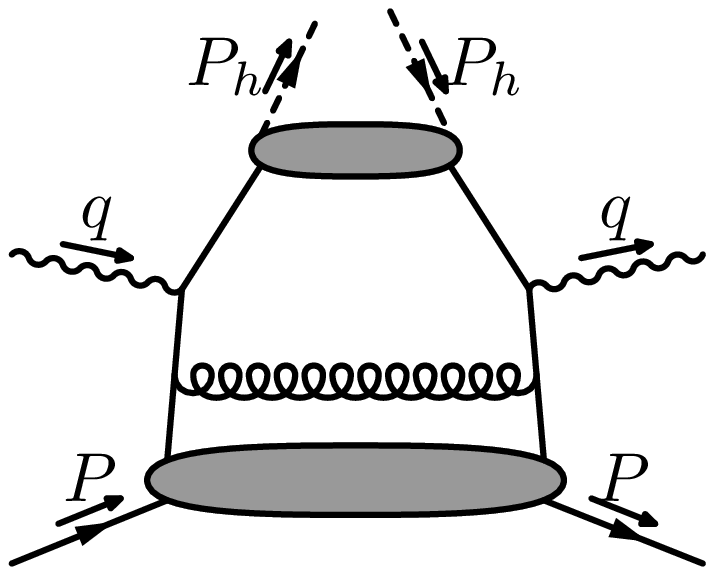} &&
\includegraphics[width=3.3cm]{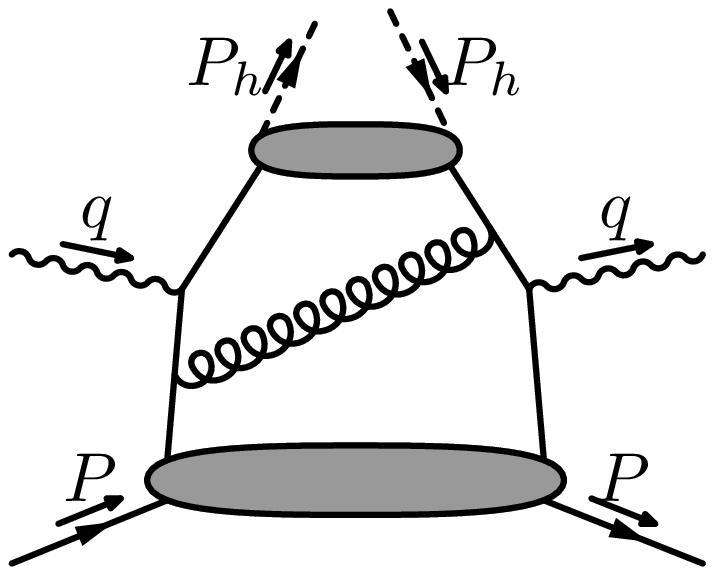}\\
(g) && (h) && (i)
\end{tabular}
\end{center}
\caption{Various interactions between the two jets and the
elementary scattering part in semi-inclusive DIS.
\label{quarkInter}}
\end{figure}

Applying the
diagrammatic expansion, all
leading $\alpha_S$ diagrams
which can contribute to the processes need to be included.
For semi-inclusive DIS, the lowest order (in $g$) diagram
is given in Fig.~\ref{quarkInter}a.
Interactions which can be absorbed, like the one in Fig.~\ref{quarkInter}b,
are
already present in the correlator definition and should not be included
separately.
Diagrams
which have several jets are $\alpha_s$
suppressed compared to diagrams which contain less jets.
The restriction
to leading order in $\alpha_s$
leads to only two outgoing jets for semi-inclusive DIS,
electron-positron annihilation, and
Drell-Yan. Further, the interactions
between the jets as illustrated in Fig.~\ref{quarkInter}c
are considered to be weak
and will be neglected. This assumption is directly
related to the question on whether factorization holds
for the correlators. This will be discussed in the next chapter.

Another possible contribution is that a
gluon-line will connect a jet with the elementary scattering
diagram. Since there can be interferences, such as depicted
in Fig.~\ref{quarkInter}d,
such an interaction is proportional to the coupling constant, $g$, and
is not $\alpha_s$ suppressed.
We shall see in the next section that one can resum
all collinear gluon-lines connecting the correlator and the scattering diagram to all
orders in $g$.
The other possibility
of additional collinear quarks, such as in Fig.~\ref{quarkInter}e,
is not taken into account here.
Ellis, Furmanski, and Petronzio~\cite{Ellis:1982wd,Ellis:1982cd}, found
that 
such contributions
are suppressed with $M^2/Q^2$ and are therefore beyond the aimed accuracy
of this chapter.

Besides additional parton-lines from the correlator one
might wonder whether there are other corrections in the scattering
diagram itself which need to be included at leading order $\alpha_s$.
Such corrections consist of loops and gluon radiation and
can partly be absorbed in the correlator definitions.
For instance, if the loop diagram in Fig.~\ref{quarkInter}f
contains a far off shell gluon, then it cannot be absorbed. However,
if the gluon is approximately on shell and collinear to one of the jets,
as illustrated in Fig.~\ref{quarkInter}g, then it
is absorbed in the correlator definition up to some scale. For gluon radiation
a similar situation applies.
If the gluon in Fig.~\ref{quarkInter}h has a small transverse momentum
with respect to the parent hadron ($P$),
then it is already present in the correlator definition. The
gluon in Fig.~\ref{quarkInter}i cannot be absorbed and represents
an $\alpha_S$ correction.

\sloppy
Absorbing interactions in the correlator introduces a
particular scale. There were a few attempts to calculate this
scale dependence (evolution equations),
but including effects from intrinsic
transverse momentum this has not been achieved
so far. For a discussion on the
scale dependence and various other difficulties
the reader is referred to
Ellis, Furmanski, Petronzio~\cite{Ellis:1982cd},
Collins, Soper, Sterman~\cite{Collins:1981tt,Collins:1982wa},
Hagiwara, Hikasa, Kai~\cite{Hagiwara:1982cq}, Ahmed,
Gehrmann~\cite{Ahmed:1999ix},
Boer~\cite{Boer:2001he}, and
Henneman~\cite{thesisAlex}.

\fussy
The conclusion of this section
is that, restricted to leading order in $\alpha_s$, one
must resum over all possible gluons which connect the correlators
and the scattering
diagram to obtain the cross section
(generalizations of Fig.~\ref{quarkInter}.d).
Such resummations will be performed
for several processes in this chapter.

\section{Semi-inclusive deep-inelastic scattering\label{hiro12}}

In the first two subsections
the expressions for the hadronic tensor including
$M/Q$ corrections will be derived. In the third subsection the
results for the hadronic tensor will be used to express some asymmetries in terms
of distribution and fragmentation functions. We will begin here by
introducing the calculation of the hadronic tensor.

Applying the general rules of the diagrammatic expansion, the hadronic
tensor is found to be a sum over various diagrams.
Considering only leading order in $\alpha_s$,
all those diagrams (to all orders in $g$) will be considered
that are not already absorbed in the correlators.
In order to structure the calculation we shall treat here
the lowest and one higher order contribution in $g$.
Those contributions in combination with a number of other diagrams will
be studied in detail in the next subsections.
The final result, Eq.~\ref{eq1},
will consist of correlators containing gauge links
for which parametrizations have been given in the previous chapter
(see Eq.~\ref{intDistr},~\ref{unintDistr},~\ref{intFrag},
~\ref{unintFrag}).

At lowest order in $g$ (see also Fig.~\ref{quarkInter}a), the
contribution to the hadronic tensor
is a convolution between the scattering
diagram, the fragmentation correlator, and the distribution correlator.
In this section arguments connected to the parent hadron will be often
omitted
for notational convenience, so $\Phi(p,P,S) \rightarrow \Phi(p)$.
The result for the hadronic tensor reads
\begin{equation}
2 M W^{\mu\nu} = \int \rmd{4}{p} \rmd{4}{k} \delta^4(p+q-k)
\tr^{D,C} \left[ \Phi(p) \gamma^\mu \Delta(k) \gamma^\nu \right] + \mathcal{O}(g),
\label{treelevel}
\end{equation}
where $\Phi$ and $\Delta$ were defined in the previous chapter
(Eq.~\ref{defPhi}, ~\ref{defDelta}).
Note that the Heisenberg fields in the correlator contain interactions
to all orders in $g$, see for example the interaction in
Fig.~\ref{quarkInter}b.

In order to evaluate the result,
a Sudakov-decomposition is made. A set of light-like vectors
($\{n_-,n_+\}$ with $n_-\cdot n_+ = 1$ and $\bar{n}_- \sim n_+$, where
the bar denotes reversal of spatial components) is introduced
such that\nopagebreak
\myBox{
\begin{align}
P &= \frac{x_B M^2}{\tilde{Q}\sqrt{2}}\ n_- + \frac{\tilde{Q}}{x_B \sqrt{2}}\ n_+,\nonumber\\
P_h &= \frac{z_h \tilde{Q}}{\sqrt{2}}\ n_- + \frac{M_h^2}{z_h \tilde{Q} \sqrt{2}}\ n_+,
\nonumber\\
q &= \frac{\tilde{Q}+\mathcal{O}(M^2/\tilde{Q})}{\sqrt{2}}\ n_-
- \frac{\tilde{Q}+\mathcal{O}(M^2/\tilde{Q})}{\sqrt{2}}\ n_+ + q_T,
\label{quarkFrame}
\end{align}
\begin{flushright}
\emph{Sudakov-decomposition for semi-inclusive DIS}
\end{flushright}
}
where $\tilde{Q}^2 \equiv Q^2 + q_T^2 = Q^2 + \mathcal{O}(M^2)$.
Using the parton model assumptions the correlators in Eq.~\ref{treelevel}
can be
integrated over the small momentum components
\begin{equation}
\begin{split}
2 M W^{\mu\nu} &= \int \rmd{2}{p_T} \rmd{2}{k_T} \delta^2(p_T+q_T-k_T)
\\
&
\eqnIndent
\times \frac{1}{3}
\tr^D \left[ \tr^C \left( \int \rmd{1}{p^-} \Phi(p) \right)_{p^+ = -q^+} \gamma^\mu
\tr^C \left( \int \rmd{1}{k^+} \Delta(k) \right)_{k^- = q^-} \gamma^\nu \right]
\\
& \eqnIndent
\times \left(1+  \MQQ \right) + \mathcal{O}(g)+ \mathcal{O}(\alpha_S),
\end{split}
\end{equation}
where the relation $\Phi_{ab} = \delta_{ab} \tr^C \Phi/3$ was applied.

At higher orders in $g$ one needs to include gluonic diagrams in which
gluon-lines are connected to the correlators.
If there is more than one soft correlator present, as is studied here,
the inclusion of such
interactions becomes more complex.
A simplification can be obtained by assuming the expressions for the
cross section to be color
gauge invariant, allowing for a suitable gauge choice. A
convenient gauge turns out to be
the light-cone gauge with retarded boundary conditions.
In such gauges,
see for example Ref.~\cite{Belitsky:2002sm},
one of the light-cone components is set to zero together
with the transverse polarizations at light-cone minus infinity, or
$A^-(\eta) = A^\alpha_T(\eta^-,-\infty,\eta_T) = 0$.

In the calculation of the hadronic tensor,
we shall employ the equations of motion.
It has been put forward by Politzer~\cite{Politzer:1980me} and subsequently
by
Boer~\cite{Boer:1998im} that the classical equations of motion,
$(i \slashii{D} - m)\psi(x) = 0$, hold within
physical matrix elements.
They lead to the following identity for the correlators
$\Delta$ and $\Delta_A^\alpha$ (see also Eq.~\ref{defDelta} and
Eq.~\ref{defDeltaA} and where $\Delta_A^\alpha
\equiv \Delta_{A_l}^\alpha t_l$)
\begin{equation}
\Delta(k) (\slashi{k}-m) = -g
\int
\rmd{4}{k_1} \Delta_A^\alpha(k,k_1) \gamma_\alpha.
\end{equation}
We will briefly come back to the validity of these equations
in the next chapter (subsection~\ref{hiro5}, Drell-Yan).

When including gluons from the
fragmentation and distribution correlator, 
the standard treatment shows that
gluons which are backwardly polarized ($S^\mu
\sim \bar{p}^\mu$)
lead to $\MQQ$ suppressed matrix elements and can therefore
be safely neglected. This means that the $A^+$-gluons from the fragmentation
correlator and the $A^-$-gluons from the distribution correlator can be
discarded. Together with the chosen light-cone gauge this means that only
the transversely polarized gluons of the fragmentation correlator
and the
longitudinally and transversely polarized gluons from
the distribution correlator
need to be considered.

We shall continue the calculation by considering a simple higher order
interaction.
Taking a gluon from the distribution correlator as displayed in
Fig.~\ref{quarkInter}d gives the following contribution
to the hadronic tensor
\begin{align}
2 M W^{\mu\nu} &= 2M W^{\mu\nu}_{\text{Fig.~}\eqref{quarkInter}d} + \text{other diagrams},\\
2M W^{\mu\nu}_{\text{Fig.~}\eqref{quarkInter}d} &=
\int \rmd{4}{p} \rmd{4}{k} \rmd{4}{p_1}
\delta^4(p+q-k)
\nonumber\\
& \eqnIndent
\times \tr^{D,C} \left[ \Phi_{A_l}^\alpha(p,p_1) \gamma^\mu \Delta(k)
\left( ig \gamma_\alpha t_l \right)
i \frac{\slashi{k} - \slashiv{p}_1 + m}{(k-p_1)^2 - m^2 + i \epsilon}
 \gamma^\nu \right],
\label{xgluon}
\end{align}
where $\Phi_A$ is now the quark-gluon-quark correlator
as defined in Eq.~\ref{defPhiA}. In the
next equation the above expression
is rewritten to indicate how the various parts of the quark-propagator in
combination with the gluon polarization contribute to the cross section
\begin{align}
2MW^{\mu\nu}_{\text{Fig~}\eqref{quarkInter}d} &=
\int \rmd{4}{p} \rmd{4}{k} \delta^4(p+q-k) \nonumber
\\
& \eqnIndent \times \left(
2Mw^{\mu\nu}_{L,\slashi{k}+m} +
2Mw^{\mu\nu}_{L,-\slashiv{p}_{1T}} +
2Mw^{\mu\nu}_{T,\slashi{k}-\slashiv{p}_{1T}+m} +
2Mw^{\mu\nu}_{T,-p_1^+\gamma^-} \right),
\\
2Mw^{\mu\nu}_{L,\slashi{k}+m} &=
\int  \frac{\rmd{4}{p_1} (-g)}{(k{-}p_1)^2 {-}m^2{+} i \epsilon}
\tr^{D,C} \left[
	 \Phi_{A_l}^+(p,p_1) \gamma^\mu \Delta(k) \gamma^- t_l
	\left(\slashi{k} + m \right)\gamma^\nu
 \right], \label{linkPlus1}
\\
2Mw^{\mu\nu}_{L,- {\slashiv{p}_1}_T} &=
\int  \frac{\rmd{4}{p_1} (-g)}{(k{-}p_1)^2 {-}m^2{+} i \epsilon}
\tr^{D,C} \left[
 \Phi_{A_l}^+(p,p_1) \gamma^\mu \Delta(k) \gamma^- t_l
	\left( - {\slashiv{p}_1}_T \right)\gamma^\nu
\right], \label{linkPlus2}
\\
2Mw^{\mu\nu}_{T,\slashi{k}-\slashiv{p}_{1T}+m} &{=}\!\!
\int\!\!\!  \frac{\rmd{4}{p_1} (-g)}{(k{-}p_1)^2 {-}m^2{+} i \epsilon}
\! \tr^{D,C}\! \left[\!
{\Phi_{A_l}}_T^\alpha(p,\!p_1) \gamma^\mu \Delta(k) \gamma_\alpha t_l
	\left(\slashi{k} {-} {\slashiv{p}_1}_T {+} m \right) \gamma^\nu\!
\right], \label{linkT1}
\\
2Mw^{\mu\nu}_{T,-p_1^+\gamma^-} &{=}\!\!
\int\!\!\!  \frac{\rmd{4}{p_1} (-g)}{(k{-}p_1)^2 {-}m^2{+} i \epsilon}
\tr^{D,C} \left[
{\Phi_{A_l}}_T^\alpha(p,p_1) \gamma^\mu \Delta(k) \gamma_\alpha t_l
	\left(  {-} p_1^+ \gamma^- \right)
	\gamma^\nu
\right]. \label{linkT2}
\end{align}
As we will see later in this section, the various terms contribute to the following orders
in $M/Q$:
$2Mw^{\mu\nu}_{L,\slashi{k}+m}$ contributes at leading order
to the longitudinal gauge link,
$2Mw^{\mu\nu}_{L,- {\slashiv{p}_1}_T}$ contributes at subleading order,
$2Mw^{\mu\nu}_{T,\slashi{k}-\slashiv{p}_{1T}+m}$ contributes at leading order
to the transverse gauge link, and
$2Mw^{\mu\nu}_{T,-p_1^+\gamma^-}$ contributes at subleading order.

In the following subsections we will evaluate these terms and
include higher order gluon insertions from the distribution correlator.
We will begin by analyzing the
leading order in detail which leads to the gauge link;
the next-to-leading
order in $M/Q$ will be discussed in the second subsection in which
transversely polarized gluons
from the fragmentation correlator contribute as well.

\subsection[Leading order in $M/ Q$]
{Leading order in $\boldsymbol{M}/ \boldsymbol{Q}$ \label{hiro}}

\subsubsection{Longitudinal gauge link}

\begin{figure}
\begin{center}
\begin{tabular}{cp{.001cm}cp{.001cm}c}
\includegraphics[width=3.3cm]{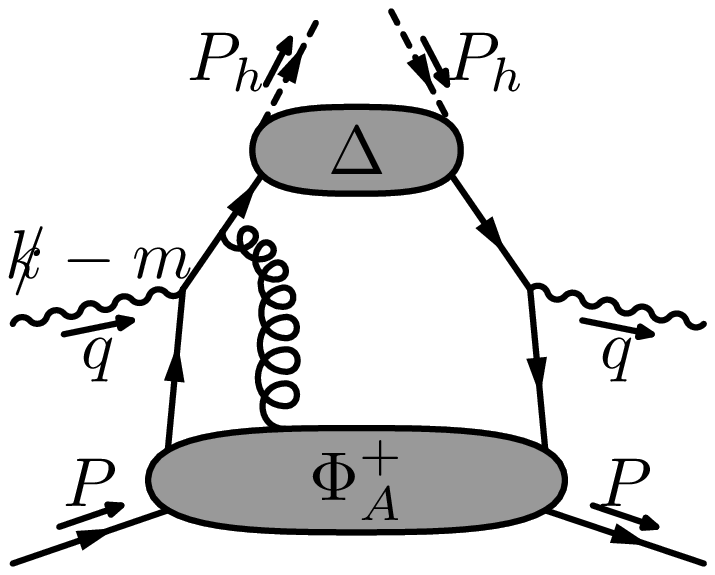} &&
\includegraphics[width=3.3cm]{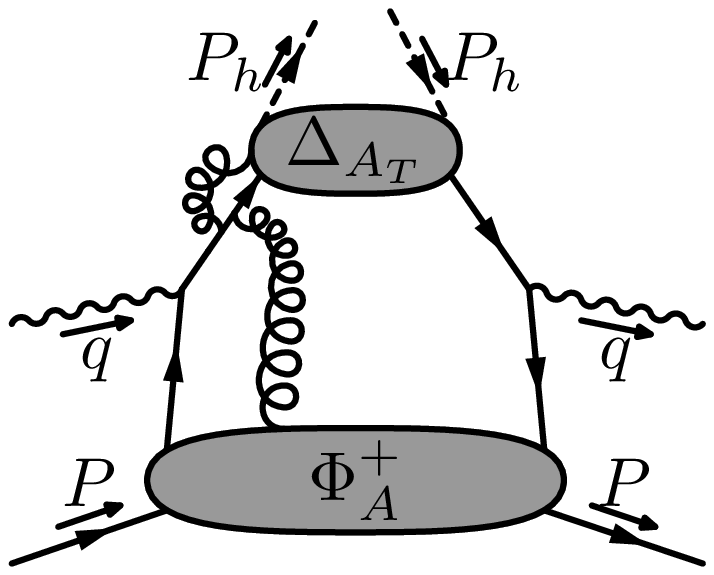} &&
\includegraphics[width=3.3cm]{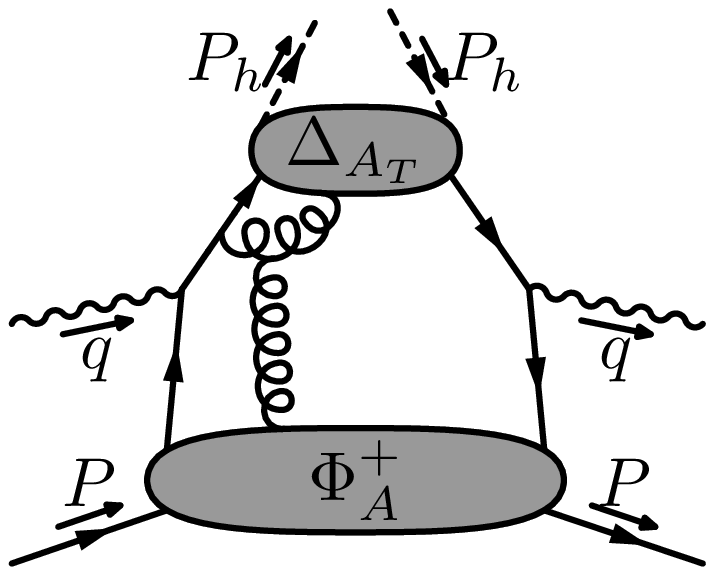}\\
(a) && (b) && (c)
\end{tabular}
\end{center}
\caption{Various contributions to the hadronic tensor.
The interactions in the figures cannot be absorbed in a single jet and
therefore need to be considered.
Figure~(c) is not of the type of figure~\ref{quarkInter}.d because
the interaction is connected with the hard part. In figure~(a)
only the indicated part of the numerator of the quark propagator is
considered. The part which is not indicated will be treated in the next
subsection.\label{plusLink}}
\end{figure}

We will study here in detail
the contribution of a longitudinally polarized
gluon inserted in the diagram. This follows the leading order in $M/Q$
calculations of
Bjorken, Kogut, Soper~\cite{Bjorken:1970ah};
Efremov, Radyushkin~\cite{Efremov:1978xm}; and
Collins, Soper, Sterman~\cite{Collins:1981uk,Collins:1981uw,Collins:1989gx}.
In Boer, Mulders~\cite{Boer:1999si}
those calculations were extended
by including $M/Q$ corrections in the diagrammatic approach.
In that paper explicit calculations were given to order $g^2$ which were
generalized by using arguments based on Ward identities.

When studying this problem Ward identities should be handled 
with care~\cite{Pijlman:2004mr}.
The considered gluon (with momentum $p_1$)
is approximately longitudinally polarized and is inserted at
various places in an amplitude,
seducing one to use $p_1^\mu \mathcal{M}_\mu (p_1) = 0$.
However, in the limit of $p_1 \rightarrow 0$
the Ward identity does not contain any information ($0=0$).
Since this issue especially becomes relevant in more complicated diagrams 
(chapter~\ref{chapter4})
explicit calculations will be performed in this chapter.

%However, Ward
%identities should be handled here with care~\cite{Pijlman:2004mr}.
%In the following we shall see that
%the gauge link is obtained in the limit
%where $p_1$ vanishes, in which case the
%Ward identity does not contain any information ($0=0$).
%Therefore explicit calculations will be performed in this chapter.

Inserting the gluon
on the left-hand-side of the cut, see also Fig.~\ref{plusLink}a,
we need to evaluate Eq.~\ref{linkPlus1} and
Eq.~\ref{linkPlus2}. The contribution from Eq.~\ref{linkPlus2}
will be calculated in the next subsection and turns out to appear at subleading
order.
To perform the integral over $p_1$, the
denominator $((k{-}p_1)^2 {-} m^2 {+} i \epsilon)^{-1}$
in Eq.~\ref{linkPlus1} is simplified into
$(-2p_1^+ k^- + i \epsilon)^{-1}$. This simplification, called
the \emph{eikonal approximation}, is justified when
making the parton model assumptions which were discussed in
section ~\ref{diagram}. The integral over $p_1^+$ in
$2Mw^{\mu\nu}_{L,\slashi{k}+m}$
(Eq.~\ref{linkPlus1}) takes now the following form
\begin{equation}
\int \rmd{1}{p_1^+} \frac{e^{ip_1^+(\eta-\xi)^-}}{-2p_1^+ k^- + i \epsilon}
A_l^+(\eta).
\label{quark23}
\end{equation}
Assuming $k^-$ to be positive, the integral can be performed by calculating
the residue. This leads
to a Heaviside function ($\theta$)
\begin{equation}
\text{Eq.~}\eqref{quark23} = \frac{2\pi i}{-2k^-}\ \theta( \eta^- - \xi^- )
\ A_l^+(\eta).
\end{equation}
The Heaviside function expresses
that there will be only contributions to the hadronic tensor
when $\eta^- {>} \xi^-$, meaning that $A^+(\eta^- {\rightarrow} \infty)$
contributes, but $A^+(\eta^- {\rightarrow} {-}\infty)$ does not.
\nopagebreak

Using the equations of motion one finds for
$2Mw^{\mu\nu}_{L,\slashi{k}+m}$
($\gamma^- (\slashi{k}{+}m) {=} 2k^- {-} (\slashi{k}{-}m) \gamma^-$)
\begin{equation}
\begin{split}
2Mw^{\mu\nu}_{L,\slashi{k}+m} &=
\int\! \frac{ \rmd{4}{\xi} }{(2\pi)^4}
e^{ip\xi} \langle P,S | \bar{\psi}(0)
\gamma^\mu \tr^C\! \left[ \frac{\Delta(k)}{3} \right]\!
\gamma^\nu\!  (-ig)\!\! \int_\infty^{\xi^-}\!\!
\rmd{1}{\eta^-} \! A^+(\eta^- \! ,\xi^+,\xi_T) \psi (\xi) | P,S \rangle_\text{c}\\
&\phantom{=} \eqnIndent \times \left( 1+ \MQQ \right)
\\
& \phantom{=}
-\! g\! \int\!\! \rmd{4}{p_1}\!\!
\tr^{D,C} \left[ \Phi_{A_l}^+(p,p_1) \gamma^\mu\!\! \int\!\! \rmd{4}{k_1} \Delta_A^\beta(k,k_1)
\gamma_\beta t_l \frac{\gamma^-}{-2k^- p_1^+ + i \epsilon} \gamma^\nu \right],
\raisetag{20pt}
\end{split}
\label{longLink}
\end{equation}
where the relation $\Delta_{ab} = \delta_{ab} \tr^C \Delta(k)/3$ was used.
The first term above contributes at leading twist and
 is exactly the first order gauge link expansion
running via infinity.
The second term in Eq.~\ref{longLink}, coming from applying the
equations of motion, is canceled when
the other diagrams of Fig.~\ref{plusLink} are included.
Although this cancellation\footnote{
In order to achieve this cancellation it was assumed that the gluon
connecting the fragmentation correlator is outgoing and approximately
on its mass-shell ($k_1^- > 0$).} occurs
in the chosen
light-cone gauge, where the gluon propagator has the
numerator (see for example Ref.~\cite{Belitsky:2002sm})
\begin{equation}
d^{\mu\nu}(l) = g^{\mu\nu} - \frac{l^\mu n_+^\nu}{l^- - i \epsilon}
- \frac{n_+^\mu l^\nu}{l^- + i \epsilon},
\end{equation}
one does obtain another term proportional
to $\Delta_{AA}$.
We see that by considering the other diagrams the $\Delta_A$ term got 
replaced by a $\Delta_{AA}$ term. This cancellation was observed in
Boer, Mulders~\cite{Boer:1999si}.
Since this issue only plays a role
when considering
fragmentation correlators (a free outgoing quark does not produce
$\Delta_A$ terms), this kind of cancellation
is expected to hold to all orders. One should find
by including higher order diagrams that
the $\Delta_{AA}$ term
gets replaced by a $\Delta_{AAA}$ term and so on.
This cancellation, which is expected to hold at each order in $g$,
%Since this cancellation
has not been
proven to all orders
and deserves further investigation.
%further research is recommended on this point.

Before continuing it should be pointed out that the cancellation also occurs
when the Feynman gluon propagator is taken
in the hard part. This suggests
the hard part and the correlators to be separately gauge invariant, which
is a minimal requirement for
a factorized description.

In order to generalize the above result, Eq.~\ref{longLink}, we consider
$n$ longitudinally polarized gluon insertions on the left-hand-side
of the cut which contribute to the hadronic tensor as
\begin{align}
2 M W^{\mu\nu} &= \int \rmd{4}{p}
\rmd{4}{k} \delta^4(p+q-k)\ 2Mw^{\mu\nu}_{n A^+} + \text{other diagrams},
\\
2Mw^{\mu\nu}_{n A^+} & =
\int \rmd{4}{p_1} \ldots \rmd{4}{p_n}
\tr^{D,C} \biggl[
\Phi_{A_{l_1}\ldots A_{l_n}}^{+ \ldots +}(p,p_1{-}p_2,\ldots, p_{n-1} {-} p_n,p_n)
\gamma^\mu \Delta(k)  \nonumber\\
& \eqnIndent {\times} (-g)^n
(t_{l_n} \gamma^-) \frac{ \slashi{k} - \slashiv{p}_n+m}{(k{-}p_n)^2 {-} m^2 {+} i\epsilon}
\ldots (t_{l_1} \gamma^- )
\frac{ \slashi{k} - \slashiv{p}_1+m}{(k{-}p_1)^2 {-} m^2 {+} i\epsilon}
\gamma^\nu \biggr].
\label{long2}
\end{align}
Performing all the $p_i$-integrals one straightforwardly
obtains an ordered product
\begin{multline}
2Mw^{\mu\nu}_{nA^+} =
\int
\frac{ \rmd{4}{\xi}}{(2\pi)^4} e^{ip\xi}
\langle P,S | \bar{\psi}(0) \gamma^\mu
\tr^C \left[ \!\frac{\Delta(k)}{3}\! \right] \gamma^\nu
(-ig)^n\int_{\infty}^{\xi^-}\!
\rmd{1}{\eta_1^-}\! \int_{\eta_1^-}^{\xi^-}\!
\rmd{1}{\eta_{2}^-}\!
\ldots \! \int_{\eta_{n-1}}^{\xi^-}\! \rmd{1}{\eta_n^-}
\\
\times
A^+ (\eta_1^-,\xi^+,\xi_T)\! \ldots \! A^+ (\eta_n^-,\xi^+,\xi_T) \psi (\xi) | P,S \rangle_\text{c}
\left( 1+ \mathcal{O}(M/Q)\right),
\label{calina1}
\end{multline}
which is the $n^\text{th}$-order longitudinal gauge link expansion.

To summarize, by including longitudinally polarized gluons from a distribution
correlator which interact with an outgoing quark,
one obtains the longitudinal
gauge link
$\mathcal{L}^{\xi_T\!,\ \xi^+}(\infty^-,\xi^-)$ (for definition see
section~\ref{sectionDistr}).
Inserting longitudinally polarized
gluons from the distribution correlator on the right-hand-side of the cut
also produces a longitudinal gauge link. Its result can be obtained by
taking the complex conjugate of the result we just obtained
(Eq.~\ref{calina1})
and interchanging $\mu$ with $\nu$. For those insertions
one finds the gauge link $\mathcal{L}^{0_T,\ \xi^+}(0^-,\infty^-)$.

\subsubsection{Transverse gauge link}

\begin{figure}
\begin{center}
\begin{tabular}{ccc}
\includegraphics[width=3.3cm]{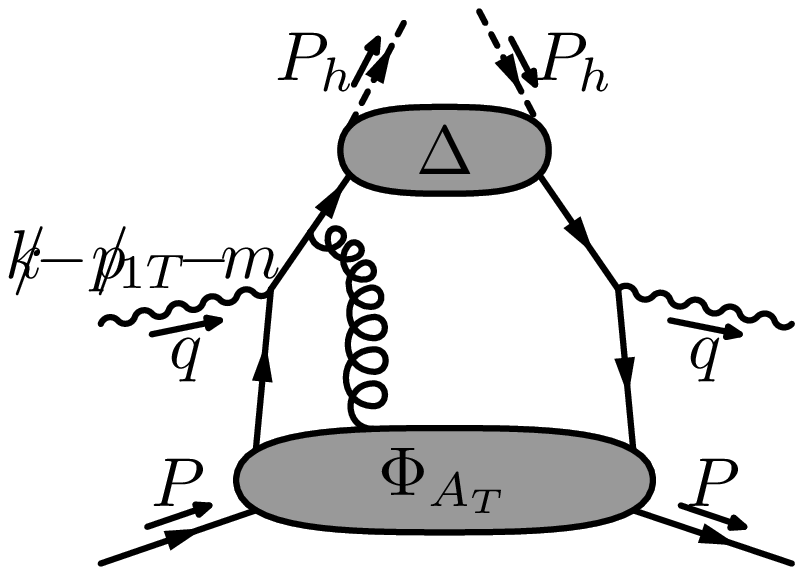} &\hspace{.1cm}
\includegraphics[width=3.3cm]{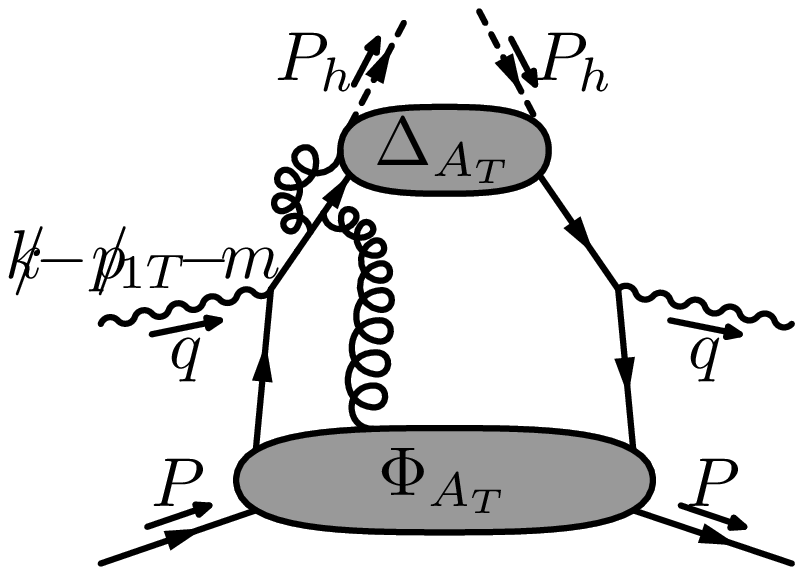} &\hspace{.1cm}
\includegraphics[width=3.3cm]{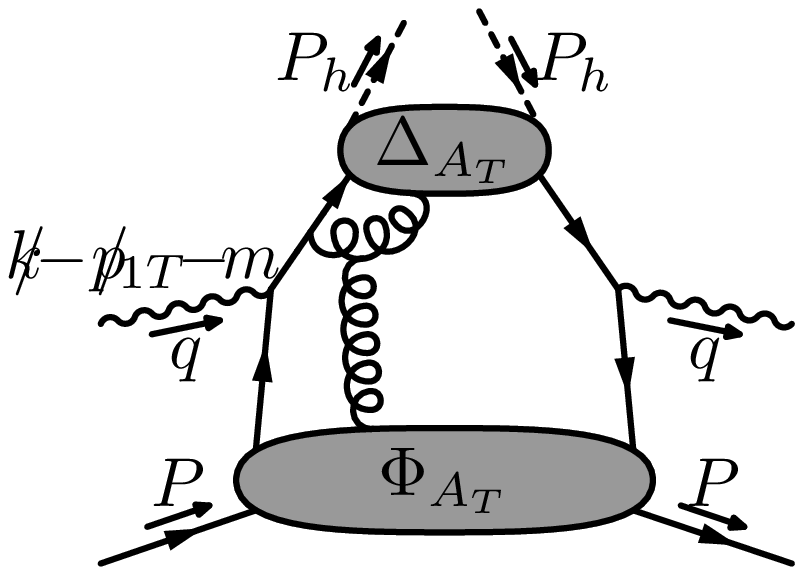}\\
(a) & \hspace{.1cm}(b) &\hspace{.1cm} (c)
\end{tabular}
\end{center}
\caption{Various contributions to the hadronic tensor. In each of the
diagrams it is indicated which part of the numerator
of the quark propagator is considered.\label{TLink}}
\end{figure}

We continue by studying the leading contributions from transversely polarized
gluons. For a long time transversely
polarized gluons were thought to appear always suppressed
in cross sections (see for instance Mulders, Tangerman~\cite{Mulders:1996dh}).
In 2002
Belitsky, Ji, and Yuan showed in Ref.~\cite{Belitsky:2002sm}
that this idea is wrong and
that transversely polarized gluons at leading
order give rise to a transverse gauge link.
Although their calculation
showed how the transverse gauge link could be derived, it also contained
a few points which deserved further clarification. Some of these points
were given attention in Ref.~\cite{Boer:2003cm}
but other points
were unintentionally left open. All issues
will be discussed here in considerable detail.

We start with
$ 2Mw^{\mu\nu}_{T,\slashi{k}-\slashiv{p}_{1T} +m}$ as given in
Eq.~\ref{linkT1} (represented in Fig.~\ref{TLink}a)
\begin{equation}
2Mw^{\mu\nu}_{T,\slashi{k}-\slashiv{p}_{1T} +m} {=}\!\!\!
\int\!\!\!  \frac{\rmd{4}{p_1}(-g)}{(k{-}p_1)^2{-}m^2 {+} i \epsilon}\!
\tr^{D,C}\! \left[\!
{\Phi_{A_l}}_T^\alpha(p,\!p_1) \gamma^\mu\! \Delta(k) g \gamma_\alpha t_l
	 \left( \slashi{k}{-}\slashiv{p}_{1T} {+}m \right)\!
	\gamma^\nu\!
\right].
\label{quark102}
\end{equation}
Performing the integral over $p_1^+$ would lead to
$\int_{\xi^-}^\infty \rmd{1}{\eta^-} A_T^\alpha (\eta)$ as was also encountered
when calculating the longitudinal link. In gauges where the gluon
fields vanish at infinity this would form a perfect matrix element, but in those
gauges where the fields do not vanish the matrix element is divergent
and needs subtractions\footnote{A similar
problem occurs also with the longitudinally polarized gluons.
In order to keep the
matrix elements containing the longitudinal gauge link convergent
one should require the
longitudinally polarized fields to vanish at infinity. This requirement
is supported in those
gauges where the gluon-fields at infinity can be described with
a scalar potential (see also Ref.~\cite{Belitsky:2002sm})
\begin{displaymath}
\phi(\eta) = \int_C^{\eta} \rmd{1}{\zeta} \cdot A(\zeta).
\end{displaymath}
In order to have the potential finite at $\eta^- = \infty$ which seems to be
physically reasonable, the $A^+$-fields need to vanish at that point.
This argument does not hold for transversely polarized gluons.}.

In order to consider the transverse fields at infinity more precisely one expresses
the $A$-fields as
\begin{equation}
A_T^\alpha (\eta) = A_T^\alpha(\infty,\eta^+,\eta_T) +
                    \left[ A_T^\alpha(\eta) -
                    A_T^\alpha(\infty,\eta^+,\eta_T) \right].
                    \label{quark101}
\end{equation}
In this decomposition the $\eta^-$-dependence has vanished in the first term.
When performing the $\eta^-$-integral over the first term
of Eq.~\ref{quark101} in Eq.~\ref{quark102}
(which is proportional to $\exp{[ip_1(\eta-\xi)]}$)
a delta-function
in $p_1^+$ emerges\footnote{The boundary terms were also
studied by Boer, Mulders, and Teryaev in
Ref.~\cite{Boer:1997bw}.}. This delta-function eliminates the large momenta
in the denominator in Eq.~\ref{quark102}, making the contribution leading
in $M/Q$.
When
performing the $p_1^+$-integral one finds that
the second term in Eq.~\ref{quark101}
contributes at order $M^2/Q^2$ and can be neglected.
The choice for taking the subtracting point in 
Eq.~\ref{quark101} at infinity
might seem arbitrary
but actually it is not. Choosing the subtraction point at 
$\eta^- {=} -\infty$
would produce, after an integration
over $p_1^+$, the divergent term $\int_\infty^{\xi^-} \!\! \rmd{1}{\eta^-}\! 
[A_T^\alpha(\eta) - A_T^\alpha({-}\infty^-,\eta^+,\eta_T)]$.
Hence it is the $i\epsilon$
prescription in Eq.~\ref{quark102} (or process)
which forces us to take the subtraction point at infinity in order
to keep the contributions from both terms in Eq.~\ref{quark101}
finite.

We continue with the first term and find~\cite{Boer:2003cm}
\begin{multline}
2Mw^{\mu\nu}_{T,\slashi{k}-\slashiv{p}_{1T}+m} =
\int \rmd{2}{p_{1T}}\rmd{1}{p_1^-}
\int \frac{ \rmd{4}{\xi} \rmd{1}{\eta^+} \rmd{2}{\eta_T} }{(2\pi)^7}
 e^{ip\xi + i p_{1T}(\eta-\xi)_T + i p_1^- (\eta-\xi)^+}
\\
{\times} \lim_{p_1^+ \rightarrow 0}
\langle P,S | \bar{\psi}(0) \gamma^\mu \Delta(k) i \gamma_\alpha
i
\frac{\slashi{k} - {\slashiv{p}_1}_T + m }{(k-p_1)^2 - m^2 + i \epsilon}
\gamma^\nu
\\
\times g
\partial_\eta^\alpha \int_C^{\eta_T} \rmd{1}{\zeta_T} \cdot A_T (\infty,
\eta^+, \zeta_T)
\psi(\xi) | P,S \rangle_\text{c}
\left(1+ \MQQ \right),
\label{sgluon}
\end{multline}
where $C$ is some constant.
To arrive at an equivalent expression
the existence of a pure gauge at infinity
was assumed in Eq.~(38) of Belitsky, Ji, Yuan~\cite{Belitsky:2002sm}.
This assumption makes the treatment for QCD less general and is avoided here.

Having the gluon collinear means that the delta-function in $p_1^+$ also
pushes the transverse momentum down. In order to proceed we needed to
interchange those limits. Therefore,
when taking the limit $p_1^+ \rightarrow 0$
we keep $p_{1T}$ finite. This
exchange of limits, which
is also present in Ref.~\cite{Belitsky:2002sm}
and seems to be unavoidable,
might lead to problems at higher orders in $\alpha_S$.

Doing now a partial integration and applying the equations of motion
one finds by performing first
the $\eta^-$-integral, the $p_1^+$-integral, and then the
rest of the integrals
($\Delta(k)[{-}\slashiv{p}_{1T}][\slashi{k}{-}\slashiv{p}_{1T} {+} m]
=[ \Delta(k)[\slashi{k}{-}\slashiv{p}_{1T} {-} m] + g\int \rmd{4}{k_1}
\Delta_A^\alpha(k,k_1) \gamma_\alpha ] [\slashi{k}{-}\slashiv{p}_{1T} {+} m]$)
\begin{align}
2 M &w^{\mu\nu}_{T, \slashi{k} - \slashiv{p}_{1T} + m}
\nonumber\\
&
=
\!\int\! \frac{ \rmd{4}{\xi}}{(2\pi)^4}
 e^{ip\xi}
\langle P,S | \bar{\psi}(0) \gamma^\mu \Delta(k)
\gamma^\nu
(-ig)\! \int_C^{\xi_T} \rmd{1}{\zeta_T}\! \cdot\! A_T (\infty,\xi^+,\zeta_T)
\psi(\xi) | P,S \rangle_\text{c}
\nonumber\\
&
\eqnIndent
\times \left( 1+ \MQQ \right)
\nonumber
\\
&
\phantom{=}
+g
\int \frac{ \rmd{4}{\xi} \rmd{1}{\eta^+} \rmd{2}{\eta_T}}{(2\pi)^7}
\lim_{p_1^+ \rightarrow 0} e^{ip\xi} e^{ip_1(\eta-\xi)}
\langle P,S | \bar{\psi}(0) \gamma^\mu
\int \rmd{4}{k_1} \Delta_A^\alpha(k,k_1) \gamma_\alpha \nonumber\\
& \eqnIndent \times
\frac{\slashi{k} - {\slashi{p_1}}_T + m}{(k-p_1)^2 - m^2 + i \epsilon}
\gamma^\nu
(-ig) \int_C^{\eta_T} \rmd{1}{\zeta_T} \cdot A_T (\infty,\eta^+,\zeta_T)
\psi(\xi) | P,S \rangle_\text{c},
\end{align}
where the first term
is exactly the first order expansion of the transverse gauge link. The second
term was not considered in Ref.~\cite{Belitsky:2002sm}, while
in Ref.~\cite{Boer:2003cm} it was
thought to be suppressed in $M/Q$.
In the case of jet-production in DIS the second term does not appear
(the outgoing quark is assumed to be free), but
when considering the fragmentation of a quark into a hadron neither
observations of Ref.~\cite{Belitsky:2002sm,Boer:2003cm} are in general valid.
The second term is
as leading
as the first term and will be considered here.

Similar terms were also encountered when deriving the longitudinal gauge link.
 Also here one finds that when including the diagrams
in Fig.~\ref{TLink}b and Fig.~\ref{TLink}c that the second term is
exchanged with a term proportional to
$\Delta_{AA}$.
Similarly as for the longitudinal gauge link, it
will be assumed that this behavior can be generalized
to all orders. When
taking the Feynman gauge for the hard
part, one finds that
the diagram in Fig.~\ref{TLink}c has an additional contribution at
next-to-leading order in $M/Q$.
In the next subsection where we will discuss
the subleading
order in $M/Q$,
it will be pointed out where this term gets canceled.

Generalizing the result above to all orders
by inserting $n$ transversely polarized gluons, one obtains for the
hadronic tensor
\begin{align}
2 MW^{\mu\nu} &=
\int \rmd{4}{p} \rmd{4}{k} \delta^4(p+q-k)\ 2Mw^{\mu\nu}_{nA_T}
+
\text{other diagrams},\\
2Mw^{\mu\nu}_{nA_T} &=
\int \rmd{4}{p_1} \ldots \rmd{4}{p_n}
\frac{ \rmd{4}{\xi} \rmd{4}{\eta_1} \ldots \rmd{4}{\eta_n} }{(2\pi)^{4(n+1)}}
e^{i(p{-}p_1)\xi} e^{i(p_1{-}p_2) \eta_1} \ldots
e^{i (p_{n-1}{-}p_n)\eta_{n-1}}
e^{ip_n\eta_n}
\nonumber\\
& \phantom{\times}\ \times(-g)^n
\langle P,S | \bar{\psi}(0) \gamma^\mu
\Delta(k)
 \gamma_{\alpha_n}t_{l_n}
\frac{ \slashi{k} - {\slashiv{p}_n}_T+m}{(k{-}p_n)^2 {-} m^2 {+} i\epsilon}
\ldots
\gamma_{\alpha_1}t_{l_1}
\frac{ \slashi{k} - {\slashiv{p}_1}_T +m}{(k{-}p_1)^2 {-} m^2 {+} i\epsilon}
\nonumber\\
& \phantom{\times}\ \times
\gamma^\nu
A^{\alpha_n}_{l_n,T} (\eta_n) \ldots A^{\alpha_1}_{l_1,T} (\eta_1) \psi (\xi) | P,S \rangle_\text{c}
\left( 1 + \mathcal{O}(M/Q) \right).
\label{quark78}
\end{align}
Following exactly the same steps as done
for the single transversely polarized gluon
one obtains
\begin{equation}
\begin{split}
2Mw^{\mu\nu}_{nA_T} &{=}\!\!\!
\int\!\!\!
\rmd{4}{p_1}\!\! \ldots \rmd{4}{p_{n{-}1}}\!\!
\frac{\! \rmd{4}{\xi}\! \rmd{4}{\eta_1}\! \ldots\! \rmd{4}{\eta_{n{-}1}}\! }{(2\pi)^{4n}}
e^{i(p{-}p_1)\xi}  e^{i (p_1{-}p_2)\eta_1}\!\ldots
e^{i(p_{n{-}2}{-}p_{n{-}1})\eta_{n{-}2}}
e^{ip_{n{-}1}\eta_{n{-}1}}
\\
& \phantom{=} \eqnIndent
\times
\langle P,S | \bar{\psi}(0) \gamma^\mu
\Delta(k)
\gamma_{\alpha_{n-1}}
\frac{ \slashi{k} - {\slashiv{p}_{n-1}}_T+m}{(k{-}p_{n-1})^2 {-} m^2 {+} i\epsilon}
\ldots
\gamma_{\alpha_1}
\frac{ \slashi{k} - {\slashiv{p}_1}_T +m}{(k{-}p_1)^2 {-} m^2 {+} i\epsilon}
\gamma^\nu
\\
& \phantom{=} \eqnIndent
\times
(-ig) \int_C^{{\eta_{n-1}}_T} \rmd{1}{\zeta_n} \cdot A(\infty,\xi^+,\zeta_{nT})
(-g)^{n-1}
A^{\alpha_{n-1}}(\eta_{n-1}) \ldots A^{\alpha_1} (\eta_1) \psi (\xi) | P,S \rangle_\text{c}
\\
& \phantom{=} \eqnIndent
\times \left(1+ \mathcal{O}(M/Q) \right).
\end{split}
\raisetag{14pt}
\label{quark79}
\end{equation}
Using now an identity which holds for non-Abelian
fields\footnote{In contrast, Eq.~(47) of Ref.~\cite{Belitsky:2002sm}
does not hold
in QCD.}~\cite{Boer:2003cm}\nopagebreak
\myBox{
\vspace{-.4cm}
\begin{multline}
\int_C^{\eta_T} \rmd{1}{\zeta_{1T}} \cdot A(\zeta_{1})
\int_{\zeta_{1T}}^{\eta_T} \rmd{1}{\zeta_{2T}} \cdot A(\zeta_2) \ldots
 \int_{\zeta_{n-1,T}}^{\eta_T} \rmd{1}{\zeta_{nT}} \cdot A(\zeta_n)\ A^\alpha(\eta) \\
= \partial_\eta^\alpha \int_C^{\eta_T} \rmd{1}{\zeta_{1T}} \cdot A(\zeta_1)
\int_{\zeta_{1T}}^{\eta_T} \rmd{1}{\zeta_{2T}} \cdot A(\zeta_{2}) \ldots
 \int_{\zeta_{nT}}^{\eta_T} \rmd{1}{\zeta_{n+1,T}} \cdot A(\zeta_{n+1}),
\end{multline}
\vspace{-.8cm}
\begin{flushright}
\emph{identity for non-Abelian fields}
\end{flushright}
}
one can subsequently perform all the integrals and obtain
%\begin{multline}
%2M w^{\mu\nu}_{nA_T}
%=
%\int \frac{ \rmd{4}{\xi}}{(2\pi)^4}
%e^{ip\xi}
%\langle P,S | \bar{\psi}(0) \gamma^\mu
%\Delta(k)  \gamma^\nu
%(-ig)^n\!\int_C^{\xi_T}\! \rmd{1}{ {\zeta_{n}}_{\alpha_n} }\!
%\!\int_C^{{\zeta_n}_T}\! \rmd{1}{ {\zeta_{n-1}}_{\alpha_{n-1}} }\!
%\ldots
%\!\int_C^{{\zeta_2}_T}\! \rmd{1}{{\zeta_1}_{\alpha_1}}
%\\
%\times
%A^{\alpha_1} (\infty,\xi^+,\zeta_1)
%\ldots A^{\alpha_n} (\infty,\xi^+,\zeta_n)
%\psi (\xi) | P,S \rangle_\text{c} \left(1+ \mathcal{O}(M/Q)\right).
%\label{quark80}
%\end{multline}
%equaling
\begin{multline}
2Mw^{\mu\nu}_{nA_T} =
\int \frac{ \rmd{4}{\xi}}{(2\pi)^4}
e^{ip\xi}  \langle P,S | \bar{\psi}(0) \gamma^\mu
\tr^C\! \left[ \frac{\Delta}{3} \right]\!  \gamma^\nu (-ig)^n \!
\int_C^{\xi_T}\! \rmd{1}{ {\zeta_{1}}_{\alpha_1} }\!
\!\int_{\zeta_1}^{\xi_T}\! \rmd{1}{ {\zeta_{2}}_{\alpha_2} }\! \ldots\!
\int_{\zeta_{n-1}}^{\xi_T}\! \rmd{1}{ {\zeta_{n}}_{\alpha_n} }
\\
\times
A^{\alpha_1} (\infty,\xi^+,\zeta_{1T})
\ldots A^{\alpha_n} (\infty,\xi^+,\zeta_{nT})
\psi (\xi) | P,S \rangle_\text{c}
\left(1+\mathcal{O}(M/Q) \right).
\end{multline}
This is the $n^\text{th}$-order expansion
of the transverse gauge link $\mathcal{L}^{\infty^-,\ \xi^+}(C,\xi_T)$.

The gluon insertions in the conjugate part of the diagram can be evaluated
as we did for the longitudinal gauge link. Taking the complex conjugate and
interchanging $\mu$ and $\nu$ one obtains $\mathcal{L}^{\infty^-,\ \xi^+}(0,C)$.

\subsubsection{The complete leading order result}

In the previous paragraphs we resummed gluons from the distribution correlator
on one side of the cut being either longitudinally or transversely polarized.
When taking the combination
of longitudinally and transversely polarized gluons on one side of the
cut, one finds in leading order only contributions from
those diagrams in which the gluons couple in a specific order.
This can be seen as follows.

\begin{figure}
\begin{center}
\begin{tabular}{cp{1cm}c}
\includegraphics[width = 2cm]{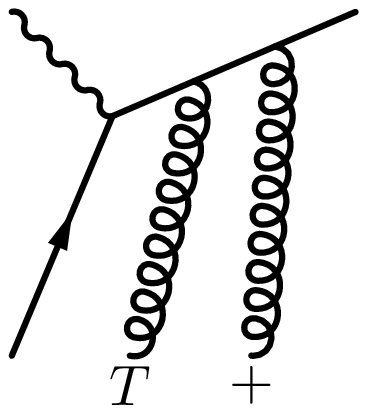} &&
\includegraphics[width = 2cm]{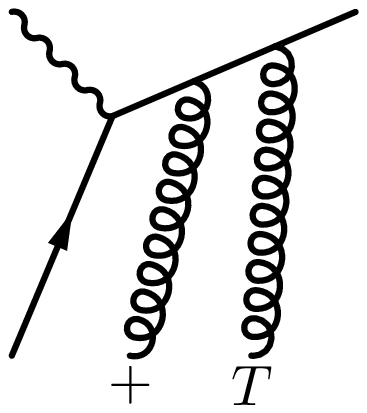} \\
&&\\
(a) && (b)
\end{tabular}
\end{center}
\caption{Two contributions in the diagrammatic approach. The parton-lines
entering from the bottom of the graphs belong to the distribution correlator.
\label{orderLink}}
\end{figure}

Consider a diagram containing a single $A_T$-insertion. The sum over all
possible longitudinally polarized gluon
insertions between the inserted vertex of the transversely polarized gluon
and the fragmentation correlator (see for instance Fig.~\ref{orderLink}.a) 
yields a longitudinal
gauge link between infinity and
the $A_T$-field which is at some coordinate $\eta$. The leading contribution
of this diagram comes when this $A_T$-field is at infinity and therefore this
longitudinal gauge link vanishes.
Only the longitudinally polarized
gluons, which couple between the vertex
of the $A_T$-gluon and the virtual photon (for instance Fig.~\ref{orderLink}.b),
contribute at leading order and provide the longitudinal gauge link.
It is
exactly this order which also appears in the full gauge link.

Taking combinations on the left-hand-side and on the right-hand-side of the
cut is relatively easy. The contributions on each side can be evaluated
without using information from the other side of the cut. One simply
obtains a product of both insertions, so
$\mathcal{L}^{0_T\!,\ \xi^+}(0^-,\infty^-)
\mathcal{L}^{\infty^-\!,\ \xi^+}(0_T,\infty_T)
\mathcal{L}^{\infty^-\!,\ \xi^+}(\infty_T,\xi_T)
\mathcal{L}^{\xi_T\!,\ \xi^+}(\infty^-,\xi^-) = \mathcal{L}^{[+]}(0,\xi^-)$.

Up to now we inserted gluons from a distribution correlator
to an amplitude and looked at their final
contribution in the cross sections. It turned out that the transversely
polarized gluons at infinity
contribute at leading order and yield the transverse gauge link.
The same calculation can be done for transversely
polarized gluons from the fragmentation correlator.
The only leading contribution comes when those fields are at minus infinity,
but those fields do not contribute
in the chosen gauge ($A^-(\eta)=A_T^\alpha(\eta^-,-\infty,\eta_T)=0$). 
This means that the transversely polarized gluons from
the fragmentation correlator always appear at subleading order in this gauge.
In order to find the gauge invariant fragmentation correlator we assume
that the expression for the cross section 
is gauge invariant. In that case, there
is only one link operator possible for the fragmentation correlator 
which vanishes
in this particular gauge. That is a gauge link running via minus infinity.
This gauge link is also found when one would have chosen to work in
$A^+ = A_T(\infty,0,\xi_T) = 0$ gauge.

In this process we have found that inserting gluons on an outgoing
quark yields a gauge link via infinity, while coupling gluons
to an incoming quark gives a gauge link via minus infinity. In
other processes we will encounter the same behavior.
Coupling to outgoing parton-lines leads to gauge links via plus infinity, while
coupling to incoming parton-lines leads to gauge links via minus infinity.

The hadronic tensor at leading order in $M/Q$ can now be expressed as
(the color factor $1/3$ has been absorbed in the fragmentation correlator)
\begin{multline}
2MW^{\mu\nu} = \int \rmd{2}{p_T} \rmd{2}{k_T} \delta^2(p_T+q_T-k_T)
\tr^D \bigg[ \Phi^{[+]}(x_B, p_T) \gamma^\mu \Delta^{[-]}(z^{-1},k_T) \gamma^\nu \bigg]
\\
\times \left(1+\mathcal{O}(M/Q)\right) + \mathcal{O}(\alpha_S).
\label{leadingOrder}
\end{multline}
This result includes an infinite amount of
diagrams at leading
order in $M/Q$ and is almost equivalent to the result of
Boer, Mulders~\cite{Boer:1997nt}. The only difference is that the correlators
above are fully gauge invariant (without assuming $A_T$-fields
at $\infty^-$ to vanish inside matrix elements), forming
a natural explanation for transverse
momentum dependent T-odd distribution functions. By integrating the hadronic
tensor over $q_T$,
these T-odd distribution
functions disappear, because the resulting 
$p_T$-integrated distribution correlator contains a gauge link of which its
path
is just a straight-line on the light-cone and does not produce T-odd effects.
The
integrated distribution functions $e_L(x)$, $f_T(x)$, and $h(x)$, as introduced
in Ref.~\cite{Boer:1997nt}, are therefore zero within this approach.

\begin{figure}
\begin{center}
\begin{tabular}{cc}
\includegraphics[width=5.5cm]{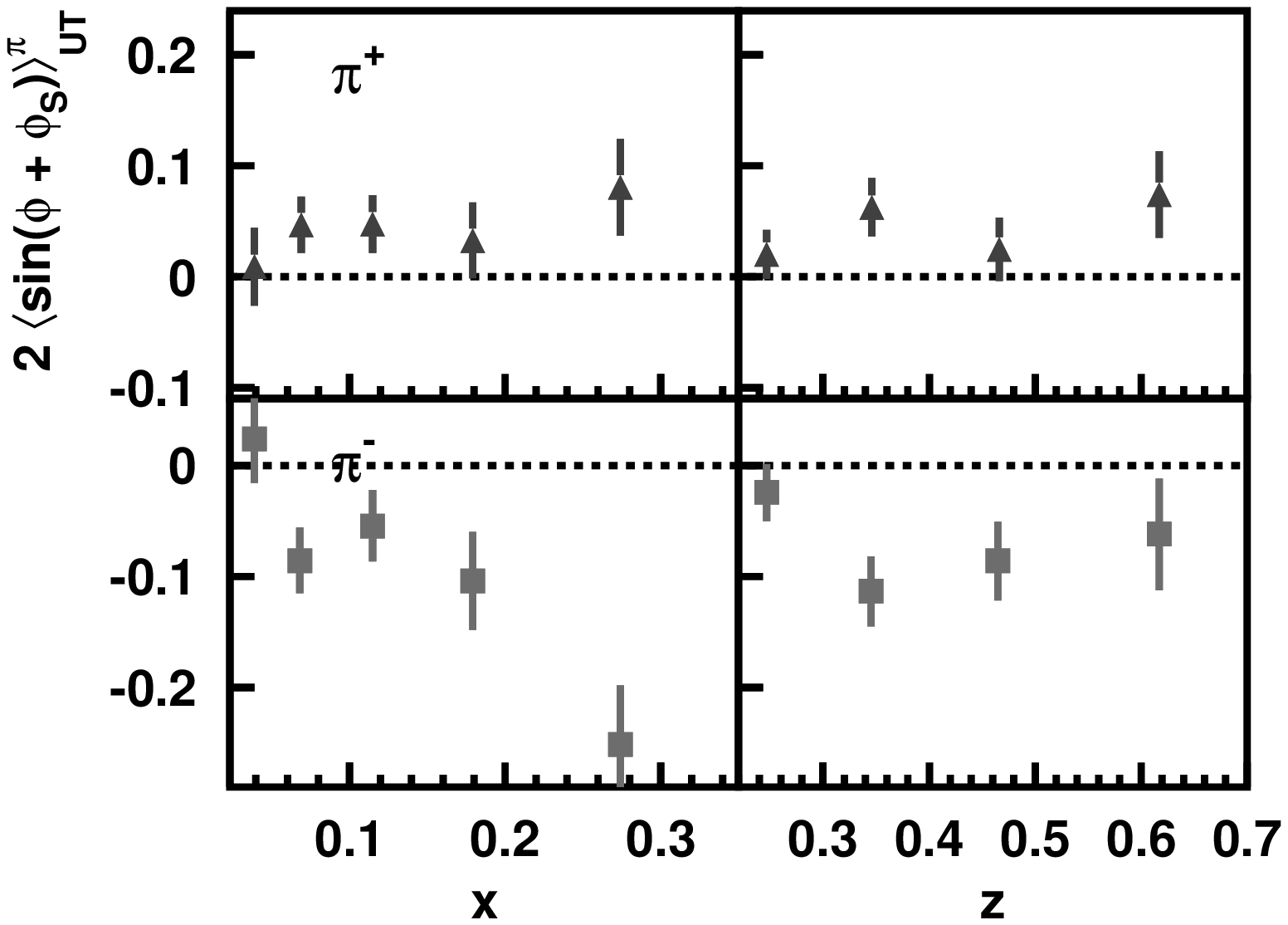} &
\includegraphics[width=5.5cm]{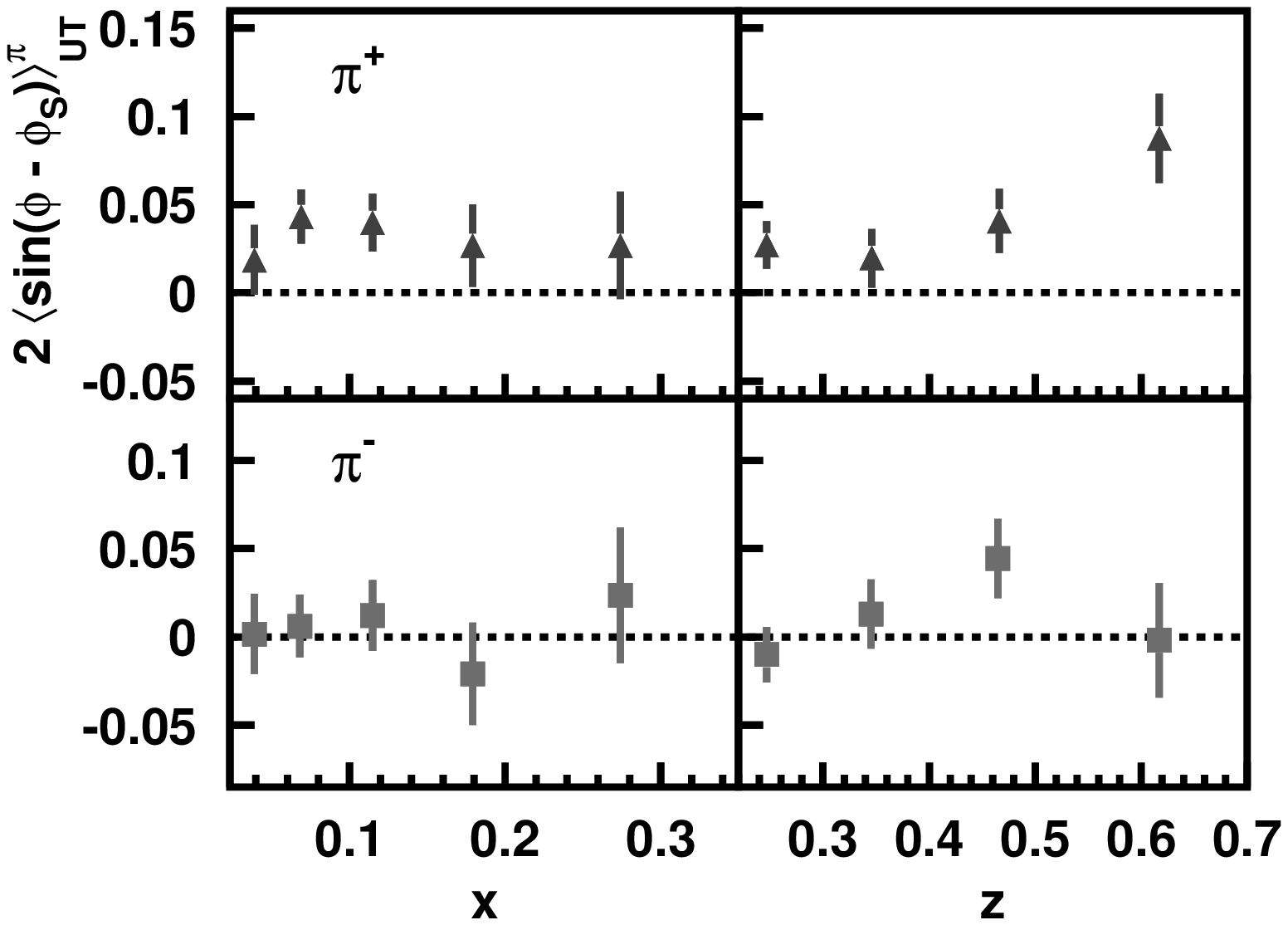}\\
(a) & (b)
\end{tabular}
\end{center}
\caption{Two projected single transverse target-spin asymmetries
measured by HERMES~\cite{Airapetian:2004tw}.
The following cuts were applied: $Q^2 > 1\ \text{GeV}^2$,
$W^2 > 10\ \text{GeV}^2$,
and $0.1< y <0.85$.
According to Boer,
Mulders~\cite{Boer:1998nt} the asymmetry
in figure~(a) is proportional to $h_1 H_1^{\perp(1)}$
while the asymmetry in figure~(b) is proportional $f_{1T}^{\perp(1)} D_1$ (in the scaling limit).
\label{quarkData}}
\end{figure}

From the hadronic tensor, like the above expression,
several asymmetries were calculated in Ref.~\cite{Boer:1997nt}. Two of these
asymmetries were recently measured by HERMES.
In the scaling limit ($Q^2 \rightarrow \infty$) the asymmetries given
in Fig.~\ref{quarkData}a and Fig.~~\ref{quarkData}b become proportional to
the
Sivers function, $f_{1T}^{\perp(1)}$, and the Collins function
$H_1^{\perp(1)}$.

\subsection[Next-to-leading order in $M/Q$]
{Next-to-leading order in $\boldsymbol{M}/\boldsymbol{Q}$}

In the previous subsection we restricted ourselves to leading order in $M/Q$.
Here
the calculation will be continued to next-to-leading order.
We will start
by analyzing in detail the contributions from a single gluon insertion
at order in $M/Q$
(Eq.~\ref{linkPlus2} and Eq.~\ref{linkT2}) and include the other contributing
diagrams later on.

\begin{figure}
\begin{center}
\begin{tabular}{ccccc}
\includegraphics[width=3.3cm]{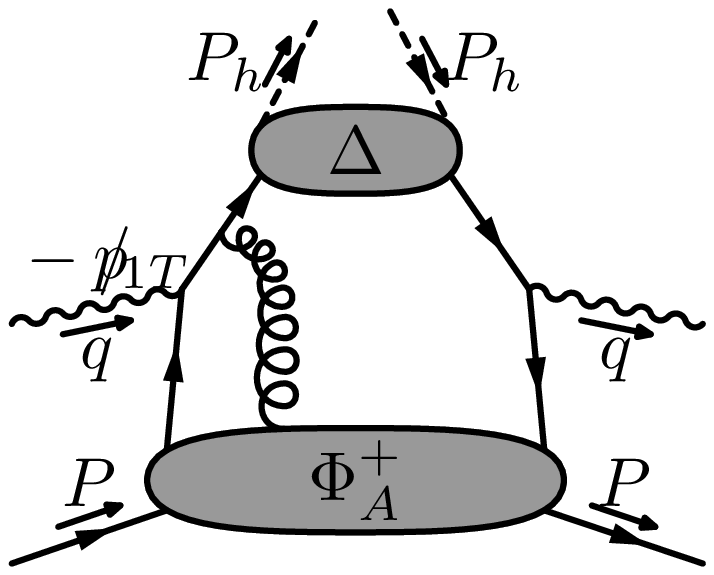} & &
\includegraphics[width=3.3cm]{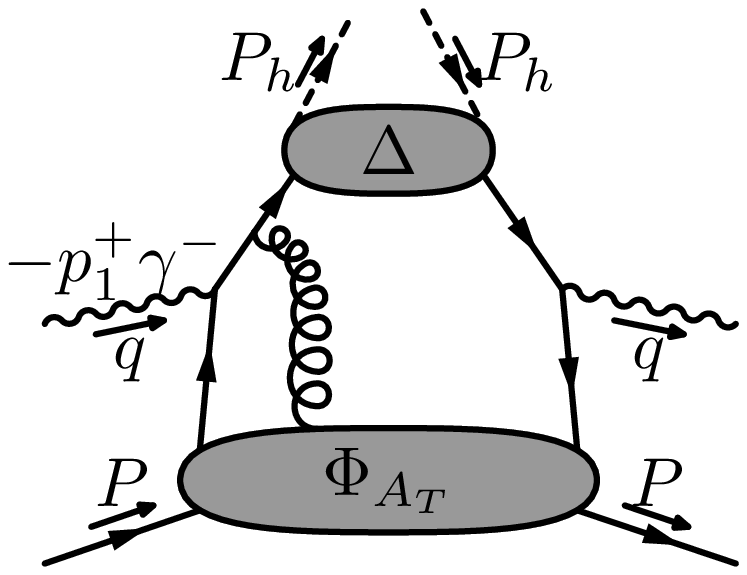} &&
\includegraphics[width=3.3cm]{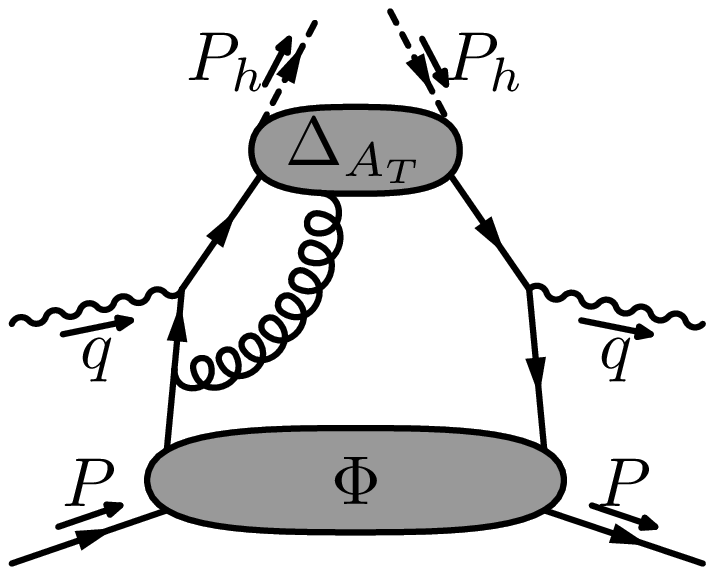}\\
(a) && (b) && (c)
\end{tabular}
\end{center}
\caption{Various diagrams contributing at next-to-leading
order in $M/Q$.\label{quarkMQ}}
\end{figure}

Evaluating Eq.~\ref{linkPlus2} (see also Fig.~\ref{quarkMQ}a) one finds
\begin{multline}
2Mw^{\mu\nu}_{L,-\slashiv{p}_{1T}} =
\int \frac{\rmd{4}{\xi}}{(2\pi)^4} e^{ip\xi}
\langle P,S| \bar{\psi}(0) \gamma^\mu \tr^C \left[ \frac{\Delta(k)}{3}\right]
\frac{\gamma^- \gamma_\alpha }{2k^-}
\gamma^\nu
\\
\times
g \int_{\xi^-}^\infty \rmd{1}{\eta^-}
\left[ \partial^\alpha_T A^+(\eta^-,\xi^+,\xi_T) \right]
\psi(\xi) | P,S \rangle_\text{c} \left(1+\mathcal{O}(M/Q)\right).
\label{sub1}
\end{multline}
This contribution was discovered in Boer, Mulders~\cite{Boer:1999si}.
Another contribution at subleading order comes from a transversely
polarized gluon, see Eq.~\ref{linkT2} and Fig.~\ref{quarkMQ}b. 
Using the relation
$p_1^+/({-}p_1^+ {+} c {+} i\epsilon) = 
{-}1 + (c{+}i\epsilon)/({-}p_1^+ {+} c {+} i\epsilon)$
and a decomposition like Eq.~\ref{quark101}, that
term gives
\begin{multline}
2Mw^{\mu\nu}_{T,-p_1^+ \gamma^-} =
\int \frac{ \rmd{4}{\xi} }{(2\pi)^4}
 e^{ip\xi}
\langle P,S | \bar{\psi}(0) \gamma^\mu \tr^C \left[ \frac{\Delta(k)}{3} \right]
\frac{-\gamma_\alpha \gamma^-}{2k^-} \gamma^\nu
\\
\times
\left( gA_T^\alpha(\xi) - gA_T^\alpha(\infty,\xi^+,\xi_T) \right)
\psi(\xi) | P,S \rangle_\text{c} \left(1 + \mathcal{O}(M/Q) \right).
\label{sub2}
\end{multline}
The contributions from
Eq.~\ref{sub1} and Eq.~\ref{sub2}
can be combined into
\begin{multline}
2Mw^{\mu\nu}_{L,-\slashiv{p}_{1T}} + 2M w^{\mu\nu}_{T,-p_1^+ \gamma^-}\\
\hspace{-2cm}
 =
\int \frac{ \rmd{4}{\xi} }{(2\pi)^4}
 e^{ip\xi}
\langle P,S | \bar{\psi}(0) \gamma^\mu \tr^C \left[ \frac{\Delta(k)}{3} \right]
\frac{-{\gamma_\alpha}_T \gamma^-}{2k^-}
 \gamma^\nu
\\
\times
g \int^{\xi^-}_\infty \rmd{1}{\eta^-} G^{+\alpha}_T(\eta^-,\xi^+,\xi_T)
\psi(\xi) | P,S \rangle_\text{c}\left(1+\mathcal{O}(M/Q)\right)
 + \mathcal{O}(g^2).
\end{multline}
The above result will be generalized to all orders in $g$
in the following paragraphs.

The sum over all possible insertions on both sides
of the cut yields in \nolinebreak \mbox{leading}
\mbox{order} the full gauge link.
Some of the insertions, like $2Mw^{\mu\nu}_{L,-\slashiv{p}_{1T}}$
encountered in Eq.~\ref{linkPlus2}
and $2Mw^{\mu\nu}_{T,-p_1^+ \gamma^-}$ in Eq.~\ref{linkT2}, give a suppression in $M/Q$. In the sum
over all insertions and considering the next-to-leading order in
$M/Q$ such terms appear only
once, either on the left or on the right-hand-side of the cut.

Consider the case
that this term is on the left-hand-side and that the
gluon from the distribution correlator, which is
related to the suppression, is longitudinally polarized
(similar to $2Mw^{\mu\nu}_{L,-\slashiv{p}_{1T}}$ as in Eq.~\ref{linkPlus2}). In that case
the other gluon insertions in the diagram
will contribute to the gauge link. Those gluons cannot have their vertices
between the vertex of
the $A^+$-gluon giving the suppression and the vertex of the
virtual photon.
The reason is that the suppressed term is proportional to $\gamma^-$
which prevents a longitudinal link between it, and that
the suppressed
term vanishes when fields at light-cone infinity couple
between the suppressed
term and the virtual photon.
The conclusion is that the suppressed term always
couples directly next to the photon vertex and the other leading insertions
are adjacent to the fragmentation correlator. This contributes to the
hadronic tensor as
\begin{equation}
\begin{split}
&
2Mw^{\mu\nu}_{L,-\slashiv{p}_{1T}}\text{ (generalized)} \\
& \eqnIndent
=
\int  \rmd{4}{p_1}
\int \frac{ \rmd{4}{\xi} \rmd{4}{\eta} }{ (2\pi)^8 } e^{ip\xi+ip_1(\eta-\xi)}
\\
& \eqnIndent \eqnIndent
\times
\langle P,S | \bar{\psi}(0) \gamma^\mu \Delta(k)
\mathcal{L}^{[+]}(0,\eta^-) i \gamma^- i \frac{- {\slashiv{p}_1}_T}{-2k^- p_1^+
+ i \epsilon} \gamma^\nu gA^+(\eta) \psi (\xi) | P,S \rangle_\text{c} .
\end{split}
\label{sub3}
\end{equation}
By performing the integral over $p_1^+$, using the relation
$\int_\infty^{\xi^-}\!\!\!\rmd{1}{\eta^-}\!
i \partial_{\eta_T}^\alpha \mathcal{L}^{[+]}(0,\eta^-)
A^+(\eta) =
\int_\infty^{\xi^-}\!\!\! \rmd{1}{\eta^-} i\partial_\eta^+
i \partial_{\eta_T}^\alpha \mathcal{L}^{[+]}(0,\eta^-)$, and applying
the identities of the previous chapter (Eq.~\ref{theoryIdentity}) to work out
the transverse derivative
(note that $\langle G^{+\alpha}(\infty,\eta^+,\eta_T)\rangle = 0$), one finds
\begin{equation}
\begin{split}
&
2Mw^{\mu\nu}_{L,-\slashiv{p}_{1T}}\text{ (generalized)}
\\
&
{=}  \!
\int\! \frac{ \rmd{4}{\xi}  }{ (2\pi)^4 } e^{ip\xi}
\langle P,S | \bar{\psi}(0) \gamma^\mu \Delta(k)
\frac{\gamma_\alpha \gamma^-}{2k^-}
\biggl[
- \mathcal{L}^{0_T\!,\ \xi^+}(0^-,\infty^-) \mathcal{L}^{\infty^-\!,\ \xi^+}
(0_T,\xi_T)
gA_T^\alpha(\infty,\xi^+\!,\xi_T)
\\
& \phantom{=}
\
+ \mathcal{L}^{[+]}(0,\xi^-) gA_T^\alpha(\xi)
 -
g \int_{\infty}^{\xi^-} \rmd{1}{\eta^-}
\mathcal{L}^{[+]}(0,\eta^-) G^{+\alpha}_T(\eta)
\mathcal{L}^{\xi_T\!,\xi^+}(\eta^-,\xi^-)
 \bigg]
 \gamma^\nu  \psi (\xi) | P,S \rangle_\text{c}
 \big|_{\substack{\eta^+ = \xi^+\\ \eta_T = \xi_T}}
\\
& \phantom{=} \
\times \left(1+ \mathcal{O}(M/Q) \right).
\end{split}
\raisetag{14pt}
\label{sub4}
\end{equation}

If the suppression term comes
from a transversely polarized gluon like
$2Mw^{\mu\nu}_{T,-p_1^+ \gamma^-}$ in
Eq.~\ref{linkT2}, then
one generally has (following the same arguments)
\begin{equation}
\begin{split}
&
2Mw^{\mu\nu}_{T,-p_1^+ \gamma^-}\text{ (generalized)}
\\
& \eqnIndent
=
\int \rmd{4}{p_1}
\int \frac{ \rmd{4}{\xi} \rmd{4}{\eta} }{ (2\pi)^8 } e^{ip\xi+ip_1(\eta-\xi)}
\\
& \eqnIndent \eqnIndent
\times \langle P,S | \bar{\psi}(0) \gamma^\mu \Delta(k)
\mathcal{L}^{[+]}(0,\eta^-) i \gamma_\alpha i \frac{- p_1^+ \gamma^-}{-2k^- p_1^+
+ i \epsilon} \gamma^\nu gA^\alpha_T(\eta) \psi (\xi) | P,S \rangle_\text{c},
\end{split}
\label{sub5}
\end{equation}
which can be rewritten into
\begin{equation}
\begin{split}
&
2Mw^{\mu\nu}_{T,-p_1^+ \gamma^-}\text{ (generalized)}
\\
& \eqnIndent
=
\int \frac{ \rmd{4}{\xi}  }{ (2\pi)^4 } e^{ip\xi}
\langle P,S | \bar{\psi}(0) \gamma^\mu \tr^C \left[ \frac{\Delta(k)}{3} \right]
\frac{-\gamma_\alpha \gamma^-}{2k^-}
\gamma^\nu
\\
& \eqnIndent \eqnIndent
\times
g
\biggl[
\mathcal{L}^{[+]}(0,\xi^-) A^\alpha_T(\xi)
{-}
\mathcal{L}^{0_T\!,\ \xi^+}(0^-\!,\infty^-)
\mathcal{L}^{\infty^-\!,\ \xi^+}(0_T,\xi_T)
A^\alpha_T(\infty,\xi^+,\xi_T)
\biggr] \psi (\xi) | P,S \rangle_\text{c}
\\
& \eqnIndent \eqnIndent
\times \left(1+ \mathcal{O}(M/Q) \right) .
\end{split}
\raisetag{10pt}
\label{sub6}
\end{equation}

The sum of Eq.~\ref{sub4} and Eq.~\ref{sub6} gives
\begin{equation}
\begin{split}
&
2Mw^{\mu\nu}_{L,-\slashiv{p}_{1T}} + 2Mw^{\mu\nu}_{T,-p_1^+\gamma^-}
\text{  (generalized)}\\
&
=
\int \frac{ \rmd{4}{\xi} }{(2\pi)^4}
 e^{ip\xi}
 \langle P,S | \bar{\psi}(0) \gamma^\mu \tr^C \left[ \frac{\Delta(k)}{3} \right]
\frac{-{\gamma_\alpha}_T \gamma^-}{2k^-}
 \gamma^\nu
\\
&
\phantom{=} \ \times
g \int^{\xi^-}_\infty \rmd{1}{\eta^-}
\mathcal{L}(0,\eta^-)
G^{+\alpha}(\eta) \mathcal{L}^{\xi_T\!,\ \xi^+} (\eta^-,\xi^-)
\psi(\xi) | P,S \rangle_\text{c} \Big|_{\substack{\eta^+ = \xi^+\\ \eta_T = \xi_T}}
\left(1+ \mathcal{O}(M/Q) \right).
\end{split}
\end{equation}
If the suppression is on the right-hand-side of the cut, one obtains
the complex conjugate of this expression with $\mu$ and $\nu$ interchanged.

Integrated over $p^-$, the
matrix elements as above appear often in cross sections, leading to
the following abbreviation for it
\begin{equation}
\begin{split}
& \left( {\Phi^{[\pm]}_{\partial^{-1}G}}\right)_{ij}^{\alpha}
(x,p_T,P,S)
\\
&
\equiv
\int \frac{\rmd{2}{\xi_T}\rmd{1}{\xi^-}}{(2\pi)^3}
e^{ip\xi}
\\
&\phantom{=}\ \times
\langle P,S| \bar{\psi}_j(0)
g\int_{\pm \infty}^{\xi^-}
 \mathrm{d}\eta^- \mathcal{L}^{[\pm]}(0,\eta^-)
G_T^{+\alpha}(\eta) \mathcal{L}^{\xi_T\!,\ \xi^+}(\eta^-,\xi^-) \psi_i(\xi) |P,S \rangle_\text{c}
 \bigg|_{\begin{subarray}{l}
\eta^+ = \xi^+=0 \\ \eta_T = \xi_T \\ p^+ = xP^+ \end{subarray}}.
\end{split}
\label{quarkPhiG}
\end{equation}
Note that ${\Phi^{[\pm]}_{\partial^{-1}G}}_{ij}^{\alpha}$ has only
nonzero transverse components.
Using the identities of the previous chapter, Eq.~\ref{theoryIdentity},
one can show that
\begin{equation}
{\Phi^{[\pm]}_{\partial^{-1}G}}^{\alpha}(x,p_T,P,S)
= {\Phi_D^{[\pm]}}_T^\alpha(x,p_T,P,S) - p_T^\alpha\ \Phi^{[\pm]}(x,p_T,P,S),
\label{hiro6}
\end{equation}
where
\begin{equation}
{\Phi_D^{[\pm]}}_{ij}^\alpha(x,p_T,P,S) {\equiv}\!\!
\int\!\! \frac{\rmd{2}{\xi_T}\!\!\rmd{1}{\xi^-}\!\!}{(2\pi)^3} e^{ip\xi}
\langle P,S | \bar{\psi}_i(0) \mathcal{L}^{[\pm]}(0,\xi^-)
iD_\xi^\alpha \psi_j(\xi) | P,S \rangle_\text{c}
\Big|_{\begin{subarray}{l} \xi^+=0 \\ p^+ = xP^+ \end{subarray}}\!\!.
\end{equation}

At subleading order in $M/Q$ one also encounters a contribution
from the interaction of a transversely polarized
 gluon from the fragmentation correlator
with the hard part (see Fig.~\ref{quarkMQ}c).
 Such a contribution gives
\begin{align}
2MW^{\mu\nu} &=
\int \rmd{4}{p} \rmd{4}{k} \delta^4(p+q-k)\ 2Mw^{\mu\nu}_{T,frag}
+ \text{other diagrams},\\
2Mw^{\mu\nu}_{T,frag} &=
-g
\int \rmd{4}{k_1}
\tr^{D,C} \left[ \Phi \gamma^\mu \Delta_{A_T}^\alpha (k,k_1) \gamma^\nu
\frac{\slashiv{p} + \slashi{k}_1 + m}{(p+k_1)^2 - m^2 + i \epsilon}
\gamma_\alpha t_l \right] \nonumber\\
&= -g
\int \rmd{4}{k_1}
\tr^{D,C} \left[ \tr^C \left[ \frac{\Phi}{3} \right]
 \gamma^\mu \Delta_{A_T}^\alpha (k,k_1) \gamma^\nu
\frac{\gamma^+}{2p^+}
\gamma_\alpha \right].
\label{quark81}
\end{align}
This result can be generalized by considering gluons from the
distribution correlator as well, yielding the gauge link
in the distribution correlator. When applying
the Feynman gauge for the elementary part, the gluon from the
distribution correlator which
couples to the gluon going to the fragmentation correlator contains
an additional term. This term cancels exactly the term
which arises when calculating the leading order transverse
gauge link with the Feynman gauge for the hard part (see the previous
subsection).
The result in the applied light-cone gauge reads
\begin{equation}
2Mw^{\mu\nu}_{T,frag}\text{ (generalized)} =
\frac{-g}{3}\!\!
\int\!\! \rmd{4}{k_1}
\tr^D \left[ \Phi^{[+]} \gamma^\mu
\tr^C \left( \Delta_{A_T}^\alpha (k,k_1) \right) \gamma^\nu
\frac{\gamma^+}{2p^+}
\gamma_\alpha \right]
\end{equation}
Because this result is obtained in the light-cone gauge with retarded boundary
conditions, the fragmentation correlator appears to be
non-gauge-invariant. Assuming gauge invariance of the expression
for the cross section one finds that the
correlator $\int \rmd{1}{k^+}\rmd{4}{k_1} \Delta_{A_T}^\alpha (k,k_1)$
corresponds to the correlator $\gamma^0 \left[
{\Delta_{\partial^{-1}G}^{[-]}}^\alpha(z^{-1},k_T,P_h,S_h)
\right]^\dagger \gamma^0$,
which is gauge invariant and defined as
\begin{align}
&
{\Delta_{\partial^{-1}G}^{[\pm]}}^{\alpha}(z^{-1},k_T,P_h,S_h)
\nonumber\\
&\equiv
\! \frac{1}{3}\!\! \sum_X\!\!\! \int\!\!\! \phaseFactor{P_X}\!\!
\int\!\! \frac{\rmd{2}{\eta_T}\rmd{1}{\eta^+}}{(2\pi)^3} e^{ik\eta}
{}_\text{out}\langle P_h;P_X | \bar{\psi}(0)
\mathcal{L}^{0_T\!,\ \eta^-}\!(0,\pm \infty^+)
\mathcal{L}^{\pm \infty^+\!,\ \eta^-}\!(0,\eta_T)
| \Omega \rangle_\text{c}
\nonumber\\
&\phantom{\equiv}
{\times}
\langle \Omega | g \!\!\!
\int_{\pm \infty}^{\eta^+}\!\!\! \rmd{1}{\zeta^+}\!\!
\mathcal{L}^{\eta_T\!,\ \eta^-}\!(\pm \infty^+\!,\zeta^+\!)
G_T^{-\alpha}\!(0,\zeta^+\!,\eta_T\!)
\mathcal{L}^{\eta_T\!,\ \eta^-}\!(\zeta^+\!,\eta^+\!) \psi(\eta)
| P_h;\! P_X \rangle_\text{out,c}
\Big|_{\begin{subarray}{l} \eta^-=0\\ k^-=z^{-1}P_h^- \end{subarray}}
\label{quarkDeltaG}
\end{align}
Also here one has the relation
\begin{equation}
{\Delta_{\partial^{-1}G}^{[\pm]}}^{\alpha}(z^{-1},k_T,P_h,S_h)
= {\Delta_{D}^{[\pm]}}_T^{\alpha}(z^{-1},k_T,P_h,S_h)-
k_T^\alpha\ \Delta^{[\pm]}(z^{-1},k_T,P_h,S_h),
\label{hiro7}
\end{equation}
where
\begin{equation}
\begin{split}
&
{\Delta_{D}^{[\pm]}}^\alpha_{ij}(z^{-1},k_T,P_h,S_h)
\\
&
\eqnIndent
\equiv \frac{1}{3} \sum_X \int \phaseFactor{P_X}
\int \frac{\rmd{2}{\xi_T}\rmd{1}{\xi^+}}{(2\pi)^3} e^{ik\xi}
{}_\text{out} \langle P_h;P_X | \bar{\psi}_j(0)
\mathcal{L}^{0_T\!,\ \xi^-}(0,\pm \infty^+) | \Omega \rangle_\text{c}
\\
&
\eqnIndent \eqnIndent
\times
\langle \Omega | \mathcal{L}^{\pm \infty^+\!,\ \xi^-}(0_T,\xi_T)
\mathcal{L}^{\xi_T\!,\ \xi^-}(\pm \infty^+,\xi^+ )
iD_\xi^\alpha \psi_i(\xi) | P_h;P_X \rangle_\text{out,c}
\Big|_{\begin{subarray}{l} \xi^-=0\\ k^-=z^{-1}P_h^- \end{subarray}}\!\!.
\end{split}
\raisetag{20pt}
\end{equation}

Considering now all insertions on both sides of the cut one finds that the
hadronic tensor including $M/Q$ corrections
now (finally) reads~\cite{Boer:2003cm} 
(for notational reasons the contribution from
$\gamma^0 [
\Delta_{\partial^{-1}G}^{-\alpha}(z^{-1},k_T,P_h,S_h)
]^\dagger \gamma^0$ was put in $(\mu\rightarrow\nu)^*$)
\myBox{
\vspace{-.4cm}
\begin{equation} \begin{split}
2M W^{\mu\nu} &{=}  \int
\rmd{2}{p_T} \rmd{2}{k_T} \delta^2(p_T+q_T-k_T)
\Bigg[
\tr^D \biggl[ \Phi^{[+]}(x_B,p_T)
\gamma^\mu \Delta^{[-]}(z^{-1},k_T)
\gamma^\nu \biggr]
\\
& \phantom{=}
+ \tr^D \biggl[
- \gamma_\alpha \frac{\slashi{n}_+}{Q\sqrt{2}} \gamma^\nu
{\Phi^{[+]}_{ \partial^{-1}G}}^{\alpha}
(x_B,p_T)
\gamma^\mu \Delta^{[-]}
(z^{-1},k_T)
\\
& \phantom{=} \phantom{\quad \times \mathrm{Tr}\biggl[}
 - {\Delta^{[-]}_{\partial^{-1}G}}^{\alpha}(z^{-1},k_T) \gamma^\nu \Phi^{[+]}(x_B,p_T)
 \gamma_\alpha \frac{\slashii{n}_-}{Q\sqrt{2}} \gamma^\mu
+ (\mu \leftrightarrow
\nu)^* \biggr] \Bigg]
\\
& \phantom{=}\times \left(1+ \MQQ + \mathcal{O}(\alpha_S) \right).
\label{eq1}
\end{split}
\raisetag{10pt}
\end{equation}
\vspace{-.8cm}
\begin{flushright}
\emph{hadronic tensor for semi-inclusive DIS including}$M/Q$
\emph{corrections}
\end{flushright}
}
This expression completes the
descriptions of Mulders, Tangerman~\cite{Mulders:1996dh} and
Boer, Mulders~\cite{Boer:1999si} by including the
transverse gauge link. Apart from diagrams which are connected to the equations
of motion, all possible gluon-interactions between the correlators and the
elementary part have been included.
Although
an infinite set of diagrams was calculated
including $M/Q$ corrections,
the final result still
looks remarkably simple. It is basically expressed in two types of correlators,
$\Phi$ and $\Phi_{\partial^{-1}G}$, and similar ones for fragmentation.
For $\Phi$ one can simply plug in the parametrizations of the previous
chapter. We could in principle parametrize $\Phi_{\partial^{-1}G}$ as well
but it will turn out that this correlator only appears in certain kind of traces.
Using the equations of motion those traces can be rewritten in terms
of the already defined functions of $\Phi$.

\subsection{Some explicit cross sections and asymmetries}

In the previous subsection the hadronic
tensor was derived including next-to-leading order
corrections in $M/Q$. This tensor was expressed in
correlators which have been parametrized in terms of distribution and
fragmentation functions.
Inserting these para\-metrizations, the expressions for the
longitudinal target-spin and beam-spin asymmetries,
as published in Ref.~\cite{Bacchetta:2004zf}, will be given.
Three other papers considered recently
beam-spin asymmetries as well and
motivated the publication of Ref.~\cite{Bacchetta:2004zf}.
Using a model calculation,
Afanasev and Carlson estimated in Ref.~\cite{Afanasev:2003ze}
the beam-spin asymmetry
and compared their results to CLAS data. Yuan obtained in
Ref.~\cite{Yuan:2003gu} an expression for
the asymmetry and made an estimation
by using a chiral quark model and a bag model.
Metz and Schlegel claimed in Ref.~\cite{Metz:2004je},
which also considers longitudinal target-spin
asymmetries,
that the calculation of Ref.~\cite{Afanasev:2003ze} is incomplete. They
completed the model by including other diagrams which, as the authors point
out themselves, are not compatible with the parton model at order $M/Q$.
In
Ref.~\cite{Bacchetta:2004zf} the analysis of Yuan~\cite{Yuan:2003gu}
is completed
by including quark-mass effects and the new function $g^\perp$.

Explicit leading order cross sections can easily be obtained by
replacing the correlators $\Phi^{[+]}$ and $\Delta^{[-]}$ in
Eq.~\ref{eq1} with the explicit parametrizations of the previous
chapter
(see Eq.~\ref{intDistr},~\ref{unintDistr},~\ref{intFrag},~\ref{unintFrag}).
Note that our choice of light-like vectors, Eq.~\ref{quarkFrame}, is
easily related to the frame in which the correlators were defined,
Eq.~\ref{frameDistr},~\ref{frameFragm}.
At order $M/Q$ the other correlators
${\Phi^{[+]}_{\partial^{-1}G}}^{\alpha}$ and
${\Delta^{[-]}_{\partial^{-1}G}}^{\alpha}$ need to be included. They
will be handled here by
making a \emph{Fierz-decomposition} which
%Given that $A {=} a_1 {+} a_2 \gamma_5 {+} a_3^\alpha \gamma_\alpha {+}
%a_4^\alpha i \gamma_\alpha \gamma_5 {+} a_5^{\rho\sigma} \sigma_{\rho\sigma}
%\gamma_5$ and similarly for $B$,
relies on the identity
\begin{align}
&
\tr^D \bigl[ A B \bigr] = a_1 b_1 + a_2 b_2 + a_3^\alpha {b_3}_\alpha +
a_4^\alpha {b_4}_\alpha + a_5^{\alpha\beta} {b_5}_{\alpha\beta},
\nonumber\\
&
\begin{aligned}
&\text{where}& a_1 &= \frac{1}{2}\tr^D \bigl[ A \bigr],&
a_2 &= \frac{1}{2} \tr^D \bigl[ A \gamma_5 \bigr],&
a_3^\alpha &= \frac{1}{2} \tr^D \bigl[ A \gamma^\alpha \bigr],
\nonumber\\
&&
a_4^\alpha &= \frac{1}{2} \tr^D \bigl[ A i \gamma^\alpha \gamma_5 \bigr],&
a_5^{\alpha\beta} &= \frac{1}{2\sqrt{2}} \tr^D \bigl[ A \sigma^{\alpha\beta} \gamma_5 \bigr]&
&\text{and similarly for $b_i$.}
\end{aligned}
\end{align}

Using the Fierz-decomposition one encounters $\Phi_{\partial^{-1}G}$
and $\Delta_{\partial^{-1}G}$ only in particular
combinations of traces which allow for a simplification by the use of
the equations of motion.
Showing only the argument connected to the polarization of the parent
hadron, the correlators $\Phi_{\partial^{-1}G}$
and $\Delta_{\partial^{-1}G}$ are decomposed into
\begin{align}
{\Phi^{[+]}_{\partial^{-1}G}}^\alpha(%x,p_T,P,
S)
&=
{\Phi^{[+]}_{\partial^{-1}G}}^\alpha(%x,p_T,P
0) +
{\Phi^{[+]}_{\partial^{-1}G}}^\alpha(%x,p_T,P,
S_L) +
{\Phi^{[+]}_{\partial^{-1}G}}^\alpha(%x,p_T,P,
S_T),
\\
{\Delta^{[-]}_{\partial^{-1}G}}^\alpha(%z^{-1}\!,k_T,P_h,
S_h)
&=
{\Delta^{[-]}_{\partial^{-1}G}}^\alpha(%z^{-1}\!,k_T,P_h
0) +
{\Delta^{[-]}_{\partial^{-1}G}}^\alpha(%z^{-1}\!,k_T,P_h,
S_{hL}) +
{\Delta^{[-]}_{\partial^{-1}G}}^\alpha(%z^{-1}\!,k_T,P_h,
S_{hT}).
\end{align}
Using the relation $[i\slashii{D}{-}m]\psi =0$, which straightforwardly
gives
$[iD^\mu {+} \sigma^{\mu\nu}D_\nu {-m}\gamma^\mu]\psi =
[i \gamma^\mu D^\nu {-} i \gamma^\nu D^\mu {+} im \sigma^{\mu\nu} {+} i
\epsilon^{\mu\nu\rho\sigma}\gamma_\sigma \gamma_5 iD_\rho]\psi=0$,
one finds for an unpolarized target
(the functions depend on $x$ and $p_T^2$)
\begin{align}
&
\frac{1}{2} \mathrm{Tr} \big[ {\Phi^{[+]}_{\partial^{-1}G}}^{\alpha}(0)
 \sigma_\alpha^{\phantom{\alpha} +}
\big]
= i M x\, e - i m\, f_1 + M x\, h - \frac{p_T^2}{M}\,
h_1^\perp,\\
&
\frac{1}{2} \mathrm{Tr}
\big[ {\Phi^{[+]}_{\partial^{-1}G}}^{\alpha}(0)
 \, i \sigma_\alpha^{\phantom{\alpha}+} \gamma_5 \big] = 0,
\\
&
\frac{1}{2} \mathrm{Tr} \big[ {\Phi^{[+]}_{\partial^{-1}G}}^{\alpha}(0) \,
 \gamma^+ \big] {-}
\frac{1}{2}\, \epsilon_T^{\alpha\beta} \mathrm{Tr}
\big[ {\Phi^{[+]}_{\partial^{-1}G}}_\beta (0) \,
i\gamma^+ \gamma_5 \big]
=
p_T^\alpha \bigl( x \,f^\perp {+}
i \frac{m}{M}\, h_1^\perp {+} i x\, g^\perp {-}  f_1 \bigr),
\end{align}
and for unpolarized observed hadrons (where the functions
depend on $z$ and $z^2 k_T^2$)
\begin{align}
&
\frac{1}{4}
\tr \bigl[ {\Delta_{\partial^{-1}G}^{[-]}}^\alpha(0) \sigma_\alpha^{\phantom{\alpha}-}
\bigr]
= i M_hE^{[-]} - im z D_1^{[-]} + M_h H^{[-]} -\frac{z k_T^2}{M_h}
H_1^{\perp[-]},
\displaybreak[0]\\
&
\frac{1}{4} \tr \bigl[ {\Delta_{\partial^{-1}G}^{[-]}}^\alpha(0) i\sigma_\alpha^{
\phantom{\alpha}+} \gamma_5 \bigr] = 0,
\displaybreak[0]\\
&
\frac{1}{4} \tr \bigl[ {\Delta_{\partial^{-1}G}^{[-]}}^\alpha (0)
 \gamma^- \bigr] {+}
\frac{1}{4}
{\epsilon_T^{\alpha}}_\beta \tr \bigl[
{\Delta_{\partial^{-1}G}^{[-]}}^\beta(0) i \gamma^-
\gamma_5 \bigr]  \nonumber
\\
& \eqnIndent \eqnIndent  \eqnIndent \eqnIndent
\eqnIndent \eqnIndent  \eqnIndent \eqnIndent
= k_T^\alpha \bigl(
D^{\perp[-]} {+} iz\frac{m}{M_h}  H_1^{\perp[-]}
{-} i G^{\perp[-]} {-} zD_1^{[-]} \bigr).
\end{align}
For longitudinal target-spin asymmetries one has the relations
\begin{align}
&
\frac{1}{2} \mathrm{Tr} \big[ {\Phi^{[+]}_{\partial^{-1}G}}^{\alpha}(S_L)
\sigma_\alpha^{\ +} \big] = 0
\\
&
\frac{1}{2} \mathrm{Tr}
\big[ {\Phi^{[+]}_{\partial^{-1}G}}^{\alpha}(S_L)
 \, i \sigma_\alpha^{\phantom{\alpha}+} \gamma_5 \big] =
- m S_L \,g_{1L}  + i M x S_L e_L +
M x S_L\, h_L - \frac{p_T^2}{M} S_L\, h_{1L}^\perp,\\
&
\frac{1}{2} \mathrm{Tr} \big[ {\Phi^{[+]}_{\partial^{-1}G}}^{\alpha}(S_L) \,
 \gamma^+ \big] -
\frac{1}{2}\, \epsilon_T^{\alpha\beta} \mathrm{Tr}
\big[ {\Phi^{[+]}_{\partial^{-1}G}}_\beta(S_L) \,
i\gamma^+ \gamma_5 \big]
\nonumber
\\
&
\eqnIndent \eqnIndent \eqnIndent \eqnIndent
 \eqnIndent \eqnIndent
=
- \epsilon_T^{\alpha\beta} p_{T\beta} \bigl(
x S_L \,f_L^\perp - i \frac{m}{M} S_L\, h_{1L}^\perp + ixS_L\, g_L^\perp
-i S_L \,g_{1L} \bigr).
\end{align}
It turns out\footnote{In the way the calculation is presented here,
it is remarkable that the equations of motion contain sufficient information
to rewrite the appearing correlators $\Phi_{\partial^{-1}G}^\alpha $ and
$\Delta_{\partial^{-1}G}^\alpha$ in terms of $\Phi$,
$\Phi_\partial^\alpha$, $\Delta$, and
$\Delta_\partial^\alpha$.
It has been pointed out by Qiu in Ref.~\cite{Qiu:1988dn}
that one can avoid the quark-gluon-quark correlators by using the equations
of motion at an earlier stage in the calculation. That could make the result
less surprising.}
 that these relations are sufficient to rewrite the appearing
correlators $\Phi_{\partial^{-1}G}$
and $\Delta_{\partial^{-1}G}$ in terms of the distribution and
fragmentation functions.

The vectors in which the cross section is expressed need to be related to
the vectors of the Cartesian basis as introduced in the previous chapter.
The following relations can be found for the vectors $P$, $q$, and
$P_h$ ($e_\pm \equiv ( e_t \pm e_z)/\sqrt{2}$)
\begin{eqnarray}
q &=& \frac{Q}{\sqrt{2}}\ e_- - \frac{Q}{\sqrt{2}}\ e_+,\\
P &=& \frac{x_B M^2 + \mathcal{O}(M^4/Q^2)}{Q\sqrt{2}}\ e_- +
      \frac{Q + \mathcal{O}(M^2/Q)}{x_B \sqrt{2}}\ e_+,\\
P_h &=& \frac{z_h Q+ \mathcal{O}(M_h^2/Q)}{\sqrt{2}}\ e_- +
        \frac{M_h^2 - {P_h}_\perp^2 + \mathcal{O}(M_h^4/Q^2)}{z_h \sqrt{2}Q}
       \  e_+ + {P_h}_\perp.
\end{eqnarray}
Comparing these relations with Eq.~\ref{quarkFrame} (and neglecting
$M^2/Q^2$ corrections)
one finds
\begin{align}
n_+ &= e_+,&
n_- &=  e_- + \frac{\sqrt{2}}{z_h \tilde{Q}} {P_h}_\perp,&
q_{T\perp} &= -P_{h\perp}/z.
\end{align}
This gives the following relations for any two
transverse vectors $m_T$ and $n_T$
\begin{align}
&
{m_T}\ = {m_T}_\perp + m_T {\cdot} e_- \ e_+ + m_T {\cdot e_+} \ e_-
= {m_T}_\perp + \frac{\sqrt{2}}{Q} q_T {\cdot} m_T \ e_+,\\
&
m_T {\cdot n_T} = {m_T}_\perp {\cdot} {n_T}_\perp .
\end{align}

So, by applying the Fierz-decomposition, using the equations of motion,
expressing the result in the Cartesian basis, and using
FORM~\cite{Vermaseren:2000nd}, the hadronic tensor can be obtained.
The tensor, also given in appendix~\ref{quarkAppen},
reads explicitly~\cite{Bacchetta:2004zf}
\nopagebreak
\myBox{
\small
\begin{align}
2 M W^{\mu\nu} &= \left[2 M W_A^{\mu\nu} + 2 M W_S^{\mu\nu}\right]
\left(1+\MQQ + \mathcal{O}(\alpha_S)\right),\label{Wsidis2}
\\
2 M W_A^{\mu\nu} &\approx
\begin{aligned}[t]
&
2 z \int \rmd{2}{p_T} \rmd{2}{k_T}
\delta^2 ({p_T}_\perp - \frac{{P_h}_\perp}{z} - {k_T}_\perp )
\\
& { \times \Biggl\{ } \phantom{+}
i \frac{2 e_t^{[ \mu} {k_T}_\perp^{\nu]}}{Q}
\left[ -\frac{M}{M_h} x_B \ eH_1^{\perp[-]} + \frac{m}{M_h} f_1 H_1^{\perp[-]} -
\frac{1}{z} f_1 G^{\perp[-]} \right]
\\
&
\phantom{\times \Biggl\{ }
{+} i \frac{2 e_t^{[ \mu} {p_T}_\perp^{\nu]}}{Q}
\left[ - \frac{m}{M} h_1^\perp D_1^{[-]} + \frac{M_h}{z M} h_1^\perp E^{[-]} -
x_B g^\perp D_1^{[-]} \right]
\\
&
\phantom{\times \Biggl\{ }
{+} i \epsilon_\perp^{\mu\nu}
\left[ S_L\ g_{1L} D_1^{[-]} - \frac{p_{T\perp}{\cdot}S_{T\perp}}{M} g_{1T} D_1^{[-]}
\right]
\\
&
\phantom{\times \Biggl\{ }
{+} iS_L \frac{2 e_t^{[ \mu} \epsilon_\perp^{\nu ]\rho} {{k_T}_\perp}_\rho}{Q}
\left[ g_{1L} \frac{D^{\perp[-]}}{z} - g_{1L} D_1^{[-]}
- \frac{m}{M} h_{1L}^\perp D_1^{[-]}
- x_B \frac{M}{M_h} e_L H_1^{\perp[-]}
\right]
\\
&
\phantom{\times \Biggl\{ }
{+} i S_L \frac{2 e_t^{[ \mu} \epsilon_\perp^{\nu ]\rho} {{p_T}_\perp}_\rho}{Q}
\left[ x_B g_L^\perp D_1^{[-]} +
\frac{M_h}{M} h_{1L}^\perp \frac{E^{[-]}}{z} -
\frac{m}{M} h_{1L}^\perp D_1^{[-]} \right] \biggr\}
\end{aligned}
\nonumber\\
2 M W_S^{\mu\nu} &\approx
\begin{aligned}[t]
&
2 z \int \rmd{2}{p_T} \rmd{2}{k_T}
\delta^2 ({p_T}_\perp - \frac{{P_h}_\perp}{z} - {k_T}_\perp )
\\
& { \times \Biggl\{ }
- g_\perp^{\mu\nu} f_1 D_1^{[-]}
{+} \frac{g_\perp^{\mu\nu}{k_T}_\perp \cdot {p_T}_\perp - {k_T}_\perp^{\{ \mu} {p_T}_\perp^{\nu \} }}{M M_h}
h_1^\perp H_1^{\perp[-]}
\\
&
\phantom{\times \Biggl\{ }
{+} \frac{2 e_t^{\{ \mu} {k_T}_\perp^{\nu \} } }{Q}
\left[ -f_1 D_1^{[-]} + f_1 \frac{D^{\perp[-]}}{z}
+ x_B \frac{M}{M_h} h \frac{H_1^{\perp[-]}}{z}
\right]
\\
&
\phantom{\times \Biggl\{ }
{+} \frac{2 e_t^{\{ \mu} {p_T}_\perp^{\nu \} } }{Q}
\left[ x_B f^\perp D_1^{[-]} + \frac{M_h}{M} h_1^\perp \frac{H^{[-]}}{z}
+\frac{\threeVec{k}_T^2}{M M_h} h_1^\perp H_1^{\perp[-]}
\right]
\\
&
\phantom{\times \Biggl\{ }
{-} \frac{ {k_T}_\perp^{\{ \mu} \epsilon_\perp^{\nu \} \rho}
{{p_T}_\perp}_\rho + {p_T}_\perp^{\{ \mu} \epsilon_\perp^{\nu \} \rho}
{{k_T}_\perp}_\rho}{2 M M_h} \left[
S_L h_{1L}^\perp H_1^{\perp[-]} - \frac{{k_T}_\perp \cdot
{S_T}_\perp}{M} h_{1T}^\perp  H_1^{\perp[-]} \right]
\\
&
\phantom{\times \Biggl\{ }
{-}
\frac{ {k_T}_\perp^{\{ \mu} \epsilon_\perp^{\nu \} \rho}
{{S_T}_\perp}_\rho + {S_T}_\perp^{\{ \mu} \epsilon_\perp^{\nu \} \rho}
{{k_T}_\perp}_\rho}{2 M_h} h_{1T} H_1^{\perp[-]}
\\
&
\phantom{\times \Biggl\{ }
{+}S_L \frac{2 e_t^{\{ \mu} \epsilon_\perp^{\nu \} \rho} {{k_T}_\perp}_\rho}{Q}
\left[ \frac{M}{M_h} x_B  h_L H_1^{\perp[-]}
- \frac{m}{M_h}  g_{1L} H_1^{\perp[-]} +  g_{1L} \frac{G^{\perp[-]}}{z}
\right]
\\
&
\phantom{\times \Biggl\{ }
{+}S_L \frac{2 e_t^{\{ \mu} \epsilon_\perp^{\nu \} \rho} {{p_T}_\perp}_\rho}{Q}
\left[ \frac{M_h}{M}  h_{1L}^\perp \frac{H^{[-]}}{z}
+ \frac{{\threeVec{k}_T}_\perp^2}{M M_h} h_{1L}^\perp H_1^{\perp[-]}
- x_B  f_L^\perp D_1^{[-]} \right] \biggr\},
\nonumber
\end{aligned}
\end{align}
\begin{flushright}
\emph{the hadronic tensor of semi-inclusive DIS}
\end{flushright}
}
where the approximation-signs indicate that transverse target polarization has
only been taken into account at leading order and any polarizations in the
final state have been discarded. The distribution functions depend
here on $x_B$ and $p_T^2$ while the fragmentation functions have the
arguments $z$ and $z^2 k_T^2$.
It should be noted that the hadronic tensor satisfies
 $q_\mu W^{\mu\nu} = q_\nu W^{\mu\nu} = 0$,
expressing electromagnetic gauge invariance.

In the above equation
the hadronic tensor is expressed in terms of
the various distribution and fragmentation functions. In order to obtain
the cross section, Eq.~\ref{crossSIDIS}, the hadronic tensor still needs to be
 contracted
with the leptonic tensor $L_{\mu\nu}$. Those contractions are
facilitated by the use of table~\ref{table1}.
\begin{table}
\begin{center}
\begin{tabular}{l|l}
 & result of contraction with $L_{\mu\nu}$ \\
\hline
$g_\perp^{\mu\nu}$ & $\frac{4Q^2}{y^2} (-1+y -y^2/2)$
\\
$a_\perp^{\{ \mu} b_\perp^{\nu \} } - a_\perp \cdot b_\perp g_\perp^{\mu\nu}
$ & $\frac{4Q^2}{y^2} (1-y)
\bigl( a\cdot e_x\ b \cdot e_x - a\cdot e_y\ b\cdot e_y \bigr)$
\\
$\frac{1}{2} \left[ a_\perp^{\{ \mu} \epsilon_\perp^{\nu \} \rho} b_{\perp \rho} +
b_\perp^{\{ \mu} \epsilon_\perp^{\nu \} \rho} a_{\perp \rho} \right]$ &
$\frac{4Q^2}{y^2} (-1+y) \bigl( a\cdot e_x\ b\cdot e_y + a\cdot e_y\ b\cdot e_x
\bigr)$
\\
$e_t^{\{ \mu} a_\perp^{\nu \} }$ & $\frac{4Q^2}{y^2} (2-y)\sqrt{1-y}\
a\cdot e_x$
\\
$e_t^{ \{ \mu} \epsilon_\perp^{\nu \} \rho} a_{\perp \rho}$ &$
\frac{4Q^2}{y^2} (y-2) \sqrt{1-y}\ a\cdot e_y$
\\
$i \epsilon_\perp^{\mu\nu}$ & $\frac{4Q^2}{y^2} \lambda_e (y-y^2/2)$
\\
$i a_\perp^{ [ \mu } b_\perp^{\nu ] }$ & $ \frac{4Q^2}{y^2} \lambda_e
(y-y^2/2) \bigl( a\cdot e_x\ b\cdot e_y + a\cdot e_y\ b \cdot e_x \bigr)$
\\
$i e_t^{ [ \mu } b_\perp^{\nu ] }$ & $\frac{4Q^2}{y^2} \lambda_e y \sqrt{1-y}\
a\cdot e_y$\\
$ e_t^{[\mu} \epsilon_\perp^{\nu]\rho} a_{\perp \rho}$ &
$\frac{4Q^2}{y^2} \lambda_e y \sqrt{1-y}\ a\cdot e_x$
\end{tabular}
\end{center}
\caption{Various contractions given frame-independently
(see also Ref.~\cite{Mulders:1996dh}).\label{table1}}
\end{table}
After having made these contractions the result is not yet
a simple product
between the various distribution and fragmentation functions,
but rather a convolution
\mbox{($\int \rmd{2}{p_T}\rmd{2}{k_T}$}\mbox{$ \delta^2({p_T}_\perp
{-} P_{h\perp}/z {-} {k_T}_\perp)$} $ f_i(x_B,p_T^2) D_j(z,z^2k_T^2)$).
The convolution
can be made into a simple product by multiplying the cross sections with some
factors ${P_h}_\perp$ and subsequently integrate
over $\rmd{2}{{P_h}_\perp}$. Note that the leptonic tensor is independent
of $P_{h\perp}$.

The unpolarized ${P_h}_\perp$-integrated cross section is given by
($L_{\mu\nu}^{(U)}$ denotes the unpolarized part of $L_{\mu\nu}$)\nopagebreak
\myBox{
\begin{equation}
\begin{split}
\int \rmd{2}{{P_h}_\perp}E_{\threeVec{P}_h} E_{ \threeVec{l}'}
\frac{  \rmd{6}{\sigma} }
{\rmd{3}{\threeVec{l}'} \rmd{3}{\threeVec{P}_h}} &=
\frac{1}{4s} \frac{ \alpha^2 }{Q^4}
L^{(U)}_{\mu\nu}
\int \rmd{2}{{P_h}_\perp}
2 M W_{U}^{\mu\nu}
\\
&=
\frac{2 \alpha^2}{sQ^2}\ \frac{1-y + y^2/2}{y^2}\ z\ f_1(x_B)\ D_1(z)
\\
& \phantom{=} \ \times
\left(1+\mathcal{O}(M/Q) +
\mathcal{O}(\alpha_S)\right).
\end{split}
\end{equation}
\begin{flushright}
\emph{unpolarized cross section for semi-inclusive DIS at leading order}
\end{flushright}
}

\begin{figure}
\begin{center}
\includegraphics[width=5.5cm]{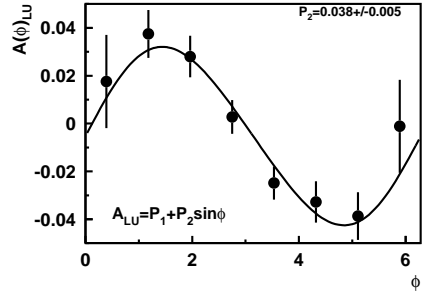}
\end{center}
\caption{Azimuthal beam-spin asymmetry measured by CLAS~\cite{Avakian:2003pk}.
The applied cuts are $Q^2 > 1\ \text{GeV}^2$, $W^2 > 4\ \text{GeV}^2$, and
$0.5 < z < 0.8$.
\label{quarkData2}}
\end{figure}

One of the single spin asymmetries we will study here is the beam-spin
asymmetry. This asymmetry has been measured by
CLAS~\cite{Avakian:2003pk} and
HERMES~\cite{Avetisyan:2004uz}.
The asymmetry will be defined as
\begin{equation}
A_{LU} = \frac{\big(L_{\mu\nu}^{\lambda_e=1}-L_{\mu\nu}^{\lambda_e=-1}\big)\
               2M W^{\mu\nu}_{\mathrm{U}}}
              {\int \mathrm{d}^2 P_h^\perp \
              \big(L_{\mu\nu}^{\lambda_e=1}+L_{\mu\nu}^{\lambda_e=-1}\big)\
              2MW^{\mu\nu}_{\mathrm{U}}},
\end{equation}
where $\lambda_e = 1$ corresponds to a polarization pointing towards the target.
For a general weighted azimuthal asymmetry the following definition is introduced
\begin{equation}
A_{( \ldots )}^{{P}_{h\perp} \cdot \hat{{a}}} \equiv
\int \mathrm{d}^2 P_{h\perp}\; {P}_{h\perp} \cdot \hat{{a}}
\ A_{( \ldots )}.
\end{equation}

Weighting the beam-spin asymmetry with ${P_h}_\perp \cdot e_y$
and integrating over $\rmd{2}{P_{h\perp}}$ gives~\cite{Bacchetta:2004zf}
\nopagebreak
\myBox{
\begin{equation}
\begin{split}
A_{LU}^{{P_h}_\perp \cdot e_y} =
\frac{{-}2y\sqrt{1{-}y}}{(1{-}y{+}y^2/2)\, f_1 D_1}\ \frac{M M_h}{Q}
\bigg[ & \frac{m}{M}z\, f_1 H_1^{\perp[-](1)} - \frac{M_h}{M}\,
f_1 G^{\perp[-](1)}
- x_B z\, e H_1^{\perp[-](1)} \\
&+ \frac{m}{M_h}z\, h_1^{\perp(1)} D_1 -  h_1^{\perp(1)} E
+ \frac{M}{M_h}x_B z\, g^{\perp(1)} D_1 \bigg]
\\
& \times \left(1+ \mathcal{O}(M/Q) + \mathcal{O}(\alpha_S)\right),
\end{split}
\raisetag{10pt}
\label{e:Aluw}
\end{equation}
\begin{flushright}
\emph{azimuthal beam-spin asymmetry for semi-inclusive DIS}
\end{flushright}
}
where the functions depend on $x_B$ or $z$.

The asymmetry given above contains besides the contributions
given in Mulders, Tangerman~\cite{Mulders:1996dh}
and Yuan~\cite{Yuan:2003gu},
two additional terms: the term proportional to $h_1^\perp D_1$ and
the term
proportional to $g^\perp D_1$. Presently, all six contributions are
unknown, making this asymmetry unsuited for studying one function in
particular. To extract information one should get a handle on some of the
contributions, either through phemenological studies
(e.g. see Efremov et al.~\cite{Efremov:2000ar,Efremov:2004tp}) or model
calculations.
For a review on the Collins functions, $H_1^\perp$,
the reader is referred to
Amrath, Bacchetta, and Metz~\cite{Amrath:2005gv} and the references therein.
Models of the Boer-Mulders function, $h_1^\perp$, have been studied by
Gamberg, Goldstein, and
Oganessyan~\cite{Goldstein:2002vv,Gamberg:2002rp,Gamberg:2003ey},
Boer, Brodsky, and Hwang~\cite{Boer:2002ju},
and Lu and Ma~\cite{Lu:2004hu,Lu:2005rq}. Hwang has attempted
to construct a model for
$g^\perp$, but encountered problems with factorization
in the model~\cite{Hwang}.

More
easy to interpret is the
asymmetry of the produced jet which can be obtained
from Eq.~\ref{e:Aluw} by summing over all possible final-state hadrons and
integrating over their phase space.
Neglecting
quark mass contributions the
azimuthal asymmetry of the jet is directly proportional to $g^\perp$,
\nopagebreak
\myBox{
\begin{equation}
A_{LU,j}^{{P_j}_\perp \cdot e_y} =  -\frac{M^2}{Q}
\frac{2 y \sqrt{1-y}}{(1-y+y^2/2)}\
\frac{x_B\,g^{\perp(1)}(x_B)}{f_1(x_B)}
\left(1+ \mathcal{O}(M/Q) + \mathcal{O}(\alpha_S)\right).
\end{equation}
\begin{flushright}
\emph{beam-spin asymmetry for jet production in DIS}
\end{flushright}
}
As explained in the previous chapter
(see subsection~\ref{subsecUnintDistr}), the function $g^\perp$ exists owing
to the directional dependence of the
gauge link which is also connected with the non-validity of the
Lorentz invariance relations. This makes the asymmetry
of increasing interest for the understanding of the theoretical description
(are such functions really non-vanishing?) even though this function does not
have a partonic interpretation (like all the other T-odd functions).
One way to access the asymmetry for the jet-momentum 
is to make an extrapolation by
first considering the leading hadron, then the sum of the leading and
next-to-leading hadron, etc.

\begin{floatingfigure}{3.3cm}
\begin{center}
\includegraphics[width=3.3cm]{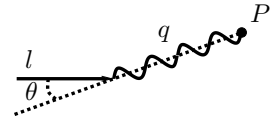}
\end{center}
\caption{Illustration of the lab-frame.\label{labframe}}
\end{floatingfigure}

\sloppy
Next we consider target-spin asymmetries
which have been measured by HERMES~\cite{Airapetian:1999tv,
Airapetian:2001eg,Airapetian:2002mf} and COMPASS~\cite{Pagano:2005jx}.
Considering these asymmetries
one needs to express the
polarization in the lab-frame into the Cartesian basis, Eq.~\ref{hiro13}.
In an experimental setup the target is usually polarized along
or perpendicular to the electron beam. Both directions are combinations
of $e_x$, $e_y$ and $e_z$.
If the target is longitudinally polarized then it has nonzero
$S_L$ and
${S_T}_\perp$.
The contribution of the latter was suggested
by Oganessyan et al.~\cite{Oganessian:1998ma,Oganessyan:2002pc}
to access transversity
via longitudinal target-spin asymmetries
(see for related work Diehl, Sapeta~\cite{Diehl:2005pc}).

\fussy
Defining $\theta$ to be the angle between $\threeVec{q}$ and $\threeVec{l}$
in the target rest frame (see also Fig.~\ref{labframe}),
the following relation can be obtained
\begin{equation}
\theta = \frac{2Mx_B}{Q} \sqrt{1-y} + \MQQ.
\end{equation}
Polarizing the target along the beam leads for
${S_T}_\perp$ and $S_L$ to
\begin{align}
S_L &= \pm 1 + \MQQ,\\
| {\threeVec{S}_T}_\perp | &= \frac{2Mx_B}{Q} \sqrt{1-y} + \MQQ.
\end{align}
Defining the longitudinal target-spin asymmetry as
\begin{equation}
A_{UL'} = \frac{L_{\mu\nu}^{\mathrm{U}}\ \big( 2MW^{\mu\nu}_{S_L'=1} -
                                               2MW^{\mu\nu}_{S_L'=-1}\big)}
                {\int \mathrm{d}^2 P_{h\perp}\ L_{\mu\nu}^{\mathrm{U}}\
                 \big( 2MW^{\mu\nu}_{S_L'=1} +
                                               2MW^{\mu\nu}_{S_L'=-1}\big)
                                           },
\end{equation}
where $S_L' = 1$ denotes a polarization against the beam-direction,
one obtains~\cite{Bacchetta:2004zf}\nopagebreak
\myBox{
\vspace{-.2cm}
\small
\begin{equation} \begin{split}
A_{UL'}^{{P_h}_\perp \cdot e_y} =
\frac{-2\sqrt{1-y}}{(1-y+y^2/2)\, f_1 D_1}\
\frac{M M_h}{Q} \bigg[ & \begin{aligned}[t] ( 2-y ) \bigg( &
\frac{m}{M}z\ g_1 H_1^{\perp[-](1)} {-} \frac{M_h}{M} g_1 G^{\perp[-](1)} {-}
x_Bz\ h_L H_1^{\perp[-](1)} \\
& + h_{1L}^{\perp(1)} \left[ H + 2z H_1^{\perp[-](1)} \right]
 - \frac{M}{M_h}x_Bz\ f_L^{\perp(1)} D_1 \bigg)
\end{aligned} \\
&+ ( 1-y ) \bigg( x_Bz h_{1} H_1^{\perp[-](1)} \bigg) \\
&-( 1-y+y^2/2 ) \bigg( \frac{M}{M_h}x_Bz\ f_{1T}^{\perp(1)} D_1 \bigg)  \bigg]
\\
&\! \times\left(1+\mathcal{O}(M/Q)+\mathcal{O}(\alpha_S)\right),
\end{split}
\raisetag{10pt}
\label{e:Aulw}
\end{equation}
\vspace{-.6cm}
\begin{flushright}
\emph{longitudinal target-spin azimuthal asymmetry in semi-inclusive DIS}
\end{flushright}
}
where the functions depend on $x_B$ or $z$. The functions
$f_L^{\perp(1)}$ and $G^{\perp(1)}$,
which were neglected in previous analyses, have been
included.

Compared to the beam-spin asymmetry there are more terms contributing to the
asymmetry. The sizes of the contributions are at present unknown but the
asymmetry is certainly not dominated by the contribution from transversity
$h_1$
(see also Hermes~\cite{Airapetian:2005jc}).
On the other hand, by 
using the different $y$-dependence one can in principle extract
the contributions from the 
Sivers function, $f_{1T}^{\perp(1)}$, and transversity.
For this extraction different collision energies are required ($y \approx
Q^2/(x_B s)$). 
Experimental statistics could be improved
by integrating over all other external variables.

\section{The Drell-Yan process\label{sectDrellYan}}

The calculation of the hadronic tensor in Drell-Yan is very similar
to the calculations we performed in the previous section. Only the most
important steps will be listed.

In the Drell-Yan process the set of light-like vectors ($\{n_-,n_+\}$ with
$n_- \sim \bar{n}_+$ and $n_-\cdot n_+=1$) is chosen
such that\nopagebreak
\myBox{
\vspace{-.2cm}
\begin{align}
P_1 &= \frac{x_1 M_1^2}{Q\sqrt{2}} n_- + \frac{Q}{x_1 \sqrt{2}} n_+,\nonumber\\
P_2 &= \frac{Q}{x_2 \sqrt{2}} n_- + \frac{x_2 M_2^2}{Q \sqrt{2}} n_+,\nonumber\\
q &= \frac{Q+\mathcal{O}(M^2/Q)}{\sqrt{2}} n_- +
     \frac{Q+\mathcal{O}(M^2/Q)}{\sqrt{2}} n_+ + q_T.
\label{SudaDY}
\end{align}
\vspace{-.6cm}
\begin{flushright}
\emph{Sudakov-decomposition for Drell-Yan}
\end{flushright}
}

\begin{floatingfigure}{3.3cm}
\begin{center}
\includegraphics[width=3.3cm]{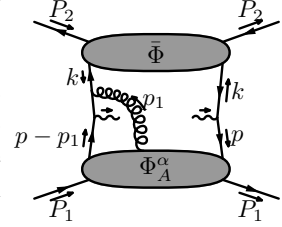}
\end{center}
\caption{A gluon coupling to the incoming antiquark.\label{linkDY}}
\end{floatingfigure}
In this process one can also resum the contributions of gluons coming from
the distribution correlator and connecting with the hard part. The difference with
respect to semi-inclusive DIS is that the quark-fragmentation correlator
is interchanged with an antiquark-distribution correlator describing
an incoming parton.
This results in a gauge link running via minus infinity for the distribution
correlator. The first order calculation illustrates this difference in direction
very clearly.
Inserting a longitudinally polarized gluon from the distribution correlator,
like Fig.~\ref{linkDY}, gives the following contribution to the hadronic tensor
(arguments connected to parent hadrons will be suppressed in this section)
\begin{align}
2MW^{\mu\nu} &=
\int \rmd{4}{p} \rmd{4}{k} \delta^4(p+k-q) \left[
2Mw^{\mu\nu}_{\text{Fig.~}\eqref{linkDY}} + \text{other diagrams} \right],
\label{zondag11}
\\
2Mw^{\mu\nu}_{\text{Fig.~}\eqref{linkDY}} &=
\int \rmd{4}{p_1}
%\nonumber\\
%& \eqnIndent \times
g \tr^{D,C} \bigg[
\Phi_{A_l}^+(p,p_1) \gamma^\mu \bar{\Phi}(k)
(i \gamma^- t_l) i
\frac{-(\slashi{k} + \slashiv{p}_1) + m}{(k{+}p_1)^2 {-} m^2 {+} i \epsilon}
\gamma^\nu \bigg].
\label{DY1}
\end{align}
In order to perform the integral over $p_1^+$ the denominator,
$(k+p_1)^2 - m^2 + i \epsilon$,
is again simplified into
$2p_1^+ k^- + i \epsilon$ (eikonal approximation).
The integral over $p_1^+$ takes now the form
\begin{equation}
\int \rmd{1}{p_1^+} \frac{e^{ip_1^+ (\eta-\xi)^-}}{2p_1^+ k^- + i \epsilon}
A_l^+(\eta)
= \frac{-2\pi i}{2k^-} \theta(\xi^- - \eta^-) A_l^+(\eta).
\end{equation}
This illustrates that there is a contribution when $\eta^- \rightarrow - \infty$,
but not when $\eta^- \rightarrow \infty$.
The pole in $p_1^+$ is now on the other side of the real axis with
respect to semi-inclusive DIS, leading to a gauge link via minus
infinity. One finds for $2Mw^{\mu\nu}_{\text{Fig.~}\eqref{linkDY}}$
\begin{multline}
2Mw^{\mu\nu}_{\text{Fig.~}\eqref{linkDY}}
=
\int \frac{\rmd{4}{\xi}}{(2\pi)^4} e^{ip\xi}\langle P_1,S_1 |
\bar{\psi}(0) \gamma^\mu \bar{\Phi}(k) \gamma^\nu
\\
\times
(-ig) \int_{-\infty}^{\xi^-} \rmd{1}{\eta^-} A^+(\eta^-,\xi_T) \psi(\xi) | P_1,S_1
\rangle_\text{c} \left( 1+ \mathcal{O}(M/Q) \right)
+ \mathcal{O}(\bar{\Phi}_A),
\end{multline}
which is exactly the first order expansion
of a gauge link via minus infinity, and an additional term,
$\bar{\Phi}_A$, which gets canceled
similarly as in semi-inclusive DIS.

We note here that via the coupling of gluons to an incoming parton
we have derived a gauge link via minus infinity. This is similar
to the gauge link for the fragmentation correlator which arose from
coupling gluons to an incoming quark.

The subleading corrections are
also very similar\footnote{
The two terms at subleading order appear opposite in sign. This sign
difference comes from having a quark or antiquark propagator.}.
The complete result
reads
\begin{equation} \begin{split}
2M W^{\mu\nu} &= \frac{1}{3} \int
\rmd{2}{p_T} \rmd{2}{k_T} \delta^2(p_T+k_T-q_T)
\mathrm{Tr}^D \Bigg[ \biggl[ \Phi^{[-]}(x_1,p_T)
\gamma^\mu \bar{\Phi}^{[-]}
(x_2,k_T)
\gamma^\nu \biggr]
\\
&\eqnIndent \phantom{\times \mathrm{Tr}^D \Bigg[}
+ \biggl[ \gamma_\alpha \frac{\slashiv{n}_+}{Q\sqrt{2}} \gamma^\nu
\Phi^{[-]\alpha}_{ \partial^{-1}G}
(x_1,p_T)
\gamma^\mu \bar{\Phi}^{[-]}
(x_2,k_T)
\\
&
\eqnIndent \phantom{\times \mathrm{Tr}^D \Bigg[ + \biggl[}
 - \gamma^\alpha \frac{\slashiv{n}_-}{Q\sqrt{2}} \gamma^\mu
{{\bar{\Phi}}^{[-]\alpha}_{\partial^{-1}G}}
(x_2,k_T)
 \gamma^\nu \Phi^{[-]}
(x_1,p_T)
+  (\mu \leftrightarrow
\nu)^* \biggr] \Bigg]
\\
& \eqnIndent
\times \left(1+ \MQQ + \mathcal{O}(\alpha_S)\right),
\end{split}
\label{maandag1}
\raisetag{14pt}
\end{equation}
where $\Phi_{\partial^{-1}G}^\alpha$ was defined in Eq.~\ref{quarkPhiG} and
(the contraction over color indices has been made explicit)
\begin{align}
{\bar{\Phi}_{ij}}^{[ \pm ]} (x_2,k_T,P_2,S_2) &=
\int \frac{\rmd{2}{\xi_T}\rmd{1}{\xi^+}}{(2\pi)^3} e^{-ik\xi}
\nonumber\\
& \ \ \times
\langle P_2,S_2 | \psi_{i,a}(\xi)
\mathcal{L}_{ba}^{[\pm]}(0,\xi^+) \bar{\psi}_{j,b}(0)
| P_2,S_2 \rangle_{\text{c}}
\big|_{\begin{subarray}{l} \xi^- = 0\\ p^- = x_2 P^-_2\end{subarray}}\!\!,
\\
{{\bar{\Phi}}^{[\pm]\alpha}_{\partial^{-1}G}}(x_2,k_T,P_2,S_2) &=
\bar{\Phi}_{D, T}^{[ \pm ]\alpha}(x_2,k_T,P_2,S_2) + k_T^\alpha\
\bar{\Phi}^{[ \pm ]}(x_2,k_T,P_2,S_2), \label{hiro8}\\
{\bar{\Phi}_{D,ij}}^{[ \pm ]\alpha} (x_2,k_T,P_2,S_2) &=
\int \frac{\rmd{2}{\xi_T}\rmd{1}{\xi^+}}{(2\pi)^3} e^{-ik\xi}
\nonumber\\
& \phantom{\equiv} {\times}
\langle P_2,S_2 |\! \left[iD^\alpha_\xi \psi_{i}(\xi)\right]_a \!
\mathcal{L}_{ba}^{[\pm]}(0,\xi^+) \bar{\psi}_{j,b}(0)
| P_2,S_2 \rangle_{\text{c}}
\big|_{\begin{subarray}{l} \xi^- = 0\\ p^- = x_2 P^-_2\end{subarray}}\!\!.
\end{align}
The hadronic tensor in Eq.~\ref{maandag1} has been derived by employing
the diagrammatic approach. Effects from instantons are for instance not
included. Such effects
could be relevant for the angular distribution of the lepton pair and
have recently been studied by
Boer, Brandenburg, Nachtmann, and Utermann~\cite{Boer:2004mv}.

In order to make a connection with the Cartesian basis
the external momenta are expressed as
($x_1 x_2 s = Q^2 + \mathcal{O}(M^2)$)
\begin{align}
q &= \frac{Q}{\sqrt{2}}\ e_- + \frac{Q}{\sqrt{2}}\ e_+,\\
P_1 &= \frac{x_1 M_1^2 + \mathcal{O}(M_1^4/Q^4)}{Q \sqrt{2}}\ e_- +
     \frac{Q + \mathcal{O}(M_1^2/Q)}{x_1 \sqrt{2}}\ e_+ + P_{1\perp},\\
P_2 &= \frac{Q + \mathcal{O}(M_2^2/Q)}{x_2 \sqrt{2}}\ e_- +
\frac{x_2 M_2^2 + \mathcal{O}(M_2^4/Q^4)}{Q \sqrt{2}}\ e_+ + P_{2\perp}.
\end{align}
Comparing these relations with Eq.~\ref{SudaDY}, one can
derive that (neglecting $M^2/Q^2$ corrections)
\begin{align}
%n_- &= \left(1+\MQQ\right)\ e_- - \left(1+\MQQ\right)\ \frac{q_{T\perp}}{Q\sqrt{2}}
%       + \MQQ\ e_+,\\
%n_+ &= \left(1+\MQQ\right)\ e_+ - \left(1+\MQQ\right)\ \frac{q_{T\perp}}{Q\sqrt{2}}
%       + \MQQ\ e_-,                                                  \\
%q_{T\perp} &= -2x_1 P_{1\perp} \left(1+\MQQ\right) = -2 x_2 P_{2\perp} \left(1+\MQQ\right).
n_- &=  e_- - \frac{q_{T\perp}}{Q\sqrt{2}}, \nonumber\\
n_+ &=  e_+ - \frac{q_{T\perp}}{Q\sqrt{2}}, \nonumber\\
q_{T\perp} &= -2x_1 P_{1\perp}  = -2 x_2 P_{2\perp}.
\end{align}
To express the hadronic tensor in the Cartesian basis the following identities
can be used
\begin{align}
\Phi_\text{twist 2}^{[-]}(x_1,p_T) &=
\frac{\slashi{n}_+ \slashi{n}_-}{2} \Phi_\text{twist 2}^{[-]}(x_1,p_T)
\frac{\slashi{n}_- \slashi{n}_+}{2}
\nonumber\\
&= \biggl[
\Phi_\text{twist 2}^{[-]}(x_1,p_{T\perp})
- \frac{\slashi{q}_{T\perp} \slashi{e}_-}{2\sqrt{2}Q}
 \Phi_\text{twist 2}^{[-]}(x_1,p_{T\perp})
\nonumber
\\
& \phantom{=}
\eqnIndent
-
\Phi_\text{twist 2}^{[-]}(x_1,p_{T\perp})
\frac{\slashi{e}_- \slashi{q}_{T\perp} }{2\sqrt{2}Q} \biggr]
\bigg|_{n_\pm \rightarrow e_\pm} + \MQQ,
\\
\bar{\Phi}_\text{twist 2}^{[-]}(x_2,k_T) &=
\frac{\slashi{n}_- \slashi{n}_+}{2} \bar{\Phi}_\text{twist 2}^{[-]}(x_2,k_T)
\frac{\slashi{n}_+ \slashi{n}_-}{2} \nonumber\\
&= \biggl[
\bar{\Phi}_\text{twist 2}^{[-]}(x_2,k_{T\perp})
- \frac{\slashi{q}_{T\perp} \slashi{e}_+}{2\sqrt{2}Q}
\bar{\Phi}^{[-]}_\text{twist 2}(x_2,k_{T\perp})
\nonumber
\\
& \phantom{=}
\eqnIndent  -
\bar{\Phi}_\text{twist 2}^{[-]}(x_2,k_{T\perp})
\frac{\slashi{e}_+ \slashi{q}_{T\perp} }{2\sqrt{2}Q}
\biggr]
\bigg|_{n_\pm \rightarrow e_\pm}+ \MQQ.
\end{align}

\pagebreak

\noindent
This yields for the hadronic tensor~\cite{Boer:2003cm}\nopagebreak
\myBox{
\small
\begin{equation} \begin{split}
2M W^{\mu\nu} &= \frac{1}{3} \int
\rmd{2}{p_T} \rmd{2}{k_T} \delta^2(p_{T\perp}+k_{T\perp}-q_{T\perp})
\tr^D \Bigg[ \biggl[ \Phi^{[-]}(x_1,p_{T\perp})
\gamma^\mu \bar{\Phi}^{[-]}
(x_2,k_{T\perp})
\gamma^\nu \biggr]
\\
& \
+ \biggl[
 - \frac{\slashiv{p}_{T\perp} \slashi{e}_-}{2\sqrt{2}Q} \Phi^{[-]}(x_1,p_{T\perp})
\gamma^\mu \bar{\Phi}^{[-]}(x_2,k_{T\perp} \gamma^\nu
- \frac{\slashi{k}_{T\perp} \slashi{e}_-}{2\sqrt{2}Q} \Phi^{[-]}(x_1,p_{T\perp})
\gamma^\mu \bar{\Phi}^{[-]}(x_2,k_{T\perp}) \gamma^\nu
\\
& \ \ \ - \Phi^{[-]}(x_1,p_{T\perp})
\gamma^\mu \bar{\Phi}^{[-]}(x_2,k_{T\perp})
\frac{\slashi{e}_+ \slashi{k}_{T\perp}}{2\sqrt{2}Q} \gamma^\nu
- \Phi^{[-]}(x_1,p_{T\perp}) \gamma^\mu
\bar{\Phi}^{[-]}(x_2,k_{T\perp})
\frac{\slashi{e}_+ \slashiv{p}_{T\perp}}{2\sqrt{2}Q} \gamma^\nu
\\
& \ \ \ + \frac{\gamma_\alpha \slashi{e}_+}{Q\sqrt{2}} \gamma^\nu
\Phi^{[-]\alpha}_{ \partial^{-1}G}(x_1,p_{T\perp})
\gamma^\mu \bar{\Phi}^{[-]}(x_2,k_{T\perp})
%\\
%& \ \ \
 - \frac{\gamma_\alpha\slashi{e}_-}{Q\sqrt{2}} \gamma^\mu
{{\bar{\Phi}}^{[-]\alpha}_{\partial^{-1}G}}(x_2,k_{T\perp})
 \gamma^\nu \Phi^{[-]}(x_1,p_{T\perp})
\\
%&
&\ \ \
+ (\mu \leftrightarrow
\nu)^* \biggr] \Bigg] \left(1+ \MQQ + \mathcal{O}(\alpha_S)\right)
\Bigg|_{n_{\pm} \rightarrow e_\pm}.
\end{split}
\raisetag{20pt}
\label{crossDY}
\end{equation}
\begin{flushright}
\emph{the hadronic tensor for Drell-Yan}\\
\emph{in the Cartesian basis including $M/Q$ corrections}
\end{flushright}
}
In the parametrizations of the correlators, which are $n_\pm$-dependent, 
the $n_\pm$ are replaced by $e_\pm$ (that is where $|_{n_{\pm} \rightarrow e_\pm}$
stands for). The functions in the parametrizations
depend on $x_1,\! p_{T\perp}^2$ or $x_2,\! k_{T\perp}^2$. Integrated
and weighted hadronic tensors are given in appendix~\ref{quarkAppen}

\indent
The
above result for the hadronic tensor is similar to the leading order
result of Ralston, Soper~\cite{Ralston:1979ys}. The only actual
difference is the presence of gauge links in the parton distributions.
The Drell-Yan process can be studied experimentally
at RHIC~\cite{GrossePerdekamp:2000wc}
and by the proposed PAX-experiment at GSI.
This has recently generated several \mbox{theoretical}
studies~\cite{Boer:2004mv,Efremov:2004qs,
Efremov:2004tp,Bianconi:2005px,Gamberg:2005ip}.

\section{Semi-inclusive electron-positron annihilation\label{ePluseMin}}

For electron-positron annihilation the calculation is also
very similar. By
including the possible gluon insertions
on the outgoing partons
one obtains gauge links which
are running via plus infinity.

The calculation is set up by
choosing the light-like vectors ($\{n_+,n_-\}$ with $n_-\sim \bar{n}_+$ and
$n_- \cdot n_+ = 1$) such that\nopagebreak
\myBox{
\vspace{-.4cm}
\begin{align}
P_{h_2} &= \frac{M_{h_2}^2}{z_2 Q\sqrt{2}} n_- + \frac{z_2 Q}{\sqrt{2}} n_+,
\nonumber
\\
P_{h_1} &= \frac{z_1 Q}{\sqrt{2}} n_- + \frac{M_{h_1}^2}{z_1 Q \sqrt{2}} n_+,
\nonumber
\\
q &= \frac{Q+\mathcal{O}(M^3/Q)}{\sqrt{2}} n_- + \frac{Q+\mathcal{O}(M^3/Q)}{\sqrt{2}} n_+ + q_T.
\end{align}
\vspace{-.8cm}
\begin{flushright}
\emph{Sudakov-decomposition for electron-positron annihilation}
\end{flushright}
}
The complete result for the hadronic tensor
reads (suppressing arguments connected to parent hadrons)
\begin{equation}
\begin{split}
2M W^{\mu\nu} &= 3    \int
\rmd{2}{p_T} \rmd{2}{k_T} \delta^2(p_T+k_T-q_T)
\tr^D \Bigg[ \biggl[ \Delta^{[-]}(z_1^{-1},p_T)
\gamma^\nu \bar{\Delta}^{[-]}
(z_2^{-1},k_T)
\gamma^\mu \biggr]
\\
&\eqnIndent \quad
+ \biggl[ \gamma_\alpha \frac{\slashi{n}_-}{Q\sqrt{2}} \gamma^\mu
\Delta^{[-]\alpha}_{ \partial^{-1}G}
(z_1^{-1},p_T)
\gamma^\nu \bar{\Delta}^{[-]}
(z_2^{-1},k_T)
\\
&\eqnIndent \qquad
 - \gamma^\alpha \frac{\slashi{n}_+}{Q\sqrt{2}} \gamma^\nu
{{\bar{\Delta}}^{[-]\alpha}_{\partial^{-1}G}}
(z_2^{-1},k_T)
 \gamma^\mu \Delta^{[-]}
(z_1^{-1},p_T)
\\
%&
&\eqnIndent \qquad
+  (\mu \leftrightarrow
\nu)^* \biggr] \Bigg]\left(1+ \MQQ + \mathcal{O}(\alpha_S)\right),
\end{split}
\raisetag{15pt}
\end{equation}
where $\Delta^{[-]\alpha}_{\partial^{-1}G}$ was defined in
Eq.~\ref{quarkDeltaG} and
\begin{equation}
\begin{split}
&
\bar{\Delta}_{ij}^{[\pm]} (z^{-1},k_T,P_h,S_h)
\\
& \ \
\equiv\! \frac{1}{3}\! \sum_X\!\! \int\! \phaseFactor{P_X}\!\!
\int\! \frac{\rmd{2}{\eta_T} \rmd{1}{\eta^-}}{(2\pi)^3}
e^{-ik\eta}
\langle \Omega |
 \bar{\psi}_j(0)
\mathcal{L}^{0_T\!,\ \eta^+}(0^-,\pm \infty^-)
 |P_X; P_h,S_h \rangle_{\text{out,c}}
\\
& \ \ \eqnIndent
\times
{}_\text{out}\langle P_X; P_h,S_h |
\mathcal{L}^{\pm \infty^-\!,\ \eta^+}(0_T,\eta_T)
{\mathcal{L}^{\eta_T\!,\ \eta^+}} (\pm \infty^-,\eta^-)
 \psi_i(\eta)
 | \Omega \rangle_{\text{c}}
\Big|_{\begin{subarray}{l} \eta^+=0\\ k^+ = z^{-1} P_h^+ \end{subarray}},
\end{split}
\raisetag{20pt}
\end{equation}
\begin{equation}
\bar{\Delta}^{[\pm]\alpha}_{\partial^{-1}G}(z^{-1},k_T,P_h,S_h)
\equiv
\bar{\Delta}_{D, T}^{[\pm]\alpha} (z^{-1},k_T,P_h,S_h) + k_T^\alpha\
\bar{\Delta}^{[\pm]} (z^{-1},k_T,P_h,S_h) \label{hiro9},
\end{equation}
\begin{equation}
\begin{split}
&
\bar{\Delta}_{D,ij}^{[\pm]\alpha} (z^{-1},k_T,P_h,S_h)
\\
&\
\equiv\! \frac{1}{3} \! \sum_X\!\! \int\! \phaseFactor{P_X}\!\!
\int\! \frac{\rmd{2}{\eta_T} \rmd{1}{\eta^+}}{(2\pi)^3}
e^{-ik\eta}
\langle \Omega |
 \bar{\psi}_j(0)
\mathcal{L}^{0_T\!,\ \eta^+}(0^-,\pm \infty^-)
 | P_h,S_h \rangle_{\text{out,c}}
\\
& \ \ \eqnIndent
\times
{}_\text{out}\langle P_h,S_h |
\mathcal{L}^{\pm \infty^-\!,\ \eta^+}(0_T,\eta_T)
{\mathcal{L}^{\eta_T\!,\ \eta^+}} (\pm \infty^-,\eta^-)
iD^\alpha_\eta \psi_i(\eta)
 | \Omega \rangle_{\text{c}}
\Big|_{\begin{subarray}{l} \eta^+=0\\ k^+ = z^{-1} P_h^+ \end{subarray}},
\end{split}
\end{equation}

The external momenta in the Cartesian basis read
\begin{align}
q &= \frac{Q}{\sqrt{2}} e_- + \frac{Q}{\sqrt{2}} e_+,\\
P_{h_2} &= \frac{M_{h_2}^2 + \mathcal{O}(M^4_{h_2}/Q^2)}{z_2 Q\sqrt{2}}\ e_-
+ \frac{z_2 Q+ \mathcal{O}(M_{h_2}^2/Q)}{\sqrt{2}}\ e_+,\\
P_{h_1} &= \frac{z_1 Q+ \mathcal{O}(M_{h_1}^2/Q)}{\sqrt{2}}\ e_-
+ \frac{M_{h_1}^2+ \mathcal{O}(M^4_{h_1}/Q^2)}{z_1 Q \sqrt{2}}\ e_+
+ P_{h_1\perp},
\end{align}
giving the relations (neglecting $M^2/Q^2$ corrections)
\begin{align}
n_- &=  e_- - \frac{\sqrt{2}\ q_{T\perp}}{Q},&
n_+ &= e_+,&
q_{T\perp} &= - P_{h_1\perp} / z_1.
\end{align}
Using the following identities to express the hadronic tensor in the
Cartesian basis
\begin{align}
\Delta_\text{twist 2}^{[-]}(z_1^{-1},p_T)
&= \frac{\slashi{n}_- \slashi{n}_+}{2}
\Delta_\text{twist 2}^{[-]}(z_1^{-1},p_T)
\frac{\slashi{n}_+ \slashi{n}_-}{2}
\nonumber\\
&= \biggl[ \Delta_\text{twist 2}^{[-]}(z_1^{-1},p_{T\perp}) -
\frac{\slashi{q}_{T\perp} \slashi{e}_+}{\sqrt{2}Q}
\Delta^{[-]}_\text{twist 2}(z_1^{-1},p_{T\perp})
\nonumber\\
& \phantom{=}
 -
\Delta_\text{twist 2}^{[-]}(z_1^{-1},p_{T\perp})
\frac{\slashi{e}_+ \slashi{q}_{T\perp} }{\sqrt{2}Q} \biggr]
\bigg|_{n_\pm \rightarrow e_\pm} + \MQQ,
\\
\bar{\Delta}_\text{twist 2}^{[-]}(z_2^{-1},k_T) &=
\frac{\slashi{n}_+ \slashi{n}_-}{2} \bar{\Delta}_\text{twist 2}^{[-]}(z_2^{-1},k_T)
\frac{\slashi{n}_- \slashi{n}_+}{2}
\nonumber
\\
&= \bar{\Delta}_\text{twist 2}^{[-]}(z_2^{-1},k_{T\perp})
\bigg|_{n_\pm \rightarrow e_\pm} +\MQQ,
\end{align}
one finds~\cite{Boer:2003cm}
\nopagebreak
\myBox{
\small
\begin{equation} \begin{split}
2M W^{\mu\nu} &= 3
\int
\rmd{2}{p_T} \rmd{2}{k_T} \delta^2(p_{T\perp}+k_{T\perp}+P_{h_1\perp}/z_1)
 \Bigg[ \tr^D \biggl[ \Delta^{[+]}(z_1^{-1},p_{T\perp})
\gamma^\nu \bar{\Delta}^{[+]}
(z_2^{-1},k_{T\perp})
\gamma^\mu \biggr]
\\
& \quad
+ \biggl[
- \tfrac{\slashiv{p}_{T\perp} \slashi{e}_+}{Q\sqrt{2}}
\Delta^{[+]}(z_1^{-1}\!,p_{T\perp})
\gamma^\nu \bar{\Delta}^{[+]}(z_2^{-1}\!,k_{T\perp}) \gamma^\mu
- \tfrac{\slashi{k}_{T\perp} \slashi{e}_+}{Q\sqrt{2}} \Delta^{[+]}(z_1^{-1}\!,p_{T\perp})
\gamma^\nu \bar{\Delta}^{[+]}(z_2^{-1}\!,k_{T\perp}) \gamma^\mu
\\
& \quad \ + \tfrac{\gamma_\alpha\slashi{e}_-}{Q\sqrt{2}} \gamma^\mu
\Delta^{[+]\alpha}_{ \partial^{-1}G}(z_1^{-1},p_{T\perp})
\gamma^\nu \bar{\Delta}^{[+]}
(z_2^{-1},k_{T\perp})
- \tfrac{\gamma_\alpha\slashi{e}_+}{Q\sqrt{2}} \gamma^\nu
\bar{\Delta}^{[+]\alpha}_{\partial^{-1}G}(z_1^{-1},k_{T\perp})
 \gamma^\mu \Delta^{[+]}
(z_1^{-1},p_{T\perp})
\\
%&
& \quad \
+ (\mu \leftrightarrow
\nu)^* \biggr] \Bigg] \left(1+ \MQQ + \mathcal{O}(\alpha_S)\right)
\Bigg|_{\substack{n_\pm \rightarrow e_\pm}}.
\end{split}
\raisetag{16pt}
\label{crossEpEm}
\end{equation}
\begin{flushright}
\emph{hadronic tensor for electron-positron annihilation}\\
\emph{including next-to-leading order in }$M/Q$
\end{flushright}
}
The integrated and weighted hadronic tensors are given in
appendix~\ref{quarkAppen}.

Subleading order effects in electron-positron annihilation
have already been studied by Boer, Jakob, and Mulders
in Ref.~\cite{Boer:1997mf}.
The only real difference of the above result
with Ref.~\cite{Boer:1997mf}
is that the fragmentation functions contain here a fully closed gauge
link.

\section{Fragmentation and universality\label{fragUni}}

\begin{figure}
\begin{center}
\begin{tabular}{cp{2cm}c}
\includegraphics[width=2.7cm]{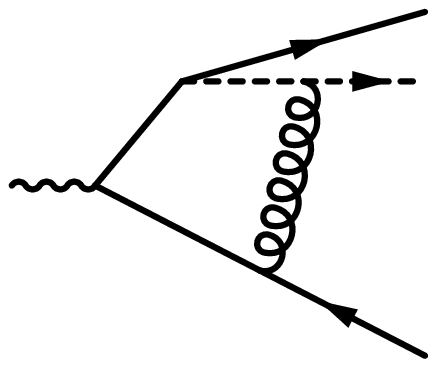} &&
\includegraphics[width=2.7cm]{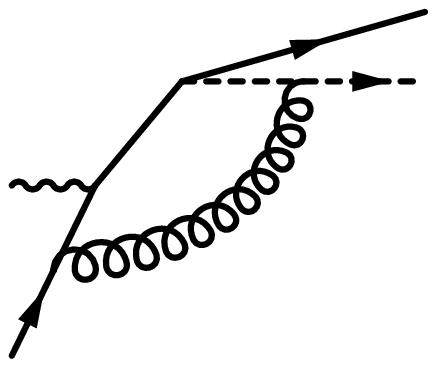}\\
(a) && (b)
\end{tabular}
\end{center}
\caption{Two typical diagrams which appear in model calculations and
produce \mbox{T-odd} effects. Diagram~(a) appears
in electron-positron annihilation and diagram~(b) appears in semi-inclusive
DIS. The dashed-lines represent some kind of particle (for instance
a scalar diquark).\label{fragUniFig}}
\end{figure}

In transverse momentum dependent correlators we encountered
gauge links via plus or minus infinity.
In subsection~\ref{hiro2} we found that
due to the possible interplay
between the two mechanisms for T-odd effects (the gauge link and
final-state interactions) , it is
not possible to relate those correlators which have
different link structures.
However, if one of the mechanisms turns out to be absent or heavily suppressed,
then
relations between fragmentation functions having different gauge links can
be derived.

%Based on model calculations it has been claimed that fragmentation
%functions should
%be universal in general.
In 2002, an article written by Metz~\cite{Metz:2002iz} appeared in
which universality of fragmentation was argued.
In that paper the effect of the
gauge link is modeled through a set of diagrams
in semi-inclusive DIS and electron-positron annihilation, see
Fig.~\ref{fragUniFig}.
Within the model it is found that the Collins function appears with the
same sign in both processes.
In order to achieve this result
the momentum in the loop in Fig.~\ref{fragUniFig}
is not integrated over the full integration domain.
This restriction is justified to make the model
consistent with the parton model but also makes it unclear whether the
universality result still holds if this restriction is lifted.

The result of Ref.~\cite{Metz:2002iz} was formalized in a
subsequent paper by Collins and Metz~\cite{Collins:2004nx}
in which it is claimed that the gauge link
is taken into account.
Having similar diagrams as in Fig.~\ref{fragUniFig},
the authors \emph{eikonalize} the quark-propagator
next to the exchanged gluon, meaning that its
transverse momentum is assumed to be small. This procedure is similar to
the restriction made in Ref.~\cite{Metz:2002iz} because it effectively
limits the
integration domain over the
momenta. When comparing the two
processes, it was found in Ref.~\cite{Collins:2004nx}
that the
pole of the eikonal propagator does not contribute\footnote{
Gamberg, Goldstein, and Oganessyan find in
Ref.~\cite{Gamberg:2003eg} in a slightly different formulation that the
eikonal pole does contribute in T-odd effects, giving a sign dependent
fragmentation function if only the gauge link is considered (no final-state
interactions). This result has been challenged by Amrath, Bacchetta,
and Metz in Ref.~\cite{Amrath:2005gv}. However, if the result
of Ref.~\cite{Gamberg:2003eg} holds then,
together with the final-state interactions, this model would illustrate the
universality problems with fragmentation.}
in the
cross section and
consequently the
difference between the processes vanishes\footnote{After a single
momentum
integration all poles fall on one
side of the real axis. This allows one to close the contour on the other
side of this axis and one finds that the difference between
the processes vanishes.}.
According to the authors this
shows that fragmentation functions are
universal in QCD.

A first comment on both papers is that
restricting the integration domain of the momenta in the loop makes
the interesting result less rigorous. It is not clear whether the
obtained result finds its origin in the analytical properties of Feynman
diagrams or whether it results from the applied restriction.
Secondly, it is not clear whether the 
presence of vertices, 
through which the production of hadrons from quarks is modeled
(like the quark-hadron-diquark vertex), introduce
additional analytical structures which might be the origin for the result.

When comparing the results of Ref.~\cite{Collins:2004nx} with the analyses
in this thesis one observes several differences. For instance, the eikonal
pole turns out not
to contribute in the analysis of Ref.~\cite{Collins:2004nx} while 
we found in
subsection~\ref{hiro} that this pole leads to the gauge link.
In addition,
the diagrams in Fig~\ref{fragUniFig}, 
which should give the effect of the gauge link
(although not proven in Ref.~\cite{Collins:2004nx}), 
give the same sign for T-odd functions
in both processes. This is in contrast with 
the analysis in subsection~\ref{hiro2}
where it was found that there should be a sign-difference
if final-state interactions were to be absent. From that point of view the
effect of the diagrams in Fig~\ref{fragUniFig} should be considered as a
final-state interaction and not as an effect of the gauge link.

%Finally, although the diagrams appear to
%contribute to the gauge link, it is not proven in Ref.~\cite{Collins:2004nx}
%that the diagrams
%actually do. Since the eikonal
%pole - which provided the gauge link in our analyses -
%does not contribute in the cross section
%and the fragmentation functions have the same sign, the interactions
%as depicted in Fig.~\ref{fragUniFig}
%can also be interpreted as a kind of
%final-state interaction (although the diagrams look similar to the diagrams
%which give rise to the gauge link).

The idea of not having a gauge link in the fragmentation functions is
by itself interesting. Fragmentation functions are defined through
matrix elements involving the vacuum which is invariant under translations.
It might be possible to use that property to show that fragmentation functions
are already gauge invariant without the presence of a gauge link.
% the correlator
%would transform as
%(showing only the relevant parts)
%\begin{multline}
%{}_\text{out} \langle P_h,P_X | \bar{\psi}(0) | \Omega \rangle_\text{c}\
%\langle \Omega | \psi(\xi) | P_h,P_X \rangle_\text{out,c}\\
%\overset{?}{\rightarrow}
%{}_\text{out} \langle P_h,P_X | \bar{\psi}(0) | \Omega \rangle_\text{c}
%\ \langle \Omega | \psi(\xi) | P_h,P_X \rangle_\text{out,c}.
%\end{multline}
If a gauge link is not needed to establish gauge invariant fragmentation
correlators then
the derived gauge link in fragmentation functions for
semi-inclusive DIS and electron-positron annihilation
might not influence the expectation value of the
matrix elements. If this turns out be true then
this would make the fragmentation functions sign-independent universal
which would also explain 
the results obtained in Ref.~\cite{Collins:2004nx}. In addition, 
in that case the
Lorentz-invariance relations for fragmentation functions
might be valid as well
(see also the discussion in subsection~\ref{subsecUnintDistr}).

Summarizing, the arguments in Ref.~\cite{Collins:2004nx} lead to an interesting
result although several issues need clarification.
Another point we discussed was
that gauge links might not be relevant for defining
gauge invariant fragmentation functions. This could allow for a nonperturbative
proof of universality of fragmentation. Experimental and theoretical
studies on universality of fragmentation are recommended.

\section{Deeply virtual Compton scattering}

Up to now we derived in this chapter the gauge links in the unintegrated
correlators for various high-energy
scattering processes.
In all these processes interactions
between the correlators and
the hard diagram were included in the form of gluon-lines.
These gluons coupled to external quarks (or fields)
on which the equations of motion were applied to derive the gauge links.
Studying a different kind of process in which gluons are coupled to internal
instead of
external partons is therefore interesting.
The process studied in this section is deeply virtual Compton
scattering (DVCS). Although the gauge link in this process has been derived
in several ways and is perfectly known,
the method which was applied in  previous sections seems
not to be presented for DVCS in the literature.
After some introductory remarks that calculation will be presented here
to illustrate the consistency of the applied approach. Note that
in this section the considered diagrams are amplitude diagrams.

In the DVCS-process an electron scatters, via the
exchange of a virtual photon, off a hadron,
giving, besides the production of a physical photon,
the hadron a momentum change $\Delta$ (with $\Delta^2 \sim -M^2$).
This process gained significant popularity after Ji showed
its connection with quark angular momentum~\cite{Ji:1996ek,Ji:1996nm,Ji:1997pf}.
Together with the intrinsic spin of the quarks, this provides indirect
access to the quark orbital angular momentum as well.
Besides the contributions of Ji considerable progress
has been made by many others. For an overview and introduction the reader
is referred to
Goeke, Polyakov, Vanderhaeghen~\cite{Goeke:2001tz},
Diehl~\cite{Diehl:2003ny}, and
Belitsky, Radyushkin~\cite{Belitsky:2005qn}.

\begin{figure}
\begin{center}
\begin{tabular}{cp{-.1cm}cp{-.1cm}c}
\includegraphics[width=3.3cm]{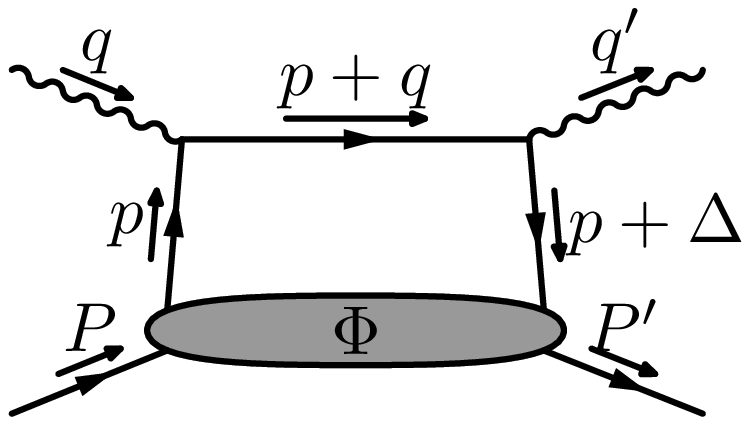} &&
\includegraphics[width=3.3cm]{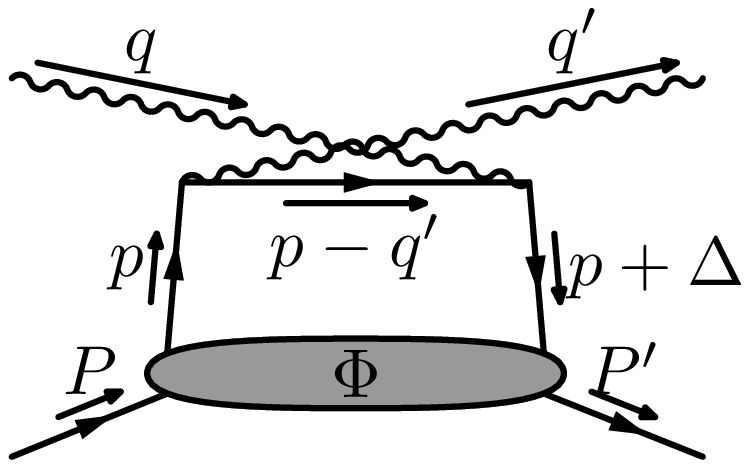} &&
\includegraphics[width=3.3cm]{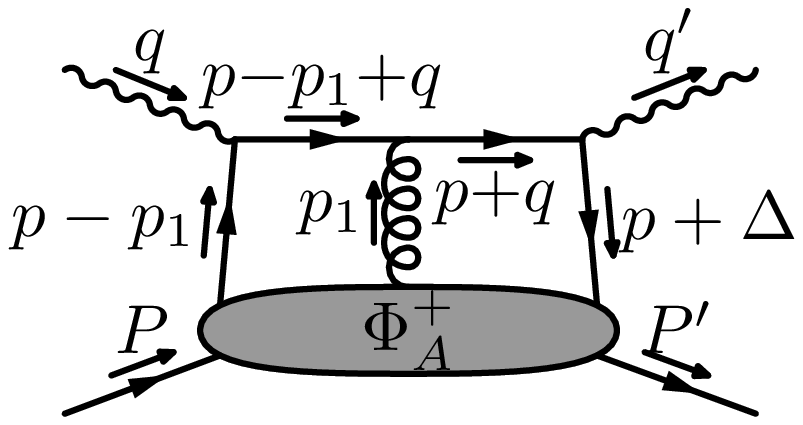}\\
(a) && (b) && (c)
\end{tabular}
\end{center}
\caption{Various contributions in DVCS.\label{figDVCS}}
\end{figure}

In order
to describe the DVCS-process the diagrammatic approach can be employed.
We will slightly deviate from
the more common notations of DVCS in order to follow the notations of this thesis.
Restricting to the
lowest order in $g$, $\alpha_S$, and $\alpha_\text{e.m.}$, there are
two contributions to the amplitude which are
represented in Fig.~\ref{figDVCS}a and Fig.~\ref{figDVCS}b. They are
given by
\begin{align}
i \mathcal{M} &= \bar{u}(k',\lambda') \gamma_\nu u(k,\lambda)
\frac{-i(-ie)(iQ_f e)^2}{q^2}
 (q') (i) T^{\nu\mu}\epsilon_\mu (q'),\\
T^{\nu\mu} &= (i)
\int \rmd{4}{p} \tr^D \bigl[ \Phi(p,P,P') \gamma^\mu i
\frac{\slashiv{p} + \slashi{q}+m}{(p+q)^2 + i \epsilon} \gamma^\nu \nonumber\\
& \phantom{=} \eqnIndent \eqnIndent \eqnIndent \eqnIndent\eqnIndent \eqnIndent+
\Phi(p,P,P') \gamma^\nu i
\frac{\slashiv{p} - \slashi{q'}+m}{(p-q')^2 + i \epsilon} \gamma^\mu \bigr]
+ \mathcal{O}(g),
\label{DVCS1}
\end{align}
where $\Phi(p,P,P')$ is now an off-forward correlator defined as
\begin{equation}
\Phi_{ij}(p,P,P') \equiv \int \frac{\rmd{4}{\xi}}{(2\pi)^4} e^{ip\xi}
{}_\text{out}\langle P',S' |\ \bar{\psi}_j(0)
\ \psi_i(\xi)\ | P,S \rangle_\text{in,c} .
\end{equation}
In the above definition the spin-labels of the hadrons and color indices
of the quark-fields have been suppressed.

Introducing a Sudakov-decomposition to facilitate the expansion in $M/Q$
\begin{align}
P &\equiv \frac{x M^2}{Q \sqrt{2}} n_- + \frac{Q}{x \sqrt{2}} n_+,&
q &\equiv \frac{Q}{\sqrt{2}}n_- - \frac{Q}{\sqrt{2}}n_+,\\
g_\perp^{\mu\nu} &\equiv g^{\mu\nu} - n_+^{\mu} n_-^\nu - n_+^{\nu} n_-^\mu,&
\epsilon_\perp^{\mu\nu} &\equiv \epsilon^{\rho\sigma\mu\nu} {n_+}_\rho {n_-}_\sigma,
\end{align}
where $x = x_B + \MQQ$, and restricting us to $\Delta^+ \sim Q$ and
$\Delta_\perp^2 \sim - M^2$, one obtains in leading order in $M/Q$
(applying also a Fierz-decomposition)\nopagebreak
\myBox{
\begin{align}
T^{\nu\mu} &= \left[ T^{\nu\mu}_{\text{Fig.~}\eqref{figDVCS}a} +
                T^{\nu\mu}_{\text{Fig.~}\eqref{figDVCS}b} \right]
              \left[ 1+ \mathcal{O}(M/Q) \right] + \mathcal{O}(g),
\\
T^{\nu\mu}_{\text{Fig.~}\eqref{figDVCS}a} &=
\int \rmd{1}{p^+}
\frac{1}{p^+ + q^+ + i \epsilon}
\left( g_\perp^{\nu\mu} F(p^+) + (i)\epsilon_\perp^{\nu\mu} \tilde{F}(p^+) \right),
\label{DVCS10}\\
T^{\nu\mu}_{\text{Fig.~}\eqref{figDVCS}b} &=
\int \rmd{1}{p^+} \frac{1}{p^+ - q^+ + \Delta^+ - i \epsilon}
\left( g_\perp^{\nu\mu} F(p^+) - (i)\epsilon_\perp^{\nu\mu} \tilde{F}(p^+)\right),
\label{DVCS2}
\\
\intertext{where}
F(p^+) &= \frac{1}{2} \int \frac{\rmd{1}{\xi^-}}{2\pi} e^{ip^+ \xi^-}
{}_\text{out}\langle P' | \bar{\psi}(0) \gamma^+
\psi(\xi^-) | P \rangle_\text{in,c},
\label{DVCS3}
\\
\tilde{F}(p^+) &= \frac{1}{2}\int \frac{\rmd{1}{\xi^-}}{2\pi} e^{ip^+ \xi^-}
{}_\text{out}\langle P' | \bar{\psi}(0) \gamma^+ \gamma_5
\psi(\xi^-) | P \rangle_\text{in,c}.
\label{DVCS4}
\end{align}
\begin{flushright}
\emph{amplitude for DVCS in leading order in }$M/Q$
\end{flushright}
}
Since all the correlators are on the light-cone the only other possible
leading contributions come from the resummation of $A^+$-gluons.
The above result is therefore complete in the light-cone gauge
(to all orders in $g$) at leading order
in $M/Q$ and corresponds to the result obtained by
Ji in Ref.~\cite{Ji:1996nm}.
Assuming the expression for the cross section to be gauge invariant leads
to a straight gauge link in a covariant gauge
between the two quark-fields in the correlators
$F$ and $\tilde{F}$ above. These correlators can be parametrized
into functions (generalized parton distribution functions) which can be
measured in experiments or be predicted by theory.

Instead of assuming the cross section to be gauge invariant we will consider
now a covariant gauge and derive the gauge link by including
the longitudinally polarized gluons explicitly. The insertion of a single gluon,
connecting the correlator and the hard part (see Fig.~\ref{figDVCS}c),
contributes to the amplitude as
\begin{align}
T^{\nu\mu} &= T_{\text{Fig.~}\eqref{figDVCS}c}^{\nu\mu} + \text{other diagrams},\\
T_{\text{Fig.~}\eqref{figDVCS}c}^{\nu\mu} &=
(i) \int \rmd{4}{p} \rmd{4}{p_1}
\tr^{D,C} \Big[ \Phi_{A_l}^+(p,p_1,P,P') \gamma^\mu
i \frac{\slashiv{p} + \slashi{q} + m}{(p+q)^2 - m^2 + i \epsilon}
 \nonumber
\\
& \phantom{=} \eqnIndent \eqnIndent  \eqnIndent \eqnIndent 
\eqnIndent \eqnIndent\eqnIndent\times
(ig \gamma^- t_l)
i \frac{\slashiv{p} - \slashiv{p}_1
+ \slashi{q} + m}{(p-p_1+q)^2 - m^2 + i \epsilon}
\gamma^\nu \Big] ,
\label{DVCSsingle}
\end{align}
where
\begin{equation}
{\Phi_{A_l}}_{ij}^+(p,p_1,P,P') \equiv
\int \frac{\rmd{4}{\xi} \rmd{4}{\eta}}{(2\pi)^8} e^{ip\xi} e^{i p_1(\eta-\xi)}
{}_\text{out} \langle P' |\ \bar{\psi}_j(0) \ A_l^+(\eta)\ \psi_i(\xi)\ | P \rangle_\text{in,c} .
\end{equation}
Analyzing the $p_1^+$-dependence of Eq.~\ref{DVCSsingle}
and making the parton model assumptions
one finds a pole at
$p_1^+ \approx p^+ + q^+ \neq 0$. Calculating the integral over $p_1^+$
by taking the residue one finds
\begin{multline}
T_{\text{Fig.~}\eqref{figDVCS}c}^{\nu\mu} =
(i) \int \rmd{1}{p^+}
\frac{\rmd{1}{\xi^-}}{2\pi} e^{-iq^+ \xi^-}
{}_\text{out} \langle P' | \bar{\psi}(0)
\gamma^\mu (i) \frac{q^- \gamma^+}{2q^- (p^+ + q^+) + i \epsilon}
\\
\times
(-i) \int_{\infty}^{\xi^-} \rmd{1}{\eta^-} A^+(\eta^-)e^{i(p^+ + q^+)\eta^-}
\gamma^\nu \psi(\xi^-) | P \rangle_\text{in,c}
\left(1+ \mathcal{O}(M/Q)\right).
\end{multline}
We see in the above expression that 
the first order expansion of the gauge link runs between
$\infty$ and $\xi^-$ instead of between $0$ and $\xi^-$. 
When evaluating the $p^+$-integral one obtains
the Heaviside function $\theta(-\eta^-)$. From the two conditions,
$\infty > \eta^- > \xi^-$ and $0>\eta^-$, 
it turns out that of the gauge link
only the part between $\xi^-$ and $0$
remains. This result can be rewritten as follows
\begin{multline}
T_{\text{Fig.~}\eqref{figDVCS}c}^{\nu\mu} =
\int \rmd{1}{p^+} \frac{1}{p^+ + q^+ + i \epsilon}\
\frac{1}{2}\
\int \frac{\rmd{1}{\xi^-}}{2\pi} e^{ip^+ \xi^-}
\\
\hspace{-1cm}
\times \Biggl(
g_\perp^{\nu\mu} {}_\text{out}\langle P' | \bar{\psi}(0) \gamma^+ (-i) \int_0^{\xi^-}
\rmd{1}{\eta^-} A^+(\eta^-) \psi(\xi^-) | P \rangle_\text{in,c}
\\
+ (i) \epsilon_\perp^{\nu\mu}
{}_\text{out}\langle P' | \bar{\psi}(0) \gamma^+ \gamma_5 (-i) \int_0^{\xi^-}
\rmd{1}{\eta^-} A^+(\eta^-) \psi(\xi^-) | P \rangle_\text{in,c} \Biggr)
\\
\ \
\times \left( 1+ \mathcal{O}(M/Q) \right).
\end{multline}
The equation above is a copy of Eq.~\ref{DVCS10} including the first
order gauge link expansion.

Although the momentum of the gluon-line did not vanish
(no pole at $p_1^+ = 0$),
the procedure of including longitudinally polarized gluons from the correlator
as
illustrated here still provides the gauge link. The presented calculation
can be straightforwardly generalized to all orders in $g$.
In leading order in $M/Q$ one obtains Eq.~\ref{DVCS2}
and the matrix elements in Eq.~\ref{DVCS3} and Eq.~\ref{DVCS4} containing
the path-ordered exponentials.

\newpage

\section{Summary and conclusions}

The diagrammatic approach was applied to various
scattering processes.
Following the ideas of Boer,
Mulders, Tangerman~\cite{Mulders:1996dh,Boer:1999si},
and Belitsky, Ji, Yuan~\cite{Belitsky:2002sm}, we assumed
factorization for the correlators and
produced the leading order in $\alpha_S$ cross section
for semi-inclusive DIS, electron-positron annihilation, and Drell-Yan.
In Ref.~\cite{Mulders:1996dh} cross sections were obtained
by choosing effectively the
$A^+ + A^- = 0$ gauge, leaving only the $A_T$-fields from the
correlators to consider. That result,
which includes subleading order corrections in $M/Q$, was studied
in Ref.~\cite{Boer:1999si}
in more detail by considering the $A^- = 0$ gauge.
In that gauge longitudinally polarized gluons between the correlators and
the elementary scattering diagram need to be considered as well.
In Ref.~\cite{Boer:1999si} they were explicitly calculated to order $g^2$
of which the result was generalized to all orders in $g$ by using Ward identities.
Afterwards it was pointed out
in Ref.~\cite{Belitsky:2002sm} that there is an additional
contribution which was not identified before (the transverse gauge link).
The theoretical description was here
extended by including all leading order $\alpha_S$ corrections
and including $M/Q$ corrections. The obtained result is similar to the results of
Ref.~\cite{Mulders:1996dh} except that T-odd distribution functions are now
also included and the approach is fully color gauge invariant.

The hadronic tensors were expressed in
gauge invariant correlators containing
gauge links. These gauge links arise from including
leading-order-$\alpha_S$-interactions which consist of gluons interacting between
correlators and the elementary scattering diagram.
The paths of these
gauge links run via plus infinity if the gluons are coupled to
an outgoing parton,
and via minus infinity if the gluons are coupled to an incoming
parton.
Following the same approach we also considered the
DVCS-process in which gluons couple to internal parton-lines.
The path of the obtained gauge link for this process is just
a straight line between the two
quark-fields as also encountered in DIS.

Using the expressions for the hadronic tensor in semi-inclusive DIS,
some explicit asymmetries were derived.
One of the more interesting asymmetries is the beam-spin asymmetry
for jet-production in DIS, which is proportional to the distribution
function $g^\perp$. This function was discussed in chapter~\ref{chapter2}
and originates from 
the directional dependence of the gauge link. Whether or not
such functions are really non-vanishing is at present an open question. Its measurement
would contribute to the understanding of the theoretical description.

In the second part
of this chapter, it was pointed out that the interesting
arguments for universality of fragmentation as given in
Collins, Metz~\cite{Collins:2004nx}
contain some
issues which need clarification. Therefore,
the difference in paths of the gauge links in the fragmentation functions
appearing in semi-inclusive DIS and electron-positron annihilation
could lead to different fragmentation functions for the two processes.
This might create a problem for extracting transversity
via the Collins effect. Alternative ways to access transversity,
besides Drell-Yan, are for instance $J/ \psi$ production in
$pp$-scattering~\cite{Anselmino:2004ki}, and via interference fragmentation
functions~\cite{Jaffe:1997hf,Bacchetta:2003vn,Bacchetta:2004it,vanderNat:2005aq}
(in which the nonlocality of the operators is light-like).

\newpage

\begin{subappendices}

\section{Hadronic tensors\label{quarkAppen}}

%\vspace{-.2cm}

Based on Ref.~\cite{Boer:2003cm}
some hadronic tensors will be given.
Contributions from antiquarks can be included
by interchanging $q$ with $-q$ and $\mu$ with $\nu$.
The following definitions were used
(suppressing arguments of parent hadrons):
\begin{align}
\Phi_\partial^{[\pm]\alpha} (x)&= \int \rmd{2}{p_T}\ p_T^\alpha\ \Phi^{[\pm]}(x,p_T),&
\bar{\Phi}_\partial^{[\pm]\alpha} (x)&= -\int \rmd{2}{p_T}\ p_T^\alpha\ \bar{\Phi}^{[\pm]}(x,p_T),
\nonumber\\
\Delta_\partial^{[\pm]\alpha} (z^{-1})&= \int \rmd{2}{k_T}\ k_T^\alpha\ \Delta^{[\pm]}(z^{-1},k_T),&
\bar{\Delta}_\partial^{[\pm]\alpha} (z^{-1})&= -\int \rmd{2}{k_T}\ k_T^\alpha\ \bar{\Delta}^{[\pm]}(z^{-1},k_T),
\nonumber\\
\Phi_D^\alpha (x)&= \int \rmd{2}{p_T} \Phi_D^{[\pm]\alpha}(x,p_T),&
\bar{\Phi}_D^\alpha (x)&= \int \rmd{2}{p_T} \bar{\Phi}_D^{[\pm]\alpha}(x,p_T),
\nonumber\\
\Delta_D^\alpha (z^{-1})&= \int \rmd{2}{k_T} \Delta_D^{[\pm]\alpha}(z^{-1},k_T),&
\bar{\Delta}_D^\alpha (z^{-1})&= \int \rmd{2}{k_T} \bar{\Delta}_D^{[\pm]\alpha}(z^{-1},k_T).
\end{align}

%\vspace{-.2cm}

\subsubsection{Semi-inclusive DIS}

The
translation of Eq.~\ref{eq1} into the Cartesian basis
is similar as was done for semi-inclusive electron-positron annihilation
or Drell-Yan
(see Ref.~\cite{Boer:2003cm} for details). 
The unintegrated hadronic tensor reads
\vspace{-.1cm}
\begin{equation}
\begin{split}
2MW^{\mu\nu} = & \int \rmd{2}{p_{T}} \rmd{2}{k_{T}}
\delta^2(p_{T\perp} - P_{h\perp}/z - k_{T\perp}) \tr^D \Big[
\left(
\Phi(x_B,p_{T\perp}) \gamma^\mu \Delta(z_1^{-1},k_{T\perp}) \gamma^\nu
\right)
\\
&+ \Big(
\Phi(x_B,p_{T\perp}) \gamma^\mu \tfrac{\slashiv{p}_{T\perp}
\slashii{e}_+}{Q\sqrt{2}} \Delta(z_1^{-1},k_{T\perp}) \gamma^\nu
-
\Phi(x_B,p_{T\perp}) \gamma^\mu
\tfrac{\slashiv{k}_{T\perp} \slashii{e}_+}{Q\sqrt{2}} \Delta(z_1^{-1},k_{T\perp}) \gamma^\nu
\\
%&+ \Phi(x_B,p_{T\perp}) \gamma^\mu  \Delta(z_1^{-1},k_{T\perp})
%\frac{\slashii{e}_+\slashiv{p}_{T\perp} }{Q\sqrt{2}} \gamma^\nu
%- \Phi(x_B,p_{T\perp}) \gamma^\mu  \Delta(z_1^{-1},k_{T\perp})
%\frac{\slashii{e}_+\slashiv{k}_{T\perp} }{Q\sqrt{2}} \gamma^\nu
%\\
&- \tfrac{\gamma_\alpha \slashii{e}_+}{Q\sqrt{2}} \gamma^\nu
\Phi_{\partial^{-1}G}^{[+]\alpha}(x_B,\!p_{T\perp}) \gamma^\mu\!
\Delta^{[-]}(z^{-1}\!\!\!,k_{T\perp})
{-}
\Delta_{\partial^{-1}G}^{[+]\alpha}(z^{-1}\!\!\!,k_{T\perp})
\gamma^\nu
\Phi^{[+]}(x_B,\!p_{T\perp}) \tfrac{\gamma_\alpha \slashii{e}_-}{Q\sqrt{2}}
\gamma^\mu
\\
&
+(\mu \leftrightarrow \nu)^* \Big) \Big]
\left(1 + \MQQ + \mathcal{O}(\alpha_S) \right)
\Big|_{\substack{n_\pm \rightarrow e_\pm}}.
\end{split}
\raisetag{14pt}
\end{equation}
The integrated hadronic tensor reads 
(employing Eq.~\ref{hiro6} and Eq.~\ref{hiro7})
\vspace{-.1cm}
\begin{equation}
\begin{split}
\int \rmd{2}{P_{h\perp}} 2MW^{\mu\nu} =& z^2
\tr^D \Big[
\Phi(x_B) \gamma^\mu \Delta(z^{-1})\gamma^\nu
\\
& + \Big( \frac{\slashii{e}_+ \gamma_\alpha}{Q\sqrt{2}} \gamma^\nu
\Phi_D^\alpha (x_B) \gamma^\mu \Delta(z^{-1}) +
\frac{\slashii{e}_- \gamma_\alpha}{Q\sqrt{2}} \gamma^\mu
\Delta_D^\alpha(z^{-1}) \gamma^\nu \Phi(x_B)
\\
& -  \frac{\slashii{e}_- \gamma_\alpha}{Q\sqrt{2}}
\gamma^\mu \Delta_\partial^{[-]}(z^{-1}) \gamma^\nu \Phi(x_B) -
\gamma^\mu \frac{\gamma_\alpha \slashii{e}_+}{Q\sqrt{2}}
\Delta_\partial^{[-]\alpha}(z^{-1}) \gamma^\nu \Phi(x_B)
\\
& + (\mu \leftrightarrow \nu)^* \Big) \Big]
\left(1+\MQQ + \mathcal{O}(\alpha_S) \right)
\Big|_{\substack{n_\pm \rightarrow e_\pm}}.
\end{split}
\raisetag{14pt}
\end{equation}
The single weighted hadronic tensor reads
\vspace{-.1cm}
\begin{equation}
\begin{split}
\int \rmd{2}{P_{h\perp}} P_{h\perp}^\alpha\ 2MW^{\mu\nu} = z^3 &
\tr^D \Big[
-\Phi_\partial^{[+]\alpha}(x_B) \gamma^\mu \Delta(z^{-1})\gamma^\nu
+ \Phi(x_B) \gamma^\mu \Delta_\partial^{[-]\alpha}(z^{-1})\gamma^\nu \Big]
\\
& \times \left(1+\mathcal{O}(M/Q) + \mathcal{O}(\alpha_S) \right)
\Big|_{\substack{n_\pm \rightarrow e_\pm}}.
\end{split}
\raisetag{14pt}
\end{equation}

\subsubsection{The Drell-Yan process}

The integrated hadronic tensor reads 
(employing Eq.~\ref{hiro6} and Eq.~\ref{hiro8})
\begin{equation}
\begin{split}
\int \rmd{2}{q_{T\perp}}& 2MW^{\mu\nu} = \frac{1}{3}
\tr^D \Big[
\Phi(x_1) \gamma^\mu \bar{\Phi}(x_2)\gamma^\nu
\\
& + \Big( - \frac{\slashii{e}_+ \gamma_\alpha}{Q\sqrt{2}} \gamma^\nu
\Phi_D^\alpha (x_1) \gamma^\mu \bar{\Phi}(x_2) +
\frac{\slashii{e}_- \gamma_\alpha}{Q\sqrt{2}} \gamma^\mu
\bar{\Phi}_D^\alpha(x_2) \gamma^\nu \Phi(x_1)
\\
& + \frac{1}{2} \frac{\slashii{e}_+ \gamma_\alpha}{Q\sqrt{2}}
\gamma^\nu \Phi_\partial^{[-]\alpha}(x_1) \gamma^\mu \bar{\Phi}(x_2) -
\frac{1}{2} \gamma^\nu \frac{\gamma_\alpha \slashii{e}_-}{Q\sqrt{2}}
\Phi_\partial^{[-]\alpha}(x_1) \gamma^\mu \bar{\Phi}(x_2)
\\
& - \frac{1}{2} \frac{\slashii{e}_- \gamma_\alpha}{Q\sqrt{2}}
\gamma^\mu \bar{\Phi}_\partial^{[-]\alpha}(x_2) \gamma^\nu \Phi(x_1) +
\frac{1}{2} \gamma^\mu \frac{\gamma_\alpha \slashii{e}_+}{Q\sqrt{2}}
\bar{\Phi}_\partial^{[-]\alpha}(x_2) \gamma^\nu \bar{\Phi}(x_1)
\\
& + (\mu \leftrightarrow \nu)^* \Big) \Big]
\left(1+\MQQ + \mathcal{O}(\alpha_S) \right)
\Big|_{\substack{n_\pm \rightarrow e_\pm}}.
\end{split}
\end{equation}
The single weighted hadronic tensor reads
\begin{equation}
\begin{split}
\int \rmd{2}{q_{T\perp}} q_{T\perp}^\alpha\ 2MW^{\mu\nu} = &\frac{1}{3}
\tr^D \Big[
\Phi_\partial^{[-]\alpha}(x_1) \gamma^\mu \bar{\Phi}(x_2)\gamma^\nu
- \Phi(x_1) \gamma^\mu \bar{\Phi}_\partial^{[-]\alpha}(x_2)\gamma^\nu \Big]
\\
& \times \left(1+\mathcal{O}(M/Q) + \mathcal{O}(\alpha_S) \right)
\Big|_{\substack{n_\pm \rightarrow e_\pm}}.
\end{split}
\end{equation}

\subsubsection{Semi-inclusive electron-positron annihilation}

The integrated hadronic tensor reads
(employing Eq.~\ref{hiro7} and Eq.~\ref{hiro9})
\begin{equation}
\begin{split}
\int \rmd{2}{q_{T\perp}}& 2MW^{\mu\nu} = 3
\tr^D \Big[
\bar{\Delta}(z_2^{-1}) \gamma^\mu \Delta(z_1^{-1})\gamma^\nu
\\
& + \Big( \frac{\slashii{e}_+ \gamma_\alpha}{Q\sqrt{2}} \gamma^\nu
\bar{\Delta}_D^\alpha (z_2^{-1}) \gamma^\mu \Delta(z_1^{-1}) -
\frac{\slashii{e}_- \gamma_\alpha}{Q\sqrt{2}} \gamma^\mu
\Delta_D^\alpha(z_1^{-1}) \gamma^\nu \bar{\Delta}(z_2^{-1})
\\
& + \frac{\slashii{e}_- \gamma_\alpha}{Q\sqrt{2}}
\gamma^\mu \Delta_\partial^{[+]\alpha}(z_1^{-1}) \gamma^\nu \bar{\Delta}(z_1^{-1})-
\gamma^\mu \frac{\gamma_\alpha \slashii{e}_+}{Q\sqrt{2}}
\Delta_\partial^{[+]\alpha}(z_1^{-1}) \gamma^\nu \bar{\Delta}(z_2^{-1})
\\
& + (\mu \leftrightarrow \nu)^* \Big) \Big]
\left(1+\MQQ + \mathcal{O}(\alpha_S) \right)
\Big|_{\substack{n_\pm \rightarrow e_\pm}}.
\end{split}
\end{equation}
The single weighted hadronic tensor reads
\begin{multline}
\int \rmd{2}{q_{T\perp}} q_{T\perp}^\alpha\ 2MW^{\mu\nu} = 3
\tr^D \Big[
\bar{\Delta}(z_2^{-1}) \gamma^\mu \Delta_\partial^{[+]\alpha}(z_1^{-1}) \gamma^\nu
- \bar{\Delta}_\partial^{[+]\alpha}(z_2^{-1}) \gamma^\mu \Delta(z_1^{-1})\gamma^\nu \Big]
\\
\times \left(1+\mathcal{O}(M/Q) + \mathcal{O}(\alpha_S) \right)
\Big|_{\substack{n_\pm \rightarrow e_\pm}}.
\end{multline}

\end{subappendices}

\chapter{Color gauge invariance in hard scattering processes
\label{chapter4}}

\vspace{-.8cm}

In the previous chapter several
processes were studied at tree-level (leading order $\alpha_S$). 
The hard scale was set
by an electromagnetic interaction
involving two hadrons and, assuming factorization, cross sections
were expressed
in nonlocal scale-dependent correlators. These correlators contain
gauge links of which the path depends on the process.
As first noted by Collins~\cite{Collins:2002kn}, this produces
a sign-flip
for T-odd distribution functions when comparing Drell-Yan
with semi-inclusive DIS.

Following the same approach in this chapter,
scattering processes will be analyzed
in which the hard scale is set by an QCD-interaction (besides the participating
hadrons).
We will find that the gauge link does not only depend on the
process, but even within a process different gauge links appear;
the gauge links will depend on the hard part or
subprocess. How awkward this at first
may seem, the procedure appears to be consistent. A prescription
for deducing gauge links
will be given together with some results for gluon-gluon correlators.

The fact that the gauge link depends on the diagram and not only
on the process
immediately raises questions on factorization and universality.
This particular topic will be discussed in the last part of this chapter.

\section{Gauge links in tree-level diagrams}

In this section
gauge links in QCD-scattering processes will be considered. Having
an $\alpha_S$-interaction in the hard part at lowest order in $g$,
the number of ways in which the gluons can be inserted
is richer than for the previously discussed
electromagnetic processes.
Not only will we find more complex gauge links than in the electromagnetic
processes, but we will also observe
that the path of the
gauge link depends on the elementary scattering diagram within the process.

In the next subsection a simple QCD-scattering process (quark-quark
scattering) will be worked out. Using the results of the previous chapter
it will be possible to derive the gauge links very quickly. The calculation
will elaborate upon the presence of complex link structures in QCD.

In the second and third subsection we will consider processes in which gluons
as external partons are present in the form of jets. 
Although the same approach is followed, the calculation in the second
subsection
is technically more involved. We will obtain some results for Drell-Yan
and semi-inclusive DIS in which an additional
gluon-jet is being produced.

The structure of the obtained results suggests a general
prescription for deriving
gauge links appearing in diagrams. 
This prescription
will be given together with some examples in the last subsection.

As a final remark, most of the calculations will be done at the amplitude level.
We will often anticipate by only keeping 
terms which contribute to the cross section. 
The intermediate results will be the amplitude diagrams in which
path-ordered exponentials appear. However, in order to absorb these path-ordered
exponentials into the correlator, one needs to consider the calculation at
the cross section level. It is then also essential to sum over the colors of
the other external partons.

\subsection{Gauge links in quark-quark scattering}

The quark-quark scattering subprocess is relevant for hadron-hadron collisions
and will be studied in this subsection. 
The gauge links are derived for the correlators which are 
attached to the elementary scattering subprocess (squared
amplitude diagram). All external partons are assumed to be separated by
large momentum differences. We shall start by 
considering gluon insertions from 
a single distribution correlator and derive the gauge link.
Results for the
other correlators will be given at the end of this subsection.

\begin{figure}
\begin{center}
\begin{tabular}{ccccc}
\includegraphics[width=3.3cm]{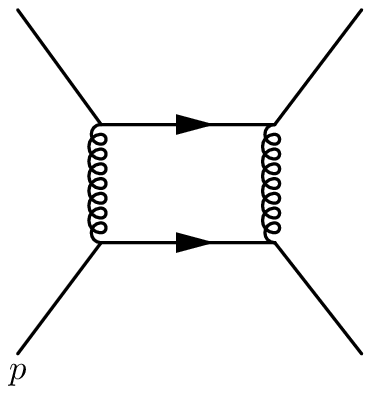}&&
\includegraphics[width=3.3cm]{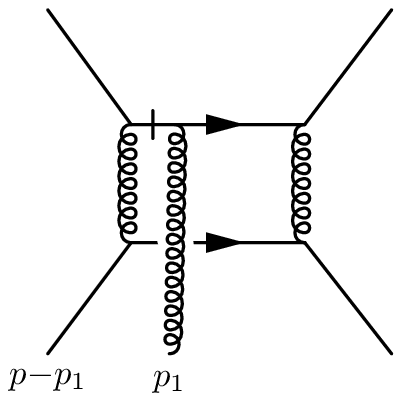}&&
\includegraphics[width=3.3cm]{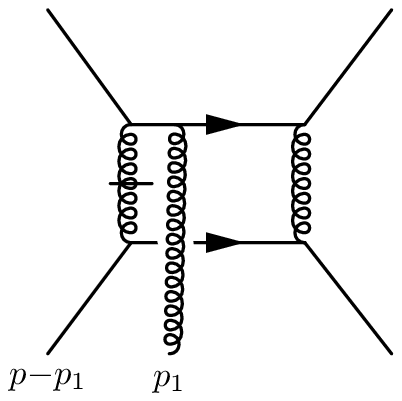}
\\
(a) && (b) && (c)
\end{tabular}
\end{center}
\caption{Figure~(a) represents
a squared Feynman diagram contributing to quark-quark scattering. In figure~(b)
and figure~(c) a gluon is inserted from the correlator into the scattering
diagram. In figure~(b) the dash indicates a pole at $p_1^+ \approx 0$,
while in figure~(c) the dash indicates a pole at $p_1^+ \neq 0$.
\label{devFigFloat}}
\end{figure}

Let us begin by considering the subprocess in 
Fig.~\ref{devFigFloat}a (the resulting gauge invariant correlator will be
given in Eq.~\ref{devExampleGauge}).
Including a longitudinally polarized gluon ($A^+$ with momentum $p_1\sim n_+$) 
from a distribution
correlator (belonging to the quark entering the graph from the bottom),
that gluon can be coupled in
four different places: to the incoming quark, 
to the outgoing quarks, and to
the exchanged virtual gluon. The inserted gluon introduces an extra integral
over $p_1$ which needs to be performed. Analyzing
the $p_1^+$-dependence one observes two classes of poles
in $p_1^+$. The first class consists of poles
at $p_1^+ \approx 0$
which arise from gluon insertions on external partons
in which the introduced propagator (by the inserted gluon) goes
on shell (see for instance Fig.~\ref{devFigFloat}b).
The second class contains the poles at $p_1^+\neq 0$ corresponding
to other internal parton-lines going on shell (see for instance
Fig.~\ref{devFigFloat}c).

In the previous chapter we have seen that the poles
at $p_1^+ \approx 0$ give a contribution to the gauge link but
the meaning of the poles
at $p_1^+ \neq 0$ is unclear.
Fortunately, in the sum over all possible gluon insertions it turns out that
the poles of the second class cancel each other. For instance,
in the case of a single
gluon insertion one can show the following cancellation
by doing the calculation
explicitly
\begin{equation}
\parbox{2cm}{\includegraphics[width=2cm]{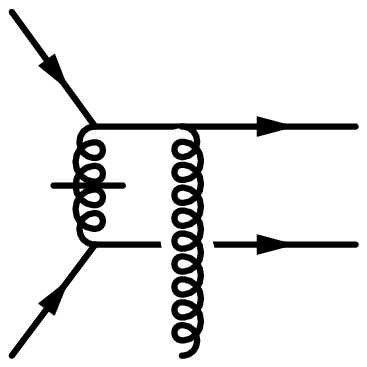}}
\eqnIndent + \eqnIndent
\parbox{2cm}{\includegraphics[width=2cm]{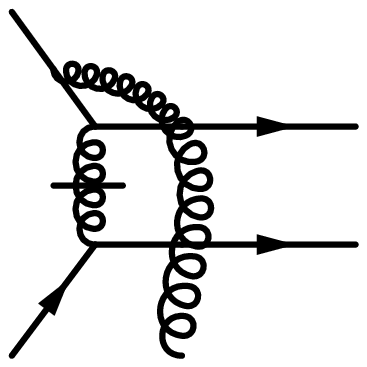}}
\eqnIndent + \eqnIndent
\parbox{2cm}{\includegraphics[width=2cm]{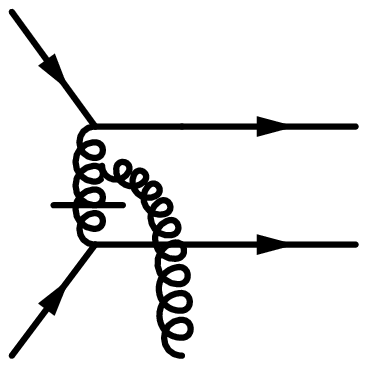}}
\eqnIndent = 0,
\end{equation}
where the dashes indicate the taken residue.
This result
can be generalized to all order insertions\footnote{By choosing the momenta
in a convenient way (use $\Phi(p,p_1{-}p_2,\ldots,p_{n-1}-p_n,p_n)$), the
general proof is not much more complicated to show than the equation
above.}
and was noted in Ref.~\cite{Bomhof:2004aw}.

The cancellation as above is observed in various processes suggesting it to
hold for amplitudes in general. Unfortunately, such a proof does not seem to
exist (not even for QED) and is not simple to establish.
Therefore, two general arguments in favor of this cancellation
will be presented, but it should be stressed that via an
explicit calculation of the cases considered in this thesis the same results
can be obtained.

The first argument is connected with a Ward identity, which states that
a longitudinally polarized gluon (on shell) with nonzero momentum
 does not
couple to an amplitude which has only physical external partons. Since
the inserted gluon (with $p_i^+ \neq 0$) is coupled to all possible
places, except for one particular interaction which is already absorbed in the
correlator definition (see for example Fig.~\ref{quarkInter}b),
such gluons should not be able to contribute in the sum.
The second argument in favor is that in the present case
(see for instance Fig.~\ref{devFigFloat}c)
the pole at $p_i^+ \neq 0$
corresponds to the internal gluon-line being on shell. However,
this configuration is
physically not possible, because the coupling between
two on shell quarks (with unequal momenta)
and an on shell gluon is forbidden by momentum
conservation. This is also connected to the fact that all
tree-level Feynman amplitudes
are finite as long as the external partons are well separated.
According to this argument the cancellation in the above equation takes place
in a nonphysical area.
This argument for a single gluon insertion
can be extended to an arbitrary amount
of insertions.

Having only the poles at $p_i^+ \approx 0$, the task to perform
is to evaluate those
poles which arise from inserting gluons on the
external parton legs.
The calculation of inserting longitudinally polarized gluons to a
single external parton
leg is similar to the calculations performed in the previous chapter.
An explicit example will be shown below.
By inserting the gluons on a single quark-line and
following the same steps as in the previous chapter,
one can derive the gauge link to all orders. There is
only one important difference: the color matrices of the inserted
vertices (of the inserted gluons) cannot be simply pulled into the considered
correlator because of the
presence of the
virtual gluon inside the graph.
Therefore, those color matrices are ``standing''
on the quark-line on which the gluons were inserted.
Taking
combinations of insertions on several
external legs is relatively straightforward because the manipulations
of the inserted
gluons on a certain external leg can be performed
without using information from the
other external legs. As a result,
the combination of insertions gives a product
of gauge links of which their color matrices remain on the quark-lines
on which the gluons were inserted.

\begin{figure}
\begin{center}
\begin{tabular}{ccc}
\includegraphics[width=4cm]{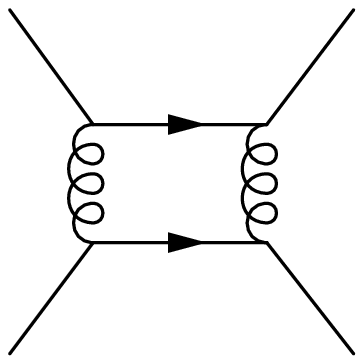} &&
\includegraphics[width=4cm]{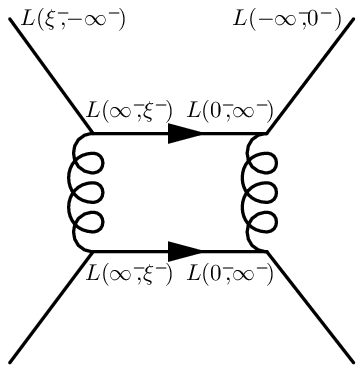}\\
(a) && (b)
\end{tabular}
\end{center}
\caption{Illustration of gauge links in
a squared Feynman diagram contributing to quark-quark scattering.
Figure~(a) represents an ordinary squared Feynman diagram.
In figure~(b) the result of the gluon insertions is presented.
It is shown on which quark-lines
the color matrices of the indicated gauge links
($L$) are acting. The superscripts $\xi^+$, and $\xi_T$ or $0_T$, have
been omitted.
%Note that $\mathcal{L}^{0_T\!,\ \xi^+}(0^-,\infty^-)
%\mathcal{L}^{\xi_T\!,\ \xi^+}(\infty^-,\xi^-)$ equals
%$\mathcal{L}^{[+]}(0,\xi^-)$ in gauges where transverse gluons fields at
%infinity can be neglected. Similarly one also has
%$\mathcal{L}^{0_T\!,\ \xi^+}(0^-,\infty^-)
%\mathcal{L}^{\xi_T\!,\ \xi^+}(\infty^-,\xi^-)$ equaling
%$\mathcal{L}^{[+]}(0,\xi^-)$
\label{dev13}}
\end{figure}

As an example, let us reconsider the subprocess given
in Fig.~\ref{devFigFloat}a
(also given in Fig.~\ref{dev13}a).
Summing over the colors of the incoming and outgoing partons,
its contribution to the cross section is proportional to
(where $\Phi$ was defined in Eq.~\ref{defPhi})
\begin{equation}
\begin{split}
\sigma_{\text{Fig.~}\eqref{dev13}a} &= K\  \tr^C \left[ t_b t_a \Phi(p) \right] \tr^C \left[ t_b t_a \right]\\
&= \frac{2K}{3} \tr^C \left[ \Phi(p) \right] \eqnIndent
\qquad\qquad \text{(where $K$ is some constant)}.
\label{dev4}
\end{split}
\end{equation}
Using the machinery of the previous chapter\footnote{Issues related to
the equations of motion will be discarded in this section.}, the inserted longitudinally
polarized gluons
from the correlator $\Phi$ produce in the amplitude diagram (left-hand-side
of the cut) the following gauge links:
$\mathcal{L}^{\xi_T\!,\ \xi^+}\!\!(\infty^-\!\!,\!\xi^-\!)$ on the outgoing quarks and
$\mathcal{L}^{\xi_T\!,\ \xi^+}\!\!(\xi^-\!\!,\!-\infty^-\!)$ on the incoming quarks
(the coordinate $\xi$ is related to the field $\psi(\xi)$ in the correlator
$\Phi(p)$).
Insertions on the right-hand-side of the cut yield similar gauge links.
The result of the all order insertions is graphically
illustrated in Fig.~\ref{dev13}b
in which it is shown that the color matrices of the
gauge links are at this moment on the
external parton-lines. The result of the insertions
reads (the notation will be a bit sloppy; the gauge links appear under
the $\xi$ integral which is present in the definition of $\Phi$)
\begin{multline}
\sigma_{\text{Fig.~}\eqref{dev13}b}
= K \tr^C \left[ t_b \mathcal{L}^{0_T\!,\xi^+}(0^-\!,\infty^-)
\mathcal{L}^{\xi_T\!,\xi^+}(\infty^-,\xi^-)
t_a \Phi(p) \right]
\\
\times \tr^C \left[ t_b \mathcal{L}^{0_T\!,\xi^+}(0^-\!,\!\infty^-)
\mathcal{L}^{\xi_T\!,\xi^+}(\infty^-\!\!,\!\xi^-) t_a
\mathcal{L}^{\xi_T\!,\xi^+}(\xi^-\!\!,\!-\infty^-)
\mathcal{L}^{0_T\!,\xi^+}(-\infty^-\!\!,\!0^-)
\right].
\end{multline}
Using now the identity ($N=3$)
\begin{equation}
(t^a)_{ij} (t^a)_{kl} = \frac{1}{2} \delta_{il} \delta_{kj} -
\frac{1}{2N} \delta_{ij}\delta_{kl},
\end{equation}
one finds
\begin{multline}
\sigma_{\text{Fig.~}\eqref{dev13}b}
= \frac{2K}{3} \Bigg(
\frac{-1}{4}
\tr^C \left[ \Phi(p) \mathcal{L}^{[+]}(0,\xi^-)
             \mathcal{L}^{[\square]}(0,\xi^-)\right]
             \\
+
\frac{5}{12} \tr^C \left[ \Phi(p) \mathcal{L}^{[+]}(0,\xi^-) \right]
\tr^C \left[ \mathcal{L}^{[\square]}(0,\xi^-)\right] \Bigg),
\end{multline}
where the transverse gauge links were added for completeness.
This addition can be performed uniquely
by assuming consistency (gauge invariance).
Using Eq.~\ref{dev4} for the normalization,
the gauge invariant quark-quark
correlator is obtained~\cite{Bomhof:2004aw,Bacchetta:2005rm}
\begin{equation}
\begin{split}
&
\Phi_{\text{Fig.~}\eqref{dev13},ij}(x,p_T,P,S)
\\
& \ {=}
\int \frac{ \rmd{1}{\xi^-} \rmd{2}{\xi_T} }{(2\pi)^3}
e^{ip\xi}\ \langle P,S | \bar{\psi}_j (0)
\\
&
\ \phantom{=}
\times
\left[
\tfrac{5}{12} \mathcal{L}^{[+]}(0,\xi^-) \tr^C \mathcal{L}^{[\square]}(0,\xi^-)
-\tfrac{1}{4} \mathcal{L}^{[+]}(0,\xi^-) \mathcal{L}^{[\square]}(0,\xi^-)
\right]
\psi_i (\xi) | P,S \rangle \bigg|_{\begin{subarray}{l}
\xi^+ = 0 \\ p^+ = x P^+ \end{subarray}}\!\!\!.
\end{split}
\label{devExampleGauge}
\end{equation}
In the correlator for this diagram a combination of path-ordered exponentials
appear. Such gauge links were not discussed before and
although this new result looks rather awkward
at first sight, \emph{it is} actually invariant
under color gauge transformations. It should also be noted that the
appearance of these complex
gauge link structures is not just a peculiarity of QCD, but also appears
in the QED-version of Fig.~\ref{dev13}. In that case the gauge link
is $\mathcal{L}^{[+]}(0,\xi^-) \mathcal{L}^{[ \square ]}(0,\xi^-)$
which was obtained in Ref.~\cite{Bomhof:2004aw}.

\begin{figure}
\begin{center}
\begin{tabular}{cccc}
\includegraphics[width=2.5cm]{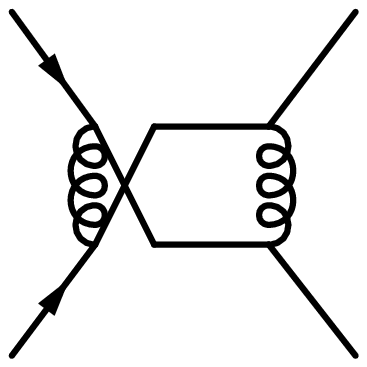}&
\includegraphics[width=2.5cm]{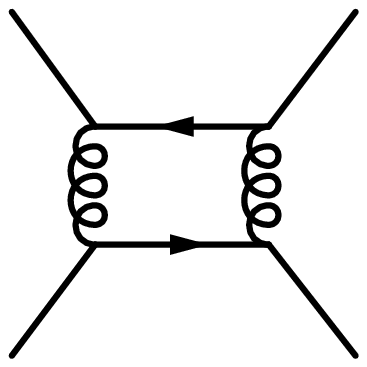}&
\includegraphics[width=2.5cm]{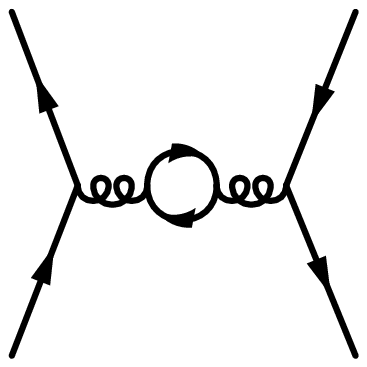}&
\includegraphics[width=2.5cm]{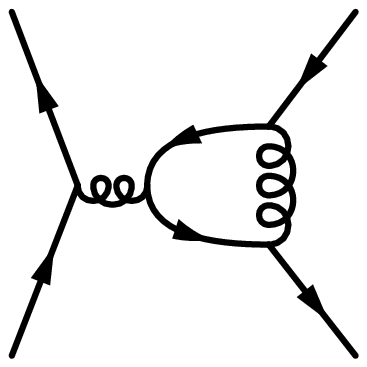}
\\
(a) & (b) & (c) & (d)
\end{tabular}
\end{center}
\caption{Various quark-(anti)quark scattering subprocesses.\label{dev8}}
\end{figure}

\begin{table}
\begin{center}
\begin{tabular}{l|c|c}
\hspace{.5cm} $\backslash$ correlator & quark  & quark \\
diagram & distribution & fragmentation \\
\hline
&&\\
Fig.~\ref{dev13}a
& $\tfrac{5}{12} \mathcal{L}^{[+]} \tr^C \mathcal{L}^{[\square]}-\tfrac{1}{4} \mathcal{L}^{[+]} \mathcal{L}^{[\square]}$
  & $\tfrac{5}{12} \mathcal{L}^{[-]} \tr^C \mathcal{L}^{[\square]\dagger}-\tfrac{1}{4} \mathcal{L}^{[-]} \mathcal{L}^{[\square]\dagger}$
  \\
&&\\
Fig.~\ref{dev8}a
& $\tfrac{3}{4} \mathcal{L}^{[+]} \tr^C \mathcal{L}^{[\square]}-\tfrac{5}{4} \mathcal{L}^{[+]} \mathcal{L}^{[\square]}$
  & $\tfrac{3}{4} \mathcal{L}^{[-]} \tr^C \mathcal{L}^{[\square]\dagger}-\tfrac{5}{4} \mathcal{L}^{[-]} \mathcal{L}^{[\square]\dagger}$
  \\
&&\\
Fig.~\ref{dev8}b
& $\tfrac{1}{24} \mathcal{L}^{[+]} \tr^C \mathcal{L}^{[\square]\dagger}+\tfrac{7}{8} \mathcal{L}^{[-]}$
  & $\tfrac{1}{24} \mathcal{L}^{[-]} \tr^C \mathcal{L}^{[\square]}+\tfrac{7}{8} \mathcal{L}^{[+]}$
  \\
&&\\
Fig.~\ref{dev8}c
&$\tfrac{3}{8} \mathcal{L}^{[+]} \tr^C \mathcal{L}^{[\square]\dagger}-\tfrac{1}{8} \mathcal{L}^{[-]}$
  &$\tfrac{3}{8} \mathcal{L}^{[-]} \tr^C \mathcal{L}^{[\square]}-\tfrac{1}{8} \mathcal{L}^{[+]}$
  \\
&&\\
Fig.~\ref{dev8}d
& $\tfrac{3}{8} \mathcal{L}^{[+]} \tr^C \mathcal{L}^{[\square]\dagger}-\tfrac{1}{8} \mathcal{L}^{[-]}$
  &$\tfrac{3}{8} \mathcal{L}^{[-]} \tr^C \mathcal{L}^{[\square]}-\tfrac{1}{8} \mathcal{L}^{[+]}$
\end{tabular}
\end{center}
\caption{The gauge links in the correlators which connect the external
parton legs
appearing
in Fig.~\ref{dev13},~\ref{dev8}. The gauge links in the
antiquark correlators
in the figures are this case the
Hermitean conjugates of the gauge links in the quark correlators.\label{dev10}}
\end{table}

Using the above procedure
one can derive the gauge links which appear in
other quark-quark and quark-antiquark scattering diagrams,
see Fig.~\ref{dev8}.
There is only one issue which still needs to be addressed.
When gluons are inserted from several correlators at the
same time, interactions \emph{between} the different insertions need to be included.
In the previous chapter we discussed processes in which two correlators were
present.
The interactions between the insertions from the two different correlators
were simplified by choosing a suitable gauge. In the present case where
four correlators are present, this trick cannot be applied.
However, the distribution gauge links as given here are exact when
two quark-jet production is considered, giving confidence in the results
in which the fragmentation process is included.
The results for quark-quark scattering were presented in
Ref.~\cite{Bacchetta:2005rm}
and are given in table~\ref{dev10}. Note that in the limit of $\xi_T\rightarrow
0$, all gauge links in table~\ref{dev10} reduce to a straight gauge link,
$\mathcal{L}^{0_T\!,\ \xi^+}(0^-,\xi^-)$,
between the two quark-fields, see also Fig.~\ref{plaatjeLinks}a.

\subsection{Gauge links in semi-inclusive
DIS and Drell-Yan with an additional gluon in the final state
\label{alphaSDY}}

\subsubsection{Introduction}

\begin{figure}
\begin{center}
\begin{tabular}{cp{.5cm}c}
\includegraphics[width = 5cm]{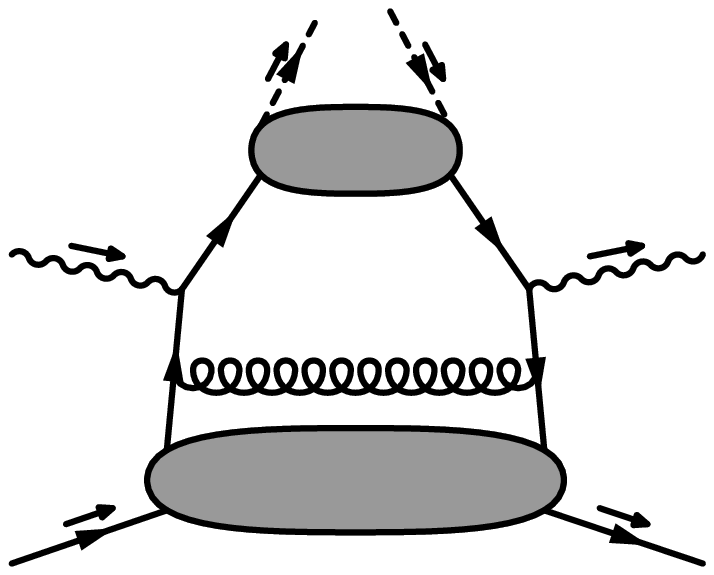}&&
\includegraphics[width = 5cm]{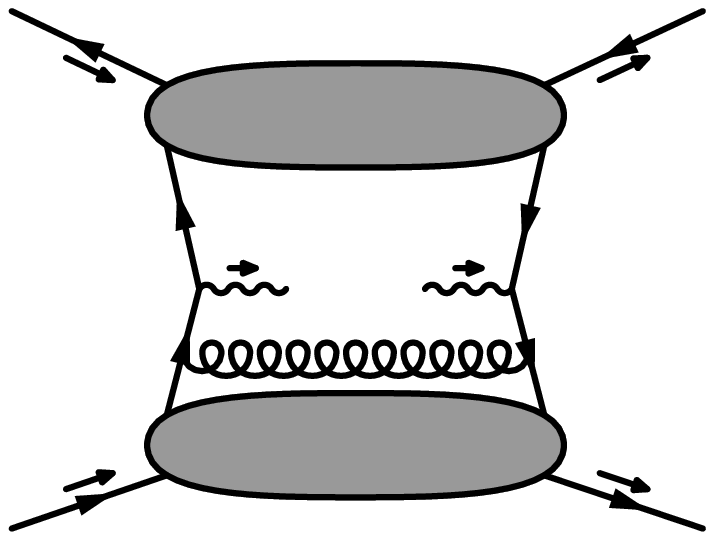} \\
(a) && (b)
\end{tabular}
\end{center}
\caption{Figure~(a) and figure~(b) represent contributions to
respectively semi-inclusive
DIS and Drell-Yan in which an additional jet is being produced.
\label{dev23}}
\end{figure}

We will consider semi-inclusive DIS and
Drell-Yan in which an additional gluon is radiated with some transverse
momentum (at least more than the hadronic scale).
The additional gluon is assumed to be observed and therefore its momentum
is not
integrated over.
Possible contributions to the processes are given
in Fig.~\ref{dev23}.

In order to obtain the gauge link in the distribution correlator all insertions
must be taken. The poles at $p_i^+ \neq 0$ cancel each other
for all possible insertions
which is compatible with the arguments presented in
the previous subsection.
Only the poles
$p_i^+ \approx 0$, coming from insertions on
the external quark and outgoing gluon, will be considered here in detail.

Inserting gluons only on the outgoing quark in semi-inclusive
DIS makes the cross section
proportional to (see also previous subsection and Fig.~\ref{dev23}a)
\begin{equation}
\sigma \sim
\mathrm{Tr}^C \bigl[ t_b\ \mathcal{L}^{[+]}(0,\xi^-)\ t_a\ \Phi(p)\ \bigr]
 \ \delta_{ab},
\label{devGluonLinks}
\end{equation}
where the gauge link appears under the $\xi$ integral in $\Phi(p)$
(sloppy notation).
The above expression is by itself not gauge invariant, the possibility
of inserting gluons on the radiated gluon simply \emph{has to be} included.
From this point of view there can be
no reason to neglect such contributions.
The interesting point here is that
one can simply argue what the gauge link should
be. If the procedure is to be
consistent then the only possible gauge link which can appear in the correlator
is a gauge link via plus infinity. Gauge links via minus infinity or
loops,
as encountered in the previous subsection,
cannot arise because the gluons are inserted on partons which are all
outgoing (incoming partons are needed to produce such gauge links).
When taking insertions on the radiated gluon into
account, there is only one possible
way in how those insertions can contribute
to obtain the proper gauge link.
Inserting gluons on the radiated gluon should lead us
to the following replacements
\begin{equation}
\begin{split}
t_a &\rightarrow \mathcal{L}^{\xi_T\!,\ \xi^+}(\xi^-,\infty^-) t_a
                 \mathcal{L}^{\xi_T\!,\ \xi^+}(\infty^-,\xi^-), \\
t_b &\rightarrow \mathcal{L}^{0_T\!,\ \xi^+}(0^-,\infty^-) t_b
                 \mathcal{L}^{0_T\!,\ \xi^+}(\infty^-,0^-). \label{dev6}
\end{split}
\end{equation}
Together with the insertions on the outgoing quark this provides us
with a gauge link via plus infinity in the quark-quark
correlator~\cite{Bomhof:2004aw}, Eq.~\ref{devGluonLinks} becomes
\begin{equation}
\begin{split}
\sigma &\sim
\mathrm{Tr}^C \bigl[ \mathcal{L}^{0_T\!,\ \xi^+}(0^-\!\!,\infty^-) t_b
                 \mathcal{L}^{0_T\!,\ \xi^+}(\infty^-\!\!,0^-)
                 \mathcal{L}^{[+]}(0,\xi^-)
                 \mathcal{L}^{\xi_T\!,\ \xi^+}(\xi^-\!\!,\infty^-) t_a
                 \mathcal{L}^{\xi_T\!,\ \xi^+}(\infty^-\!\!,\xi^-)
                 \Phi \bigr] \delta_{ab}
\\
& = \frac{4}{3}
\mathrm{Tr}^C \bigl[ \mathcal{L}^{[+]}(0,\xi^-) \Phi \bigr].
\end{split}
\raisetag{14pt}
\end{equation}

With simple arguments, the effect
of inserting gluons on a radiated gluon in semi-inclusive DIS was derived.
Using the same replacement rules, we can now also
study the gauge link structure of Drell-Yan plus an additional outgoing gluon
(see Fig.~\ref{dev23}b). Using
the same rules, one finds in that case
for the gauge link~\cite{Bomhof:2004aw}
\begin{equation}
\frac{3}{8} \mathcal{L}^{[+]}(0,\xi^-) \tr^C
\left[ \mathcal{L}^{[\square \dagger]}(0,\xi^-) \right]
- \frac{1}{8} \mathcal{L}^{[-]}(0,\xi^-).
\end{equation}
The result is here a combination of gauge links via plus and minus
infinity. Although the result is not very appealing, it is
gauge invariant. The appearance of gauge links via plus and minus infinity
appear because insertions were taken on incoming \emph{and} outgoing
partons. Based on the previous subsection, such a result was to be expected.

The replacements in Eq.~\ref{dev6} were derived by assuming the procedure
to be consistent with color gauge invariance.
In the following it will be argued that the replacement rules are
in fact the proper rules.

\subsubsection{The explicit calculation}

The derivation of gauge links
will be presented which underlies the results of
Ref.~\cite{Bomhof:2004aw}.
The amplitude in semi-inclusive DIS and Drell-Yan in which an additional
gluon is produced will be considered at order $\alpha_S$, and the gauge link
will be derived by including gluon insertions.
Before we begin with the technical derivation we note
that if the polarization vector of the outgoing
gluon is replaced by its momentum,
then the amplitude
vanishes which is a result of a Ward identity (at this order in
$\alpha_S$ there are no ghost contributions).

For the coupling of an inserted gluon on the
external gluon-line some work needs to be performed.
Consider
the outgoing gluon with momentum $l$
for a given diagram in a particular amplitude (for instance
the left-hand-side of the cut in Fig.~\ref{dev23}b).
In order to indicate the presence of the outgoing gluon, the polarization
vector $\epsilon_{a\beta}^{(c)\dagger}(l)$ will be introduced, where
$c$ is the ``color charge'' of the gluon ($c\in 1,\ldots,8$),
$a$ is the color index ($a\in 1,\ldots,8$), and
$\beta$ is the usual Lorentz-index. Although $\epsilon_{a\beta}^{(c)\dagger}(l)$
is proportional to $\delta_{ac}$ (as $e_1^\alpha \sim \delta_1^\alpha$), this
property will not be used because it is preferred here to show explicitly
the presence of this external outgoing gluon.
Using these definitions,
the considered diagram can be expressed as
($i,j,k,l \in 1,2,3$, and the Dirac indices are not explicitly shown)
\begin{equation}
%t^a_{ij}\ g^{\alpha'\beta}\ \epsilon_{a\beta}^{(c)\dagger}(l)\ \otimes \ \Phi(p) ,
t^a_{ij}\ g_{\alpha'}^{\phantom{\alpha'}\beta}\ \epsilon_{a\beta}^{(c)\dagger}(l)\
M^{\alpha'}_{ijkl}(p,l)\
\Phi_{kl}(p),
\label{eq56}
\end{equation}
in which $M$ denotes the remainder of the expression for the diagram 
and $t^a$ is a
$3\times3$ color matrix. Although the calculation will be performed at the
amplitude level, the full correlator $\Phi$ is present in the above expression
for economics of notation. We will not rely on the part of $\Phi$ which
is in the conjugate amplitude.

Inserting a gluon with momentum $p_1\sim n_+$ and polarization
$n_+$ on the radiated gluon
replaces $\Phi$ by $\Phi_A$ and changes the above expression into
(in the Feynman gauge)
\begin{equation}
\begin{split}
&
\int \mathrm{d}^4 p_1\
t^b_{ij}
\frac{(-i) g_{\alpha'\alpha } }{(l-p_1)^2 + i \epsilon}
f^{bsa} \\
&
\times \left( g^{\alpha-} (l-2p_1)^\beta + g^{-\beta}
(p_1 + l)^\alpha - 2l^- g^{\alpha \beta} \right)
\epsilon_{a\beta}^{(c)\dagger} (l) \ M^{\alpha'}_{ijkl}(p{-}p_1,l)\ \Phi_{A_s,kl}^+(p,p_1)
\\
& \eqnIndent
{=} \int \mathrm{d}^4 p_1\
t^b_{ij}
\frac{(-i) g_{\alpha'\alpha } }{(l-p_1)^2 + i \epsilon}
f^{bsa}\\
& \eqnIndent \phantom{=}
\times \left(  - 2l^- g^{\alpha \beta} + (l-p_1)^\alpha n_+^\beta +
n_+^\alpha l^\beta\right)
\epsilon_{a\beta}^{(c)\dagger} (l) \ M^{\alpha'}_{ijkl}(p{-}p_1,l)\ \Phi_{A_s,kl}^+(p,p_1).
\end{split}
\label{eq54}
\end{equation}
When referring to this equation, we will refer to the right-hand-side.
The first term of Eq.~\ref{eq54} does not
change the original Lorentz structure of the diagram and will
yield the first order expansion of the gauge link.
The second and third term have a less clear meaning.
In  the processes studied in this subsection
the second term vanishes in the
sum over all diagrams in the amplitude, and
the third term does not couple to the Hermitean conjugate amplitude
and vanishes therefore in the cross section. Both terms do therefore not
contribute which is
a result of a Ward identity which can be applied since
$l-p_1 \neq 0$.

\begin{floatingfigure}{3.5cm}
\begin{center}
\includegraphics[width=3cm]{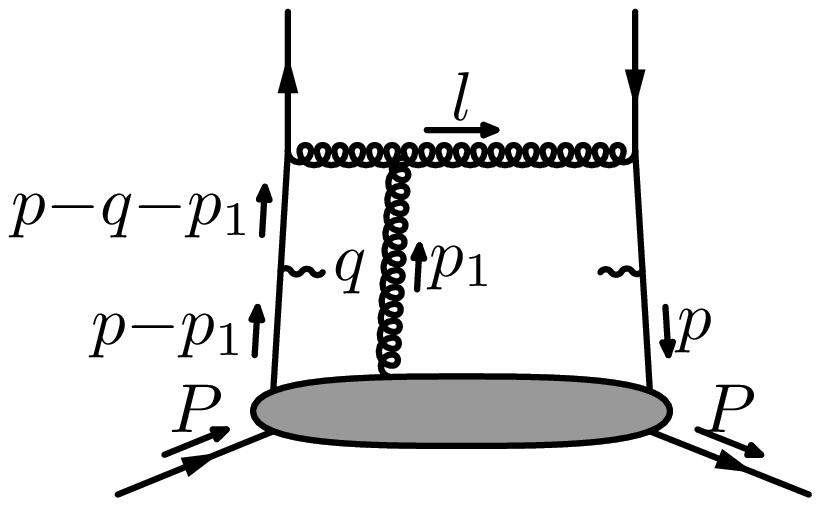}
\end{center}
\caption{Example\label{hiro10}}
\end{floatingfigure}

In other processes like two-gluon production the second and third
term do not
vanish. As will be shown
later, it turns out that
these terms together with the ghost contributions exactly cancel the nonphysical
polarizations of the radiated gluon in the Feynman gauge.
This cancellation is far from being
trivial and it should be considered as a firm consistency check.

When continuing
with the above expression, Eq.~\ref{eq54},
one should realize that there could be an additional
$p_1^+$-dependence in the rest of the diagram (in $M$), see for instance
Fig.~\ref{hiro10}.
Before closing the contours and
evaluating the integrals by taking their residues,
the powers of $p_1^+$ in the denominator and in the
numerator should be counted. Using that
$\Phi \sim \slashiv{p}$ it is possible to
show that the denominator is one order higher
than the numerator which justifies the closing of the contour at infinity.
Using this result the $p_1^+$-dependence
in the possible present internal quark propagator (in $M$)
can be analyzed more closely.
The $p_1^+$ in the numerator
can be discarded because
that term removes the pole at $p_1^+ = 0$ leaving
the pole at $p_1^+ \neq 0$ behind. Such poles do not contribute
which can be checked explicitly.
The $p_1^+$-dependence in the denominator can be
discarded as well because only the poles at $p_1^+ = 0$ will be considered.

Continuing with the non-vanishing terms of
Eq.~\ref{eq54} and performing the integrations leads
to (using that $if^{bsa} t_b = [t_s,t_a]$)
\begin{equation}
\begin{split}
&
\text{first term of Eq.~}\eqref{eq54}
\\
& \hspace{.2cm}
=  g_{\alpha'}^{\phantom{\alpha'} \beta}\ \epsilon^{(c)\dagger}_{a\beta}(l)\
  (t_s\ t_a)_{ij}\ M^{\alpha'}_{ijkl}(p,l)
\\
& \hspace{.4cm} \phantom{=}\ \   \times
  \int \frac{\mathrm{d}^4\xi}{(2\pi)^4} e^{ip\xi}
  \langle P,S | \bar{\psi}_l(0)
  (-ig)\int_{\xi^-}^\infty \mathrm{d}\eta^- A^+_s (\eta^-,0,\xi_T) \psi_k(\xi) | P,S \rangle
  \\
& \hspace{.2cm} \phantom{=}
  + g_{\alpha'}^{\phantom{\alpha'} \beta}\ \epsilon^{(c)\dagger}_{a\beta}(l) \
  (t_a\ t_s)_{ij}\ M^{\alpha'}_{ijkl}(p,l)\\
& \hspace{.4cm} \phantom{=}\ \ \times
 \int \frac{\mathrm{d}^4\xi}{(2\pi)^4} e^{ip\xi}
  \langle P,S | \bar{\psi}_l(0)
  (-ig)\int_\infty^{\xi^-} \mathrm{d}\eta^- A^+_s (\eta^-,0,\xi_T)
  \psi_k(\xi) | P,S \rangle,
\end{split}
\end{equation}
which is exactly the first order expansion of
replacement stated in
Eq.~\ref{dev6}. The above equation expresses that two link operators
are produced for insertions on outgoing gluons.
In the following we will generalize this result to all possible insertions.

\begin{figure}
\begin{center}
\begin{tabular}{ccccc}
\includegraphics[width=3cm]{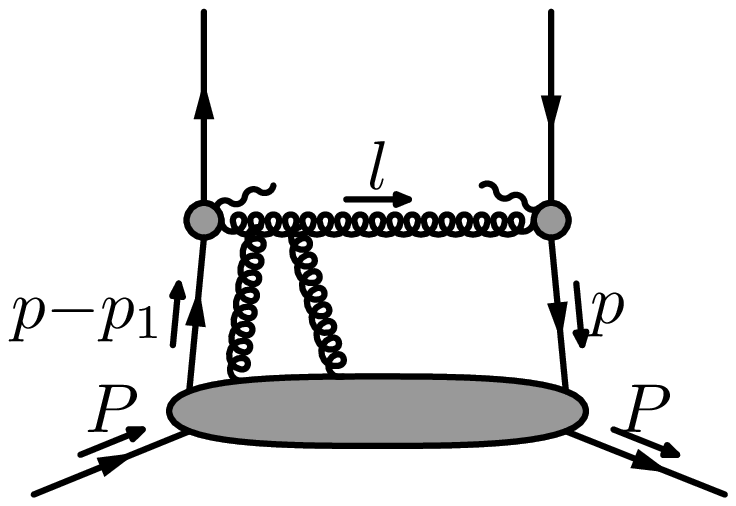} &&
\includegraphics[width=3cm]{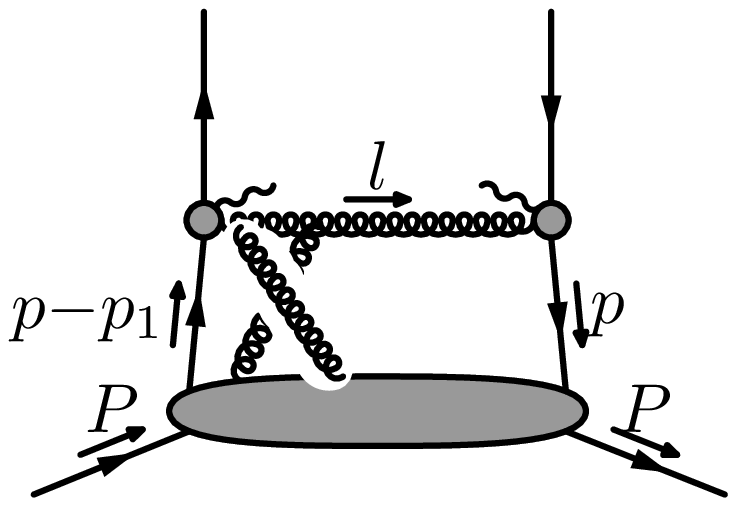} &&
\includegraphics[width=3cm]{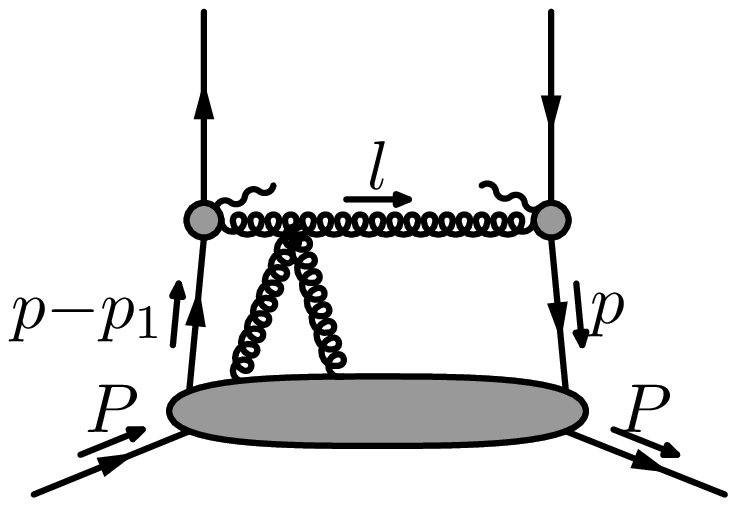}\\
(a) && (b) && (c)
\end{tabular}
\end{center}
\caption{Various insertions in the squared amplitude.
The discussed
interference terms in figure~(a) and figure~(b) cancel the contribution of
figure~(c). For the sum of these three diagrams one remains with the
non-interference contributions (the part of the inserted vertices consisting
of metric tensors) of figure~(a) and figure~(b).
\label{devGluonInsert}}
\end{figure}

When attaching more than a single gluon we have to consider the
Lorentz structure of the
inserted
vertices a bit more closely (see the three terms in Eq.~\ref{eq54}).
The inserted vertex which is the closest to the electromagnetic interaction
will be called
the first vertex (see for instance the left triple-gluon vertex in
Fig.~\ref{devGluonInsert}a), and so on.
When inserting a number of gluons the second term of the first
inserted vertex still vanishes by a Ward identity in the sum
over all amplitude diagrams.
The combination of the first term of the first
vertex with the second term of the second vertex also vanishes because
of a Ward identity, and so on. The same holds for the third term of the last
vertex which does not couple to the conjugate amplitude.
The conclusion is that the
only non-vanishing
contributions come from the
part of the inserted vertices which are products of
metric tensors and interference terms. These interference terms
consist of contractions between terms like the second
and third in Eq.~\ref{eq54} belonging to different insertions.

Consider two adjacent insertions with momenta $p_1-p_2$ and $p_2$ on
a gluon-line with momentum $l-p_1$, giving
the additional gluon propagators: $(l-p_1)^2$ and $(l-p_2)^2$
(see Fig.~\ref{devGluonInsert}a).
The interference term between those two insertions
produce a term in the numerator
proportional to $(l-p_2)^2$ canceling one of the gluon propagators.
This interference term is canceled if one includes the contributions from
the insertion
in which the considered vertices interchanged (see Fig.~\ref{devGluonInsert}b)
and the insertion in which the
two gluons couple into a four-gluon
vertex\footnote{
Up to now we have not considered insertions in which the gluons from the
correlator cross.
When allowing
for $N$ such insertions
one should actually add for all diagrams a symmetry factor
$1/N$. The diagram which has the four-gluon vertex counts in that case double
and should be multiplied by two.
}
(see Fig.~\ref{devGluonInsert}c).
When an interference term does not come from adjacent insertions, then
this cancellation occurs in a slightly more complicated manner.
It will be assumed that similar cancellations also appear when more than
two of those
terms interfere.

Attaching $N$ gluons to the outgoing gluon and
considering now only the metric tensors of the insertions (like the first
term of Eq.~\ref{eq54})
modifies in the Feynman gauge Eq.~\ref{eq56} into
\begin{multline}
\int \left( \prod_i \mathrm{d}^4 p_{i} \right)
t^b_{ij} \frac{(-2l^-)(-i)}{(l-p_{1})^2 +i \epsilon} f^{b l_1 c_{2}}
%\\
\times
\frac{(-2l^-) (-i) }{(l-p_{2})^2 +i \epsilon} f^{c_{2} l_{2} c_{3}} \ldots
\frac{(-2l^-)(-i) }{(l-p_{N})^2 +i \epsilon} f^{c_{N} l_{N} a}\\
\times
\ g_{\alpha'}^{\phantom{\alpha'}\beta}\
\epsilon^{(c)\dagger}_{a\beta}(l)\ M^{\alpha'}_{ijkl}(p{-}p_1,l)\
\Phi_{A_{l1}\ldots A_{lN},kl}^{+\ldots+}
(p,p_{1}{-}p_2,\ldots,p_{N-1}{-}p_N,p_{N}),
\label{ff2}
\end{multline}
where we already anticipated on leaving out terms which do not contribute to
the cross section.
Considering the color
structure explicitly, Eq.~\ref{ff2} reads (see Eq.~\ref{zondag12} for
clarification)
\begin{multline}
\text{Eq.~}\eqref{ff2}=
\int \rmd{4}{p_1} \ldots \rmd{4}{p_N}
  \frac{1}{- p_1^+ + i \epsilon} \ldots
\frac{1}{-p_N^+ + i \epsilon}
g_{\alpha'}^{\phantom{\alpha'}\beta}
\\
\times
\left[
\begin{array}{l}
\text{A sum over all possible combinations of
$t_a$ and $N$ color matrices $t_{l_i}$ with the}\\
\text{index $i$ on the right of $t_a$ going down (from left to right)
and on the left of $t_a$}\\
\text{going up. The sign of
each combination is } (-1)^{\text{number of terms right of }t_a}
\end{array}
\right]_{ij}\\
\times
\epsilon^{(c)\dagger}_{a\beta}(l)\ M_{ijkl}^{\alpha'}\
\Phi_{A_{l1}\ldots A_{lN},kl}^{+\ldots+}
(p_{1},\ldots,p_{N}).
\label{ffa3}
\end{multline}
The factor in the middle contains the following terms
in the case of four gluon insertions,
\begin{equation}
t_{l_1} t_{l_2} t_a t_{l_4} t_{l_3} - t_{l_2} t_a t_{l_4} t_{l_3} t_{l_1}
+t_{l_1} t_{l_4} t_a t_{l_3} t_{l_2} + \ldots
\label{zondag12}
\end{equation}

Performing the integrations over $p_i$ leads to a path-ordered product.
Using
the following identity for non-commuting $a$'s and $b$'s\footnote{So
$[a(x_1),a(x_2)] \neq 0 $ and $[a(x_1),b(x_2)] \neq 0$.}
which can be proven by using induction
\myBox{
\begin{equation}
\begin{split}
&
\int_\xi^C \mathrm{d}\eta_1 \int_{\eta_1}^C \mathrm{d}\eta_2
\ldots \int_{\eta_N}^C \mathrm{d}\eta_N
\bigg[ b(\eta_1) \ldots b(\eta_m) a(\eta_{N}) \ldots a(\eta_{m+1})\\
&
+
\text{all possible coordinate permutations such that the $b$'s are}\\
&
\phantom{+ \bigg[}
\text{
path-ordered and the $a$'s are anti-path-ordered} \bigg] \\
&
\eqnIndent \eqnIndent
{=}\int_{\xi}^C \mathrm{d}\eta_1 \int_{\eta_1}^C \mathrm{d}\eta_2
\ldots \int_{\eta_{m-1}}^C \mathrm{d} \eta_m \
b(\eta_1) \ldots b(\eta_m) \\
& \phantom{\eqnIndent \eqnIndent =}
\times \int_\xi^C \mathrm{d} \eta_{m+1}
\int_{\eta_{m+1}}^C \mathrm{d} \eta_{m+2} \ldots
\int_{\eta_{N-1}}^C \mathrm{d} \eta_N \ a(\eta_N) \ldots a(\eta_{m+1}),
\end{split}
\label{kaka}
\end{equation}
\begin{flushright}
\emph{identity to obtain a product of ordered exponentials}
\end{flushright}
}
one obtains for the part of Eq.~\ref{ffa3},
which consists of
$m$ $t$-matrices left of $t_a$ and $n$ $t$-matrices right of $t_a$
(with $n+m=N$),
the following expression
\begin{equation}
\begin{split}
&
\left[t_{l_1} \ldots t_{l_{m}} t_a t_{l_{N}} \ldots t_{l_{m+1}}\right]_{ij}
g_{\alpha'}^{\phantom{\alpha'}\beta}\ \epsilon^{(c)\dagger}_{a\beta}(l)\
M^{\alpha'}_{ijkl}(p,l)
\\
&
\times \int \frac{\rmd{4}{\xi}}{(2\pi)^4} e^{ip\xi}
\langle P,S | \bar{\psi}_l(0)
(-i)^{m+n} \int_{\xi^-}^\infty \rmd{1}{\eta_{1}^-}
\int_{\eta_{1}^-}^\infty \rmd{1}{\eta_{2}^-}
\ldots \int_{\eta_{m-1}^-}^\infty \rmd{1}{\eta_{m}^-}
A^+_{l_{1}}(\eta_{1}^-) \ldots A^+_{l_{m}}(\eta_{m})
\\
&
\times\!\!\! \int_\infty^{\xi^-}\!\! \rmd{1}{\eta_{N}^-}\! A^+_{l_{N}}(\eta_{N})
\!\!\int_{\eta_N^-}^{\xi^-}\!\! \rmd{1}{\eta_{N-1}^-}\! A^+_{l_{N-1}}(\eta_{N-1})
 \ldots\!
\!\!\int_{\eta_{m+2}^-}^{\xi^-}\!\! \rmd{1}{\eta_{m+1}^-}\! A^+_{l_{m+1}}(\eta_{m+1})
\psi_k(\xi) | P,S \rangle.
\end{split}
\end{equation}
This expression represents the $m^\text{th}$ order expansion
of $\mathcal{L}^{\xi_T\!,\xi^+}(\xi^-,\infty^-)$ multiplied with
the $n^\text{th}$ order expansion of $\mathcal{L}^{\xi_T\!,\xi^+}(\infty^-,\xi^-)$.

Including now all terms of Eq.~\ref{ffa3} by
summing the above expression
over all possible values of $n$ and $m$,
and summing over all $N$, one obtains
\begin{multline}
\epsilon_{a\beta}^{(c)\dagger}(l)\ g_{\alpha'}^{\phantom{\alpha'}\beta}\
       M^{\alpha'}_{ijkl}(p,l)\
\\
\times
\int \frac{\mathrm{d}^4\xi}{(2\pi)^4} e^{ip\xi}
\langle P,S | \bar{\psi}_l(0) \left[
\mathcal{L}^{\xi_T\!,\ \xi^+}(\xi^-,\infty^-) t_a
\mathcal{L}^{\xi_T\!,\ \xi^+}(\infty^-,\xi^-) \right]_{ij} \psi_k(\xi) | P,S \rangle,
\end{multline}
which is exactly the replacement rule as stated in Eq.~\ref{dev6}.

By
taking all possible insertions on a radiated gluon
the derivation of the gauge link
was shown in considerable detail in the Feynman gauge.
The color matrices of these gauge links are
convoluted with the rest of the diagram and are not directly acting
on the fields in the considered correlator (see the next subsection for
how this expression gets handled further).
The calculation for an incoming gluon is very similar and its result
will be stated in the next subsection.

\subsection{Prescription for deducing gauge
links in tree-level diagrams\label{rules}}

The previous subsections suggest a general
procedure for evaluating gauge links
appearing in tree-level squared
amplitude diagrams.
Although not rigorously proven\footnote{In all considered cases the prescription
yielded the correct result. The structure of the
calculations in the previous subsections indicates that
a general proof by induction should be feasible.},
a prescription will be given
together with some examples. We note that the derived gauge links appear
under the $\xi$ integral of the considered correlator (the notation
is a bit sloppy).

In order to determine the gauge link for a correlator (its parent
hadron is moving mainly in the $n_+$-direction), which is connected to
a specific elementary tree-level squared amplitude diagram having
physical external partons,
the following set of rules can be applied:
\begin{enumerate}
\item Write down the contribution of this diagram to the cross section
keeping only the color factors.
Rewrite this expression
into a product
of the amplitude diagram (left-hand-side of the squared diagram)
and the
conjugate amplitude diagram (right-hand-side of the squared diagram).
Color charges of external partons must be made explicit. For the considered
correlator we simply use $\Phi_{ab}$,
for the other incoming and outgoing quarks we apply
the vectors ($u^{(i)}_j$ and
${u^{(i)\dagger}_j}$) indicating
their color charge $i$ and component $j$
and similarly for antiquarks (${v^{(i)\dagger}_j}$ and $v^{(i)}_j$).
For the gluons we introduce $\epsilon^{(a),\text{in/out}}_{b}$ where $a$ is the color
charge and
$b$ is the color component.
Any structure constant,
$f^{abc}$ has to be expressed in terms of
the basic color matrices, $t^{(i)}_{jk}$ via the relation
$f^{abc} = (-2i) \tr^C [ t_c [t_a,t_b] ]$.
\item
Taking all the insertions from a quark distribution or antiquark
fragmentation correlator
on the external partons
in the amplitude diagram (except for
the parton connected to the considered correlator),
can be translated into the following replacements
(outgoing refers to crossing the cut):
        \begin{itemize}
        \item an outgoing quark:
        $u^{(k)\dagger}_i \rightarrow
         u^{(k)\dagger}_j  \,\mathcal{L}^{\xi_T\!,\ \xi^+}_{ji}(\infty^-,\xi^-)$
        \item an outgoing antiquark:
        $v^{(k)}_i \rightarrow \mathcal{L}^{\xi_T\!,\ \xi^+}_{ij}(\xi^-,\infty^-) \,v^{(k)}_j$
        \item an outgoing gluon:
        $ \epsilon^{(c),\text{out}\dagger}_{a} t^{(a)}_{ij} \rightarrow
        \mathcal{L}_{ik}^{\xi_T\!,\ \xi^+}(\xi^-,\infty^-) \, 
        \epsilon^{(c),\text{out}\dagger}_{a} t^{(a)}_{kl}
        \,\mathcal{L}_{lj}^{\xi_T\!,\ \xi^+} (\infty^-,\xi^-) $
        \item an incoming quark:
        $u^{(k)}_i \rightarrow \mathcal{L}^{\xi_T\!,\ \xi^+}_{ij}(\xi^-,-\infty^-) \,u^{(k)}_j$
        \item an incoming antiquark:
        $v^{(a)\dagger}_i \rightarrow v^{(a)\dagger}_j \,\mathcal{L}^{\xi_T\!,\ \xi^+}_{ji}(-\infty^-,\xi^-)$
        \item an incoming gluon:
        $\epsilon^{(a),\text{in}}_b t^{(b)}_{ij} \rightarrow
	\mathcal{L}^{\xi_T\!,\ \xi^+}_{ik}(\xi^-,-\infty^-) \,\epsilon^{(a),\text{in}}_b
         t^{(b)}_{kl} \,\mathcal{L}^{\xi_T\!,\ \xi^+}_{lj}(-\infty^-,\xi^-)$
        \end{itemize}
The coordinate $\xi$ is the argument of the parton-field
in the quark distribution or antiquark fragmentation correlator.
If the considered
correlator is quark fragmentation or antiquark distribution correlator,
the same rules apply but with $\xi$ replaced by $0$.
See also the correlators given in section~\ref{sectDrellYan} and
section~\ref{ePluseMin} for further clarification.
\item
For the insertions in the conjugate amplitude diagram one makes the following
replacements when considering a quark distribution or antiquark
fragmentation correlator
(outgoing refers to crossing the cut):
        \begin{itemize}
        \item an outgoing quark:
        $u^{(k)}_i \rightarrow
        \mathcal{L}^{0_T\!,\ \xi^+}_{ij}(0^-,\infty^-)\, u^{(k)}_j$
        \item an outgoing antiquark:
        $v^{(k)\dagger}_i \rightarrow
        v^{(k)\dagger}_j \mathcal{L}^{0_T\!,\ \xi^+}_{ji}(\infty^-,0^-)$
        \item an outgoing gluon:
        $ \epsilon^{(c),\text{out}}_{a} t^{(a)}_{ij} \rightarrow
        \mathcal{L}_{ik}^{0_T\!,\ \xi^+}(0^-,\infty^-) \,
        \epsilon^{(c),\text{out}}_{a} t^{(a)}_{kl}
        \,\mathcal{L}_{lj}^{0_T\!,\ \xi^+} (\infty^-,0^-) $
        \item an incoming quark:
        $u^{(k)\dagger}_i \rightarrow u^{(k)\dagger}_j
        \mathcal{L}^{0_T\!,\ \xi^+}_{ji}(-\infty^-,0)$
        \item an incoming antiquark:
        $v^{(a)}_i \rightarrow \mathcal{L}^{0_T\!,\ \xi^+}_{ij}(0^-,-\infty^-)
        v^{(a)}_j$
        \item an incoming gluon:
        $\epsilon^{(a)\text{in}\dagger}_b t^{(b)}_{ij} \rightarrow
	\mathcal{L}^{0_T\!,\ \xi^+}_{ik}(0^-\!\!,-\infty^-)
	\epsilon^{(a)\text{in}\dagger}_b
         t^{(b)}_{kl} \mathcal{L}^{0_T\!,\ \xi^+}_{lj}(-\infty^-\!\!,0^-)$
        \end{itemize}
and replaces $0$ by $\xi$ when considering a quark
fragmentation or antiquark distribution correlator.
\item
Using that $u_b^{(r)} = \delta_{rb}$, $\epsilon^{(r)}_b = \delta_{rb}$,
etc., simplify the obtained expression. At this point it is essential
to sum over the colors of the other external partons.
In the final expression the transverse gauge link pieces are
included which can be done uniquely.
\item Divide the expression by the normalization which can be obtained
by replacing all $\mathcal{L}$'s by $1$ and $\Phi$ by $1/3$.
\item The result is the gauge link of the considered
correlator.
\end{enumerate}
To illuminate this set of rules
three examples with explicit results will be given.

\subsubsection{Example: Drell-Yan}

We will reconsider the gauge link in the distribution correlator in Drell-Yan.
The parent hadron is assumed to be moving mainly in the $n_+$-direction.
The result of each step of the prescription is as follows:
\begin{flalign}
1:\ &
\sigma = K \tr^C \big[ \Phi [ v^{(r)}(k) v^{(r)\dagger}(k) ] \big],
\text{where $K$ is some constant. In analogy with Dirac}
\nonumber\\
& \text{spinors one has in this trace }
[ v^{(r)}(k) v^{(r)\dagger}(k) ]_{ij} \equiv
v_i^{(r)}(k) v_j^{(r)\dagger}(k) .
\nonumber \phantom{\Bigg[}\\  % & \nonumber\displaybreak[0]\\
2:\ &
\sigma = K \tr^C \big[ \Phi [ v^{(r)}(k) v^{(r)\dagger}(k) ]
\mathcal{L}^{\xi_T\!,\ \xi^+}(-\infty^-,\xi^-) \big].
\phantom{\Bigg[} \nonumber\displaybreak[0]\\
3:\ &
\sigma = K \tr^C \big[ \Phi \mathcal{L}^{0_T\!,\ \xi^+}(0^-,-\infty^-)
[v^{(r)}(k) v^{(r)\dagger}(k)]
\mathcal{L}^{\xi_T\!,\ \xi^+}(-\infty^-,\xi^-) \big].
\phantom{\Bigg[}\nonumber\displaybreak[0]\\
4:\ &
\sigma = K \tr^C [ \Phi \mathcal{L}^{0_T\!,\ \xi^+}(0^-,-\infty^-)
\mathcal{L}^{\xi_T\!,\ \xi^+}(-\infty^-,\xi^-) ] = K
\tr^C [ \Phi \mathcal{L}^{[-]}(0,\xi^-) ],
\nonumber\displaybreak[0]\\
& \text{where in the last step the transverse gauge link was included.}
\phantom{\Bigg[}\nonumber\displaybreak[0]\\
5:\ & \text{normalization is } K .
\phantom{\Bigg[}\nonumber\displaybreak[0]\\
6:\ & \text{the gauge link is } \mathcal{L}^{[-]}(0,\xi^-) \nonumber.
\end{flalign}

\subsubsection{Example: Drell-Yan and gluon-jet}

In this example the gauge link is calculated
for a distribution correlator (with parent hadron along $n_+$) in
Drell-Yan with an extra gluon-jet
in the final state,
see for instance Fig.~\ref{dev23}b. The results
read:
\begin{flalign}
1:\ &
\sigma = K\tr^C \big[ \Phi t_b [v^{(r)}(k) v^{(r)\dagger}(k)] t_a \big]
\epsilon_b^{(s)\text{out}} \epsilon_a^{(s)\text{out}\dagger}.
\phantom{\Bigg[}\nonumber\displaybreak[0]\\
2:\ &
\sigma = K \tr^C \big[ \Phi t_b [v^{(r)}(k) v^{(r)\dagger}(k)]
\mathcal{L}^{\xi_T\!,\ \xi^+}(-\infty^-,\xi^-) \nonumber\displaybreak[0]\\
& \eqnIndent
\times
\mathcal{L}^{\xi_T\!,\ \xi^+}(\xi^-,\infty^-) t_a
\mathcal{L}^{\xi_T\!,\ \xi^+} (\infty^-,\xi^-) \big]
\epsilon_b^{(s)\text{out}} \epsilon_a^{(s)\text{out}\dagger}.
\phantom{\Bigg[}\nonumber\displaybreak[0]\\
3:\ &\sigma  = K \tr^C \big[ \Phi
\mathcal{L}^{0_T\!,\ \xi^+}(0^-,\infty^-) t_b
\mathcal{L}^{0_T\!,\ \xi^+} (\infty^-,0^-)
\mathcal{L}^{0_T\!,\ \xi^+}(0^-,-\infty^-) \nonumber\\
& \eqnIndent \times
[ v^{(r)}(k) v^{(r)\dagger}(k)]
\mathcal{L}^{\xi_T\!,\ \xi^+}(-\infty^-,\xi^-)
\mathcal{L}^{\xi_T\!,\ \xi^+}(\xi^-,\infty^-) t_a
\mathcal{L}^{\xi_T\!,\ \xi^+} (\infty^-,\xi^-) \big]
\epsilon_b^{(s)\text{out}} \epsilon_a^{(s)\text{out}\dagger}.
\phantom{\Bigg[}\nonumber\displaybreak[0]\\
4:\ &
\sigma = (K/2)  \tr^C [ \Phi \mathcal{L}^{[+]}(0,\xi^-) ]
\tr^C [ \mathcal{L}^{[ \square ]\dagger}(0,\xi^-) ]
- \frac{K}{6} \tr^C [\Phi \mathcal{L}^{[-]}(0,\xi^-) ].
\phantom{\Bigg[}\nonumber\displaybreak[0]\\
5:\ &\text{normalization is } \frac{4}{3}\ K.
\phantom{\Bigg[}\nonumber\displaybreak[0]\\
6:\ &\text{the gauge link is }
\frac{3}{8} \mathcal{L}^{[+]}(0,\xi^-)
\tr^C [ \mathcal{L}^{[ \square ]\dagger}(0,\xi^-) ]
-\frac{1}{8}\mathcal{L}^{[-]}(0,\xi^-).\nonumber
\end{flalign}

\subsubsection{Example: Three jet-production in hadron-hadron collisions}

\begin{figure}
\begin{center}
\includegraphics[width=3cm]{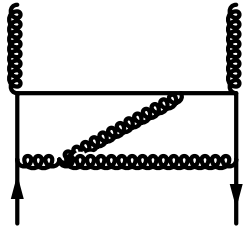}
\end{center}
\caption{A tree-level diagram contributing to three jet-production.
\label{hallo}}
\end{figure}

In this example we consider
the gauge link for the distribution
correlator, with parent hadron along $n_+$,
attached to the lower quark-lines in
the diagram in Fig.~\ref{hallo}. This diagram contributes to three
jet-production in hadron-hadron collisions.
The results read:
\begin{flalign}
1:\ & {\sigma} = iK \tr^C [ \Phi t_a t_b t_c u^\dagger u t_d t_e ]
f^{efg}
\epsilon^{(h)\text{out}\dagger}_f
\epsilon^{(i)\text{out}\dagger}_g
\epsilon^{(j)\text{in}}_d
\epsilon^{(h)\text{out}}_a
\epsilon^{(i)\text{out}}_c
\epsilon^{(j)\text{in}\dagger}_b
\nonumber\\
& \phantom{\sigma} =K
\tr^C \big[ \Phi t_a t_b t_c [u^\dagger u] t_d [t_f,t_g] \big]
\epsilon^{(h)\text{out}\dagger}_f
\epsilon^{(i)\text{out}\dagger}_g
\epsilon^{(j)\text{in}}_d
\epsilon^{(h)\text{out}}_a
\epsilon^{(i)\text{out}}_c
\epsilon^{(j)\text{in}\dagger}_b .
\phantom{\Bigg[}\nonumber\displaybreak[0]\\
2:\ &\sigma =K
\tr^C [ \Phi t_a t_b t_c [u^\dagger u]
\mathcal{L}^{\xi_T\!,\ \xi^+}(\infty^-,\!-\infty^-)
t_d \mathcal{L}^{\xi_T\!,\ \xi^+}(-\infty^-,\!\infty^-) [t_f,t_g]
\mathcal{L}^{\xi_T\!,\ \xi^+}(\infty^-,\xi^-)]
\nonumber\\
& \eqnIndent \times
\epsilon^{(h)\text{out}\dagger}_f
\epsilon^{(i)\text{out}\dagger}_g
\epsilon^{(j)\text{in}}_d
\epsilon^{(h)\text{out}}_a
\epsilon^{(i)\text{out}}_c
\epsilon^{(j)\text{in}\dagger}_b ,
\phantom{\Bigg[}\nonumber\displaybreak[0]\\
3:\ &
\sigma =K
\tr^C [ \Phi \mathcal{L}^{0_T\!,\ \xi^+} (0^-,\infty^-)
t_a \mathcal{L}^{0_T\!,\ \xi^+}(\infty^-,-\infty^-)
t_b \mathcal{L}^{0_T\!,\ \xi^+}(-\infty^-,\infty^-)
t_c [u^\dagger u]
\nonumber\\
& \eqnIndent \times
\mathcal{L}^{\xi_T\!,\ \xi^+}(\infty^-,-\infty^-)
t_d \mathcal{L}^{\xi_T\!,\ \xi^+}(-\infty^-,\infty^-) [t_f,t_g]
\mathcal{L}^{\xi_T\!,\ \xi^+}(\infty^-,\xi^-)]
\nonumber\\
& \eqnIndent
\times
\epsilon^{(h)\text{out}\dagger}_f
\epsilon^{(i)\text{out}\dagger}_g
\epsilon^{(j)\text{in}}_d
\epsilon^{(h)\text{out}}_a
\epsilon^{(i)\text{out}}_c
\epsilon^{(j)\text{in}\dagger}_b .
\phantom{\Bigg[}\nonumber \displaybreak[0]\\
4:\ &
\sigma = \frac{-K}{24} \tr^C \left[ \Phi \mathcal{L}^{[+]}(0,\xi^-) \right]
+ \frac{K}{8}
\tr^C \left[ \Phi \mathcal{L}^{[+]}(0,\xi^-) \mathcal{L}^{[ \square ]}
(0,\xi^-)
\right]
\tr^C \left[ \mathcal{L}^{[\square] \dagger}(0,\xi^-) \right].
\phantom{\Bigg[}\nonumber\\
5:\ &\text{normalization is }K/3 .
\phantom{\Bigg[}
\nonumber\displaybreak[0]\\
6:\ &\text{gauge link is }
\frac{-1}{8}  \mathcal{L}^{[+]}(0,\xi^-)
+ \frac{3}{8}
 \mathcal{L}^{[+]}(0,\xi^-) \mathcal{L}^{[ \square ]}(0,\xi^-)
\tr^C \left[ \mathcal{L}^{[\square] \dagger}(0,\xi^-) \right].\nonumber
\end{flalign}

\section{Relating correlators with different gauge links\label{devLinks2}}

The gauge invariant
correlators, appearing in the diagrams, will here
be related to the correlators introduced in chapter~\ref{chapter2}. We
will find that the T-odd functions, appearing in different diagrams, can in
general differ by more than just a sign.

The gauge link structure consists in general
of several terms of which each term consists of a link
($\mathcal{L}^{[+]}$ or $\mathcal{L}^{[-]}$) multiplied
by some traces of gauge loops. After an integration over the transverse momentum
of a correlator these gauge links collapse on the light-cone, one remains
with a simple straight gauge link. The transverse momentum integrated
distribution and fragmentation
functions
are therefore universal\footnote{In a fragmentation correlator the gauge links
appear in two different matrix elements. To show the universality of integrated
fragmentation correlators with an arbitrary gauge link, one can
choose the light-cone gauge.}.

To treat the first transverse moment of T-even
distribution functions, one can use time-reversal
to project out the T-even structure of the unintegrated
correlator. This enables
one to write the T-even part
of the correlator as
the average of itself and the correlator having the time-reversed gauge link
($\infty \leftrightarrow -\infty$, as done in Eq.~\ref{theoryTeven}).
Using the identities in Eq.~\ref{theoryIdentity}, which also gives the
relation (based on $\tr^C t_a = 0$)
\begin{equation}
\partial_{\xi_T}^\alpha \tr^C \left[ \mathcal{L}^{[ \square ]}(0,\xi^-)
\right] \Big|_{\xi_T=0} = 0,
\label{traceLoop}
\end{equation}
one can show that the
first transverse moment of the unintegrated correlator
is link-independent. The first transverse moment
of T-even distribution functions is therefore universal.

The first transverse moment of T-odd distribution
functions having some gauge link is treated analogously.
The T-odd part of the correlator is projected out by taking the difference
of itself and the time-reversed one
(see for example Eq.~\ref{theoryTodd}).
Using the identities in Eq.~\ref{theoryIdentity} and Eq.~\ref{traceLoop}
one can show that the first transverse moment of a T-odd distribution
correlator
equals the T-odd correlator as defined in Eq.~\ref{theoryTodd} up to a constant.
Therefore one can relate the first transverse moment of T-odd distribution
functions to the ones appearing in semi-inclusive DIS.

For example, let us consider the first transverse moment of the T-odd
part of a distribution correlator containing the gauge link:
$\mathcal{L}' =  \frac{3}{8} \mathcal{L}^{[+]}(0,\xi^-)
\tr^C [ \mathcal{L}^{[ \square ]\dagger}(0,\xi^-) ]
-\frac{1}{8}\mathcal{L}^{[-]}(0,\xi^-)$. Using time
reversal the first transverse moment
reads as follows~\cite{Bomhof:2004aw}
\begin{equation}
\begin{split}
\Phi_{\partial}^{[\text{T-odd}]\mathcal{L}'\alpha}
&= \frac{1}{2}
\int \rmd{2}{p_T} p_T^\alpha \int \frac{\rmd{1}{\xi^-}\rmd{2}{\xi_T}}{(2\pi)^3}
e^{ip\xi}
\\
\times&
\Bigl( \langle P,S | \bar{\psi}(0)
\big[
\tfrac{3}{8} \mathcal{L}^{[+]}(0,\xi^-)
\tr^C [ \mathcal{L}^{[ \square ]\dagger}(0,\xi^-) ]
-\tfrac{1}{8}\mathcal{L}^{[-]}(0,\xi^-)
\big] \psi(\xi) | P,S \rangle_{\text{c}}
\\
&
-
\langle P,S | \bar{\psi}(0)
\big[
\tfrac{3}{8} \mathcal{L}^{[-]}(0,\xi^-)
\tr^C [ \mathcal{L}^{[ \square ]}(0,\xi^-) ]
-\tfrac{1}{8}\mathcal{L}^{[+]}(0,\xi^-) \big]
\psi(\xi) | P,S \rangle_{\text{c}} \Bigr) 
\Big|_{\begin{subarray}{l}\xi^+=0 \\ p^+ = xP^+ \end{subarray}}
\\
&= \frac{5}{4} \Phi_{\partial}^{[\text{T-odd}] \alpha}
\end{split}
\raisetag{14pt}
\end{equation}
The above expression is equivalent to $\frac{5}{4}$
times the first transverse moment of the T-odd correlator as
defined in Eq.~\ref{theoryTodd}. The first transverse moment
of the T-odd functions (like the Sivers function $f_{1T}^{\perp(1)}$)
is therefore $\frac{5}{4}$-times the T-odd function which is measured in
semi-inclusive DIS.

For fragmentation functions the interplay of the two possible
mechanisms for \mbox{T-odd} effects
created the problem for comparing the fragmentation
functions in semi-inclusive DIS with the functions in
electron-positron annihilation.
The first transverse moment of those fragmentation functions read
\begin{multline}
\Delta^{[\pm]\alpha}_\partial (z^{-1}) = \frac{1}{3}\!
\sum_X\!\! \int\! \phaseFactor{P_X}\!
\int \frac{\rmd{1}{\xi^+}}{2\pi} e^{iP_h^- \xi^+ /z}
{}_\text{out}
\langle P_h,P_X | \bar{\psi}(0) \mathcal{L}^{0_T\!,\xi^-}(0^+,\pm \infty^+)
| \Omega \rangle_\text{c}
\\
\times \langle \Omega | \Bigg[
(-g) \int_{\pm \infty}^{\xi^+} \rmd{1}{\eta^+}
\mathcal{L}^{0_T\!,\ \xi^-}(\pm \infty^+,\eta^+) G_T^{-\alpha}(\eta)
\mathcal{L}^{0_T\!,\ \xi^+}(\eta^+,\xi^+)
\\
+
\mathcal{L}^{0_T\!,\ \xi^-}(\pm \infty^+,\xi^+) iD_T^\alpha \Bigg]
\psi(\xi) | P_h, P_X \rangle_{\text{out,c}}
\Big|_{\begin{subarray}{l}
\eta^- = \xi^- = 0\\ \eta_T = \xi_T = 0 \end{subarray}}.
\end{multline}
By making comparisons in the light-cone gauge,
the first transverse moment of a fragmentation correlator with some
gauge link can be expressed in the transverse moments
of the fragmentation functions appearing in
semi-inclusive DIS and electron-positron
annihilation.

\begin{table}[t]
\begin{center}
\begin{tabular}{cc}
a: &        \begin{tabular}{c|c|c|c|c}
        Fig.~\ref{dev13}a & Fig.~\ref{dev8}a & Fig.~\ref{dev8}b
        & Fig.~\ref{dev8}c & Fig.~\ref{dev8}d
        \\
        \hline
        $1/2$ & $ -3/2 $ & $-3/4$ & $5/4$ & $5/4$
        \end{tabular}
\\
& \\
b: &
        \begin{tabular}{c|c|c|c|c}
        Fig.~\ref{dev13}a & Fig.~\ref{dev8}a & Fig.~\ref{dev8}b
        & Fig.~\ref{dev8}c & Fig.~\ref{dev8}d
        \\
        \hline
        $-1/2$ & $ 3/2 $ & $3/4$ & $-5/4$ & $-5/4$
        \end{tabular}
\end{tabular}
\end{center}
\caption{The comparison of functions having complex gauge links
with the functions measured in semi-inclusive DIS and
electron-positron annihilation. The factors in the table are obtained
by joggling with color matrices. The factors
are thus effectively a function of $N_c$.
\newline
a:
The factors for T-odd distribution functions.
The first moment
of a
T-odd function appearing
Fig.~\ref{dev13} and Fig.~\ref{dev8} is the T-odd
function of semi-inclusive DIS multiplied by the given factor. So
${\Phi^{[\text{T-odd}]}_\text{Fig.(...)}}^\alpha_\partial =
\text{factor} \times {\Phi^{[\text{T-odd}]}}^{\alpha}_\partial$.
\newline
b: The factors for T-even and T-odd
fragmentation functions, the first transverse moment of the functions in
Fig.~\ref{dev13}, and Fig.~\ref{dev8} can be written as a linear combination of the
functions appearing in electron-positron annihilation
and semi-inclusive DIS. So
${\Delta_{\text{Fig.(...)}}}_\partial^\alpha =
( {\Delta^{[+]}}_\partial^\alpha + {\Delta^{[-]}}_\partial^\alpha )/2
+ \text{factor}\times
( {\Delta^{[+]}}_\partial^\alpha - {\Delta^{[-]}}_\partial^\alpha )/2$.
\label{devFactors}}
\end{table}

As an example, let us consider the correlator with the
gauge link structure
$\mathcal{L}'=
\mathcal{L}^{[ \square ]}(0,\xi^+) \mathcal{L}^{[+]}(0,\xi^+)$. When
taking its first transverse moment one obtains
\begin{equation}
\begin{split}
\Delta^{\mathcal{L}' \alpha}_\partial (z^{-1})
&= \frac{1}{3} \sum_X \int \phaseFactor{P_X}
\int \frac{\rmd{1}{\xi^+}}{2\pi} e^{iP_h^- \xi^+ /z}
\\
& \phantom{=} \times
{}_\text{out}
\langle P_h,P_X | \bar{\psi}_a(0) \mathcal{L}_{ab}^{0_T\!,\ \xi^-}(0^+,\infty^+)
\mathcal{L}^{0_T\!,\ \xi^-}_{ef}(-\infty^+,\infty^+)| \Omega \rangle_\text{c}
\\
& \phantom{=}
\times \langle \Omega | \Bigg[
(-g) \int_{\infty}^{-\infty} \rmd{1}{\eta^+}
\mathcal{L}^{0_T\!,\ \xi^-}_{bc}(\infty^+,\eta^+) G_{cd}^{-\alpha}(\eta)
\mathcal{L}_{de}^{0_T\!,\ \xi^-}(\eta^+,-\infty^+)
\mathcal{L}_{fg}^{0_T\!,\ \xi^-}(\infty^+,\xi^+)
\\
& \phantom{=}
+
\mathcal{L}_{be}^{0_T\!,\ \xi^-}(\infty^+,-\infty^+)
(-g) \int_{\infty}^{\xi^+} \rmd{1}{\eta^+}
\mathcal{L}_{fc}^{0_T\!,\ \xi^-}(\infty^+,\eta^+) G_{cd}^{-\alpha}(\eta)
\mathcal{L}_{dg}^{0_T\!,\ \xi^-}(\eta^+,\xi^+)
\\
& \phantom{=}
+
\mathcal{L}_{be}^{0_T\!,\ \xi^-}(\infty^+,-\infty^+)
\mathcal{L}_{fg}^{0_T\!,\ \xi^-}(\infty^+,\xi^+) iD^\alpha \Bigg]
\psi_g(\xi) | P_h, P_X \rangle_{\text{out,c}}
\Big|_{\begin{subarray}{l}
\eta^- = \xi^- = 0\\ \eta_T = \xi_T = 0 \end{subarray}}.
\end{split}
\raisetag{16pt}
\end{equation}
By comparing this expression to $\Delta_\partial^{[ \pm ]}$
in the light-cone gauge one finds
\begin{equation}
\Delta^{\mathcal{L}' \alpha}_\partial (z^{-1}) = 2
\Delta^{[ + ]\alpha}_\partial (z^{-1}) -
\Delta^{[ - ]\alpha}_\partial (z^{-1}).
\end{equation}
The above expression states that the functions with this particular gauge link
are equivalent to twice the functions measured
in electron-positron annihilation minus the functions which are measured
in semi-inclusive DIS. That relation holds for T-even and T-odd fragmentation
functions.

Using the method as given above, the functions
appearing in quark-quark scattering can be compared
with the functions appearing in semi-inclusive DIS and electron-positron
annihilation. The results are given
in table~\ref{devFactors}.

In this section it was shown how the integrated
functions and the first transverse moments can be compared with the
functions in semi-inclusive DIS and electron-positron annihilation.
For the higher transverse moments this comparison is not always possible. In
those higher
moments new matrix elements are encountered which do not
appear in semi-inclusive DIS or electron-positron annihilation.

\section{Unitarity in two-gluon production}

The gauge link does not only depend on the process but also
depends on the diagram, forming a potential danger for unitarity.
If a theory is unitary then the cross section for the
production of nonphysical boson polarizations is canceled by the
cross section for the production of ghosts and antighosts. The fact that
QCD is an unitary theory is not trivially to see.
In Ref.~\cite{'tHooft:1972ue,'tHooft:1971fh}
't Hooft and Veltman showed
that in the sum over all diagrams
unitarity is maintained to all orders in QCD.
Having different gauge links
(and thus different functions) for each squared amplitude diagram,
it is not clear whether the approach still obeys unitarity.
In this section it will be argued at lowest order in $\alpha_S$
that including a gauge link in the correlators does not produce
nonphysical polarizations in two-gluon production in
quark-antiquark scattering. This may be
considered as a firm consistency check.
Before starting the gauge link calculation let us review, without the
gauge link, the unitarity proof for this process to some extent.

Consider the gluon-production amplitude of the process
\begin{equation}
\parbox{2cm}{\includegraphics[width=2cm]{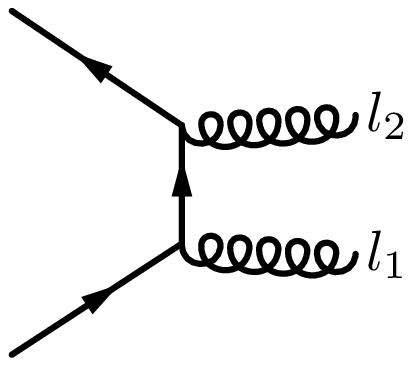}}\eqnIndent + \eqnIndent
\parbox{2cm}{\includegraphics[width=2cm]{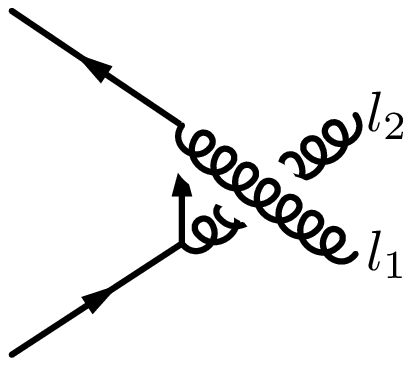}}\eqnIndent + \eqnIndent
\parbox{2cm}{\includegraphics[width=2cm]{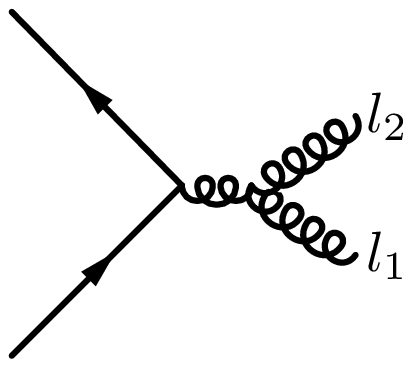}}
\eqnIndent.
\label{dev11}
\end{equation}
When squaring this amplitude there are nonphysical polarizations contributing
in the Feynman gauge.
In the complete square
those polarizations have either $\epsilon(l_1)$ being forwardly
polarized ($\epsilon^\alpha(l) \sim l^\alpha$) and
$\epsilon(l_2)$ being backwardly polarized
($\epsilon^\alpha(l) \sim \bar{l}^\alpha$, the bar denotes
reversal of spatial components), or vice versa. Together
they contribute
to the cross section as
\begin{equation}
\left| \text{Eq.~}\eqref{dev11}\right|^2 = 2
\bar{v}(k) \slashi{l}_2 t_d u(p) \frac{f^{dab}}{(l_1+l_2)^2}
\left[
\bar{v}(k) \slashi{l}_1 t_{d'} u(p) \frac{f^{d'ba}}{(l_1+l_2)^2}
\right]^\dagger .
\label{dev12}
\end{equation}
The ghost-production amplitude is
\begin{equation}
\parbox{2cm}{\includegraphics[width=2cm]{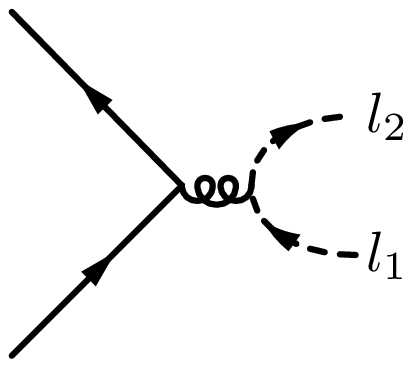}}\eqnIndent + \eqnIndent
\parbox{2cm}{\includegraphics[width=2cm]{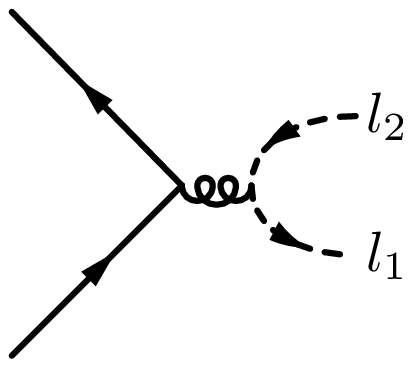}}
\eqnIndent.
\end{equation}
Taking the square of this amplitude exactly yields Eq.~\ref{dev12}
with an opposite sign\footnote{The opposite sign comes from the cut
fermion loop.}.
The cross section for the production of nonphysical polarizations
is therefore canceled by the ghost and antighost contributions. One remains
with purely transversely polarized outgoing gluons. This explicitly shows
unitarity at this order in $\alpha_S$.

\begin{figure}
\begin{center}
\begin{tabular}{p{5cm}p{1cm}p{5cm}}
\hspace{1cm}\includegraphics[width=3cm]{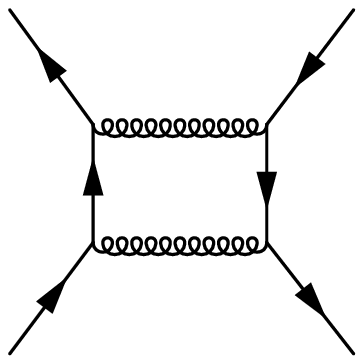} &&
\hspace{1cm}\includegraphics[width=3cm]{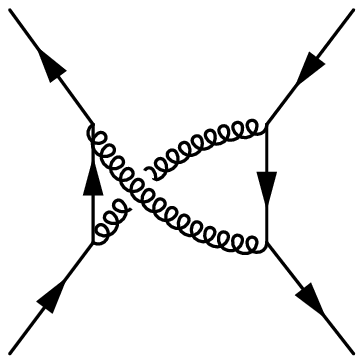}\\
&&\\
$\frac{1}{64} \mathcal{L}^{[-]}(0,\xi^-)$
&& $- \frac{5}{4} \mathcal{L}^{[-]}(0,\xi^-)$ \\
&&\\
\eqnIndent $+\frac{21}{64}
\mathcal{L}^{[+]}(0,\xi^-) \tr^C \mathcal{L}^{[\square]\dagger}(0,\xi^-)$
&&
\eqnIndent $+\frac{3}{4}
\mathcal{L}^{[+]}(0,\xi^-) \tr^C \mathcal{L}^{[\square]\dagger}(0,\xi^-)$
\end{tabular}
\end{center}
\caption{Two elementary scattering squared amplitude
diagrams and the gauge links corresponding
to the correlators attached to the quarks below (parent hadron is assumed
to be moving along the $n_+$-direction). The gauge links are derived by
applying the prescription of the previous section. As shown in this section,
the prescription gives
the correct result when the gluons crossing the cut have a physical
polarization.
\label{uni99}}
\end{figure}

\begin{figure}
%\begin{center}
\begin{tabular}{l}
\parbox{2cm}{\includegraphics[width=2cm]{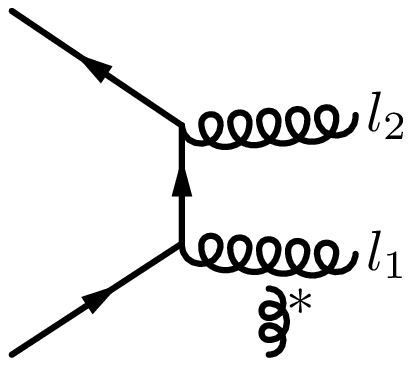}}\eqnIndent +
\parbox{2cm}{\includegraphics[width=2cm]{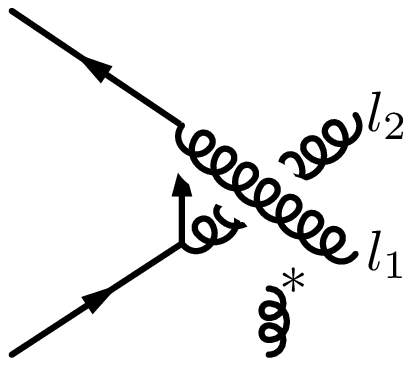}} \eqnIndent
 +
\parbox{2cm}{\includegraphics[width=2cm]{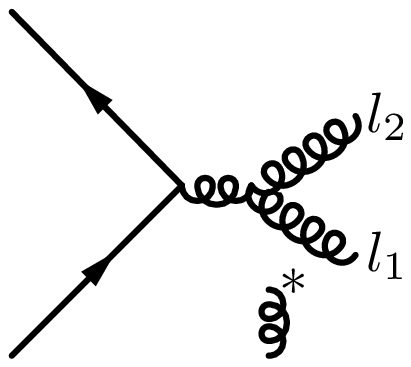}}\eqnIndent +
\parbox{2cm}{\includegraphics[width=2cm]{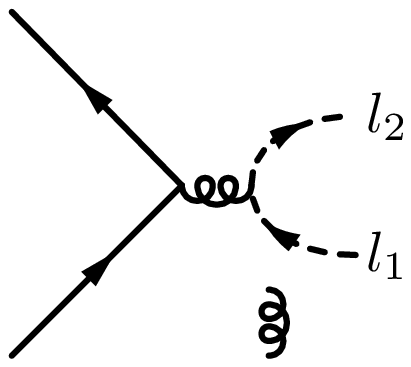}}
\\
\\
\\
\phantom{sdfkjhsdfksdsdsasdasdadfsfsfsfsfsfssfjsdfkjf}
=
\parbox{2cm}{\includegraphics[width=2cm]{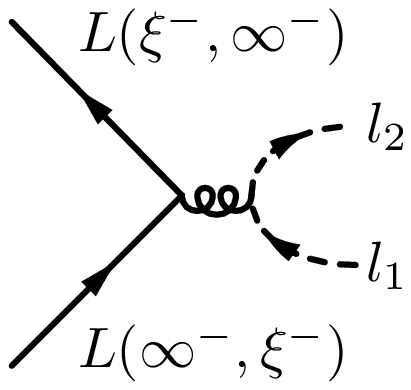}}
+
\parbox{2cm}{\includegraphics[width=2cm]{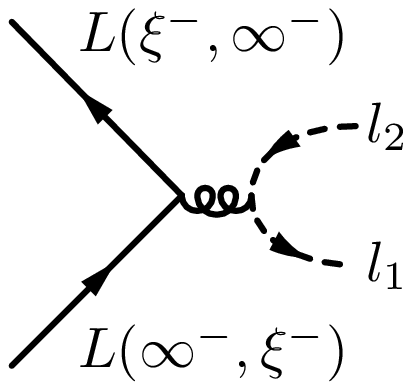}}
\end{tabular}
%\end{center}
\caption{Illustration of the calculation.
The $*$ indicates that of all the insertions on $l_1$ and $l_2$
only certain parts are included. Each
of those parts consists of several (zero or more)
terms like the first term of Eq.~\ref{dev14}, and
several terms (one or more) like the second \emph{or} third term of
Eq.~\ref{dev14}.\label{figUnitarity}}
\end{figure}

\begin{figure}
\begin{tabular}{l}
\parbox{2cm}{\includegraphics[width=2cm]{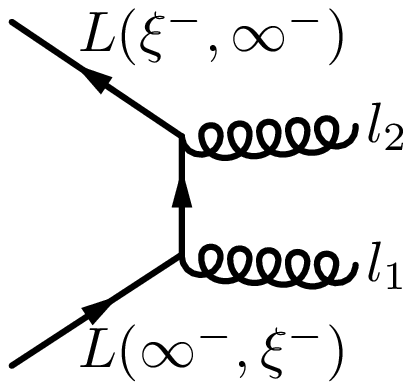}}\eqnIndent +
\parbox{2cm}{\includegraphics[width=2cm]{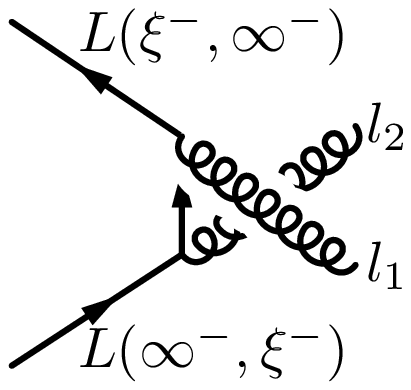}}
\\
\\
\\
\phantom{sdfkjhsdfkjsdfkjf} +
\parbox{2cm}{\includegraphics[width=2cm]{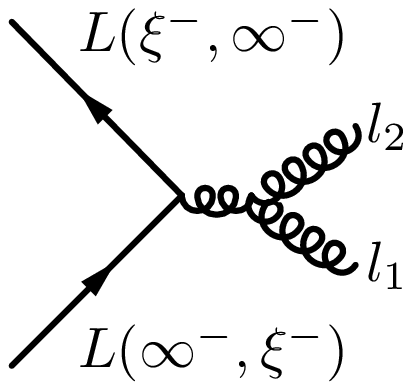}}\eqnIndent +
\parbox{2cm}{\includegraphics[width=2cm]{Developments_Figures_ghostProduction1Link.eps}}\eqnIndent +
\parbox{2cm}{\includegraphics[width=2cm]{Developments_Figures_ghostProduction2Link.eps}}
\end{tabular}
\caption{The result of the insertions on $l_1$ and $l_2$ on the
amplitude level. In obtaining this result properties of the conjugate amplitude
were used. In the diagrams
it is indicated on which quark-lines the color matrices of the gauge links
are standing.\label{uniResult2}}
\end{figure}

Including the gluon insertions for the gauge link
one finds in general different gauge links for different
diagrams (see for instance Fig.~\ref{uni99}).
Having different gauge links for the diagrams and thereby
also
having different functions, it is a priory not clear whether such a cancellation
between ghosts and nonphysical gluon polarizations still
occurs. In the remainder
of this section the rather technical
argument will be briefly outlined
that this is still the
case. Important intermediate results are illustrated in
Fig.~\ref{figUnitarity} and Fig.~\ref{uniResult2}. The reader who is not
interested in the derivation of these results can skip the next paragraphs
and can read the last paragraph
of this section.

When inserting a gluon with momentum $p_i$
on an outgoing gluon-line with momentum $l$ and carrying the
vector index $\alpha$,
the inserted vertex has the following form
\begin{equation}
%-2 l^- g^{\alpha'\alpha} + n_+^{\alpha'} l^\alpha + (l-p_i)^{\alpha'}n_+^\alpha,
-2 l^- g^{\alpha'\alpha} + (l-p_i)^{\alpha'}n_+^\alpha + n_+^{\alpha'} l^\alpha,
\label{dev14}
\end{equation}
where $\alpha'$ is contracted with the internal part of the amplitude.
When considering multiple insertions,
the terms of the inserted vertices, which are like
the first term of Eq.~\ref{dev14},
give back the same elementary
amplitude diagram with gauge links according to the
replacement rule as given in subsection~\ref{rules} step 2 (this has been shown
in subsection~\ref{alphaSDY}).
Arising from multiple insertions, we also discussed
in subsection~\ref{alphaSDY}
interference contributions
between terms like the second and third
term of Eq.~\ref{dev14}. These interference terms are expected
to cancel
against insertions containing four-gluon vertices. Of all the insertions,
this leaves us to
discuss the vertices in which combinations
between terms like the first term with the second or third term appear.
Those combinations were
canceled in semi-inclusive
DIS and Drell-Yan in which an additional gluon was radiated, but they do
not cancel in the present case.

Consider $N$ gluon insertions in which the vertices contain
$N-1$ terms are
like the first term in Eq.~\ref{dev14} and one term like
the third term in Eq.~\ref{dev14}.
This particular combination
contributes as having the outgoing gluon being backwardly
polarized (it becomes proportional to $l_1 \cdot \epsilon(l_1)$).
Coupling this term to the conjugate amplitude leads to a
forwardly polarized gluon on $l_2$ in the amplitude
(see the discussion at the beginning of this section).
In the sum over the amplitude diagrams only one non-vanishing contribution
remains. This contribution comes from
the
diagram with the
two triple-gluon vertices (like the third term of Eq.~\ref{dev11}) and
is proportional to $l_1^{\alpha'}$.
It gets contracted with the inserted vertex ($n_+^{\alpha'} l_1^\alpha$),
making this particular contribution proportional to
$(-i)l^-/(-2l^- p_1^+ + i \epsilon)$ when the inserted gluon
propagator (corresponding
to this vertex) is included.
When comparing this with the first
term of Eq.~\ref{eq54} one finds that the contribution here is opposite
in sign and is divided by a factor of 2.
Being opposite in sign this contribution is as if we would have coupled
the gluon to an antighost (with a ghost on $l_2$)
giving $1/2$ times the $N^\text{th}$ order
expansion of the gauge link.

When considering $N$ gluon insertions with $N-2$ terms
like the first term of Eq.~\ref{dev14} and $2$ terms like the third term
of Eq.~\ref{dev14}, one finds that this
contribution can also be interpreted as coupling
to an antighost multiplied with $-(1/2)^2$
times the $N^\text{th}$ order expansion of the
gauge link.
Considering now
$N$ insertions and summing over all possible third terms (number $m$)
and first terms (total number of combinations is $N!/(N!-m!)/m!$),
one finds that this contributes as if we would have inserted gluons
on an antighost
but with the following factor in front of the
$N^\text{th}$ order expansion of the gauge link
\begin{equation}
\sum_{m=1}^{N} \frac{(-1)^{m+1}}{2^m}\ \frac{N!}{(N-m)! m!} =
1 - \left( \frac{1}{2} \right)^N . \label{hiro11}
\end{equation}
Summarizing, when inserting gluons on the external gluon-line
carrying momentum $l_1$ one encounters besides the discussed replacement
rule several other terms. Some of these cancel when insertions with
the four-gluon vertices are included. The others can be interpreted as
insertions on an antighost-line (in the amplitude)
with $1-2^{-N}$ times the
$N^\text{th}$ order expansion of the gauge link. The
replacement rule is similar as for the gluons.

Inserting $N$ gluons on external ghost-lines and
antighost-lines is relatively simple. One obtains
the original amplitude
diagram times the $N^\text{th}$-order expansion of the
gauge link
multiplied by $(1/2)^N$.
The replacement rule is similar to the replacement rule as for
insertions on external gluon-lines, but the 
factor $(1/2)^N$ at each order of the expansion
prevents one forming a gauge link which would make the correlator gauge 
invariant.
However, the insertions on an antighost-line together
with the contributions of insertions on an external gluon-line,
in which terms appear
like the third of Eq.~\ref{dev14},
can be recast into a
ghost-antighost production
diagram with a full gauge
link ($(1/2)^N +$ Eq.~\ref{hiro11} $=1$).

\sloppy
The contributions, arising from insertions with terms like the
second term of Eq.~\ref{dev14}, work out in very similar way.
Those terms can be interpreted as coupling to a ghost-line.
Together with the insertions on a ghost-line the result can be written as the
production of
ghost-antighost multiplied with a full gauge link.
The
result of the above calculation has been represented in
Fig.~\ref{figUnitarity}.

\fussy

%\begin{multline}
%\parbox{2cm}{\includegraphics[width=2cm]
%{Developments/Figures/gluonProduction1ExtraTerm.eps}}\eqnIndent +
%\parbox{2cm}{\includegraphics[width=2cm]
%{Developments/Figures/gluonProduction2ExtraTerm.eps}}
%\\
%+
%\parbox{2cm}{\includegraphics[width=2cm]
%{Developments/Figures/gluonProduction3ExtraTerm.eps}}\eqnIndent +
%\parbox{2cm}{\includegraphics[width=2cm]
%{Developments/Figures/ghostProduction1ExtraTerm.eps}}\eqnIndent
%=
%\parbox{2cm}{\includegraphics[width=2cm]
%{Developments/Figures/ghostProduction1Link.eps}}
%\end{multline}
%where the $*$ indicates that of all the insertions
%only a part is included. That part consists of several (zero or more)
%terms like the first term of Eq. ~\ref{dev14}, and
%several terms (one or more) like the second \emph{or} third term of
%Eq.~\ref{dev14}.

%only those terms are included which
%have several terms (zero or more) like the first term of Eq.~\ref{dev14} and
%several terms (one or more)
%like the second term of Eq.~\ref{dev14}, or
%several terms like the first  (zero or more) and
%several terms like the third (one or more) of Eq.~\ref{dev14}.

Summarizing, when inserting gluons in the two-gluon production
amplitude, one obtains
the gauge link for each amplitude diagram plus some additional terms.
Part of these additional terms (interferences between the second and
the third term of Eq.~\ref{dev14}) should be canceled by four-gluon vertices
as explained in subsection~\ref{alphaSDY}.
The other terms together with the insertions on ghost-lines combine
into full gauge links with an elementary ghost-antighost production-diagram.
We remain now with the same set of elementary scattering diagrams
in which gauge links according to the replacement
rules appear. Those gauge links are at this moment not absorbed in the
correlator. The result is illustrated in Fig.~\ref{uniResult2}.

Insertions in the conjugate part of the diagram yield the same result.
Since the insertions do not modify the property that if one external
gluon is forwardly
polarized the other has to be backwardly polarized, one can
even take combinations of insertions on the left-hand-side  and on the
right-hand-side of the cut and obtain the same result.

To show unitarity is now relatively straightforward. On the amplitude level,
one can show that if one gluon has a nonphysical polarization, then
the other gluon must have a nonphysical polarization as well in order
to have a non-vanishing amplitude.
This contribution, which comes from the diagram with the triple-gluon vertex,
is canceled in the cross section by the ghost-antighost contributions, because
the gauge link for ghost-antighost production is similar to the gauge link
of that contribution.
This shows for this example
that the procedure of obtaining gauge links is consistent.

%In the full
%amplitude the other term constitute
%of unphysical polarizations of both gluons coming from the diagram with
%the triple gluon vertex. This contribution together with the gluon insertions
%on ghosts can be written as ghost-antighost production multiplied with
%a full gauge link. This result is for each squared diagram a full gauge link.
%Considering in this set of diagrams an unphysical polarization on one
%gluon gives only non-vanishing contributions when the other gluon has
%also an unphysical polarization. That contribution comes from the diagram
%with two triple gluon vertices. Since that guage link is the same as for
%ghost-antighost production these unphysical polarization are canceled.

\section{Gauge links in gluon-gluon correlators}

In the description of evolution and in the
treatment of hadron-hadron collisions, one encounters quark-quark correlators
and
gluon-gluon correlators. In this section the gauge links in gluon-gluon
correlators will be studied.
Defining transverse momentum dependent
gluon distribution and fragmentation functions is a
difficult task. In several articles this topic has been studied
(see for instance
Collins, Soper~\cite{Collins:1981uk,Collins:1981uw},
Rodrigues, Mulders~\cite{Mulders:2000sh},
Burkardt~\cite{Burkardt:2004ur}, and
Ji, Ma, Yuan~\cite{Ji:2005nu}),
but a
definition, including hard-part-dependent gauge links as
obtained for quark distribution functions, has not yet
been derived on the basis of the diagrammatic expansion.
Using the same techniques as for the quark
distribution correlator and assuming  the theory to be consistent, it is
possible to ``derive'' gluon distribution correlators.
Although
one would rather like to show the consistency of the theory, the approach
to be followed
here provides at least some insight.

\begin{floatingfigure}{3.3cm}
\centering
\includegraphics[width=3cm]{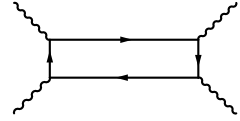}
\vspace{-.1cm}
\caption{A contribution to photon photon-distribution scattering. The correlator
is connected to the photons entering from the bottom of the graph.
\label{photonDistr}}
\end{floatingfigure}

Similarly as for the quark-quark
correlator the photon-photon correlator can be derived using
the LSZ-reduction
formalism\footnote{
Identifying $A$-fields with partons is a complicated
procedure in the LSZ-formalism in an arbitrary gauge
(see for instance Itzykson, Zuber~\cite{Itzykson:1980rh}).
However, in the Coulomb gauge it is relatively simple,
motivating the choice for this gauge.}.
In the Coulomb gauge a
photon-photon correlator without gauge link
(appearing for instance in photon photon-distribution scattering,
see Fig.~\ref{photonDistr})
reads
%\begin{equation}
\begin{multline}
\Phi_{AA}^{\alpha \beta} (x,p_T,P,S)
\! =\!\!
\int\! \frac{\mathrm{d} \xi^- \mathrm{d}^2 \xi_T}{(2\pi)^3} e^{ip\xi}
\\
\times
\langle P,S | A^\alpha (0) A^\beta (\xi)
| P,S \rangle \Big|_{\begin{subarray}{l}  \xi^+ = 0\\p^+ = xP^+\end{subarray}}
\!\!\!.
\label{PhiAA}
%\end{equation}
\end{multline}
Although this correlator is not gauge invariant, the expression
for the cross section
is actually already
gauge invariant which can be seen as follows.
Making a gauge transformation modifies Eq.~\ref{PhiAA} into
\begin{equation}
\Phi_{AA}^{\alpha \beta} (p,P,S)
{=}\!\!\!
\int\!\! \frac{\mathrm{d} \xi^- \rmd{2}{\xi_T}}{(2\pi)^3}\! e^{ip\xi}\!
\langle P,S | \Big( A^\alpha (0){+} \partial^\alpha \Xi(0) \Big)
\Big( A^\beta (\xi){+} \partial^\beta \Xi(\xi) \Big)
| P,S \rangle
\Big|_{\begin{subarray}{l} \xi^+ = 0 \\ p^+ = xP^+\end{subarray}}\!\!\!.
\label{partialIntegration}
\end{equation}
When performing a partial integration the derivatives become proportional
to the momentum $p$. These terms, however, do not show up in the cross section
because they do not couple in the sum over all diagrams (Ward identity).
The expression for the cross
section is thus shown to be gauge invariant.

Being gauge invariant, the expression for the 
cross section can be compared to any other gauge
invariant object. If they are the same in a certain gauge
(for instance $\bar{n}(p){\cdot}A{=}0$), then they are
the same in any gauge.
This justifies the use of the correlator
%\begin{multline}
\begin{equation}
\text{Eq.~}\eqref{PhiAA} \rightarrow
\int \frac{\rmd{1}{\xi^-} \rmd{2}{\xi_T}}{(2\pi)^3} e^{ip\xi}
\bar{n}_\delta(p) \bar{n}_\gamma(p)
%\\
%\times
\langle P,S | F^{\delta\alpha} (0)
 F^{\gamma\beta} (\xi)
| P,S \rangle
\Big|_{\begin{subarray}{l} \xi^+ = 0 \\ p^+ = xP^+\end{subarray}}\!\!\! ,
%\end{multline}
\end{equation}
with
$n(p)^2 = 0,\ n(p)\sim p,\ \bar{n}(p)\cdot p = 1$.
So by starting with a certain correlator connected to the cross section
we were able to derive its gauge invariant form.

In general processes
other kind of correlators appear, also containing gauge
links.
In principle these gauge links can be obtained by
following the same procedure as outlined in
the subsection~\ref{rules} for the insertions.
The result one obtains is the $A$-fields together with
the presence of gauge links. Now to show to which gauge invariant expressions
the obtained correlators correspond is more
difficult than the previously considered case. The main problem is that
one picks up a contribution of the gauge link when doing
a partial integration
as performed in Eq.~\ref{partialIntegration}.
Such contributions are similar to the gluonic pole matrix elements which
we          also
encountered for the quark-quark correlators. Since there is no reason to
neglect those terms, this forms at present the main obstacle
for obtaining gauge invariant gluon-gluon correlators.
%\footnote{We think that this problem is connected with the eikonal approximation
%one is making. In the eikonal approximation one assumes that the
%transverse momentum of the gluon is much smaller than its longitudinal
%momentum. In principle that's fine but
%whether this is also true at the pole remains to be seen.
%The pole in $p_1^+$ forces
%the transverse momentum to vanish in the eikonal approximation.
%Since the inserted gluon and the original gluon are indistinguishable
%we do not know on for hand which gluon becomes the gauge link
%for the other gluon of which the transverse momentum is not neglected.
%. It is
%this reason why we think that the eikonal approximation leads to difficulties
%concerning gauge invariance.}

So how can one proceed? Well, one could argue that although
it is not completely clear how to obtain the
correlator in terms of the $F$-fields, the longitudinal
gauge link as obtained by the presented rules should still be correct.
If the approach
is consistent then these longitudinal links should be closed at infinity
by adding transverse gauge links, which
can be done uniquely. In addition,
if the cross section is gauge invariant (as it should be),
then one can compare this expression in the light-cone gauge with any other
gauge invariant expression. Comparing with the same expression, but with
the $A$-fields replaced with the $F$-fields provides then a gauge
invariant definition.
It provides a solution for the moment,
but more research is definitely needed here.

\begin{figure}
\begin{center}
\begin{tabular}{ccc}
\includegraphics[width=5cm]{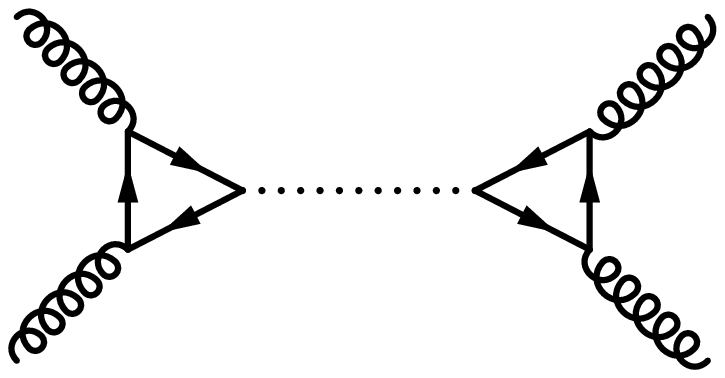}&&
\includegraphics[width=5cm]{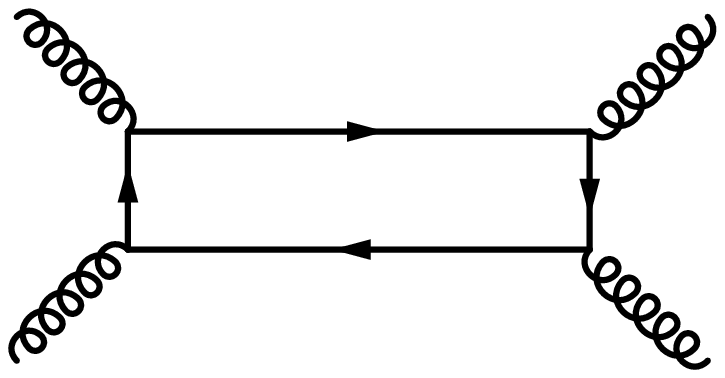}           \\
(a) && (b)
\end{tabular}
\end{center}
\caption{Two processes with gluon distribution functions attached to the
gluons entering the graphs from the bottom.
Figure~(a) contributes to Higgs-production and
figure~(b) contributes to two jet-production in hadron-hadron
scattering.\label{gluonLinks}}
\end{figure}

Using this approach one is able to obtain gauge invariant correlators. As
an example, results will be given here
for the two subprocesses illustrated
in Fig.~\ref{gluonLinks}. 
Using the prescription of subsection~\ref{rules} and factorizing the 
color matrices with which the gluons couple into the hard part, the 
gauge link for the diagram contributing to Higgs-production
can be found to be\footnote{In this diagram there is a virtual loop present which we have
not considered before. In the next section we will study gauge links and
loop corrections and we will argue that poles in the loop from insertions
can be discarded.} (Fig.~\ref{gluonLinks}a)
\begin{multline}
\Phi^{\alpha\beta}_{\text{Fig.~}\eqref{gluonLinks}a} {=} 2
\int \frac{ \rmd{1}{\xi^-}\rmd{2}{\xi_T}}{(2\pi)^3}
e^{ip\xi}
\\
\times \langle P,S | \tr^C\! \left[\! F^{+\beta}(0) \mathcal{L}^{[-]}(0,\xi^-)
F^{+\alpha}(\xi) \left[ \mathcal{L}^{[-]}(0,\xi^-)\right]^\dagger
\right] |P,S \rangle_\text{c}
\Big|_{\begin{subarray}{l} \xi^+ = 0\\p^+ = xP^+ \end{subarray}}\!\!\!.
\end{multline}
All matrices in the above expression are $3\times 3$ matrices.
This expression is equivalent to the result of Ji, Ma, and Yuan
who study in Ref.~\cite{Ji:2005nu} this particular process as a first step
to factorization of effects from intrinsic transverse momentum
in hadron-hadron collisions.
The equivalence can be seen by the following relation
$2 \tr^C [ t_a \mathcal{L}^{[-]}\!(0,\!\xi^-) t_b \mathcal{L}^{[-]\dagger}\!(0,\!\xi^-) ]
= \mathcal{L}^{[-]}_{ab}(0,\xi^-)$ in which the
$A_l$-fields on right-hand-side
are contracted with the structure constants ($i f^{alb}$)
instead of the
color matrices. In a sense this relation was obtained in
subsection~\ref{alphaSDY}.

Another example is the gauge link of a subprocess
which contributes  to two jet-production in hadron-hadron
scattering (see Fig.~\ref{gluonLinks}b).
The gauge invariant gluon-gluon correlator is found to be
\begin{align}
& \Phi^{\alpha\beta}_{\text{Fig.~}\eqref{gluonLinks}b} = 2
\int \frac{ \rmd{1}{\xi^-}\rmd{2}{\xi_T}}{(2\pi)^3}
e^{ip\xi}
\nonumber\\
& \eqnIndent \eqnIndent\eqnIndent \ \
\times
\langle P,S |
\Bigg[
\frac{3}{8} \tr^C \left[ F^{+\alpha}(\xi) [ \mathcal{L}^{[+]}(0,\xi^-)]^\dagger
F^{+\beta}(0) \mathcal{L}^{[-]}(0,\xi^-) \right]
\tr^C \left[ \mathcal{L}^{[ \square ]}(0,\xi^-) \right]
\nonumber\\
& \eqnIndent\eqnIndent\eqnIndent\eqnIndent \ \
-
\frac{1}{8}
\tr^C \left[ F^{+\alpha}(\xi) \mathcal{L}^{[+]\dagger}(0,\xi^-)
F^{+\beta}(0) \mathcal{L}^{[+]}(0,\xi^-) \right] \Bigg]
 |P,S \rangle_\text{c} \Big|_{\begin{subarray}{l}\xi^+ = 0\\ p^+ = xP^+
\end{subarray}}\!\!\!.
\end{align}
In this new result various gauge links via plus and minus infinity appear.
It is not possible to rewrite this result in terms of gauge links in which
only structure constants are used (as in the previous example).

In Ref.~\cite{Burkardt:2004ur}
Burkardt studied gluon distribution functions and suggested a sum rule
for the Sivers quark and gluon distribution functions.
It may be good to point
out that those gluon distribution
functions, containing gauge links via plus infinity, appear
in semi-inclusive
lepton-hadron scattering, like two-jet production. A contribution to that cross
section would be Fig.~\ref{gluonLinks}b in which the incoming gluons from the
top of the graph are replaced by incoming virtual photons.

In this section a method for obtaining gauge invariant
gluon-gluon correlators has been suggested although a significant
amount of work remains to be done. Note that one still needs to look for observables sensitive
to the path of the gauge link. Those
observables should be sensitive to the intrinsic transverse
momenta of the gluons. One particular observable will be discussed for
hadron-hadron scattering in the next chapter.

\section{Factorization and universality}

Up to now processes have been described by using the diagrammatic
approach in which correlators were attached
to an infinite number of
hard scattering diagrams. These correlators, like $\Phi$, $\Phi_A$, and
$\Delta_{AA}$'s as defined in
section~\ref{diagram}, contained a certain renormalization scale and
it has been assumed that these correlators could be factorized, enabling one
to calculate in principle its scale dependence process-independently.
We found at leading order in $\alpha_S$
that in several semi-inclusive processes, this infinite set
of hard scattering
diagrams and correlators combined into a finite set of diagrams convoluted
with a finite set of gauge invariant correlators (containing a gauge link).
Although the starting point was purely process-independent, it turned out that
the hard scattering part determined the path of the gauge links in the final
result.
In this section the validity of the applied factorized
approach will be discussed.

Whether processes allow for a factorized description
has been studied for several decades. Ellis, Georgi, Machacek, Politzer, and
Ross studied the issue of factorization for semi-inclusive DIS and
Drell-Yan~\cite{Ellis:1978sf,Ellis:1978ty}. By applying methods of
Libby and Sterman~\cite{Sterman:1978bi,Libby:1978qf}, Collins and Sterman
were able to show in Ref.~\cite{Collins:1981ta}
to all orders in $\alpha_S$ that
semi-inclusive electron-positron annihilation is
free of infra-red divergences (divergences appearing in separate diagrams
when momenta of virtual or radiated partons vanish, $l \rightarrow 0$).
Subsequently, Collins and Soper introduced in
Ref.~\cite{Collins:1981uk,Collins:1981uw}
infra-red free
factorization formulas for electron-positron annihilation
with two almost back-to-back hadrons being observed. Those factorization
formulas were constructed
at high $q_T$ (order $-q_T^2 \sim
q^2=Q^2\gg M^2$) and for fully integrated over $q_T$, where $q_T$ is the transverse momentum
of the virtual photon with respect to the observed hadrons.
A factorization theorem
for low $q_T$ ($q_T^2 \sim -M^2$) was also proposed. In that theorem
the infra-red part, coming from the vertex correction in which all momenta become soft,
was factorized from the jets
into a soft factor. In the same paper also an attempt was made
to describe Drell-Yan, but a factorization theorem could not be obtained
due to the interplay of final-state and
initial-state interactions.
Possible problems due to this interplay were also noticed by
Doria, Frenkel, and Taylor~\cite{Doria:1980ak}.

The study on Drell-Yan was continued in several papers among which
papers of Collins, Qiu, Soper,
and Sterman~\cite{Collins:1981tt,Collins:1983pk,Collins:1983ju,
Collins:1984kg,Collins:1985ue,Collins:1988ig,Qiu:1990xy}, and
Bodwin~\cite{Bodwin:1984hc}.
In the end it was believed that the problems of initial and final-state
interactions
were under control, because the
interactions between spectators
were expected to be on a longer time scale and
should therefore vanish by
unitarity\footnote{
Being on a longer time scale it was not expected that those interactions
could influence the short time scale
production of the virtual photon.}.
This yielded
fairly well established
factorization theorems for $q_T$-integrated Drell-Yan and small
$q_T$-unintegrated
Drell-Yan~\cite{Collins:1989gx}.
In 1992 Collins included straightforwardly
polarizations of participating hadrons into the factorization
theorems~\cite{Collins:1992xw} and suggested in Ref.~\cite{Collins:1993kk}
a factorization theorem for
semi-inclusive DIS at small $q_T$.

In 2002, Brodsky, Hwang, and Schmidt showed that final-state
interactions
between spectators lead to single spin asymmetries in semi-inclusive
DIS~\cite{Brodsky:2002cx}. Based on this surprising result,
Collins concluded that the studied interactions
are actually on a shorter time-scale than it at first sight seems.
He pointed out
in Ref.~\cite{Collins:2002kn} that
such interactions could give problems for transverse momentum
dependent factorization theorems (at small $q_T$)
in Drell-Yan in which both initial and final-state interactions are present.

The discussion on factorization and
universality for transverse momentum dependent cross sections was
recently continued
in several papers.
In 2004, Ji, Ma, and Yuan argued and claimed in
Ref.~\cite{Ji:2004wu,Ji:2004xq} to have shown
all order factorization
theorems at small $q_T$ for Drell-Yan and semi-inclusive DIS.
%including twist three
%effects.
%However, in those papers it appears that contributions from, terms like
%$\Phi_{\partial^{-1}G}^\alpha$ (see chapter~\ref{chapter3})
%and which contributes at $M/Q$, have been discarded, making it
%unclear whether the factorization theorems are complete as they stand.
A one
loop calculation illustrated the factorization theorems, which was generalized
to all orders by using \nolinebreak power \nolinebreak \mbox{counting}.
Complications from combining
final and initial-state interactions, which appear at
two loops or higher (see for instance Fig.~\ref{jiFiguur}), were not explicitly
addressed. Further, it was not discussed
whether or not the fragmentation functions are process-independent.

\begin{floatingfigure}{3cm}
\centering
\includegraphics[width=3cm]{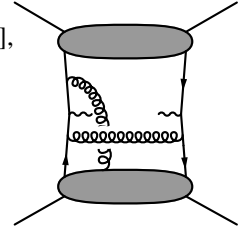}
\caption{A combined initial and final-state interaction in Drell-Yan
\label{jiFiguur}}
\end{floatingfigure}

By showing explicit results, it was subsequently
pointed out in Ref.~\cite{Bomhof:2004aw,Pijlman:2004wb,Pijlman:2004mr}
that there could be a problem for the transverse momentum dependent
factorization theorems at small $q_T$
for semi-inclusive DIS and Drell-Yan.
The problem appears due to the presence of the gauge link which
has a non-expected behavior in processes in which there are
initial and final QCD-states present
(in the language of
Ref.~\cite{Collins:2004nx,Ji:2004wu,Ji:2004xq} this issue
appears at two loops and higher).
This will be elaborated upon in the next subsections.

The study on universality and factorization
was continued
by Collins and Metz~\cite{Collins:2004nx}. Inspired by earlier
work of Metz~\cite{Metz:2002iz}, the conclusion was drawn
that the fragmentation process is
universal for which an additional argument was provided
(which was discussed in section~\ref{fragUni}).
Universality of transverse momentum dependent distribution and fragmentation
functions was consequently claimed for semi-inclusive DIS, electron-positron
annihilation, and Drell-Yan, but
the issue of final and initial-state interactions in Drell-Yan was not further
illuminated.

The mentioned references use various starting points for the
discussion of factorization, which
prevents straightforward comparisons of their results. However,
one important point,
namely the coupling of almost collinear and longitudinally
polarized gluons to various places in
elementary diagrams yielding the gauge link,
is obtained in essentially the same manner in all the approaches.
Being similar,
the results on this point should be equivalent.
Another point in the discussion of factorization are the virtual corrections
which have not been studied so far in this thesis.
Therefore, we shall consider
the vertex and self energy corrections in the next subsection.
Together with the results on gauge links appearing in tree-level diagrams,
the validity of factorization theorems
for various processes will be discussed in the second subsection.

\subsection{Virtual corrections}

In this subsection gauge links will be discussed for diagrams in which
virtual corrections appear.
Intuitively one can already guess the outcome. Since virtual corrections
do not modify the nature of the external particles
(incoming or outgoing), one does not expect
to find different gauge links for diagrams in which virtual corrections
are included. In this subsection technical arguments
for some specific cases in semi-inclusive DIS will be given to show that
this idea indeed holds.
It will be argued in the Feynman gauge
that the first order expansion of the gauge link is the
same as for the diagram in which the virtual correction is
absent~\cite{Pijlman:2004wb}. The conclusion drawn for semi-inclusive
DIS also holds for Drell-Yan and semi-inclusive electron-positron annihilation.
The first order expansion of the gauge link is sufficient
to illustrate the complications related to
factorization.

\subsubsection{Vertex correction}

We will begin by studying the problem in QED and then extend the result
to QCD.
There are three possible ways to insert
a single photon with momentum $p_1$
to the vertex correction diagram (showing the integral over $p_1$ explicitly),
\begin{equation}
\int \rmd{4}{p_1} \left[
\parbox{2cm}{\includegraphics[width=1.8cm]{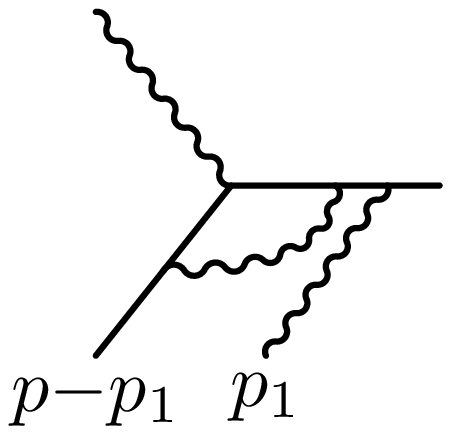}} +
\parbox{2cm}{\includegraphics[width=1.8cm]{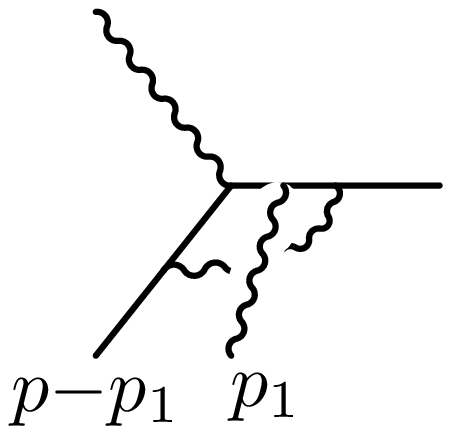}} +
\parbox{2cm}{\includegraphics[width=1.8cm]{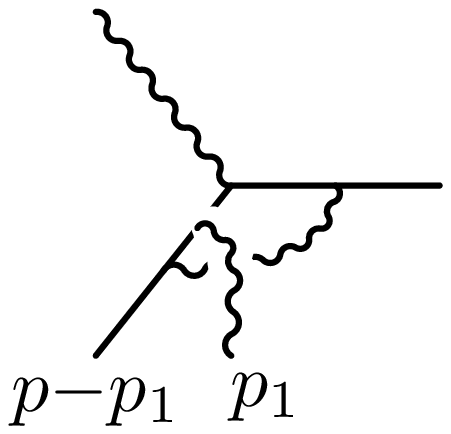}}
\right],
\label{vertexCorr}
\end{equation}
where the lines at the bottom of the graph are connected to the
relevant part of the correlator $\Phi_A^+(p,p_1)$.

The momentum dependence $p_1$ can be routed back to the
correlator via the photon propagator or via the electron propagators.
We will choose here the latter.
Having
this $p_1$-dependence, the propagators
contain poles in $p_1^+$ which can be evaluated
by taking their residues.
By introducing dashes for the taken residues,
this pole calculation is made more explicit
\begin{equation}
\begin{split}
\text{Eq.~}\eqref{vertexCorr}
&
= \eqnIndent
\parbox{1.8cm}{\includegraphics[width=1.6cm]{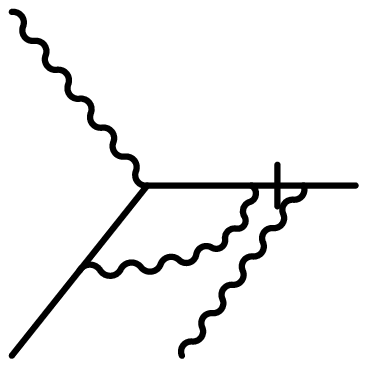}} +
\parbox{1.8cm}{\includegraphics[width=1.6cm]{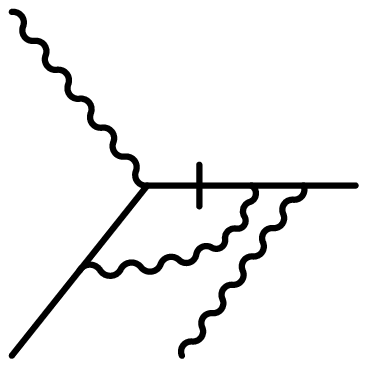}} +
\parbox{1.8cm}{\includegraphics[width=1.6cm]{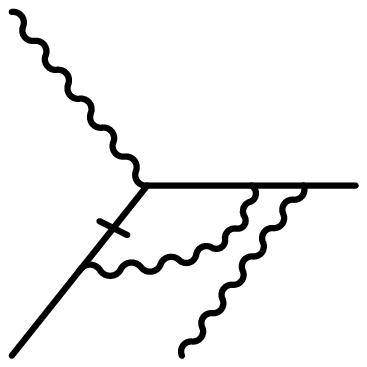}}
\\
&
 \eqnIndent
+
\parbox{1.8cm}{\includegraphics[width=1.6cm]{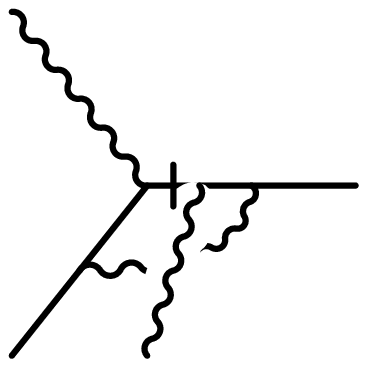}} +
\parbox{1.8cm}{\includegraphics[width=1.6cm]{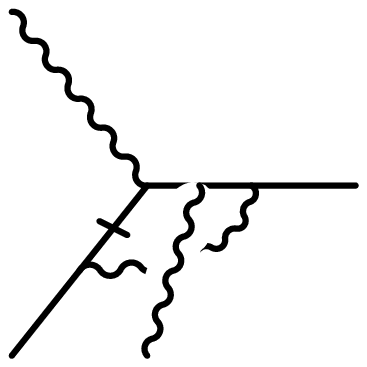}} +
\parbox{1.8cm}{\includegraphics[width=1.6cm]{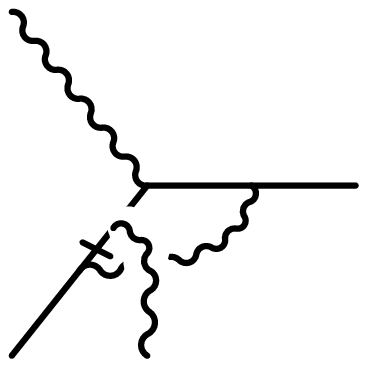}}
\\
&=
\eqnIndent
\left[
\parbox{1.8cm}{\includegraphics[width=1.6cm]{Developments_Figures_vertex113.eps}} +
\parbox{1.8cm}{\includegraphics[width=1.6cm]{Developments_Figures_vertex122.eps}} +
\parbox{1.8cm}{\includegraphics[width=1.6cm]{Developments_Figures_vertex131.eps}}
\right]\\
& \eqnIndent + \left[
\parbox{1.8cm}{\includegraphics[width=1.6cm]{Developments_Figures_vertex112.eps}} +
\parbox{1.8cm}{\includegraphics[width=1.6cm]{Developments_Figures_vertex121.eps}} \right] +
\parbox{1.8cm}{\includegraphics[width=1.6cm]{Developments_Figures_vertex111.eps}},
\end{split}
\label{vertexBracket}
\end{equation}
where the possibility that two propagators going simultaneously on shell
has been discarded.
It is fairly straightforward to show that the
bracketed terms vanish by doing the calculation explicitly.
The last diagram contains the pole on the external
parton and yields
the first order expansion of the gauge link.

This result will now be extended to QCD. The virtual photon
and the inserted photon are replaced
by gluons, and the electrons are replaced by quarks.
The calculation is in QCD slightly different, because 
in the QCD-version of Eq.~\ref{vertexBracket} the order of the
color matrices in the bracketed terms is different.
This can be solved 
by using the freedom on how to route the $p_1$-momentum through the loop,
giving us the following identity
\begin{equation}
\begin{split}
%&
\parbox{1.7cm}{\includegraphics[width=1.6cm]{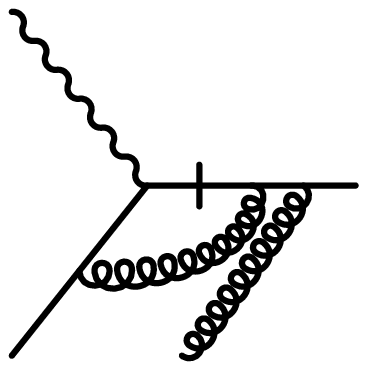}}
{+}\!\!
\parbox{1.7cm}{\includegraphics[width=1.6cm]{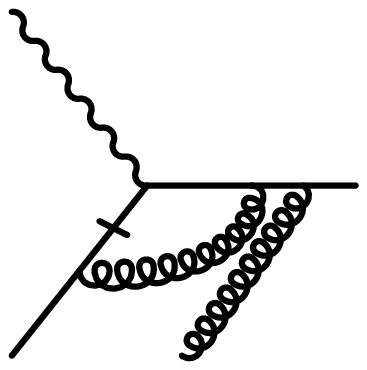}}
%\\
&
%\eqnIndent \eqnIndent \eqnIndent \eqnIndent 
{=} 
\parbox{2cm}{\includegraphics[width=1.6cm]{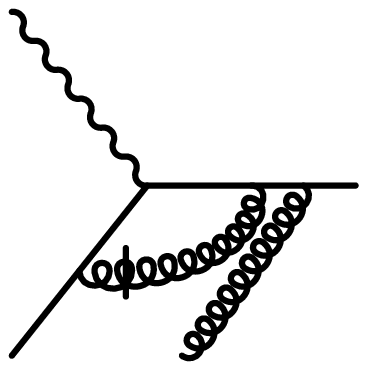}}
\\
&
% \eqnIndent \eqnIndent \eqnIndent \eqnIndent 
{=}
\frac{-1}{8}
%\left( t_b t_l t_b \right) \left( t_l t_b t_b \right)^{-1}
\left[
\parbox{1.7cm}{\includegraphics[width=1.6cm]{Developments_Figures_vertexGluon112.eps}}
{+}
\parbox{1.7cm}{\includegraphics[width=1.6cm]{Developments_Figures_vertexGluon113.eps}}
\right]
%\\
{+}
\frac{9}{8} 
%\left[1 - \left( t_b t_l t_b \right) \left( t_l t_b t_b \right)^{-1}\right]
\parbox{1.7cm}{\includegraphics[width=1.6cm]{Developments_Figures_vertexGluon114.eps}}.
\end{split}
\label{dev98}
\end{equation}
Substituting this identity in the QCD-version of Eq.~\ref{vertexBracket}
all bracketed terms (present in Eq.~\ref{vertexBracket}, \ref{dev98}) 
are canceled.
We remain with the last diagram of the QCD-version of Eq.~\ref{vertexBracket}
(giving the first order of the gauge link), the last diagram of Eq.~\ref{dev98}, and
the diagram in which the gluon is inserted on the virtual gluon.
The latter diagram contains the inserted vertex consisting
of three terms: $-2l^- g^{\alpha'\alpha} + n_+^{\alpha'} l^\alpha
+ (l-p_1)^{\alpha'} n_+^\alpha$ (similarly to Eq.~\ref{dev14}). The first term
cancels the last diagram of Eq.~\ref{dev98}, while the second and third
terms are canceled by terms appearing in a similar way
when treating the self energy corrections (see further below).
The conclusion is that the first order of the gauge link 
for the vertex corrected diagram is similar to the uncorrected diagram.

\subsubsection{Self energy correction}

\vspace{-.6cm}
\phantom{x}

\begin{floatingfigure}{3cm}
\centering
\includegraphics[width=2.5cm]{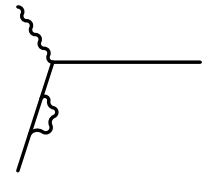}
\caption{Self energy correction on the parton 
belonging to the correlator.\label{selfOther}}
\end{floatingfigure}

\noindent
The calculation of the gauge link in a diagram in which the parton connected
to the considered
correlator has a self energy correction (see Fig.~\ref{selfOther}) 
is similar to the calculation
of the vertex correction. It can be straightforwardly
shown that the gauge link is the same as for the uncorrected
diagram. The calculation
for the diagram, in which the self energy correction is on the other external 
parton, is conceptually more difficult. That calculation will be presented
here in more detail.
We will begin with QED and then extend the result to QCD.

\indent
When taking a self energy diagram which is not directly connected to
the considered correlator there are three possible places to insert a photon
for the gauge link. This gives the following 
in terms of their residues
\begin{multline}
\int \rmd{4}{p_1} \left[
\parbox{1.8cm}{\includegraphics[width=1.6cm]{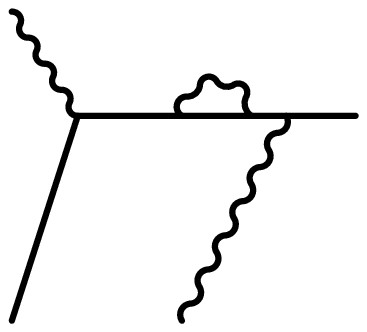}} +
\parbox{1.8cm}{\includegraphics[width=1.6cm]{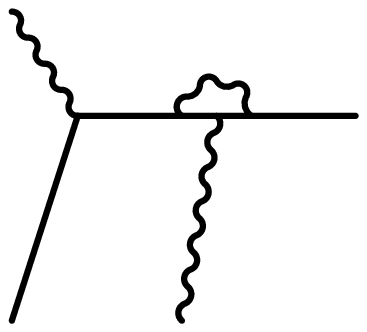}} +
\parbox{1.8cm}{\includegraphics[width=1.6cm]{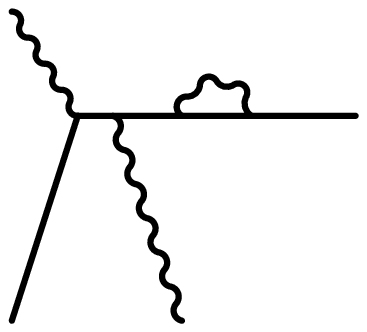}}
\right] 
\\
=
\left[
\parbox{1.8cm}{\includegraphics[width=1.6cm]{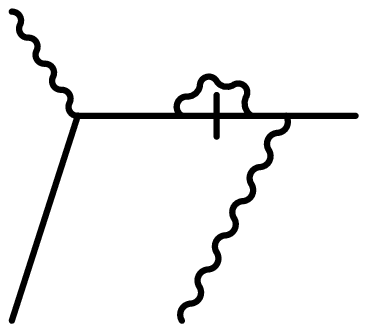}} +
\parbox{1.8cm}{\includegraphics[width=1.6cm]{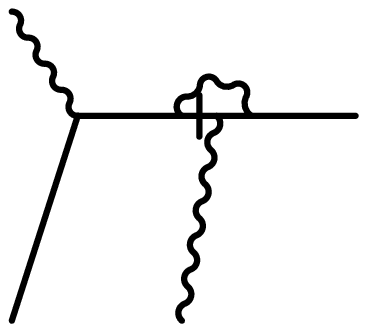}} 
\right]
\\
+ \parbox{1.8cm}{\includegraphics[width=1.6cm]{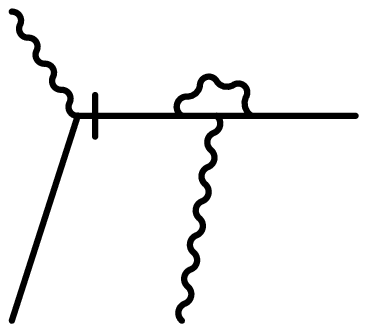}} +
\parbox{1.8cm}{\includegraphics[width=1.6cm]{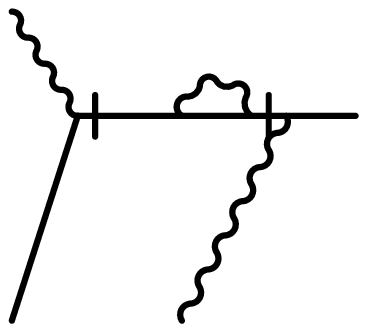}} +
\parbox{1.8cm}{\includegraphics[width=1.6cm]{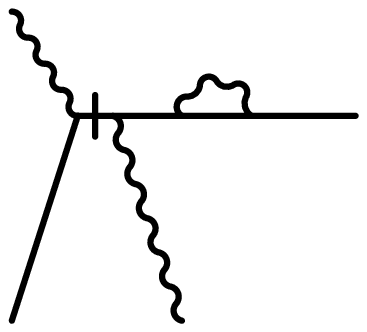}}
\end{multline}
The term between brackets on the right-hand-side again cancels.
The first of the three remaining terms 
can be calculated explicitly and yields the first order
gauge link multiplied with $(-) \delta Z_2$ ($\delta Z_2$ is of order $g^2$).
The second term is more difficult to calculate, because it contains
a double pole.
To circumvent this double pole problem
we will exponentiate this diagram together with the last diagram.

Consider $N$ (including $N=0$) consecutive self energy loops on an 
electron-line. Inserting a single photon
either before or after each loop and summing over $N$, the
result can be written as product between two geometrical series of which
one depends on $p_1$ and yields the first order gauge link expansion
multiplied by $Z_2$, and the other just gives $Z_2$. Together, this
gives the first order gauge link expansion multiplied with $Z_2^2 \approx
1 + 2 \delta Z_2$.
Note that in this sum also the first order link diagram without self
energy correction is included.
The total
result of all insertions to order $g^3$ is the first order link
expansion (order $g$) times $\delta Z_2$ (order $g^2$).

In QCD one has to route the momentum partly through the virtual gluon
as done when treating the vertex correction.
Similarly as for the vertex correction, one
obtains the first order gauge link expansion for the sum over the insertions,
but some terms remain which come from
the insertion on the virtual gluon of the self energy correction. Similar
results are achieved when the self energy correction is on the other
parton leg (see Fig.~\ref{selfOther}). Together,
the remaining terms of the self energy corrections
cancel the remaining term produced by the vertex correction.

The conclusion is that the first order of
the gauge link remains unchanged when including virtual corrections
in semi-inclusive DIS. The same conclusions can be reached for semi-inclusive
electron-positron annihilation and Drell-Yan.
\vspace{.6cm}

% This vspace is because of to compensate the vspace given in the beginning
% of this subsection.

\subsection{Evolution, factorization, and universality\label{hiro5}}

Scaling violations arise in a natural way when including higher order
corrections in $\alpha_S$. In this subsection $\alpha_S$ corrections
in combination with gauge links will be discussed. One of the corrections,
which was not discussed, is the self energy correction of the inserted gluons (or
gauge link). Since virtual corrections  are not expected
to modify the gauge link structure of the graph,
it will be assumed that its result can be rewritten as a charge renormalization
($\alpha_S$, see for related work Collins, Soper, Sterman~\cite{Collins:1982wa}).

\subsubsection{Semi-inclusive electron-positron annihilation}

\begin{figure}
\begin{center}
\begin{tabular}{cccc}
\includegraphics[width=2.5cm]{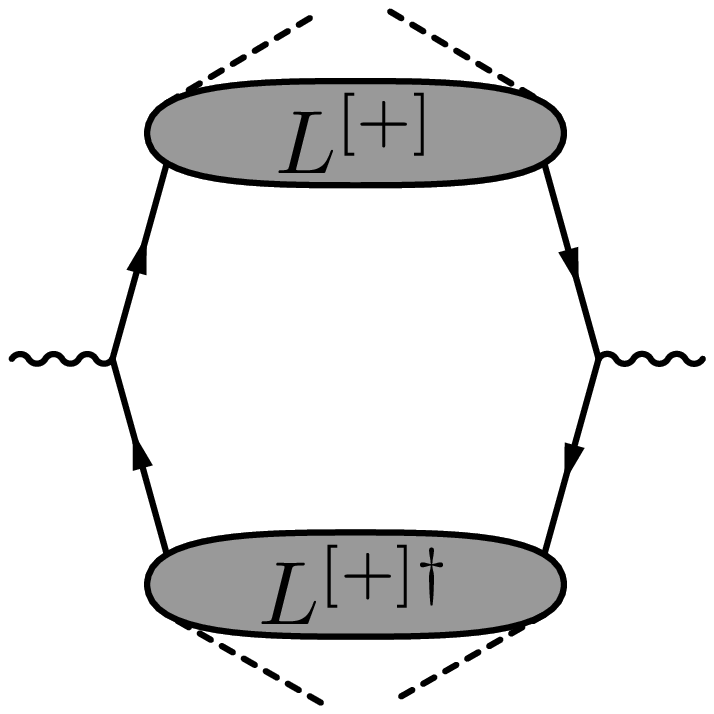} &
\includegraphics[width=2.5cm]{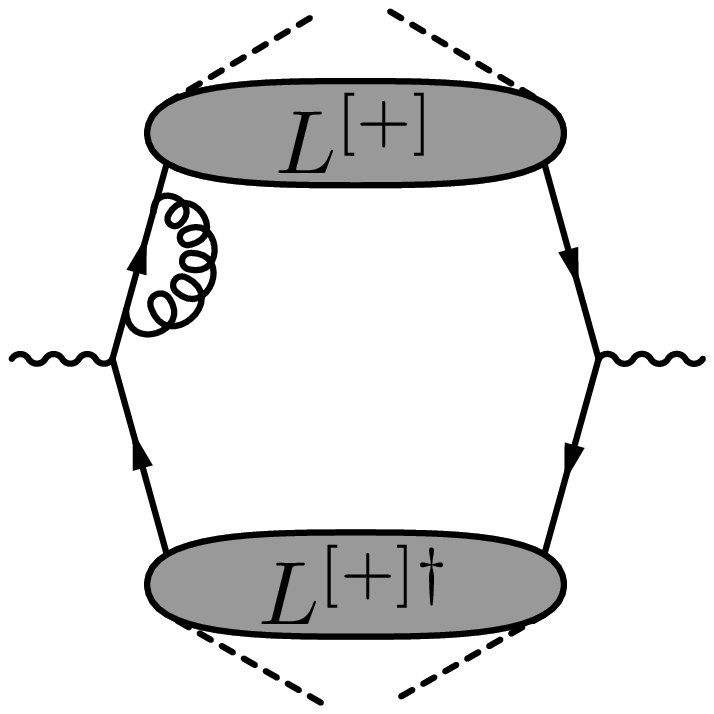} &
\includegraphics[width=2.5cm]{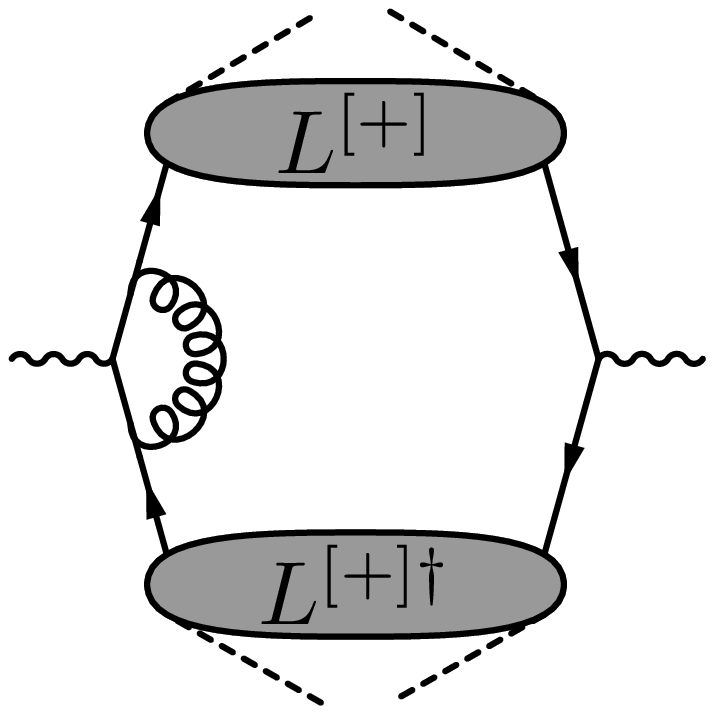} &
\includegraphics[width=2.5cm]{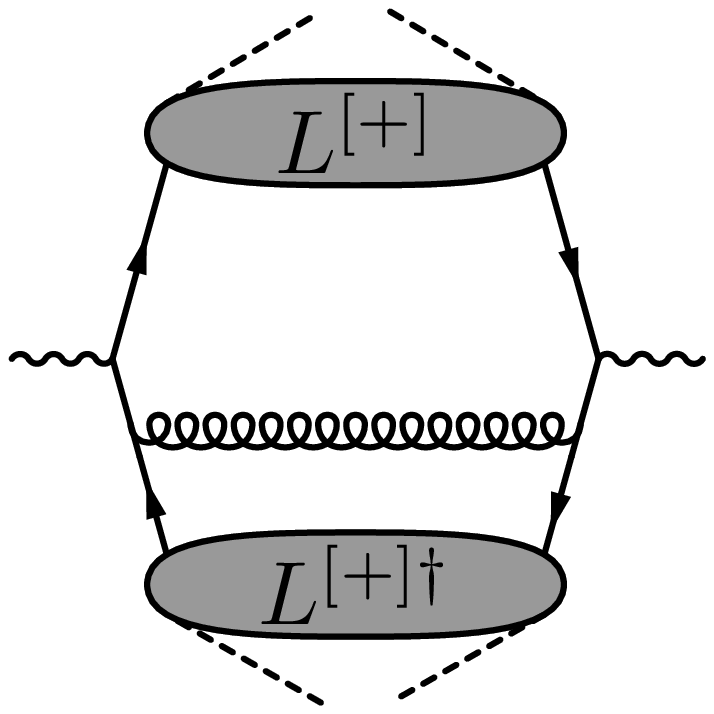}
\end{tabular}
\end{center}
\caption{Various elementary scattering diagrams convoluted with gauge invariant
correlators contributing to electron-positron annihilation. 
All corrections have correlators with gauge links
via plus infinity. The derivation of those gauge links in similar
to semi-inclusive DIS and Drell-Yan. \label{devePluseMin}}
\end{figure}

To compare the process at two different scales $Q^2$, corrections in $\alpha_S$
should be included.
At order $\alpha_S$
there are various contributions of which
some typical examples are given in Fig.~\ref{devePluseMin}. In the
figures the gauge links are indicated which result from
all gluon insertions in the elementary scattering diagram.
The gauge links are in this case all the same.
When constructing the correlators, gluon radiation was absorbed up to some
scale we will call $\mu^2\sim M^2$.
When increasing the scale of the process, additional gluons will be
radiated having a transverse momentum of at least $\mu^2$.
Since we are considering small $q_T$ their transverse
momenta cannot be to high (they should still be in the order of the hadronic
scale). Although
the gauge link was determined in subsection~\ref{alphaSDY} for gluons
having a large transverse momentum,
the technical derivation for the gauge link is expected to be the same in the
limit of $-l_T^2 \rightarrow \mu^2$.
That conclusion can also be reached
by following the general arguments as given in subsection~\ref{alphaSDY}.

When integrating over $q_T$ and over the unobserved gluon
(with momentum $l$) one finds at this order
that the infra-red
divergences\footnote{The infra-red divergences appear when $l\rightarrow0$ 
in separate (hard) scattering diagrams which are convoluted with
correlators. Following the general arguments of subsection~\ref{alphaSDY},
the gauge links depend on the nature of the external
particles, giving every correlator a gauge link via plus infinity.}
are canceled in the cross section, which is consistent with
the result of Collins and Soper~\cite{Collins:1981uk,Collins:1981uw}.
When considering finite small $q_T$ and integrating the unobserved
gluon over a restricted region $\threeVec{l}_T^2 < \Lambda^2$, where
$\Lambda^2$ is of some hadronic scale, the cancellation of infra-red
divergences can also be found. It may be good to point out that if
the radiated gluon is integrated over a restricted part of
its full phase-space, the
bilocal operators in the correlators are still off the light-cone and
transverse momentum dependent effects (like T-odd effects) are still included.
The factorized approach appears to be consistent.
When constructing factorized correlators appearing in cross sections
for small $q_T$,
no problems are encountered in perturbation theory at this order in $\alpha_S$.
Since
soft gluon radiation does not modify the original gauge links,
ladder diagrams (like the last diagram in Fig.~\ref{devePluseMin})
can be included when constructing the correlators, giving
them a scale-dependence.

\subsubsection{Drell-Yan}

\begin{figure}
\begin{center}
\begin{tabular}{cccc}
\includegraphics[width=2.5cm]{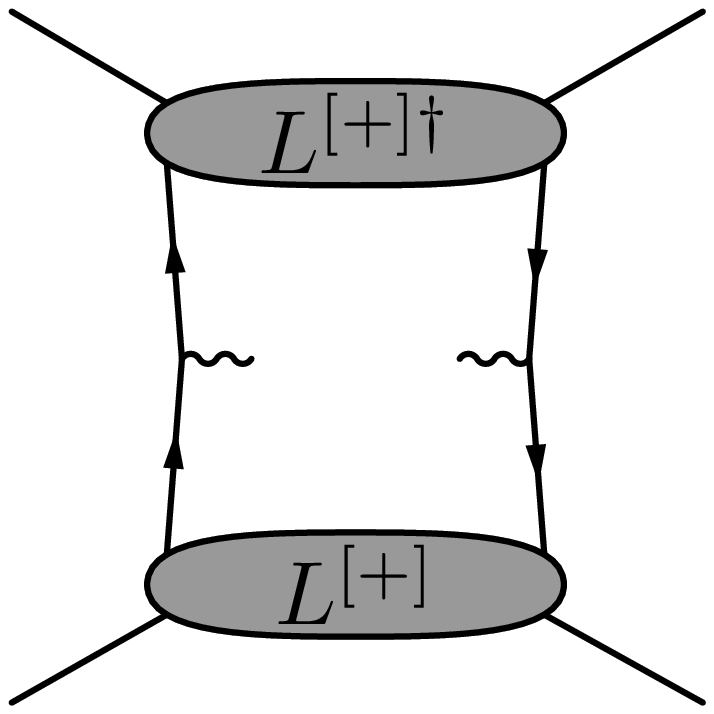} &
\includegraphics[width=2.5cm]{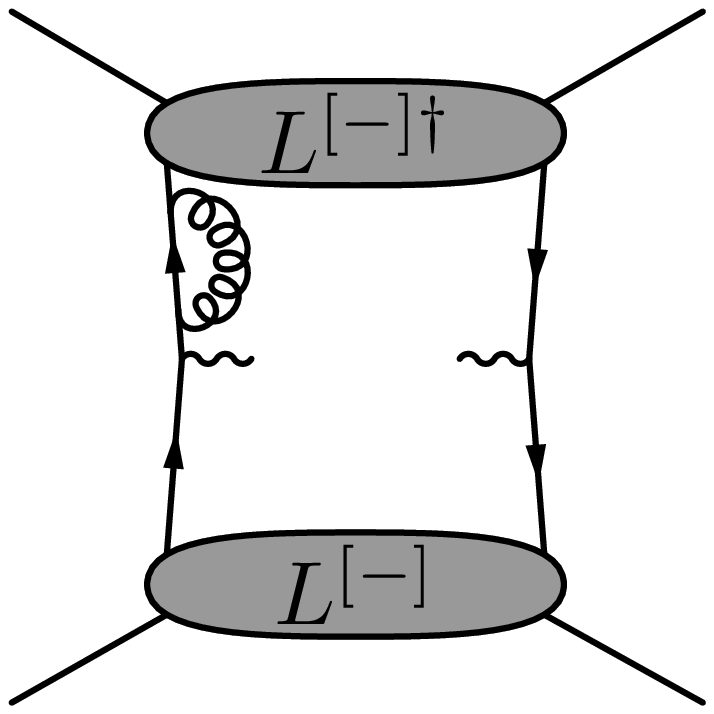} &
\includegraphics[width=2.5cm]{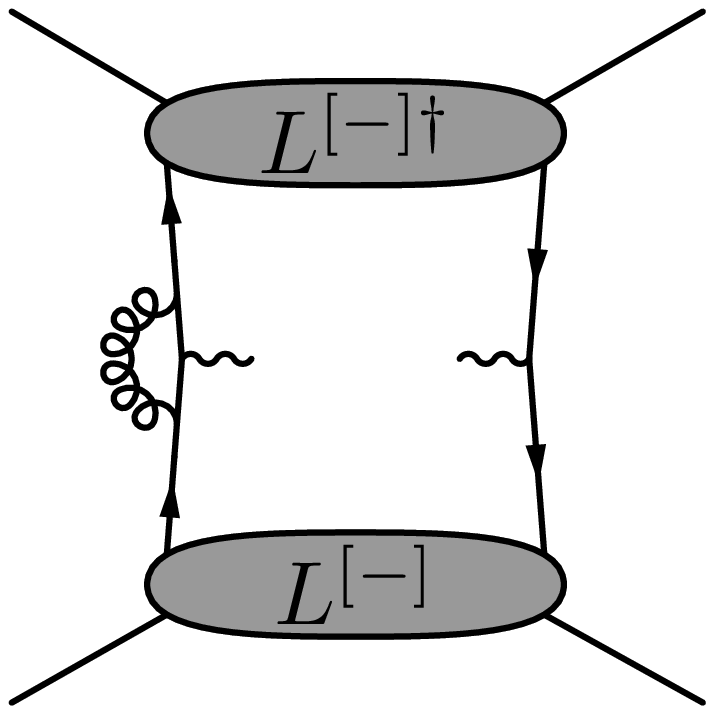} &
\includegraphics[width=2.5cm]{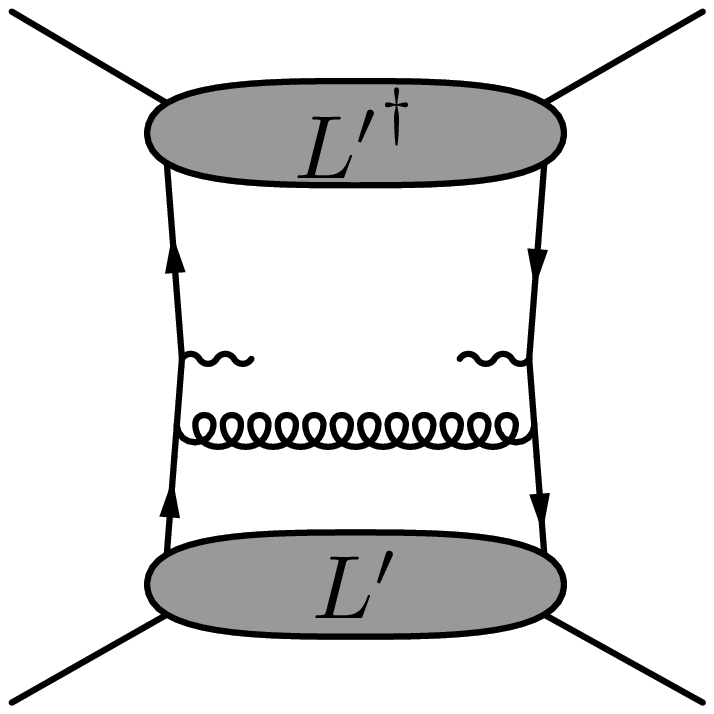}
\end{tabular}
\end{center}
\caption{Various elementary scattering diagrams convoluted with gauge invariant
correlators contributing in Drell-Yan.
At this order in $\alpha_S$, all virtual corrections have correlators
with gauge links via minus infinity, while all real
corrections have correlators with the gauge link
$\mathcal{L}' = \frac{3}{8} \mathcal{L}^{[+]}
\tr^C \mathcal{L}^{[ \square ]\dagger} - \frac{1}{8} \mathcal{L}^{[-]}$.
\label{devDrellYan}}
\end{figure}

The situation is quite different for Drell-Yan. Considering
$\alpha_S$ corrections
in a similar way (see Fig.~\ref{devDrellYan}),
one encounters various gauge links in the
correlators for the corrections.
Gluon insertions
on the radiated gluon were included to obtain gauge invariant correlators.
If the approach is to be consistent (gauge invariant)
then those contributions \emph{cannot} be neglected or circumvented
unless there are additional \mbox{analytical} properties of the correlator\footnote{
For instance if T-odd distribution functions are zero or if gauge links
are not an intrinsic property of the nucleon, but a vacuum effect instead.}.
Note that
when the transverse momentum of the radiated gluon is of a hadronic size,
the interactions appear at a similar time-scale as the insertions on the
incoming antiquark in Drell-Yan or as the insertions
on the outgoing quark in ordinary DIS.

The above results point to difficulties when calculating the scale-dependence
of the overall process. When considering $\alpha_S$ corrections to compare
different scales, the behavior of the correlators is different.
Since brehmsstralung diagrams have different correlators then the tree-level
diagram, the effect of this kind of radiation cannot be absorbed in
a correlator or other constructed objects (like for instance a soft factor)
in the approach followed here.
The scale-dependence is thus significantly more
difficult to calculate than in electron-positron annihilation.
%Further, by following the general arguments
%of subsection~\ref{alphaSDY} the gauge links presented in
%Fig~\ref{devDrellYan} should remain the same when
%the soft limit is taken.
It should be noted that when simplifying this process to QED, the
problem of the gauge links in combination with radiation does not
appear. Since the photon does not carry any charge,
the gauge links do not change when photon
radiation is
included.
QCD has a different behavior here.

In the previous chapter, section~\ref{hiro12}, the equations of motion
were applied in the calculation of gauge links.
Considering the diagrams, it seems that when
increasing the scale of the process, which produces more gluon radiation,
the gauge links in the correlators get modified. 
If radiation effects can be absorbed into the correlators appearing in the
factorized form $\Phi \otimes \bar{\Phi} \otimes H$, then
it seems
as if the equations of motion were applied in the wrong way, or that
the equations of motion are not invariant under scale
transformations\footnote{
The same problem also appears in
semi-inclusive DIS (see Fig.~\ref{plusLink}). When applying the equations
of motion we assumed analytical properties of the gluon connecting the
fragmentation
correlator (it was assumed to be outgoing). This assumption
could be invalid and may need improvement.}. Note that this problem does not
appear when discussing electron-positron annihilation.
In the following it will be indicated how this problem can be circumvented.

%A way to circumvent the application of the equations of motion is
%to expand the upper
%correlator in a set of free parton states (Fock-state decomposition).
%Since the external partons are then in essence free, one can derive the
%gauge link in the lower correlator 
%by using the results of the previous sections. 

We will try to construct the factorized correlator connected
to the external partons which enter the process from above 
in perturbation theory, although this is strictly
speaking not allowed.
Constructing the correlator, the incoming hadron and its
remnant is expanded in free parton states. Note that only the connected part
of the diagrams needs to be considered. 
Since those partons are then essentially free, the
gauge link in the lower correlator can be determined
by using the results of the previous sections.
For the subprocess
in Drell-Yan
in which partons, which cannot be absorbed in the lower correlator,
do not cross the cut,
one finds the gauge link for the
lower correlator to run via minus infinity.
However, if one of the partons, connected to the upper hadron,
radiates a gluon to the final state
(note that its energy depends of the process), then
the gauge link in the lower correlator gets modified. So for each
component of the Fock-state expansion of the upper correlator,
one can calculate the gauge link for the lower correlator. However, 
each Fock-state component will have in general \mbox{different} gauge links, making the
procedure unsuitable for constructing the upper correlator factorized from the
rest of the process. This
problem only appears for cross sections which are sensitive to intrinsic
transverse momentum. After
an integration over $q_T$, one finds at leading order in $M/Q$ that
all gauge links in the correlators are on the
light-cone and run along straight paths between the two quark-fields
(see Fig.~\ref{plaatjeLinks}a). In that case the problems with gauge links
and factorization disappear.

The above arguments illustrate that even when applying perturbation theory
and a simple Fock-state expansion one encounters difficulties with
factorization theorems sensitive to the intrinsic transverse momentum. 
This problem originates from gluons which are inserted on both initial and final-state
partons. Since this issue has not been explicitly discussed in 
Ref.~\cite{Ji:2004wu,Ji:2004xq,Collins:2004nx}, factorization for 
azimuthal asymmetries at small $q_T$ remains an open question for Drell-Yan.
These problems are not encountered
when considering fully
$q_T$-integrated
factorization theorems at leading twist.
In that case one finds that the infra-red divergences are
canceled at this order in $\alpha_S$. At subleading twist, the
$q_T$-integrated factorization theorem contains correlators off the light-cone
(see the previous chapter, appendix~\ref{quarkAppen})
and therefore it suffers from the same problems as
for the transverse momentum dependent factorization theorem.

\subsubsection{Semi-inclusive DIS}

\begin{figure}
\begin{center}
\begin{tabular}{cccc}
\includegraphics[width=2.5cm]{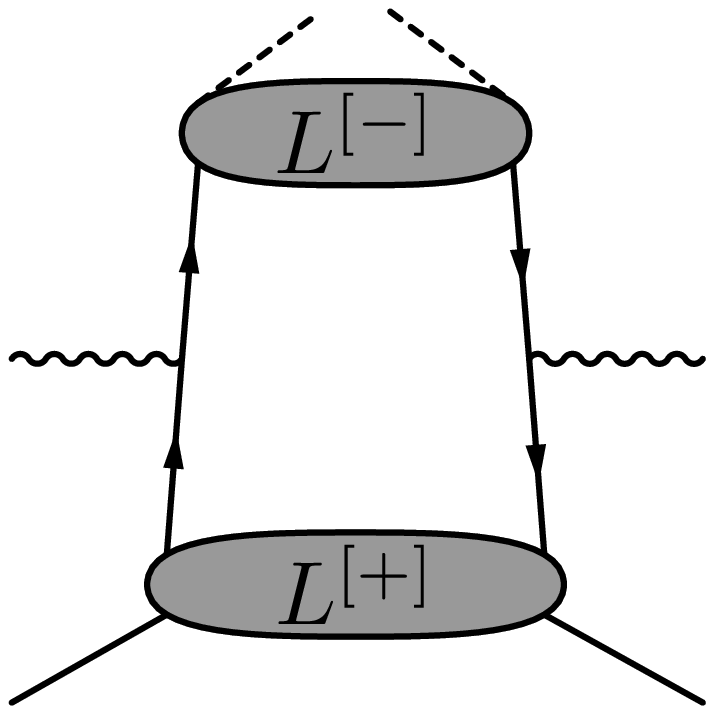} &
\includegraphics[width=2.5cm]{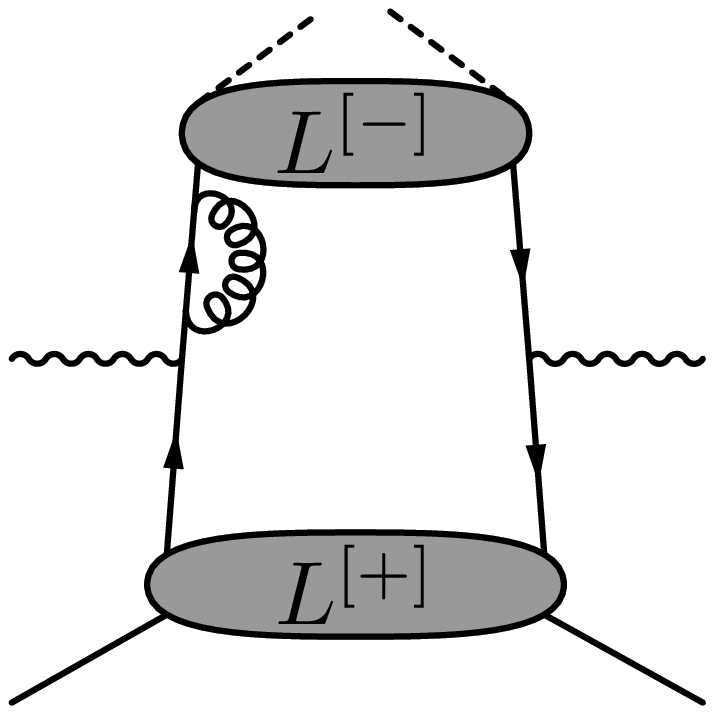} &
\includegraphics[width=2.5cm]{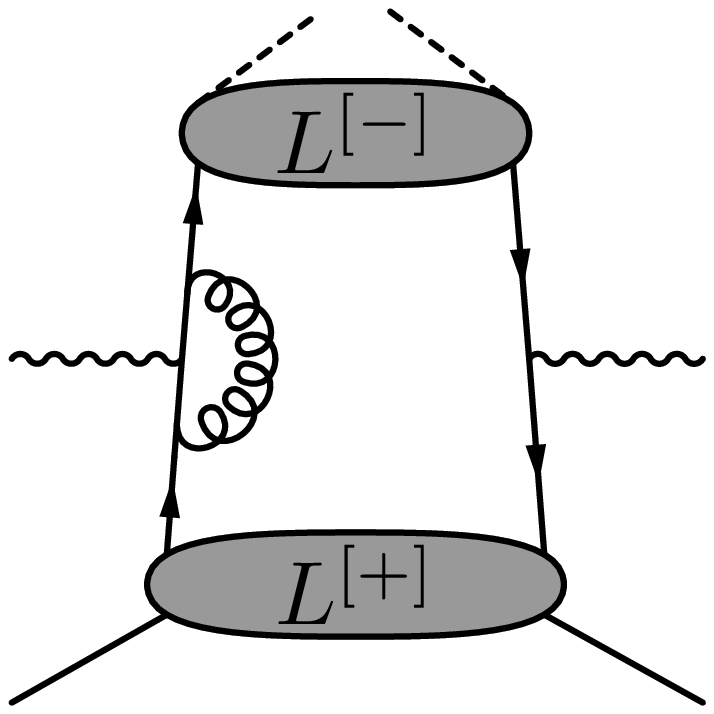} &
\includegraphics[width=2.5cm]{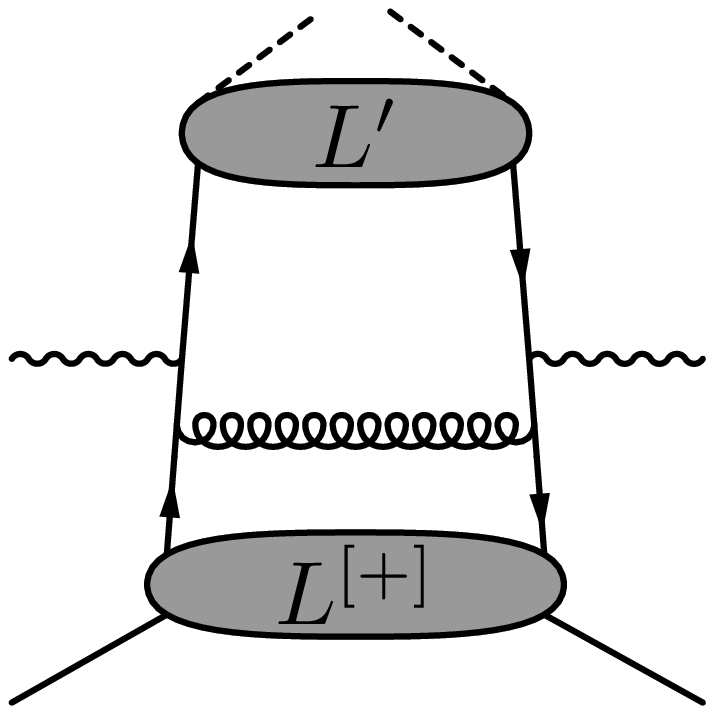}
\end{tabular}
\end{center}
\caption{Various elementary scattering diagrams convoluted with gauge
invariant correlators contributing in semi-inclusive DIS.
At this order in $\alpha_S$, the distribution 
correlator contains a gauge link via plus infinity. The virtual 
corrections have a
fragmentation correlator with a gauge link via minus infinity, while
all real corrections have a fragmentation correlator with the gauge link
$\mathcal{L}' =
\frac{3}{8}\mathcal{L}^{[+]} \mathcal{L}^{[ \square ] \dagger} -
\frac{1}{8}\mathcal{L}^{[-]}$.
\label{devSIDIS}}
\end{figure}

The situation of semi-inclusive DIS is similar  to
Drell-Yan (see also Fig.~\ref{devSIDIS}).
The real $\alpha_S$ corrections modify the original link structure of
the fragmentation correlator. This forms a problem for factorization theorems
which include effects from intrinsic transverse momentum.
The integrated cross section is
still infra-red free at leading twist, but at subleading twist problems
with factorization occur.

Concerning the construction of factorized correlators via a Fock-state expansion, 
when expanding the fragmentation
correlator in free parton states
one finds for each component the same gauge link in the
distribution correlator.
This points out that for (multiple) jet-production in DIS factorization
theorems sensitive to intrinsic transverse momentum effects might be feasible.
When expanding the distribution correlator in parton states,
several partons will cross the cut which cannot be absorbed in the fragmentation
correlator. These partons influence the gauge link structure which
forms a problem for constructing the distribution correlator in semi-inclusive
DIS. 

It may be good to point out that if gluonic-pole matrix elements for fragmentation
vanish, then the fragmentation process is universal (universality of
fragmentation functions has been
advocated by
Collins and Metz~\cite{Metz:2002iz,Collins:2004nx}). In that case the problem
with gauge links in the fragmentation correlator does not exist
and  factorization theorems could be feasible.

As a final remark, in Ref.~\cite{Ji:2004wu,Ji:2004xq} factorization has been
discussed including $M/Q$ effects. However, since contributions from
gluons at $M/Q$ have been discarded (the $\Phi_{\partial^{-1}G}$ and
$\Delta_{\partial^{-1}G}$ terms of chapter~\ref{chapter3}), it is not clear
whether the factorization theorems are complete as they stand.

\newpage

\section{Summary and conclusions}

In this chapter gauge links appearing in more complicated diagrams
were considered. 
When deriving those gauge links by considering gluon interactions between
the correlators and the hard part, two classes of poles were encountered.
The first class contains the poles $p_i^+ \approx 0$
which correspond to the limit in which the gluon momenta vanish.
This class 
yields contributions to the gauge link. The second class contains the other
poles and
in the examples shown, those poles canceled. 
Supplementing \mbox{general} arguments in favor of this cancellation
have been given and
a set of rules has been conjectured
that allows one to deduce gauge links for
arbitrary diagrams. Applying these rules for gluon-gluon correlators,
gauge invariant transverse momentum dependent
gluon distribution functions  were obtained. In a sense, these rules
factorize
interactions between a correlator and the elementary scattering part.

It was found that 
gauge links do in general not only depend on the overall process but also
depend on the elementary scattering diagram. The encountered
gauge links 
also had more complicated paths than the ones appearing in 
semi-inclusive DIS
or Drell-Yan. To compare the correlators with these different gauge links,
we considered the first transverse moments of distribution and fragmentation
functions. It was found that these transverse
moments can be related to the transverse moments appearing in semi-inclusive
DIS and electron-positron annihilation.
For the higher transverse moments no
relations were obtained.

The fact that the gauge link in the correlator depends on the subprocess instead
of the overall process
could form a potential danger for the unitarity of the approach.
However,
in an explicit nontrivial example (two-gluon production) it was
found that the factorized description in terms of gauge invariant
correlators still respects unitarity.

\sloppy
We also discussed the subject of factorization and universality
for semi-inclusive DIS, electron-positron annihilation, and Drell-Yan.
In semi-inclusive electron-positron annihilation we did not encounter any
inconsistencies because QCD-states appeared only in the final state.
For semi-inclusive DIS and Drell-Yan there was in 
Ref.~\cite{Ji:2004wu,Ji:2004xq,Collins:2004nx}
recently important
progress reported on small $q_T$-unintegrated\footnote{
$q$ is the momentum of the virtual photon and $-q_T^2 \sim M^2$} 
factorization theorems, but
one particular issue
of dealing with both final and initial-state interactions has
not satisfactory been treated. 
This issue is related to the observation that
gauge links in transverse momentum
dependent correlators can get modified when considering radiation effects.
This could create a problem for factorizing effects from intrinsic
transverse momentum appearing in
cross sections for semi-inclusive DIS and Drell-Yan.
Since fully $q_T$-integrated factorization theorems 
contain at subleading twist effects from intrinsic transverse momentum
(see appendix~\ref{quarkAppen}), it forms at that order
a potential problem as well.
%These problems are not encountered in QED because
%photon radiation does not modify the gauge link structure (although 
%complex gauge links do in general appear in QED).
Factorization
of intrinsic transverse momentum dependent effects in Drell-Yan and semi-inclusive
DIS remains therefore at present an open
issue~\cite{Pijlman:2004mr}. Similar problems also appear for
semi-inclusive hadron-hadron scattering.

\fussy
The non-Abelian character plays an essential role in the discussion of
final and initial-state interactions. For instance, in QED where photons
do not carry charge,
photon radiation does not modify the gauge link structure.
Although more complicated 
gauge links than the
ones appearing in semi-inclusive DIS and Drell-Yan are still encountered, 
factorization of effects from intrinsic transverse momentum in hard QED
scattering processes should be feasible.

It may be good to point out that in QCD the problems for semi-inclusive DIS
originate from the gauge link in the fragmentation correlator in
combination with gluon radiation. If this gauge link turns out to be
irrelevant
(which means that gluonic-pole matrix elements for
fragmentation functions would be suppressed), then the
problems with factorizing effects from intrinsic transverse momentum
disappear.
In that scenario, which is favored
by the in section~\ref{fragUni} differently given interpretation
of the results obtained
by Collins and Metz~\cite{Collins:2004nx}, factorization theorems
including effects from intrinsic transverse momentum might be obtainable.
Such factorization theorems can also be expected to hold for (multiple)
jet-production in DIS, because in those cases the problem with fragmentation
functions is not present.

As a last remark, in the
diagrammatic approach used in this thesis, no assumptions were made for
the correlators other than the assumptions similar to the parton model
(see chapter~\ref{chapter2}). Many issues might be unknown for that reason,
one of them being the evolution equations for
transverse momentum dependent functions
(for a detailed study of this problem see Henneman~\cite{thesisAlex}).
It is possible
that additional assumptions,
which should be physically justified,
might solve some of the issues.
One possibility could be
that gauge links in general
do not influence the expectation values of matrix elements.
In that case T-odd distribution functions would be zero and
proofs on factorization and universality would
be simplified significantly.
Another possibility could be that gauge links correspond to certain
interactions in the hard part in a particular kinematical limit (for
instance the vertex correction in the limit that
the gluon is collinear with one of the external quarks).
In that case the gauge link is not an intrinsic property of the nucleon,
allowing (part of) its contribution to be possibly
absorbed in other matrix elements or factors.

\newpage

\thispagestyle{empty}

\chapter{Results for single spin asymmetries in
hadronic scattering\label{chapter5}}

In the previous chapter the necessary tools were developed
to determine the Wilson
lines 
which appear in correlators in hard scattering processes.
As discussed, it is not clear whether effects from intrinsic
transverse momentum allow for a factorized description in hadron-hadron
scattering. We will assume it does and obtain results for single
spin asymmetries. In these asymmetries the effect of the gauge link appears
to be more than just a sign; it also determines the sizes of the asymmetries.
It will be shown how
the intrinsic transverse momenta of quarks can be accessed
by observing
unpolarized
hadrons in opposite jets. For the ease of the calculation, only
contributions from quark distribution and fragmentation functions
will be considered.
Gluon distribution and fragmentation functions are neglected,
but can be straightforwardly incorporated by following the same procedure.

\section{Introduction}
% for the observables

We will study the transverse momentum dependent
distribution and fragmentation functions via single spin asymmetries
in hadron-hadron scattering.
Within the applied diagrammatic approach,
an odd number of T-odd functions is needed to produce cross sections
for single spin asymmetries at leading order in $\alpha_S$.
At leading twist, T-odd functions are transverse momentum dependent.
Integrated T-odd
distribution functions are assumed to be zero and
unpolarized
integrated
T-odd fragmentation functions appear only at subleading
twist~\cite{Bacchetta:2001rb}.
Single spin asymmetries at leading order in inverse powers of the hard scale
must therefore arise \nolinebreak from \nolinebreak \mbox{intrinsic}
\nolinebreak \mbox{transverse} \nolinebreak \mbox{momenta}.

The validity of the applied theoretical description depends on two issues.
The first issue is connected to factorization. This was discussed in the
previous chapter and in the present study it will be assumed to hold.
The second issue is related to asymptotic freedom. It is expected that only
those processes can be described perturbatively in
which observed external hadrons are well separated in momentum space.
The observed outgoing hadrons must have a large
perpendicular momentum with respect to the beam-axis. For observed
hadrons close to the beam-axis, one does
not only have a problem with perturbation theory, but one will
also encounter interference effects between the hard scattering process
and the remnant of the incoming hadrons. In that case fracture functions
need to be included.

One of the most studied semi-inclusive cross section
in hadron-hadron collisions is single
hadron production. It is also this process in which the first single
spin asymmetries in inelastic collisions were observed~\cite{Bunce:1976yb}.
Since then, single spin asymmetries have been
measured in several processes (see for instance
Ref.~\cite{Adams:1991rw,Adams:1991cs,Bravar:1996ki,Apanasevich:1997hm,
Krueger:1998hz,Adler:2003pb,Adams:2003fx,Adler:2005in}), 
and extensive theoretical
studies have been made (see for instance
Ref.~\cite{Anselmino:2002pd,Qiu:1998ia,Jaffe:1996ik,Anselmino:1999pw,
Anselmino:2004nk,Jager:2004jh,D'Alesio:2004up,Anselmino:2004ky,Vogelsang:2005sq}).
The main theoretical challenge
with single hadron production at large transverse momentum is that
there is no observable which is directly connected to the intrinsic
transverse momenta of the
partons. This makes the extraction of the
transverse momentum dependent
functions complex. It could be that the only possible manner of extracting
information is to consider a specific form for the transverse momentum 
dependence of distribution or
fragmentation functions (for example $\exp [-k_T^2/M^2]$). 
Such a form can
lead to problems with the
gauge link because 
it generally contains higher order transverse
moments.
At present only the first transverse moment of functions
can be easily related to the transverse moment of functions appearing in semi-inclusive DIS and
electron-positron annihilation.
In the higher transverse moments new matrix elements are involved
which are more difficult to relate (see also section~\ref{devLinks2}). Besides, for the higher transverse moments 
convergence becomes an issue as well.

The two hadron production
process, in which the two hadrons belong to different jets and
are approximately back-to-back
in the perpendicular plane, does offer 
an observable directly sensitive to the
intrinsic transverse momenta.
The fact that the two hadrons are not completely back-to-back 
can at leading order in $\alpha_S$ be interpreted as an
effect from intrinsic transverse momentum.
This process will be studied in this chapter.
Having obtained transverse target-spin asymmetries,
a simple extension will be made to
jet-jet production by
just summing over the observed hadrons and observing the jet instead.
This latter process has
been studied by Boer and Vogelsang in Ref.~\cite{Boer:2003tx}
(see also Vogelsang, Yuan~\cite{Vogelsang:2005cs} for related work).
The study here should be considered as an extension of Ref.~\cite{Boer:2003tx}
to hadron-hadron production including a full treatment of gauge links.

To find out what observables are present, let us reconsider
the Drell-Yan process.
The extraction
of the intrinsic transverse momenta is
connected with momentum conservation in the hard scattering cross section,
which in the case of Drell-Yan is expressed by a four
dimensional
delta-function in Eq.~\ref{zondag11}.
The presence of a hard scale,
originating from the electromagnetic interaction involving
two hadrons,
allows for a Sudakov-decomposition. This enables one
to eliminate two delta-functions,
leading to the fixed light-cone momentum fractions $x_1$ and $x_2$.
The remaining two-dimensional delta-function,
$\delta^2(p_T {+} k_T {-} q_T)$ in Eq.~\ref{maandag1},
is directly connected to the intrinsic
transverse momenta of quarks. Note that 
the momenta $p_T$ and $k_T$ are transverse
with respect to their parent hadron. These momenta can
be accessed by considering
azimuthal asymmetries (see Drell-Yan
in appendix~\ref{quarkAppen}).

In the hadron-hadron production process, there are
four correlators and
two hard scales present.
If we use the large momentum difference of the
initial hadrons to fix the perpendicular plane ($\perp$) and the
light-cone momentum fractions of the initial quarks, then
the remaining two dimensional delta-function reads
$\delta^2(p_{1\perp} {+} p_{2\perp} {-} k_{1\perp} {-} k_{2\perp})$.
The momentum $p_{i\perp}$ is already transverse with
respect to its parent hadron, so $p_{1\perp} {=} {p_{1T}}_\perp$.
This is not the case for $k_{i\perp}$ because the momenta
of the outgoing hadrons, $K_i$, \nolinebreak
have \nolinebreak large perpendicular components. It is convenient
to make the decomposition $k_i {=} K_i/z_i {+} k_{iT}$, where
$k_{iT}$ is defined to be transverse with respect to $K_i$ and is of a
hadronic scale. The large momentum difference between $K_{1\perp}$ and
$K_{2\perp}$, which is still present in the two-dimensional delta-function,
can then be used to fix one of the remaining
light-cone momentum fractions, $z_i$. This leaves one
delta-function behind which contains the non-back-to-backness and forms
a natural observable as we will see.
In the case of two-jet production
there is no other light-cone momenta to fix.
In that case the sum of the
two jet momenta in the perpendicular plane
is already proportional to the intrinsic transverse momenta of the partons, providing
one a two-dimensional vector variable.

\section{Calculating cross sections for hadronic scattering}

The calculation for the cross sections will be outlined.
After considering kinematics, observables will be defined which
are sensitive to transverse momentum dependent functions.
The cross sections will be
expressed in an elementary hard scattering cross section, 
fragmentation functions, and distribution functions. Those functions
are defined through bilocal matrix elements and
contain a gauge link which depends on the squared amplitude diagram
(which we also call subprocess).
A simplification \nolinebreak of \nolinebreak the cross section will be achieved by introducing
\emph{gluonic-pole cross sections}.
This enables one to express the cross section in simple hard
scattering cross sections convoluted with the
in chapter~\ref{chapter2} defined
distribution and fragmentation
functions.

\subsubsection{Kinematics}

For the ease of the calculation we will work in the center of mass frame
of the incoming hadrons.
The incoming hadron with momentum $P_1$ fixes the direction of the
$z$-axis and the spatial momenta which are perpendicular to this axis will
carry the subscript $\perp$. The hadron with momentum $P_2$ enters the process
from the opposite direction.
A hard scale is set by $s\equiv(P_1+P_2)^2$. The pseudo-rapidity
is defined as
$
\eta_i \equiv - \ln( \tan (\theta_i / 2)),
$
%\end{equation}
where $\theta_i$ is the polar angle of an
outgoing hadron with respect to the beam-axis. 
We introduce a scaling variable $x_{i\perp}$ defined as
%, $x_{i\perp}$, is introduced
%\begin{equation}
$
x_{i\perp} \equiv 2 | \threeVec{K_{i\perp}} | / \sqrt{s},
$
%\end{equation}
where $K_i$ is the momentum of an observed outgoing hadron.
These definitions yield the following relations
\begin{align}
P_1 \cdot K_1 &= \tfrac{1}{4} s x_{1\perp} e^{-\eta_1} + \mathcal{O}(M^2),&
P_2 \cdot K_1 &= \tfrac{1}{4} s x_{1\perp} e^{\eta_1}+ \mathcal{O}(M^2),
\nonumber\\
P_1 \cdot K_2 &= \tfrac{1}{4} s x_{2\perp} e^{-\eta_2}+ \mathcal{O}(M^2),&
P_2 \cdot K_2 &= \tfrac{1}{4} s x_{2\perp} e^{\eta_2}+ \mathcal{O}(M^2).
\end{align}
The Mandelstam variables for the partons are defined as
\begin{align}
\hat{s} &\equiv (p_1+p_2)^2,&
\hat{t} &\equiv (p_1-k_1)^2,&
\hat{u} &\equiv (p_1-k_2)^2,
\end{align}
and fulfill $\hat{s}+\hat{t}+\hat{u} = p_1^2 + p_2^2 + k_1^2 + k_2^2$.
The variable $y$, which is observable and
the analogue of the $y$ variable used in semi-inclusive DIS,
is defined to be
\begin{equation}
y \equiv \frac{-\hat{t}}{\hat{s}} = \frac{1}{\exp (\eta_1 - \eta_2) + 1}
\left(1 + \mathcal{O}(M^2/s) \right).
\end{equation}
%and is observable. Another useful relation is 
%\begin{equation}
%x_1 x_2 = \frac{x_\perp^2}{4 y (1-y)}.
%\end{equation}

\subsubsection{Defining observables}

Using the diagrammatic expansion
(see also chapter~\ref{chapter2})
the cross section for two hadron
production reads ($\phi_i$'s are the azimuthal angles of the observed hadrons)
\begin{equation}
\begin{split}
\rmd{1}{\sigma}
&{=}
\frac{1}{2s}\
\frac{d^3 K_1}{(2\pi)^3\,2E_{\threeVec{K}_1}}\
\frac{d^3 K_2}{(2\pi)^3\,2E_{\threeVec{K}_2}}\
\mathcal{A}^2
\\
& {=}
\frac{x_{1\perp}x_{2\perp}s}{128(2\pi)^4}
 \rmd{1}{x_{1\perp}}  \rmd{1}{x_{2\perp}}
 \rmd{1}{\eta_1}\rmd{1}{\eta_2}
\frac{\rmd{1}{\phi_1}}{2\pi}
\frac{\rmd{1}{\phi_2}}{2\pi}
\
\mathcal{A}^2 \ \left(1+\mathcal{O} \left(\frac{M^2}{s} \right) \right),
\end{split}
\end{equation}
where $\mathcal{A}^2 = \sum_X \int \phaseFactor{P_X}
(2\pi)^4 \delta(P_1{+}P_2 {-} K_1{-}K_2{-}P_X)\
|\mathcal{M}|^2$ which is generically expressed as
\begin{multline}
\mathcal{A}^2
 = \int \rmd{4}{p_1}\rmd{4}{p_2}\rmd{4}{k_1}\rmd{4}{k_2} \
(2\pi)^4\ \delta^4(p_1{+}p_2{-}k_1{-}k_2)\\
\times
\tr^{D,C}\big\{\,\Phi(p_1) \otimes \Phi(p_2) \otimes
\Delta(k_1) \otimes \Delta(k_2)
\\
\otimes
H(p_1,p_2,k_1,k_2) \otimes H^*(p_1,p_2,k_1,k_2)\,\big\},
\end{multline}
suppressing the momenta and spins of parent hadrons.
The symbol $\otimes$ represent a convolution in color and Dirac indices, and
the hard elementary
scattering amplitudes are denoted by $H$.
The gauge links of the correlators can be derived by applying
techniques developed in the previous chapter. Those gauge links depend
in general on the subprocess.

Each of the external partons and its  parent hadron
have both a nearly light-like momentum in a more or less
common direction
(parton model assumption).
The introduction of the light-like vectors, $ n_{P_i}$,
with $n_{P_i} \cdot \bar{n}_{P_i}=1$ (bar denotes reversal of spatial components)
such that $n_{P_i}$ is proportional to $P_i$ in the asymptotic limit,
allows us to classify at which order components appear.
The components $p_i \cdot \bar{n}_{P_i}$ appear at order 
$\sqrt{s}$ (and similarly for fragmentation),
the spatial transverse components appear all at a hadronic scale. The light-like
components which are left, $p_i\cdot n_{P_i}$, appear at order $M^2/\sqrt{s}$.
Parton momentum fractions are defined as usual
\begin{align}
x_i &\equiv \frac{p_i \cdot \bar{n}_{P_i}}{P_i \cdot \bar{n}_{P_i}}, &
z_i &\equiv \frac{K_i \cdot \bar{n}_{K_i}}{k_i \cdot \bar{n}_{K_i}}.
\end{align}
The light-cone momentum components $p_i\cdot n_{P_i}$ can be simply integrated
because they are suppressed in the hard parts.
The hard scale $s$ can be used to fix the incoming
light-cone momentum fractions, $x_1$ and $x_2$, in
terms of $z_1$ and $z_2$. This gives
\begin{align}
&
\int
\rmd{4}{p_1}
\rmd{4}{p_2}
\rmd{4}{k_1}
\rmd{4}{k_2}\
\delta^4(p_1+p_2-k_1-k_2)
\nonumber\\
& \eqnIndent
{=}
\frac{2}{s}\ \int
\rmd{1}{(p_1\cdot \bar{n}_{P_1})}
\rmd{1}{(p_2\cdot \bar{n}_{P_2})}
\rmd{1}{(k_1\cdot \bar{n}_{K_1})}
\rmd{1}{(k_2\cdot \bar{n}_{K_2})}\
\delta(x_1 - \tfrac{2}{s} r\cdot P_2)\
\delta(x_2 - \tfrac{2}{s} r\cdot P_1)
\nonumber\\
& \eqnIndent
\phantom{=}\times
\int
\rmd{2}{p_{1T}}
\rmd{2}{p_{2T}}
\rmd{2}{k_{1T}}
\rmd{2}{k_{2T}}
\delta^2({q_T}_\perp - r_\perp)
\nonumber\\
& \phantom{=}\eqnIndent
\times
\int
\rmd{1}{(p_1\cdot n_{P_1})}
\rmd{1}{(p_2\cdot n_{P_2})}
\rmd{1}{(k_1\cdot n_{K_1})}
\rmd{1}{(k_2\cdot n_{K_2})} \left( 1+ \mathcal{O}(M^2/s) \right),
\end{align}
\begin{align}
&\text{with}& q_T & \equiv p_{1T} + p_{2T} - k_{1T} - k_{2T},&
r &\equiv \frac{K_1}{z_1} + \frac{K_2}{z_2}.
\end{align}
The transverse parton momenta ($\{p_{1T},\ p_{2T},\ k_{1T},\ k_{2T} \}$)
are four-vectors
and defined to be transverse with respect to their parent hadron
while the symbol $\perp$ means perpendicular with respect to $P_1$ and $P_2$. 
The vector
$q_T$ is thus of a hadronic scale and all its components are in general nonzero.
In the case of jet-jet production the integrals over $k_1$ and $k_2$ do not
appear in the expression above, $q_T = p_{1T} + p_{2T}$, and $K_i/z_i$ is 
replaced by $k_{i}^\text{jet}$. In that case the
vector $r_\perp$ gives access to the intrinsic transverse momenta of the
initial partons ($q_T$) which is similar to Drell-Yan.

Another hard scale
is formed by the scalar product of $K_1$ and $K_2$. This scale, which
is present in $\delta^2({q_T}_\perp - r_\perp)$, can be used to
express one of the light-cone momentum fractions
in terms of the others, so $z_1(x_1,x_2,z_2)$ or $z_2(x_1,x_2,z_1)$. One
remains with an integral over one momentum fraction.
Since the choice for this fraction is arbitrary, a
more symmetrical expression can be obtained by introducing an additional
integral over $x_\perp$, giving (for details see Ref.~\cite{Bacchetta:2005rm})
\begin{equation}
\begin{split}
&
\int
\rmd{4}{p_1}
\rmd{4}{p_2}
\rmd{4}{k_1}
\rmd{4}{k_2}\
\delta^4(p_1+p_2-k_1-k_2)\\
%&
%{=}
%\int
%\rmd{1}{p_1\cdot \bar{n}_{P_1}}
%\rmd{1}{p_2\cdot \bar{n}_{P_2}}
%\rmd{1}{k_1\cdot \bar{n}_{K_1}}
%\rmd{1}{k_2\cdot \bar{n}_{K_2}}
%\rmd{2}{p_{1T}}
%\rmd{2}{p_{2T}}
%\rmd{2}{k_{1T}}
%\rmd{2}{k_{2T}}\\
%&
%\phantom{=} \times
%\delta \left( p_1 \cdot \bar{n}_{P_1} -
%\tfrac{1}{2}
%\left(
%  z_1^{-1}
%  x_{1\perp} \exp (\eta_1) -
%  z_2^{-1}
%  x_{2\perp} \exp (\eta_2)
%\right)
%\right)
%\\
%&
%\phantom{=} \times
%\delta \left( p_2 \cdot \bar{n}_{P_2} -
%\tfrac{1}{2}
%\left(
%  z_1^{-1}
%  x_{1\perp} \exp (-\eta_1) -
%  z_2^{-1}
%  x_{2\perp} \exp (-\eta_2)
%\right)
%\right)
%\\
%&
%\phantom{=} \times
%\delta \left(
%z_1^{-1}
%\left| \threeVec{K_1^\perp} \right| -
%z_2^{-1}
%\left| \threeVec{K_2^\perp} \right| \right)\
%\delta \left( \frac{{q_T}_\perp \cdot K_1^\perp}{|\threeVec{K_1^\perp}|}
%- z_1^{-1}
%\left| \threeVec{K_2^\perp} \right| \sin \delta \phi \right)
%\\
%& \phantom{=}
%\times
%\int
%\rmd{1}{p_1\cdot n_{P_1}}
%\rmd{1}{p_2\cdot n_{P_2}}
%\rmd{1}{k_1\cdot n_{K_1}}
%\rmd{1}{k_2\cdot n_{K_2}}
%\\
%
& \eqnIndent
{=}
\frac{4}{s^2 x_{1\perp} x_{2\perp}}
\int \rmd{1}{x_\perp}
\rmd{1}{(p_1\cdot \bar{n}_{P_1})}
\rmd{1}{(p_2\cdot \bar{n}_{P_2})}
\rmd{1}{(k_1\cdot \bar{n}_{K_1})}
\rmd{1}{(k_2\cdot \bar{n}_{K_2})}
\\
& \eqnIndent
\phantom{=} \times
\delta \left( x_1 -
\tfrac{1}{2} x_\perp
\left( e^{\eta_1} + e^{\eta_2} \right)
\right)\
\delta \left( x_2 -
\tfrac{1}{2} x_\perp
\left(  e^{-\eta_1} + e^{-\eta_2} \right)
\right)
\\
& \eqnIndent
\phantom{=} \times
\delta \left( z_1^{-1} - \frac{x_\perp}{x_{1\perp}} \right)\
\delta \left( z_2^{-1} - \frac{x_\perp}{x_{2\perp}} \right)
\\
& \eqnIndent
\phantom{=}
\times
\int
\rmd{2}{p_{1T}}
\rmd{2}{p_{2T}}
\rmd{2}{k_{1T}}
\rmd{2}{k_{2T}}
\delta \left(
\frac{{q_T}_\perp \cdot e_{1N}}{\sqrt{s}}
- \tfrac{1}{2} x_\perp \sin \delta \phi \right)
\\
& \eqnIndent\phantom{=}
\times
\int
\rmd{1}{(p_1\cdot n_{P_1})}
\rmd{1}{(p_2\cdot n_{P_2})}
\rmd{1}{(k_1\cdot n_{K_1})}
\rmd{1}{(k_2\cdot n_{K_2})}
\left(1+\mathcal{O}(M/ \sqrt{s}) \right),
\end{split}
\label{complex2}
\end{equation}

\begin{floatingfigure}{5cm}
\vspace{-.5cm}
\begin{center}
\includegraphics[width=3.5cm]{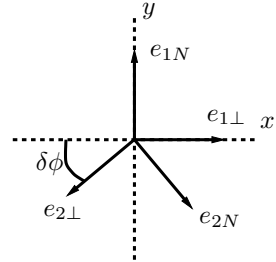}
\end{center}
\vspace{-.3cm}
\caption{Plane perpendicular to the incoming hadron momenta.
\label{complexFig1}}
\end{floatingfigure}

\noindent
where $e_{1\perp} \equiv K_{1\perp} / | \threeVec{K_{1\perp}} | $,
$e_{1N}^\sigma \equiv (-2/s) \epsilon^{\mu\nu\rho\sigma} P_{1\mu}$ $\times
P_{2\nu}{e_{1\perp}}_\rho$ and $\delta \phi$
is defined in Fig.~\ref{complexFig1} and is of order $\mathcal{O}(M/\sqrt{s})$.
The above expression illustrates that
the non-back-to-backness, $\delta \phi$, provides access to the transverse
momenta of the quarks (via ${q_T}_\perp$).
Weighting the cross section with
$\sin \delta \phi$ produces a projection of $q_T$ which leads
to the first transverse moments of distribution and fragmentation functions.

\indent
The following cross sections are now defined
\begin{align}
\langle\ \rmd{1}{\sigma} \rangle
&\equiv
\int \rmd{1}{\phi_2} \frac{\rmd{1}{\sigma}}{\rmd{1}{\phi_2}}
\nonumber\\
&= \frac{\rmd{1}{x_{1\perp}} \rmd{1}{x_{2\perp}} \rmd{1}{\eta_1}\rmd{1}{\eta_2}}
{32 \pi s} \frac{\rmd{1}{\phi_1}}{2\pi}
\int \frac{\rmd{1}{x_\perp}}{x_\perp} \Sigma(x_\perp),
\displaybreak[0]\\
\langle\ \tfrac{1}{2}\sin(\delta\phi)\ \rmd{1}{\sigma} \rangle
&\equiv
\int \rmd{1}{\phi_2}\ \tfrac{1}{2}\sin(\delta\phi)
\frac{\rmd{1}{\sigma}}{\rmd{1}{\phi_2}}
\nonumber\\
&
=
\frac{\rmd{1}{x_{1\perp}} \rmd{1}{x_{2\perp}} \rmd{1}{\eta_1}\rmd{1}{\eta_2}}
{32 \pi s^{3/2}} \frac{\rmd{1}{\phi_1}}{2\pi}
\int \frac{\rmd{1}{x_\perp}}{x_\perp^2} e_{1N} \cdot \Sigma_\partial(x_\perp),
\end{align}
where
\begin{align}
\Sigma(x_\perp) &
\equiv
(P_1\! \cdot\! \bar{n}_{P_1}) (P_2\! \cdot \! \bar{n}_{P_2})
(K_1\! \cdot\! \bar{n}_{K_1}) (K_2\! \cdot\! \bar{n}_{K_2})
\nonumber\\
& \eqnIndent  {\times}\!
\int \rmd{2}{p_{1T}}\rmd{2}{p_{2T}}\rmd{2}{k_{1T}}\rmd{2}{k_{2T}}
\Phi(x_1,p_{1T}) \otimes \Phi(x_2,p_{2T})\otimes
\Delta(z_1^{-1},k_{1T})
\nonumber\\
&\eqnIndent\eqnIndent \phantom{\equiv}\eqnIndent
 \otimes \Delta(z_2^{-1},k_{2T}) \otimes H(p_1,p_2,k_1,k_2)
\otimes H^*(p_1,p_2,k_1,k_2),
\\
\Sigma_\partial^\alpha(x_\perp)
&\equiv
(P_1\! \cdot\! \bar{n}_{P_1}) (P_2\! \cdot \! \bar{n}_{P_2})
(K_1\! \cdot\! \bar{n}_{K_1}) (K_2\! \cdot\! \bar{n}_{K_2})
\nonumber\\
& \eqnIndent  {\times}\!
\int \rmd{2}{p_{1T}}\rmd{2}{p_{2T}}\rmd{2}{k_{1T}}\rmd{2}{k_{2T}}\
\left[ q_T^\alpha \right]\  \Phi(x_1,p_{1T}) \otimes \Phi(x_2,p_{2T})\otimes
\Delta(z_1^{-1},k_{1T})
\nonumber\\
&\eqnIndent \eqnIndent \phantom{\equiv}\eqnIndent
 \otimes \Delta(z_2^{-1},k_{2T}) \otimes H(p_1,p_2,k_1,k_2)
\otimes H^*(p_1,p_2,k_1,k_2).
\end{align}
In the expressions the momentum fractions, $x_1,\ x_2,\ z_1,\text{ and } z_2$, 
are a function of $x_\perp$ via the arguments
of the first four
delta-functions in Eq.~\ref{complex2}.

\subsubsection{Calculating cross sections}

\begin{figure}[t]
\begin{center}
\begin{tabular}{ccccc}
\includegraphics[width=3cm]{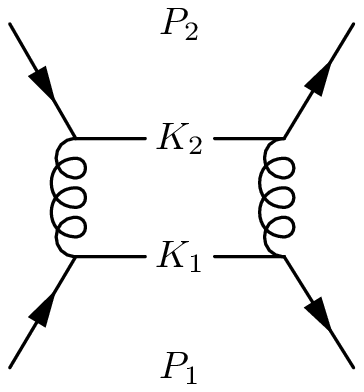} &&
\includegraphics[width=3cm]{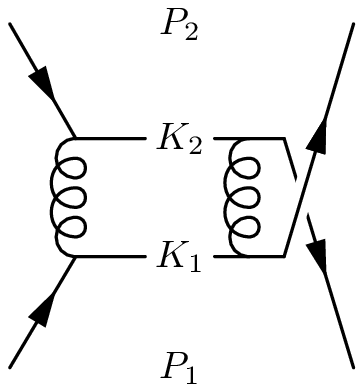} &&
\includegraphics[width=3cm]{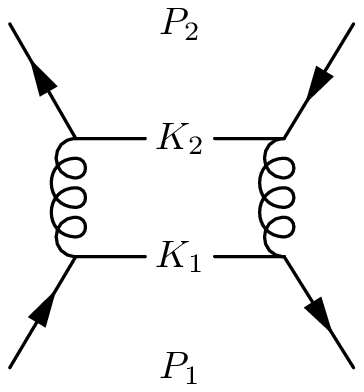}
\\
&& && \\
$D_{qq}^{[tt]}$ && $D_{qq}^{[tu]}$ && $ D_{q\bar{q}}^{[tt]}$
\\
&& && \\
\hline
&& && \\
\includegraphics[width=3cm]{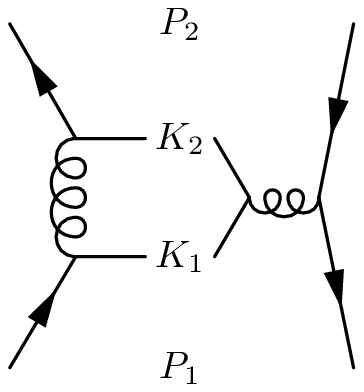} &&
\includegraphics[width=3cm]{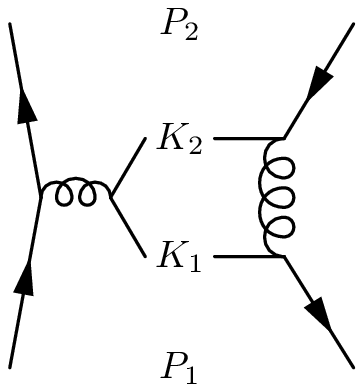} &&
\includegraphics[width=3cm]{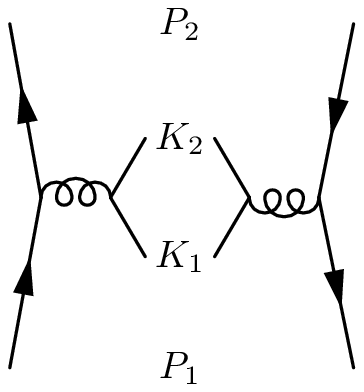}
\\
&& && \\
$D_{q\bar{q}}^{[ts]}$ && $D_{q\bar{q}}^{[st]}$ &&$ D_{q\bar{q}}^{[ss]}$
\end{tabular}
\end{center}
\caption{Contributions in hadron-hadron production. In the figures
only the hard part is shown, omitting
the correlators and parent hadrons.\label{complexFig2}}
\end{figure}

In the case studied here, gluon distribution and fragmentation functions
are neglected. Taking only the quark and antiquark correlators into account,
the forms of the expressions for $\mathcal{A}^2$, $\Sigma$ and $\Sigma_\partial$
will be presented. These were obtained in Ref.~\cite{Bacchetta:2005rm}
by using FORM~\cite{Vermaseren:2000nd}.

The following sum of diagrams contribute to
$\mathcal{A}^2$, $\Sigma$ and $\Sigma_\partial$
\begin{equation}
D_{qq}^{[tt]} + D_{qq}^{[tu]} +
   D_{q\bar{q}}^{[tt]} +
   D_{q\bar{q}}^{[ts]} +
   D_{q\bar{q}}^{[st]} + D_{q\bar{q}}^{[ss]}
+ (K_1 \leftrightarrow K_2) + (q \leftrightarrow \bar{q}),
\label{complex4}
\end{equation}
where the $D$ stands for the diagrams as displayed in Fig.~\ref{complexFig2}.
For $\mathcal{A}^2$ the following convolution can be obtained
\begin{equation}
\begin{split}
\mathcal{A}^2
&
\sim \int \rmd{1}{x_\perp} \rmd{2}{p_{1T}} \rmd{2}{p_{2T}}
\rmd{2}{k_{1T}} \rmd{2}{k_{2T}} 
\delta \left(\frac{{q_T}_\perp \cdot e_{1N}}{\sqrt{s}}
               - \tfrac{1}{2} x_\perp \sin \delta \phi \right)
\\
&
\times                              
\sum_D \sum_{i,j,k,l} f_i^{[D]}(x_1,p_{1T}^2)\ f_j^{[D]}(x_2,p_{2T}^2)
\frac{\rmd{1}{\sigma_{ijkl}^{[D]}}}{\rmd{1}{\hat{t}}}
D_k^{[D]}(z_1,z_1^2k_{1T}^2)\ D_l^{[D]}(z_2,z_2^2k_{2T}^2),
\end{split}
\end{equation}
where
the $f_i^{[D]}$ and $D_k^{[D]}$ represent some distribution and
fragmentation functions having
a diagram-dependent gauge link, the light-cone momentum fractions are
fixed by the first four delta-functions in Eq.~\ref{complex2},
and $\rmd{1}{\sigma_{ijkl}^{[D]}} / \rmd{1}{\hat{t}}$ is an elementary
parton scattering subprocess (or diagram) convoluting the functions.

In the unweighted cross section, $\langle\ \rmd{1}{\sigma} \rangle$,
the intrinsic transverse momenta can be neglected at leading
order in $M/\sqrt{s}$.
In that case all the gauge links in the correlators
are the same and on the light-cone, see Fig.~\ref{plaatjeLinks}a.
The sum over diagrams can be performed, yielding a simple parton scattering
cross section in $\Sigma$
\begin{equation}
\Sigma \sim \int \frac{\rmd{1}{x_\perp}}{x_\perp} 
 \sum_{i,j,k,l} f_i(x_1)\ f_j(x_2)\
\frac{\rmd{1}{\sigma_{ijkl}}}{\rmd{1}{\hat{t}}}
D_k(z_1)\ D_l(z_2).
\end{equation}
Also here the light-cone momentum fractions are a function of $x_\perp$ and
$\rmd{1}{\sigma_{ijkl}} / \rmd{1}{\hat{t}}$ is the elementary
parton scattering cross section.

In a weighted single spin asymmetry
each contribution consists of three integrated
correlators (T-even) containing gauge links
on the light-cone and one correlator (\mbox{T-odd})
of which a transverse moment is taken (as discussed in the introduction).
The first transverse moment of a function can be related to the
functions defined in chapter~\ref{chapter2}
(see also the previous chapter, section~\ref{devLinks2}).
The first transverse moment of a fragmentation function is in general
a combination of the ones appearing in semi-inclusive DIS and
electron positron annihilation. Containing an arbitrary gauge link, a fragmentation
function can be written as
\begin{equation}
D_i^{[ \mathcal{L}' ](1)} = C(\mathcal{L}')\ \tilde{D}_i^{(1)} 
+ D_i^{(1)},
\end{equation}
where $C$ is some calculable constant
and where definitions of Ref.~\cite{Boer:2003cm} were applied
\begin{align}
D_i^{(1)} (z) &\equiv  \frac{1}{2} \left(D_i^{[+](1)} (z) + D_i^{[-](1)} (z)
\right), &
%\nonumber\\
\tilde{D}_i^{(1)} (z) &\equiv  \frac{1}{2} \left(D_i^{[+](1)} (z) - D_i^{[-](1)} (z)
\right).
\label{maandag2}
\end{align}
The functions $D_i^{(1)}(z)$ are independent of the gauge link direction,
while the
functions $\tilde{D}_i^{(1)}(z)$ exist by the presence of the gauge link.
The tilde functions therefore vanish if the transverse momentum dependent
fragmentation functions appearing in electron-positron annihilation
are the same as in semi-inclusive DIS
(for a discussion on universality, see 
subsection~\ref{fragUni}).
For single spin asymmetries this leads for
$\Sigma_\partial^\alpha$ to
\begin{equation}
\begin{split}
\Sigma_\partial
&\sim \int \frac{\rmd{1}{x_\perp}}{x_\perp^2}
\sum_D \sum_{i,j,k,l} f_i^{[D](1)}(x_1)\ f_j^{[D]}(x_2)
\frac{\rmd{1}{\sigma_{ijkl}^{[D]}}}{\rmd{1}{\hat{t}}}
D_k^{[D]}(z_1)\ D_l^{[D]}(z_2) + \ldots
\\
&= \int \frac{\rmd{1}{x_\perp}}{x_\perp^2}
\sum_{i,j,k,l} f_i^{(1)}(x_1)\ f_j(x_2)\
\underbrace{
\sum_D {\mathcal{C}_{G,f(1)}^{[D]}}\
\frac{\rmd{1}{\sigma_{ijkl}^{[D]}}}{\rmd{1}{\hat{t}}}
}_\text{gluonic-pole cross section}
D_k(z_1)\ D_l(z_2) + \ldots,
\end{split}
\end{equation}
where terms related to the first transverse moment of the T-odd
function $f^{[D](1)}_i(x_1)$ have been
explicitly shown.
The combination $\sum_D\!  {\mathcal{C}_{G,f(1)}^{[D]}}
\rmd{1}{\sigma_{ijkl}^{[D]}} {/} \rmd{1}{\hat{t}}$ is given the name
gluonic-pole scattering.
The color factors, $C_G^{[D]}$, are obtained by comparing the diagram dependent
functions, $f_i^{[D](1)}$, with the
in chapter~\ref{chapter2} defined
functions $f_i^{(1)}$. For the
fragmentation functions $\tilde{D}_i^{(1)}$, which are similarly
defined as T-odd distribution functions
(compare Eq.~\ref{maandag2} with Eq.~\ref{theoryTodd}), one
also finds
color factors. For the functions
$D_i^{(1)}$ all the
color factors are simply $1$ because
those functions are independent of the link direction.
The functions $D_i^{(1)}$ are therefore convoluted
with ordinary parton scattering cross sections.

It may be good to point out that since the first transverse moment of a T-odd
distribution function is of order $g$ and higher,
the calculation of a single gluon insertion
in the diagrammatic approach (for the gauge link)
is already sufficient to obtain the color factors $C_G^{[D]}$.
The fact that the all order insertions provide fully gauge
invariant correlators can be seen as a consistency check.

\section{Results for cross sections and asymmetries}

In this section an explicit cross section and several single transverse
target-spin
asymmetries
will be given. The unpolarized observed hadrons, or jets, are assumed to be approximately
opposite in the perpendicular plane.
The results are based on Ref.~\cite{Bacchetta:2005rm} in which explicit
expressions as a function of $y$ can also be found.
The quark-quark subprocesses together with
the gluonic-pole subprocesses are given
appendix~\ref{complexA}.

\subsubsection{Unpolarized cross section for hadron-hadron production}

The unpolarized unweighted cross section for hadron-hadron production
is found to be a simple convolution between the integrated functions and
an elementary parton scattering cross section, reading
\myBox{
%\small
\vspace{-.3cm}
\begin{multline}
\langle\ \rmd{1}{\sigma} \rangle
= \rmd{1}{x_{1\perp}}\rmd{1}{x_{2\perp}}\rmd{1}{\eta_1}
\rmd{1}{\eta_2}\frac{\rmd{1}{\phi_1}}{2\pi}
\int \frac{\rmd{1}{x_\perp}}{x_\perp}
\\
\times
\sum_{q_1q_2q_3q_4}
f_1^{q_1}(x_1)f_1^{q_2}(x_2)\ \frac{\hat s}{2}
\,\frac{\rmd{1}{\hat \sigma_{q_1q_2\rightarrow q_3q_4}}}{\rmd{1}{\hat t}}\,
D_1^{q_3}(z_1)D_1^{q_4}(z_2)
\\
\times \left( 1 + \mathcal{O} \left( \frac{M}{\sqrt{s}} \right) +
\mathcal{O}\left( \alpha_S \right)\right),
\end{multline}
\vspace{-.7cm}
\begin{flushright}
\emph{integrated cross section for hadron-hadron production}
\end{flushright}
}
where the summation is over all quark and antiquark flavors. The arguments
of the functions are defined through
$x_1 = x_\perp (\exp (\eta_1) + \exp (\eta_2))$,
$x_2 = x_\perp (\exp (-\eta_1) + \exp (-\eta_2))$, and
$z_i = x_{i\perp} / x_\perp$. Furthermore, the variable $\hat{s}$ depends
on $x_\perp$ and $y$ through the relation
$\hat{s} = x_\perp^2 s / (4y(1-y))$.\nopagebreak

\subsubsection{Single transverse target-spin asymmetries}

In the $\sin \delta \phi$ weighted asymmetry for hadron-hadron production,
gluonic-pole scattering subprocesses are encountered. The asymmetry
reads
\myBox{
%\small
\vspace{-.3cm}
\begin{align}
& \langle  \ \tfrac{1}{2} \sin ( \delta \phi ) \rmd{1}{\sigma} \rangle
\nonumber
\\
&
=
\rmd{1}{x_{1\perp}}\rmd{1}{x_{2\perp}}
\rmd{1}{\eta_1}\rmd{1}{\eta_2}
\frac{ \rmd{1}{\phi_1} }{2\pi}\ \cos(\phi_1 - \phi_S)
\int \frac{ \rmd{1}{x_\perp} }{x_\perp}
\nonumber
\\
&
\phantom{=}
\times \biggl\{\
\frac{M_1}{x_\perp \sqrt s}
\sum_{q_1 q_2 q_3 q_4}\!\!\!{f^{q_1}}_{1T}^{\perp(1)}(x_1) f_1^{q_2}(x_2)
\frac{\hat s}{2}
\frac{\rmd{1}{\hat\sigma_{\widehat{gq}{}_1q_2\rightarrow q_3q_4}}\!} {\rmd{1}{\hat t}}
D_1^{q_3}(z_1)D_1^{q_4}(z_2) 
\nonumber
\\
&\phantom{=} \phantom{\times \bigg\{ }
+\frac{M_2}{x_\perp\,\sqrt s}
\sum_{q_1q_2q_3q_4}\!\!\!
h_1^{q_1}(x_1){h^{q_2}}_1^{\perp(1)}(x_2)\,
\frac{\hat s}{2}\,
\frac{\rmd{1}{\Delta\hat\sigma_{q_1^\uparrow \widehat{gq}{}_2^\uparrow
\rightarrow q_3q_4}}\!}{\rmd{1}{\hat t}}\,
D_1^{q_3}(z_1)D_1^{q_4}(z_2)
\nonumber
\\
&
\phantom{=} \phantom{\times\bigg\{ }
-\frac{M_{h_1}}{x_\perp\,\sqrt s}
\sum_{q_1q_2q_3q_4}\!\!\!
h_1^{q_1}(x_1)f_1^{q_2}(x_2)
\frac{\hat s}{2}
\frac{ \rmd{1}{\Delta\hat\sigma_{q_1^\uparrow q_2^{\phantom{\uparrow}}
       \rightarrow q_3^\uparrow q_4^{\phantom{\uparrow}} }}\!}
{\rmd{1}{\hat t}}
H^{q_3}{}_1^{\perp(1)}(z_1)D_1^{q_4}(z_2)
{+}\big(K_1{\leftrightarrow}K_2\big)
\nonumber
\\
&\phantom{=} \phantom{\times\bigg\{\  }
-\frac{M_{h_1}}{x_\perp\,\sqrt s}
\sum_{q_1q_2q_3q_4}\!\!\!
h_1^{q_1}(x_1)f_1^{q_2}(x_2)\,
\frac{\hat s}{2}\,
\frac{\rmd{1}{\Delta\hat\sigma_{q_1^\uparrow q_2^{\phantom{\uparrow}}
\rightarrow\widehat{gq}{}_3^\uparrow q_4^{\phantom{\uparrow}}}}\!}{\rmd{1}{\hat t}}
\widetilde H^{q_3}{}_1^{\perp(1)}(z_1)D_1^{q_4}(z_2)
{+}\big(K_1{\leftrightarrow}K_2\big) \biggr\}
\nonumber\\
&
\phantom{=}
\times
\left( 1 + \mathcal{O} \left( \frac{M}{\sqrt{s}} \right) +
\mathcal{O}\left( \alpha_S \right)\right),
\end{align}
\vspace{-.7cm}
\begin{flushright}
\emph{weighted asymmetry for hadron-hadron production}
\end{flushright}
}
where the summation is over all quark and antiquark flavors. The arguments
of the functions are defined through
$x_1 = x_\perp (\exp \eta_1 + \exp \eta_2)$,
$x_2 = x_\perp (\exp -\eta_1 + \exp -\eta_2)$, and
$z_i = x_{i\perp} / x_\perp$. Further, $\hat{s}$ depends
on $x_\perp$ and $y$ through $\hat{s} = x_\perp^2 s / (4y(1-y))$.

The asymmetry for
hadron-jet production where the hadron belongs to a different jet
is obtained by replacing $D_1(z_2)$ by $\delta(1-z_2) \delta_{j_2q}$
which fixes $x_\perp$ to be $x_{2\perp}$
(the subscript $2$ refers to the measured jet), and setting all other
functions to zero. This yields
\myBox{
%\small
\vspace{-.3cm}
\begin{equation}
\begin{split}
\langle \ \tfrac{1}{2}\sin (\delta\phi) \rmd{1}{\sigma} \rangle 
%\\
&=\rmd{1}{x_{1\perp}}\rmd{1}{x_{2\perp}}
\rmd{1}{\eta_1}\rmd{1}{\eta_2}
\frac{\rmd{1}{\phi_1}}{2\pi}\ \cos(\phi_1-\phi_S)\\
&\phantom{=}
\times\bigg\{\
\frac{M_1}{x_{2\perp}\sqrt s}
\sum_{q_1q_2q_3q_4}\!\!\!{f^{q_1}}_{1T}^{\perp(1)}(x_1)f_1^{q_2}(x_2)
\frac{\hat s}{2}
\frac{\rmd{1}{\hat\sigma_{\widehat{gq}{}_1q_2\rightarrow q_3q_4}}}
{\rmd{1}{\hat t}}
D_1^{q_3}(z_1)\\
& \phantom{= \times\bigg\{\ }
+\frac{M_2}{x_{2\perp}\,\sqrt s}
\sum_{q_1q_2q_3q_4}\!\!\!
h_1^{q_1}(x_1){h^{q_2}}_1^{\perp(1)}(x_2)
\frac{\hat s}{2}
\frac{\rmd{1}{\Delta\hat\sigma_{q_1^\uparrow \widehat{gq}{}_2^\uparrow
\rightarrow q_3q_4}}}{\rmd{1}{\hat t}}
D_1^{q_3}(z_1)\\
& \phantom{= \times\bigg\{\ }
-\frac{M_{h_1}}{x_{2\perp}\sqrt s}
\sum_{q_1q_2q_3q_4}\!\!\!
h_1^{q_1}(x_1)f_1^{q_2}(x_2)
\frac{\hat s}{2}
\frac{\rmd{1}{\Delta\hat\sigma_{q_1^\uparrow q_2^{\phantom{\uparrow}}\rightarrow
q_3^\uparrow q_4^{\phantom{\uparrow}}}}}
{\rmd{1}{\hat t}}
H^{q_3}{}_1^{\perp(1)}(z_1)\\
& \phantom{= \times\bigg\{\ }
-\frac{M_{h_1}}{x_{2\perp}\sqrt s}
\sum_{q_1q_2q_3q_4}\!\!\!
h_1^{q_1}(x_1)f_1^{q_2}(x_2)
\frac{\hat s}{2}
\frac{\rmd{1}{\Delta\hat\sigma_{q_1^\uparrow q_2^{\phantom{\uparrow}}
\rightarrow \widehat{gq}{}_3^\uparrow q_4^{\phantom{\uparrow}}}}}{\rmd{1}{\hat t}}\,
\widetilde H^{q_3}{}_1^{\perp(1)}(z_1)\ \bigg\}
\\
&
\phantom{=}
\times
\left( 1 + \mathcal{O} \left( \frac{M}{\sqrt{s}} \right) +
\mathcal{O}\left( \alpha_S \right)\right).
\end{split}
\raisetag{18pt}
\end{equation}
\vspace{-.7cm}
\begin{flushright}
\emph{weighted asymmetry for hadron-jet production}
\end{flushright}
}

The $\sin \delta \phi$ weighted
asymmetry for jet-jet production, where the jets are approximately 
back-to-back in the perpendicular plane, reads
%In jet-jet production effects from intrinsic transverse momentum
%can be studied through the vector $r_\perp^\alpha$. The asymmetry given here
%is just one projection of this vector. A more general asymmetry could
%give a better separation between the two contributions. For the
%$\sin \delta \phi$ asymmetry one obtains
\myBox{
%\small
\vspace{-.3cm}
\begin{align}
\langle \ \tfrac{1}{2}\sin(\delta\phi) \rmd{1}{\sigma}\rangle
&=
\rmd{1}{x_{1\perp}}\rmd{1}{x_{2\perp}}\rmd{1}{\eta_1}
\rmd{1}{\eta_2}
\frac{\rmd{1}{\phi_1}}{2\pi}\ \cos(\phi_1-\phi_S)\
 \delta(x_{1\perp}{-}x_{2\perp})
\nonumber\\
&\phantom{=}
\times\bigg\{\
\frac{M_1}{\sqrt s}
\sum_{q_1q_2q_3q_4}{f^{q_1}}_{1T}^{\perp(1)}(x_1)f_1^{q_2}(x_2)\,
\frac{\hat s}{2}
\,\frac{\rmd{1}{\hat\sigma_{\widehat{gq}{}_1q_2\rightarrow q_3q_4}}}{\rmd{1}{\hat t}}
\nonumber\\
& \phantom{= \times \bigg\{\ }
+\frac{M_2}{\sqrt s}
\sum_{q_1q_2q_3q_4}\,
h_1^{q_1}(x_1){h^{q_2}}_1^{\perp(1)}(x_2)\,
\frac{\hat s}{2}\,
\frac{\rmd{1}{\Delta\hat\sigma_{q_1^\uparrow \widehat{gq}{}_2^\uparrow
\rightarrow q_3q_4}}}{\rmd{1}{\hat t}}\ \bigg\}
\nonumber\\
&
\phantom{=}
\times
\left( 1 + \mathcal{O} \left( \frac{M}{\sqrt{s}} \right) +
\mathcal{O}\left( \alpha_S \right)\right).
\end{align}
\vspace{-.7cm}
\begin{flushright}
\emph{weighted asymmetry for jet-jet production}
\end{flushright}
}
As discussed in the previous section,
in jet-jet production effects from intrinsic transverse momentum
can also be studied through the vector $r_\perp^\alpha$. 
The asymmetry given here
is just one projection of this vector. An $r_\perp$-weighted asymmetry could
give a better separation between the two contributions.

\newpage

\section{Summary and conclusions}

\sloppy
Effects from intrinsic transverse momentum were studied through
single spin asymmetries in hadron-hadron
collisions. Assuming factorization, the diagrammatic expansion was applied to
derive the tree-level expressions and to illustrate effects of the gauge link.
That this effect can be more than just a sign was pointed out in 
chapter~\ref{chapter4} and is a generalization of the earlier work in
chapter~\ref{chapter3}.
Contributions to the cross sections were considered from quark and antiquark
distribution and fragmentation functions. For the contributions from gluons
a similar procedure can be followed.

\fussy
A challenge was how 
to extract the transverse momentum dependent functions
from the cross section. It was found that by weighting with a specific
angle (the non-back-to-backness), the asymmetry becomes proportional
to three integrated distribution and fragmentation functions and one
distribution or fragmentation function of which the first transverse moment
is taken. This method has the advantage that no momentum dependence of functions
needs to be assumed, offering a model-independent way of studying effects
from intrinsic transverse momentum. Whether such a method can also be applied in the case of single
hadron production remains to be seen.

The distribution and fragmentation functions contain gauge links which 
were calculated by using the prescription of the previous chapter. Those
gauge links become
\mbox{relevant} when considering transverse moments of the transverse momentum
dependent functions. The first transverse moments
of the correlators in hadron-hadron scattering also  appear (with factors)
in 
semi-inclusive DIS and electron-positron annihilation. For the higher moments
such relations have not been achieved (see chapter~\ref{chapter4}).

The single spin
asymmetries were obtained in terms of functions containing a 
subprocess-dependent (diagram-dependent)
gauge link. Consequently, the asymmetries were rewritten as a folding
of the functions defined in chapter~\ref{chapter2} 
with newly defined gluonic-pole cross sections.
Those gluonic-pole cross sections
are just partonic scattering diagrams where each diagram is weighted
with an additional
factor (compare for instance Eq.~\ref{complexa1} with Eq.~\ref{complexa2}).
This factor comes from taking the first transverse moment
of the T-odd functions. Except for the fragmentation functions $D_i^{(1)}$
where such factors do not appear, the T-odd functions are of order $g$
and higher (see for instance Eq.~\ref{theoryGluonicPole}). 
Therefore, these factors can already
be calculated by doing a single gluon insertion. The all order insertions, as
performed in the previous chapter to obtain gauge invariant results, 
are not necessary to produce these factors but
of course provide confidence on the approach we have followed.
Since single gluon insertions are sufficient to produce these
factors, comparisons
with the results of Qiu and Sterman~\cite{Qiu:1998ia} are possible.

\newpage

\begin{subappendices}

\section{Partonic cross sections\label{complexA}}

In this appendix the elementary and gluonic-pole scattering subprocesses 
will be listed.
The elementary expressions are taken from
Bacchetta, Radici~\cite{Bacchetta:2004it} in which
minus signs from interchanging
fermions are made explicit. Interchanging $k_1$ with
$k_2$ is equivalent to interchanging $\hat{u}$ with $\hat{t}$ or
$y \leftrightarrow (1-y)$. For constructing the cross sections 
a useful relation is $4 x_1 x_2 y (1-y) = x_\perp^2 [1+\mathcal{O}(M^2/s)]$
which can be employed underneath the $x_\perp$ integral.

\subsection*{Quark-quark scattering}

The unpolarized quark-quark scattering subprocesses are given by
\begin{align}
\frac{\rmd{1}{\hat \sigma^{[tt]}_{qq^\prime\rightarrow qq^\prime}}}{\rmd{1}{\hat t}}
&=\frac{4\pi\alpha_S^2}{9\,\hat s^2}
\frac{\hat s^2 + \hat u^2}{\hat t^2},
\qquad
\frac{\rmd{1}{\hat \sigma^{[uu]}_{qq^\prime\rightarrow q^\prime q}}}{\rmd{1}{\hat t}}
=\frac{4\pi\alpha_S^2}{9\,\hat s^2}
\frac{\hat s^2 + \hat t^2}{\hat u^2},
\nonumber
\\
\frac{\rmd{1}{\hat \sigma^{[tu]}_{qq\rightarrow qq}}}{\rmd{1}{\hat t}}
&=\frac{4\pi\alpha_S^2}{27 \hat s^2}\frac{\hat s^2}{\hat t\hat u},
\nonumber
\end{align}
giving for the total cross section
\begin{align}
\frac{\rmd{1}{\hat \sigma_{qq\rightarrow qq}}}{\rmd{1}{\hat t}}
&=\frac{\rmd{1}{\hat \sigma^{[tt]}_{qq^\prime\rightarrow qq^\prime}}}{\rmd{1}{\hat t}}
+\frac{\rmd{1}{\hat \sigma^{[uu]}_{qq^\prime\rightarrow q^\prime q}}}{\rmd{1}{\hat t}}
-2\,\frac{\rmd{1}{\hat \sigma^{[tu]}_{qq\rightarrow qq}}}{\rmd{1}{\hat t}}.
\label{complexa1}\\
\intertext{
For unpolarized gluonic-pole-quark
scattering (gluonic pole is associated with first quark)
the above expression is modified into
}
\frac{\rmd{1}{\hat\sigma_{\widehat{gq}q\rightarrow qq}}}{\rmd{1}{\hat t}}
&= \mathcal{C}^{[tt,qq]}_{G,f(1)}
\frac{\rmd{1}{\hat \sigma^{[tt]}_{qq^\prime\rightarrow qq^\prime}}}{\rmd{1}{\hat t}}
+
\mathcal{C}^{[uu,qq]}_{G,f(1)}
\frac{\rmd{1}{\hat \sigma^{[uu]}_{qq^\prime\rightarrow q^\prime q}}}{\rmd{1}{\hat t}}
-
2\mathcal{C}^{[tu,qq]}_{G,f(1)}\,\frac{\rmd{1}{\hat \sigma^{[tu]}_{qq\rightarrow qq}}}{\rmd{1}{\hat t}}.
\label{complexa2}
\end{align}
Consulting table~\ref{devFactors} one has the following factors:
$\mathcal{C}^{[tt,qq]}_{G,f(1)} = 1/2$, 
$\mathcal{C}^{[uu,qq]}_{G,f(1)} = 1/2$, and
$\mathcal{C}^{[tu,qq]}_{G,f(1)} = -3/2$.

The relevant polarized quark-quark subprocesses are
\begin{displaymath}
\frac{\rmd{1}{\Delta\hat\sigma^{[tt]}_{q^\uparrow q'\rightarrow q^\uparrow q'}}}{\rmd{1}{\hat t}}
=-\frac{8\pi\alpha_S^2}{9\,\hat s^2}\,\frac{\hat u\hat s}{\hat t^2},
\qquad
\frac{\rmd{1}{\Delta\hat\sigma^{[tu]}_{q^\uparrow q\rightarrow q^\uparrow q}}}
{\rmd{1}{\hat t}}
=-\frac{8\pi\alpha_S^2}{27\,\hat s^2}\,\frac{\hat s}{\hat t},
\end{displaymath}
giving for the cross sections
\begin{align}
\frac{\rmd{1}{\Delta\hat\sigma_{q^\uparrow q\rightarrow q^\uparrow q}}}{\rmd{1}{\hat t}}
&=\frac{\rmd{1}{\Delta\hat\sigma^{[tt]}_{q^\uparrow q'\rightarrow q^\uparrow q'}}}{\rmd{1}{\hat t}}
-\frac{\rmd{1}{\Delta\hat\sigma_{q^\uparrow q\rightarrow q^\uparrow q}^{[tu]}}}
{\rmd{1}{\hat t}},
\\
\frac{\rmd{1}{\Delta\hat\sigma_{q^\uparrow q^\uparrow\rightarrow qq}}}{\rmd{1}{\hat t}}
&=\frac{\rmd{1}{\Delta\hat\sigma^{[tu]}_{q^\uparrow q^\uparrow\rightarrow qq}}}{\rmd{1}{\hat t}}
=-\frac{8\pi\alpha_S^2}{27\,\hat s^2},
\\
\intertext{
For the polarized gluonic-pole scattering cross sections, the above
expressions are modified into}
\frac{\rmd{1}{\Delta\hat\sigma_{q^\uparrow q
\rightarrow\widehat{gq}{}^\uparrow q}}}{\rmd{1}{\hat t}}
&=
\mathcal{C}_{G,D(1)}^{[tt,qq]}
\frac{\rmd{1}{\Delta\hat\sigma^{[tt]}_{q^\uparrow q'\rightarrow q^\uparrow q'}}}{\rmd{1}{\hat t}}
-
\mathcal{C}_{G,D(1)}^{[tu,qq]}
\frac{\rmd{1}{\Delta\hat\sigma_{q^\uparrow q\rightarrow q^\uparrow q}^{[tu]}}}
{\rmd{1}{\hat t}}
,
\\
\frac{\rmd{1}{\Delta\hat\sigma_{q^\uparrow\widehat{gq}^\uparrow{\rightarrow}qq}}}
{\rmd{1}{\hat t}}
&=
\mathcal{C}_{G,f(2)}^{[tu,qq]}
\frac{\rmd{1}{\Delta\hat\sigma_{q^\uparrow q^\uparrow\rightarrow qq}^{[tu]}}}
{\rmd{1}{\hat t}}.
\end{align}
Table~\ref{devFactors} gives the relations:
$\mathcal{C}_{G,D(1)}^{[tt,qq]} = -1/2$,
$\mathcal{C}_{G,D(1)}^{[tu,qq]} = 3/2$,
$\mathcal{C}_{G,f(2)}^{[tu,qq]} = -3/2$.

\subsection*{Quark-antiquark scattering}

The unpolarized quark-antiquark subprocesses are given by
\begin{align}
\frac{\rmd{1}{\hat \sigma^{[tt]}_{q\bar q^\prime\rightarrow q\bar q^\prime}}}{\rmd{1}{\hat t}}
&
=\frac{4\pi\alpha_S^2}{9\,\hat s^2}\frac{\hat s^2 + \hat u^2}{\hat t^2},
\qquad
\frac{\rmd{1}{\hat \sigma^{[ss]}_{q\bar q \rightarrow q^\prime \bar q^\prime}}}{\rmd{1}{\hat t}}
=\frac{4\pi\alpha_S^2}{9\,\hat s^2}\,\frac{\hat t^2 + \hat u^2}{\hat s^2}\ ,
\nonumber
\\
\frac{\rmd{1}{\hat \sigma^{[ts]}_{q\bar q\rightarrow q\bar q}}}{\rmd{1}{\hat t}}
&
=\frac{4\pi\alpha_S^2}{27\,\hat s^2}\,\frac{\hat u^2}{\hat t\hat s},
\nonumber
\end{align}
giving for the cross section
\begin{align}
\frac{\rmd{1}{\hat \sigma_{q\bar q\rightarrow q\bar q}}}{\rmd{1}{\hat t}}
&=\frac{\rmd{1}{\hat \sigma^{[tt]}_{q\bar q^\prime\rightarrow q\bar q^\prime}}}{\rmd{1}{\hat t}}
+\frac{\rmd{1}{\hat \sigma^{[ss]}_{q\bar q \rightarrow q^\prime \bar q^\prime}}}{\rmd{1}{\hat t}}
-2 \frac{\rmd{1}{\hat \sigma^{[ts]}_{q\bar q\rightarrow q\bar q}}}{\rmd{1}{\hat t}}.
\\
\intertext{
For the unpolarized gluonic-pole scattering cross section, the above expression
is modified into}
\frac{\rmd{1}{\hat\sigma_{\widehat{gq}\bar q\rightarrow q\bar q}}}{\rmd{1}{\hat t}}
&
= \mathcal{C}^{[tt,q\bar{q}]}_{G, f(1)}
\frac{\rmd{1}{\hat \sigma^{[tt]}_{q\bar q^\prime\rightarrow q\bar q^\prime}}}{\rmd{1}{\hat t}}
+ \mathcal{C}^{[ss,q\bar{q}]}_{G, f(1)}
\frac{\rmd{1}{\hat \sigma^{[ss]}_{q\bar q \rightarrow q^\prime \bar q^\prime}}}{\rmd{1}{\hat t}}
-2 \mathcal{C}^{[ts,q\bar{q}]}_{G, f(1)}
\frac{\rmd{1}{\hat \sigma^{[ts]}_{q\bar q\rightarrow q\bar q}}}{\rmd{1}{\hat t}},
\end{align}
where 
$\mathcal{C}^{[tt,q\bar{q}]}_{G, f(1)} = -3/4$,
$\mathcal{C}^{[ss,q\bar{q}]}_{G, f(1)} = 5/4$,
$\mathcal{C}^{[ts,q\bar{q}]}_{G, f(1)} = 5/4$
(using table~\ref{devFactors}).

The polarized quark-antiquark scattering subprocesses are
\begin{align}
\frac{\rmd{1}{\Delta\hat\sigma^{[ss]}_{q^\uparrow\bar q^\uparrow\rightarrow q'\bar q'}}}
{\rmd{1}{\hat t}}
&
=-\frac{8\pi\alpha_S^2}{9\,\hat s^2}\,\frac{\hat t\hat u}{\hat s^2}, 
\qquad
\frac{\rmd{1}{\Delta\hat\sigma^{[st]}_{q^\uparrow\bar q^\uparrow\rightarrow q\bar q}}}
{\rmd{1}{\hat t}}
=-\frac{8\pi\alpha_S^2}{27\,\hat s^2}\,\frac{\hat u}{\hat s},
\nonumber
\\
\frac{\rmd{1}{\Delta\hat\sigma^{[tt]}_{q^\uparrow\bar q'\rightarrow q^\uparrow\bar q'}}}
{\rmd{1}{\hat t}}
&
=-\frac{8\pi\alpha_S^2}{9\,\hat s^2}\,\frac{\hat u\hat s}
{\hat t^2},
\qquad
\frac{\rmd{1}{\Delta\hat\sigma^{[ts]}_{q^\uparrow\bar q\rightarrow q^\uparrow\bar q}}}
{\rmd{1}{\hat t}}
=-\frac{8\pi\alpha_S^2}{27\hat s^2}\,\frac{\hat u}{\hat t},
\nonumber
\end{align}
giving for the cross sections
\begin{align}
\frac{\rmd{1}{\Delta\hat\sigma_{q^\uparrow\bar q^\uparrow\rightarrow q\bar q}}}{\rmd{1}{\hat t}}
&
=\frac{\rmd{1}{\Delta\hat\sigma^{[ss]}_{q^\uparrow\bar q^\uparrow\rightarrow q'\bar q'}}}
{\rmd{1}{\hat t}}
-\frac{\rmd{1}{\Delta\hat\sigma^{[st]}_{q^\uparrow\bar q^\uparrow\rightarrow q\bar q}}}
{\rmd{1}{\hat t}},
\\
\frac{\rmd{1}{\Delta\hat\sigma_{q^\uparrow\bar q\rightarrow q^\uparrow\bar q}}}{\rmd{1}{\hat t}}
&
=\frac{\rmd{1}{\Delta\hat\sigma^{[tt]}_{q^\uparrow\bar q'\rightarrow q^\uparrow\bar q'}}}
{\rmd{1}{\hat t}}
-\frac{\rmd{1}{\Delta\hat\sigma^{[ts]}_{q^\uparrow\bar q\rightarrow q^\uparrow\bar q}}}
{\rmd{1}{\hat t}},
\\
\frac{\rmd{1}{\Delta\hat\sigma_{q^\uparrow\bar q\rightarrow\bar q^\uparrow q}}}{\rmd{1}{\hat t}}
&
=
\frac{\rmd{1}{\Delta\hat\sigma^{[ts]}_{q^\uparrow\bar q\rightarrow\bar q^\uparrow q}}}{\rmd{1}{\hat t}}
=
-\frac{8\pi\alpha_S^2}{27\hat s^2}.
\\
\intertext{For the gluonic-pole scattering cross sections the above expression
is modified into}
\frac{\rmd{1}{\Delta\hat\sigma_{q^\uparrow\widehat{g\bar q}{}^\uparrow
\rightarrow q\bar q}}}{\rmd{1}{\hat t}}
&=
\mathcal{C}^{[ss,q\bar{q}]}_{G,\bar{f}(2)}
\frac{\rmd{1}{\Delta\hat\sigma^{[ss]}_{q^\uparrow\bar q^\uparrow\rightarrow q'\bar q'}}}
{\rmd{1}{\hat t}}
-
\mathcal{C}^{[st,q\bar{q}]}_{G,\bar{f}(2)}
\frac{\rmd{1}{\Delta\hat\sigma^{[st]}_{q^\uparrow\bar q^\uparrow\rightarrow q\bar q}}}
{\rmd{1}{\hat t}},
\\
\frac{\rmd{1}{\Delta\hat\sigma_{q^\uparrow\bar q
\rightarrow\widehat{gq}{}^\uparrow\bar q}}}{\rmd{1}{\hat t}}
&
=
\mathcal{C}^{[tt,q\bar{q}]}_{G,D(1)}
\frac{\rmd{1}{\Delta\hat\sigma^{[tt]}_{q^\uparrow\bar q'\rightarrow q^\uparrow\bar q'}}}
{\rmd{1}{\hat t}}
-
\mathcal{C}^{[ts,q\bar{q}]}_{G,D(1)}
\frac{\rmd{1}{\Delta\hat\sigma^{[ts]}_{q^\uparrow\bar q\rightarrow q^\uparrow\bar q}}}
{\rmd{1}{\hat t}},
\\
\frac{\rmd{1}{\Delta\hat\sigma_{q^\uparrow\bar q
\rightarrow\widehat{g\bar q}{}^\uparrow q}}}{\rmd{1}{\hat t}}
&
= \mathcal{C}^{[ts,q\bar{q}]}_{G,D(1)}
\frac{\rmd{1}{\Delta\hat\sigma^{[ts]}_{q^\uparrow\bar q\rightarrow\bar q^\uparrow q}}}{\rmd{1}{\hat t}},
\end{align}
where
$\mathcal{C}^{[ss,q\bar{q}]}_{G,\bar{f}(2)} = 5/4$,
$\mathcal{C}^{[st,q\bar{q}]}_{G,\bar{f}(2)} = 5/4$,
$\mathcal{C}^{[tt,q\bar{q}]}_{G,D(1)} = 3/4$,
$\mathcal{C}^{[ts,q\bar{q}]}_{G,D(1)} = -5/4$,
$\mathcal{C}^{[ts,q\bar{q}]}_{G,\bar{D}(1)} = -5/4$
(using table~\ref{devFactors}).

\newpage

\thispagestyle{empty}

\end{subappendices}

\chapter{Summary and conclusions}

\markboth{Summary and conclusions}{Summary and conclusions}

Effects from
intrinsic transverse momentum of partons
were studied in several hard scattering processes with an emphasis on
color gauge invariance.
In order to describe the processes, the diagrammatic expansion was employed
which is a field-theoretical approach.
Extending the work of Boer, Mulders~\cite{Boer:1999si} and Belitsky, Ji,
Yuan~\cite{Belitsky:2002sm}, factorization of effects from intrinsic
transverse
momentum was assumed, and by considering an infinite number of diagrams 
the tree-level expressions
including $M/Q$ corrections
were evaluated in chapter~\ref{chapter3}
for semi-inclusive DIS, Drell-Yan, and electron-positron
annihilation. 
In those processes transverse momentum dependent
distribution and fragmentation functions
were encountered which are defined through matrix elements
in which bilocal operators are folded with
a gauge link.
From a theoretical point of view, the
presence of this gauge link
(also called Wilson line) is pleasant because it makes the bilocal operator
invariant under color gauge transformations.
These gauge links are of increasing interest because by several
papers, among which Brodsky, Hwang, Schmidt~\cite{Brodsky:2002cx},
Collins~\cite{Collins:2002kn}, and Belitsky, Ji, Yuan~\cite{Belitsky:2002sm},
it has been shown that the gauge links are not the same in every process
and could lead to observable effects.

Distribution and fragmentation functions containing a gauge link were
discussed in chapter~\ref{chapter2}. These functions describe the way in
which quarks are distributed in a nucleon or how a quark decays into a jet
and a particular hadron. They form a bridge between theoretical
predictions and experimental observations, and are vital
for our understanding of the nucleon's substructure. In the diagrammatic
approach these functions naturally appear when studying hard scattering 
processes. The exact form of these functions, including the path of the
gauge link, is thus not a starting point but rather derived.

In the transverse momentum integrated
distribution functions the presence of
the gauge link did not produce new effects in contrast
to the transverse momentum dependent functions. In the latter the gauge
link does not run along a straight path between the two quark-fields and therefore
allows for the existence of T-odd distribution functions.
The existence of such functions was conjectured by
Sivers~\cite{Sivers:1989cc,Sivers:1990fh} in order to explain the observation
of single spin asymmetries. As pointed out by Collins~\cite{Collins:2002kn},
those functions have the interesting property that they appear with different
signs when comparing Drell-Yan with semi-inclusive DIS.
This interesting prediction should of course
be experimentally verified.

Another possible source for T-odd effects was uncovered by Qiu and
Sterman~\cite{Qiu:1991pp,Qiu:1991wg}. They suggested that the presence
of gluonic pole matrix elements could produce single spin asymmetries in
hadron-hadron collisions. It was shown
in chapter~\ref{chapter2} that those matrix elements have the same
form as matrix elements from which the
T-odd distribution functions are defined. The presence of the gauge link
and the gluonic pole matrix elements are therefore in essence the same
mechanism.

The path of the
gauge link in the transverse momentum dependent functions runs in
a particular direction via the light-cone boundary
 (see Fig.~\ref{plaatjeLinks}).
In chapter~\ref{chapter2} it was discussed that this direction is
a potential source for new effects or functions (see for related work
Goeke, Metz, Schlegel~\cite{Goeke:2005hb}).
While calculating the longitudinal target-spin and beam-spin asymmetries
for semi-inclusive DIS
in chapter~\ref{chapter3}, it was found that one of these new functions,
$g^\perp$, could produce a
nonzero azimuthal single spin asymmetry for jet-production
in lepton-hadron scattering. Given the reason for the existence of this
function, this prediction deserves experimental verification.

The presence of gauge links in transverse momentum dependent
fragmentation functions confronted us in chapter~\ref{chapter2} with
issues related to universality. It was found that transverse momentum dependent
fragmentation functions
could appear to be different when comparing
semi-inclusive DIS with electron-positron
annihilation. The reason for this difference is that for fragmentation
functions there are two possible
sources for T-odd effects: the gauge link and
final-state interactions. If one of the two mechanisms is suppressed,
relations between the two processes can be drawn. It should be pointed out
that
Collins and Metz obtained
in Ref.~\cite{Metz:2002iz,Collins:2004nx} in a quite general treatment
fragmentation functions which appear
with the same sign in the two different processes. This interesting
result was discussed in chapter~\ref{chapter3} and deserves further
attention. Also from the experimental side this universality issue can be
addressed. For instance, by comparing properties,
like the $z$-dependence, of
transverse momentum dependent fragmentation functions between different
processes.

In chapter~\ref{chapter4} it was discovered that gauge links can arise which are much
more complex than the gauge links in the discussed electromagnetic processes.
Besides being more complex, the path of the gauge link turned out to
not only depend on the process, but also on the subprocess (or
squared amplitude diagram).
These complex structures predict that T-odd distribution functions can in
general differ by more than just a sign when comparing different processes.
In
fact only their first transverse moment
can be straightforwardly compared between different processes. A prescription
was given in order to deduce the gauge link in squared amplitude diagrams,
and it was shown in a two gluon-production process
that nonphysical polarizations of the gluons are canceled
among diagrams although
separate diagrams are convoluted with functions having diagram-dependent
gauge links. This is a firm consistency test of the applied approach.

An illustration of these more complex gauge links
was given
in hadron-hadron production in hadron-hadron scattering. In hadron-hadron
scattering
the challenge is to extract the transverse momentum dependent functions.
In chapter~\ref{chapter5} such an observable was presented. 
This observable has the advantage that it
does not require input on the explicit form for the 
momentum dependence of the functions.
By using this observable,
it was shown that effects from the gauge link yield more than just a sign
in single spin asymmetries. The expression for the single spin asymmetry was
found to be a set of elementary scattering subprocesses convoluted with
universal
integrated functions and a function of which a transverse moment was taken.
In this latter function the gauge link depends on the subprocess.
Since the first transverse moment of functions 
can be related to a set of ``standard''
functions, which also appear in semi-inclusive
DIS and electron-positron annihilation (just a factor), 
the asymmetries were rewritten
in terms of these ``standard'' functions folded with the newly defined
gluonic-pole cross sections.
Gluonic-pole cross sections are just elementary parton scattering cross sections
in which the various subprocesses (squared amplitude diagrams) 
are weighted with a particular factor.
Contributions from gluon distribution and fragmentation functions
were discarded for the ease of the calculation, but should be included
in future studies
to make realistic estimates of single spin asymmetries. The definition
of transverse momentum dependent gluon distribution functions was addressed
in chapter~\ref{chapter4},
but needs further improvement.
Just like experiments
could verify in the
Drell-Yan process
the sign change of T-odd distribution functions,
experiments should also be able to check the more
involved appearance of T-odd distribution functions in hadron-hadron production. 
Other processes
which also contain such more complicated effects are photon-jet production,
and two jet-production.

Besides leading to new effects, the gauge link also poses
theoretical challenges among which the issue of factorization
of intrinsic transverse momentum dependent effects. This issue
underlies most treatments and results in this field
including this thesis. Recently,
significant progress was made by Ji, Ma, and Yuan in
Ref.~\cite{Ji:2004wu,Ji:2004xq}
and by Collins and Metz in Ref.~\cite{Collins:2004nx}.
They considered semi-inclusive DIS and Drell-Yan at small measured $q_T$
($q$ is the momentum of the virtual photon and $q_T^2 \sim -M^2$).
One particular issue which was not explicitly addressed is the path of the
gauge link in distribution and fragmentation functions connected to higher
order diagrams (two loops or higher). At those orders it was argued in
chapter~\ref{chapter4} that the effect of 
gauge links might endanger factorization
of intrinsic transverse momentum dependent effects in semi-inclusive DIS and
Drell-Yan. This could form a problem for azimuthal asymmetries at small $q_T$
at leading order in $M/Q$ and
at subleading order for fully $q_T$-integrated cross sections.
This issue awaits further clarification.
Model calculations could be very useful here.
Just like a model calculation uncovered the possible existence of T-odd
distribution functions, a two loop model calculation could illustrate
some of the important aspects related to factorization.

In this thesis effects from intrinsic transverse momentum were studied in
hard scattering processes. The existing theoretical formalism was further
developed and several issues have been clarified. Various cross sections
and asymmetries were obtained. Their measurement can contribute to
our understanding of the nucleon's substructure within the framework of
QCD and very likely will guide physicists in answering several remaining
questions.

\chapter*{Acknowledgements}
\addcontentsline{toc}{chapter}{Acknowledgements}
\markboth{Acknowledgements}{Acknowledgements}

Several persons are acknowledged for the support they have given
me in the last four years.
First I would like to acknowledge my supervisor Piet Mulders.
Piet, in het begin vond ik
het moeilijk om m'n eigen weg te zoeken, maar achteraf ben ik je
dankbaar
voor de zelfstandigheid die je mij hebt gegeven. Verder heb je mij erg
gemotiveerd door mij vanaf het begin als een zelfstandig
natuurkundige te behandelen en door mij de mogelijkheid te geven
om naar workshops en conferenties te gaan. Naast onze samenwerking aan
artikelen waardeer ik ook zeer dat je er altijd was op de momenten
dat ik je echt nodig had.

The second person who I would like to thank is Daniel Boer. Daniel,
de inspiratie die je kunt geven tijdens het discussieren, je snelle manier
van begrijpen, en je talent van meedenken zijn uitzonderlijk.
Ik wil je bedanken voor de samenwerking van de afgelopen jaren en voor het
nauwkeurig lezen van het proefschrift.
Mede dankzij jou kan ik nu met tevredenheid
terugkijken op de resultaten van dit proefschrift.

I also would like to acknowledge
Alessandro Bacchetta and Cedran Bomhof. I appreciate the collaboration we
had and the results we obtained. In addition, it was also
a pleasure to share an office with you.
Acknowledged are also my collaborators Ben Bakker and Miranda van Iersel.
Ook hier kijk ik met plezier terug naar de resultaten die we hebben bereikt.
En Ben, ik dank je voor de vele uiteenlopende discussies die we gehad hebben
en voor je altijd beschikbare wijze adviezen.

I also would like to thank the reading comittee, Mauro Anselmino,
Daniel Boer, Markus Diehl, Eric Laenen, and Gerard van der Steenhoven for the time
they have spent on reading my thesis and for the suggestions they have made.
Acknowledged is also Bob van Eijk for leading the useful evaluation
meetings in the last four years. Ik keek altijd naar dit soort gesprekken uit.
I also would like to thank
Philipp H\"agler, John McNamara, and Paul van der Nat for reading and
correcting parts of the thesis.

Besides the people alread mentioned,
I met in the last years several others at conferences and workshops
with whom I had nice discussions:
Umberto D'Alesio, Elke Aschenauer, Harut Avakian, Elena
Boglione, John Collins, Delia Hasch,
Leonard Gamberg, Dae Sung Hwang, Elliot Leader, Andreas Metz, Gunar
Schnell, Dennis Sivers,
and Werner Vogelsang.
I also would like to thank the members of the theory group
(including secretaries) for a stimulating and friendly atmosphere. Especially
I acknowledge Paul Becherer, Henk Blok, Hartmut Erzgr\"aber, David Fokkema,
Hugo Schouten, Harmen Warringa, and Erik Wessels for their endless discussions.

Acknowledged are also my friends for their support in the last years.
I also thank my parents, sister, and their families.
Ik dank jullie voor de gegeven mogelijkheid
om te kunnen studeren en voor de gegeven ondersteuning.

The last person I would like to thank is C\u{a}lina Ciuhu. C\u{a}lina, 
\^{\i}n\c{t}elegerea \c{s}i \mbox{sprijinul} t\u{a}u mi-au fost de mare ajutor.
Dar mai ales \^{\i}\c{t}i mul\c{t}umesc pentru dragostea 
\c{s}i motiva\c{t}ia oferit\u{a} de-a lungul acestor ani.

%your
%understanding and support have been a great help to me.
%In addition, I thank you for the love and motivation you have given me.

%Harmen, Hartmut, Hugo, Erik, Alex, Gui, P. Becherer, David,
%Allaart, Lenstra, Mackintosh,

%voetbalteam

%Familie en vrienden (Arno, Ruud, Maarten, Aernout, Usmar, John, Hartmut,
%Joost Hulshof, Daan Lenstra).
%Mogelijkheid studeren.

%Calina.

%\newpage

%\thispagestyle{empty}

% Check
% http://amath.colorado.edu/documentation/LaTeX/reference/faq/bibstyles.html#styles
%
% for style files on bibliography

\bibliographystyle{myBibliography}
\addcontentsline{toc}{chapter}{\bibname}
{\raggedright
%\begin{flushleft}
\bibliography{references}
%\end{flushleft}
}

\thispagestyle{empty}

\section*{Samenvatting}

%\chapter*{Samenvatting}
\addcontentsline{toc}{chapter}{Samenvatting}
\markboth{Samenvatting}{Samenvatting}

\begin{figure}[p]
\begin{center}
\includegraphics[width=7.6cm]{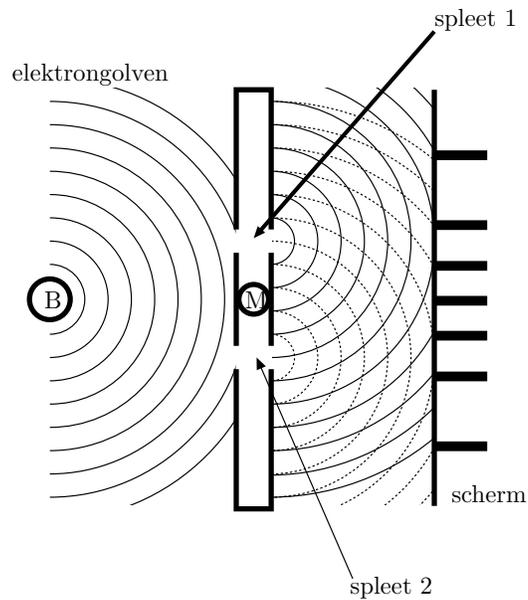}
\end{center}
\caption{Een voorstelling van een elektron uit $B$ dat
een beeld produceert op het
scherm. Doordat het elektron in een quantumtheorie niet lokaal is, golft
het elektron als het ware door de ruimte
naar het scherm toe (het scherm is bijvoorbeeld een fotografische plaat). 
Deze golven zijn aangegeven met halve cirkels en zijn ver\-ge\-lijk\-baar met
watergolven.
Doordat deze golven door twee
spleten reizen, 
ontstaan er effectief twee golfbronnen aan de rechterkant van de 
twee spleten. Aan de rechterkant van de twee spleten 
lopen de golven daarom door elkaar heen en ontstaat er op het scherm een
patroon.
Dit patroon wordt gevormd door de snijpunten van twee golflijnen en wordt
het interferentiepatroon genoemd.
Het interferentiepatroon verandert wanneer er in $M$
een lange staafmagneet wordt geplaatst die het vlak in- of uitwijst. 
Ondanks dat deze magneet geen elektrische of magnetische velden produceert
die voelbaar zijn voor de elektrongolven, verandert het interferentiepatroon
toch. Deze verandering is onder andere 
afhankelijk van de rich\-ting in welke de magneet
wijst: het vlak in of uit.\label{figSamenvatting}}
\end{figure}

\subsection*{
%\begin{flushright}
Enkelvoudige spin asymmetrie\"en en ijkinvariantie
\\ 
in harde verstrooiingsprocessen
%\end{flushright}
}

%\emph{De opbouw van hadronen uit quarks en gluonen roept 
%een aantal vragen op die in dit
%proefschrift worden aangepakt. De relevante begrippen zullen
%worden uitgelegd tezamen met de gemaakte vooruitgang.}

De materie
om ons heen is opgebouwd uit kleinere bouwsteentjes. 
De kleinste bouwsteentjes die wij kennen, worden de \emph{elementaire deeltjes}
genoemd en zijn op dit moment dus niet op te delen in nog kleinere deeltjes.
Voorbeelden van deze elementaire
deeltjes zijn het elektron en het foton. Op dit moment is de 
grootte van deze deel\-tjes niet bekend, ze worden beschouwd oneindig
klein te zijn oftewel puntachtig klein.
In de studie van elementaire deeltjes speelt een bepaalde
familie van deeltjes,
genaamd \emph{hadronen},
een belangrijke rol. In tegenstelling tot de elementaire
deeltjes hebben hadronen 
wel degelijk een zekere grootte. Een voorbeeld van een
hadron is bijvoorbeeld het proton met een afmeting van ongeveer
$0,0000000000001$~cm. Hadronen zijn echter geen elementaire deeltjes, ze
zijn opgebouwd uit kleinere puntdeeltjes die 
\emph{quarks} en \emph{gluonen} heten.
Op dit moment weten we al aardig wat over de opbouw van hadronen uit deze
quarks en gluonen,
en daar zouden we graag nog veel meer over willen weten. Niet
alleen vanwege de structuur, maar ook omdat de daaraan
gerelateerde natuurkunde
uitdagend en verrassend is. In dit proefschrift zullen dan ook de drie
pijlers van de moderne natuurkunde: de \emph{relativiteitstheorie}, de
\emph{quantumtheorie} en het concept symmetrie
(in het bijzonder \emph{ijkinvariantie}),
worden aangewend om een
beschrijving te realiseren.

De natuurkunde die nodig is voor de beschrijving van elementaire deeltjes
is in veel opzichten totaal anders dan waar wij in ons dagelijks
leven aan gewend zijn. Zo spelen op die hele kleine schaal
quantumeffecten een belangrijke rol. 
Intuititief zijn we gewend dat 
objecten \emph{gelokaliseerd} zijn, betekenende 
dat objecten altijd op een bepaalde lokatie zijn (de Dom staat bijvoorbeeld
ergens in
Utrecht). In een quantumtheorie is dit anders. De positie van een bepaald 
deeltje is wel meetbaar, maar
indien er geen meting wordt uitgevoerd, dan is het deeltje in zekere zin
overal (in dat geval is de Dom overal in Utrecht).
Dat lijkt misschien
absurd (in het geval van de Dom), 
maar voor hele kleine deeltjes is dit wel degelijk het geval en
experimenteel geverifieerd.
%Neem het volgende voorbeeld. U komt thuis na een lange werkdag en u wilt de
%deur openen. Op het moment dat u de sleutel in het slot steekt, knippert
%u echter me uw ogen. Precies op dat moment in het huis tijdelijk niet lokaal
%en heeft het in principe de tijd om te bewegen. Op het moment dat u de ogen
%weer opent, blijkt het huis ineen keer achter u te staan. Dit soort
%absurde effecten zijn de grondslag voor quantummechanica.

\sloppy
Het feit dat deeltjes niet meer gelokaliseerd zijn, leidt tot
effecten die in de klas\-sieke theorie (zonder quantumeffecten)
niet bestaan.
Een voorbeeld is het twee-spleten-experiment met elektronen, zie
figuur~\ref{figSamenvatting}. Wanneer
een
elektronenbron \'e\'en elektron per tijdseenheid produceert,
dan blijkt er een interferentiepatroon te ontstaan op het scherm achter
de twee spleten.
Dit interferentiepatroon is te ver\-kla\-ren door aan te nemen dat
het enkele elektron in plaats van door \'e\'en spleet, door beide
spleten \emph{tegelijk} is gegaan. Het niet lokaal zijn van de deeltjes
is dus de oorzaak van het effect.

\fussy
Doordat deeltjes niet gelokaliseerd zijn, is de beschrijving van de krachten
op deeltjes
anders dan in een klassieke theorie. In hetzelfde twee-spleten-experiment verandert het
interferentiepartoon bijvoorbeeld als er een lange staafmagneet 
wordt geplaatst
tussen de twee
spleten (in $M$). Ondanks dat het elektron klassiek gezien deze magneet niet voelt,
verandert het interferentiepatroon toch. Dit quantumeffect is simpelweg
ondenkbaar in de
klassieke natuurkunde en heet het \emph{Aharonov-Bohm effect}.
Volgens de klassieke natuurkunde produceert de magneet
een magnetisch veld maar in deze specifieke situatie
kunnen de elektrongolven het magnetisch veld echter niet voelen.
Om het effect te kunnen verklaren moet er dus nog iets anders zijn wat
de elektrongolven wel be\"{\i}nvloedt.
Dit andere veld wordt
het \emph{ijkveld} of \emph{fotonveld} genoemd en dit veld blijkt
inderdaad op het quantumniveau
met de elektrongolven te kunnen wisselwerken.
Het effect hangt samen met een theoretisch begrip genaamd
\emph{ijkinvariantie}.

Wanneer we de allerkleinste deeltjes bestuderen moeten we onder andere
deze effecten
in kaart brengen. Een
probleem bij de studie van de structuur van hadronen
is dat de bouwelementen, de quarks en gluonen, alleen lijken te bestaan
in de hadronen zelf. In tegenstelling tot alle andere deeltjes die wij kennen
(en dat zijn er heel wat), zijn quarks en gluonen nog nooit vrij geobserveerd.
Het zijn dus een soort legoblokjes die afzonderlijk niet lijken te bestaan.
Dit roept natuurlijk vragen op die op dit moment nog niet volledig zijn
begrepen, en dit maakt
de verdeling van quarks en gluonen
in een hadron natuurlijk des te interessanter.

De structuur van hadronen wordt bestudeerd in experimenten. E\'en
van die experimenten is \emph{elektron-hadron verstrooiing}. In dat experiment
botst een elektron met hoge snelheid op een hadron.
Door deze botsing breekt het hadron op in stukken en
dat levert een scala aan andere deeltjes en andere soorten
hadronen op, allemaal met bepaalde snelheden en rich\-tingen.
Het theoretische model wat wordt toegepast is eenvoudig in oorsprong. Het idee
is dat in bepaalde situaties
de kans dat een bepaald deeltje of hadron na een botsing
wordt gemeten evenredig is met de
kans dat een quark in een hadron wordt geraakt, vermenigvuldigt met de kans
dat ditzelfde quark vervalt in het gemeten hadron. Er wordt dus verondersteld
dat het vervalproces van een quark onafhankelijk is van de structuur van
het geraakte hadron. Deze aanname, genaamd \emph{factorizatie}, 
kan gedeeltelijk worden onderbouwd en heeft in veel experimenten
ook een goede beschrijving gegeven.

Een observatie die tot voor kort nog onbegrepen was, is het optreden
van \emph{enkel\-voudige spin asymmetrie\"en} in
elektron-hadron verstrooiing en \emph{hadron-hadron verstrooiing}.
In het laatste experiment worden
in plaats van een elektron op een hadron te schieten
twee hadronen keihard op elkaar
geschoten en wordt er een ander soort hadron na de botsing gemeten.
Hadronen kunnen in het algemeen een spin hebben. Dat betekent dat ze als het
ware tollen om hun eigen as.
Als in het experiment
blijkt dat waneer de spin van het ingaande hadron wordt omgedraaid
ook de snelheid of rich\-ting van het gemeten deeltje verandert, spreekt men van
een enkelvoudige spin
asymmetrie. Deze asymmetrie\"en treden ook op in de eerder besproken
elektron-hadron verstrooiing. In figuur~\ref{intro2b}
op bladzijde~\pageref{intro2b} is voor dit proces
de gemeten asymmetrie (langs de verticale as) weergegeven.

In dit proefschrift wordt de oorzaak van deze enkelvoudige spin asymmetrie\"en
onderzocht.
%Door bij het fundament van de theorie
%te beginnen worden aspecten samenhangend met factorizatie
%onderzocht.
In hoofdstuk~\ref{chapter2} wordt een aantal hoge energie
verstrooiingsprocessen ge\"{\i}ntro\-duceerd. De basis van de theorie wordt uitgelegd
en de \emph{distributie en fragmentatie functies} worden
gedefinieerd die later blijken op te duiken in experimentele
grootheden. Deze functies beschrijven de structuur van het hadron
en bevatten de informatie die we willen weten.
In hoofstuk~\ref{chapter2} is een nieuwe klasse van functies ontdekt
die ons begrip van de structuur kunnen vergroten. Een voorbeeld van deze
nieuwe functies is $g^\perp$, aangegeven in vergelijking~\ref{unintDistr}
(op bladzijde~\pageref{unintDistr}).

In hoofdstuk~\ref{chapter3} wordt getracht een aantal verstrooiingsprocessen te
beschrijven. In deze beschrijving blijkt
dat ijkinvariantie een grote rol speelt. Ijkinvariantie is een wiskundige
symmetrie die aanwezig zou moeten zijn in de theoretische beschrijving.
Tot voor kort was dit
niet helemaal netjes meegenomen,
maar men wist ook niet of het missen van de symmetrie
werkelijk een probleem zou kunnen zijn.
Door in dit proefschrift de theorie nauwkeurig uit te werken, wordt er een
bijdrage geleverd aan het herstellen van ijkinvariantie. Behalve 
dat dit een verbeterde 
theoretische beschrijving oplevert, blijkt er een subtiel effect te zijn
wat een natuurlijke verklaring voor het optreden
van enkelvoudige spin asymmetrie\"en geeft. 
Een verklaring in elektron-hadron verstrooiing
is de wissel\-werking tussen de geraakte quark 
en het \emph{ijkveld} (\emph{gluonveld}) van het hadron
(de gluonen zijn ge\"{\i}llustreerd in figuur~\ref{intro2a} door gekrulde lijnen).
Het effect is daardoor in een aantal opzichten 
vergelijkbaar met het Aharonov-Bohm effect. In beide gevallen wordt het
effect veroorzaakt doordat een deeltje (hetzij elektron of quark)
wisselwerkt met een ijkveld in gebieden waar de elektrische en magnetische
velden geen effect hebben.
%\footnote{In verstrooiingsexperimenten reist
%het geraakte quark vanuit het hadron in een rich\-ting naar buiten. Het heeft
%dus interacties met het ijkveld zolang het binnen het hadron is en
%mogelijkerwijs ook nog daar buiten. Merk op dat alleen de fysische velden
%opgesloten zijn. Door een bepaalde ijk te kiezen kan het effect geschreven
%worden in termen van ijkvelden op lokaties waar geen fysische velden
%kunnen bestaan. Deze ijkkeuze is nodig omdat het quark vanuit het hadron
%vertrekt. In dit opzicht verschilt het effect met het Aharanov-Bohm
%effect.}.
%Zoals de spin van het hadron, wat de rich\-ting van het gluonmagneetveld
%aangeeft, in elektron-hadron verstrooiing
%een asymmetrie veroorzaakt, zo verandert bijvoorbeeld ook
%het interferentiepatroon wanneer de
%staafmagneet wordt omgedraaid.

Deze effecten in kaart brengend,
zien we in hoofdstuk~\ref{chapter3}
dat de functies van hoofdstuk~\ref{chapter2} kunnen
worden gemeten in elektron-hadron verstrooiing.
Vergelijking~\ref{Wsidis2} is bijvoorbeeld belangrijk om de
verscheidene functies te kunnen meten.
In hoofdstuk~\ref{chapter4} wordt de theorie verder ontwikkeld om
ook hadron-hadron verstrooiing
te kunnen beschrijven. E\'en van de belangrijkste resultaten,
maar technisch van aard, wordt ge\"{\i}llustreerd in de
figuren~\ref{figUnitarity}~en~\ref{uniResult2}.
In dit hoofdstuk wordt verder ook de consistentie van de
gefactorizeerde aanpak besproken: Is het vervalproces van quarks in hadronen
wel onafhankelijk van de structuur van het geraakte hadron?
Het antwoord op deze vraag blijkt niet eenvoudig.
In hoofdstuk~\ref{chapter5}
worden de ontwikkelde technieken van hoofdstuk~\ref{chapter4} gebruikt
om voorspellingen te doen voor asymmetrie\"en in
hadron-hadron verstrooiingsprocessen. Experimenten kunnen deze voorspellingen
verifi\"eren en daarmee de geldigheid van de ontwikkelde idee\"en toetsen.

\end{document}